%% file: ew.tex
\documentclass[a4paper,twoside,notitlepage,fleqn]{report}

\usepackage{lb_rep}

\usepackage{graphicx}
\usepackage{rotating}
\usepackage{mcite}

\usepackage{PhysRep}

\input{physrep_mcite}

\input{ew_mcite}

\begin{document}

\begin{titlepage}

\noindent                
{\Large
 $\phantom{0}$        \hfill UCD-PHYC/071001\\[1mm]
 $\phantom{0}$        \hfill arXiv:0710.2838 [hep-ex]\\[2mm]
 $\phantom{0}$\bf     \hfill 15 October 2007\\[1mm]
}
\begin{center}

\vskip 4cm

\vskip 1cm
{\Huge\bf Experimental Precision Tests for the Electroweak Standard Model \\}
\vskip 1cm
{\Large {\bf Martin W. Gr\"unewald}\\[20pt]
        University College Dublin\\
        UCD School of Physics\\
        Belfield, Dublin 4\\
        Ireland\\}
\vfill
{\bf Abstract} \\[10pt]
\end{center}
{

This paper contains a review of recent precision measurements of
electroweak observables and resulting tests of the electroweak
Standard Model.
  
}
\vskip 1cm
\begin{center}
\em{ Invited Chapter for a Handbook of Particle Physics}
\end{center}

\end{titlepage}

\cleardoublepage

\setcounter{page}{1}                      

\chapter{Experimental Precision Tests for the Electroweak Standard Model}
%
%

\input{body.tex}

%% file: physrep_mcite.tex
\newcommand{\Wdisco}      {Arnison:1983rp,*Banner:1983jy}
\newcommand{\Zdisco}      {Arnison:1983mk,*Bagnaia:1983zx}
\newcommand{\GWS}         {Glashow:1961tr,*Weinberg:1967tq,*Salam:1968rm,*Veltman:1968ki,*'tHooft:1971rn,*'tHooft:1972fi,*'tHooft:1972ue}
\newcommand{\LEPacc}      {LEP,*Myers:1990sk,*Brandt:2000xk,*Assmann:2002th}
\newcommand{\ALEPHdet}    {Decamp:1990jr,*Buskulic:1995wz}
\newcommand{\DELPHIdet}   {Aarnio:1991vx,*Abreu:1996uz}
\newcommand{\Ldet}        {L3:1990kx,*Acciarri:1994yk,*Chemarin:1994bp,*Adam:1996fj}
\newcommand{\OPALdet}     {Ahmet:1991eg,*Allport:1993kp,*Allport:1994ec,*Anderson:1994ve}
\newcommand{\SLDdet}      {SLD-CDC,*SLD-LAC,*SLD-CRID,*SLD-LUM}
\newcommand{\TOPAZref}    {Montagna:1993py,*Montagna:1993ai,*Montagna:1996ja,*Montagna:1998kp}
\newcommand{\ZFITTERref}  {Bardin:1989di,*Bardin:1990tq,*Bardin:1991fu,*Bardin:1991de,*Bardin:1992jc,*Bardin:1999yd,*Kobel:2000aw,*Arbuzov:2005ma}
\newcommand{\ALEPHls}     {Decamp:1990ky,*Decamp:1992aj,*Buskulic:1993gu,*Buskulic:1994ea,*Barate:1999ce}
\newcommand{\DELPHIls}    {Abreu:1991wj,*Abreu:1994wg,*Abreu:1994ds,*Abreu:2000mh}
\newcommand{\Lls}         {Adeva:1991jh,*Adriani:1993gk,*Acciarri:1994gx,*Acciarri:2000ai}
\newcommand{\OPALls}      {Alexander:1991qw,*Acton:1993yc,*Akers:1994is,*Abbiendi:2000hu}

\newcommand{\SLDalr}      {Abe:1993sh,*Abe:1994wx,*Abe:1997nj,*Abe:2000dq}

\newcommand{\ALEPHTAU}    {ALEPHTAU,*Buskulic:1996vx,*Buskulic:1993vk,*Decamp:1991vz}
\newcommand{\DELPHITAU}   {DELPHITAU,*Abreu:1995ku}
\newcommand{\LTAU}        {L3TAU,*Acciarri:1994qt,*Adriani:1992zn}
\newcommand{\OPALTAU}     {OPALTAU,*Alexander:1996ha,*Akers:1995db,*Alexander:1991am}

\newcommand{\alasy}       {ref:alasy}
\newcommand{\dlasy}       {Abreu:1995vn,*Abdallah:2003gp}
\newcommand{\llasy}       {ref:llasy1,*ref:llasy2}

\newcommand{\SLDacbl}     {ref:SLD_AQL,*Abe:1999hb}
\newcommand{\SLDabj}      {Abe:1998wk,*Abe:2002fs}
\newcommand{\SLDabk}      {ref:SLD_ABK1}

\newcommand{\dalphaQCD}   {bib-Swartz,*bib-Zeppe,*bib-Alemany,*bib-Davier,*bib-alphaKuhn,*bib-jeger99,*bib-Erler,*bib-ADMartin,*bib-Troconiz-Yndurain,*bib-Hagiwara:2003}

%% file: ew_mcite.tex
\newcommand{\CDFdet} {Abe:1988me}
\newcommand{\DZerodet} {Abachi:1993em,*Abazov:2005pn}

%% file: body.tex
\noindent
\textsc{Martin W. Gr\"unewald\\
University College Dublin\\
UCD School of Physics\\
Belfield\\
Dublin 4\\
Ireland\\}
\texttt{e-mail: Martin.Grunewald@cern.ch}
\vspace{36 pt}

\section{Introduction}
\label{sec:ew:Intro}

The last twenty years have seen enormous progress in experimental
precision measurements in particle physics and tests of the
electroweak Standard Model.  New generations of experiments using
advanced detectors at high-energy particle colliders perform
measurements with a precision unprecedented in high-energy particle
physics.  This review summarises the major exciting experimental
results measured at the highest energies and pertaining to the
electroweak interaction.  Comparisons with the theory, the Standard
Model of particle physics ($\SM$)~\cite{\GWS}, are used to test the
theory and to constrain its free parameters.  The data sets analysed
for the precision measurements presented here have been accumulated at
the world's highest-energy particle colliders over the last two
decades.  They verify the $\SM$ as a renormalisable field theory
correctly describing nature.

Electron-positron collisions at $91~\GeV$ centre-of-mass energy were
studied by the SLD detector~\cite{\SLDdet}, operating at Stanford's
Linear Collider (SLC)~\cite{SLC} (1989-1998), and by the experiments
ALEPH~\cite{\ALEPHdet}, DELPHI~\cite{\DELPHIdet}, L3~\cite{\Ldet} and
OPAL~\cite{\OPALdet}, taking data at the Large Electron Positron
collider (LEP)~\cite{\LEPacc} (1989-1995) at the European Laboratory
for Particle Physics, CERN, Geneva, Switzerland.  This centre-of-mass
energy corresponds to the mass of the Z boson, the heavy neutral
exchange particle of the electroweak interaction, which is thus
produced in resonance, $\ee\to\mathrm{Z}$.  While SLC provided
interactions at a fixed centre-of-mass energy corresponding to the
maximum of the Z resonance cross section, the \LEPI\ collider provided
collisions at several centre-of-mass energy points, thereby scanning
the Z resonance lineshape in the range from $88~\GeV$ to $94~\GeV$.
By increasing the LEP center-of-mass energy up to $209~\GeV$ (\LEPII,
1996-2000), the precision measurements of Z-boson properties at \LEPI\
were complemented by measurements of the properties of the W-boson,
the charged carrier of the electroweak interaction, in the reaction
$\eeWW$ at \LEPII.

Proton-antiproton collisions are provided by the Tevatron collider
operating at the Fermi National Accelerator Laboratory close to
Chicago in the USA, at centre-of-mass energies of $1.8~\TeV$ (Run-I,
1992-1996) and $2.0~\TeV$ (Run-II, since 2001), and are studied by the
experiments CDF~\cite{\CDFdet} and D\O~\cite{\DZerodet}.  Results from
the Tevatron experiments important for the electroweak interaction
include measurements of the mass of the W boson, as well as the
discovery of the sixth and heaviest quark known today, the top quark,
in 1995 and the measurements of its properties, in particular its
mass.

In addition, key measurements were performed in dedicated experiments
at lower interaction energies, notably the measurement of the
anomalous magnetic moment of the muon~\cite{Bennett:2006fi}, as well
as the measurements of parity violation effects in atomic
transitions~\cite{QWCs:exp:1,QWCs:exp:2,QWCs:theo:2003:new}, in Moller
scattering~\cite{E158RunI,E158RunI+II+III} and in neutrino-nucleon
scattering~\cite{bib-NuTeV-final}.  The following sections summarise
the major experimental measurements of particular relevance to
electroweak physics, starting with electron-positron collisions at the
Z pole and above the W-pair threshold, the measurements at the
Tevatron collider, and the specialised measurements at lower momentum
transfer.  The combined set of experimental results is compared to
predictions based on the Standard Model of particle physics and used
to constrain its free parameters. In particular, the predictive power
of the $\SM$ and the precision of the electroweak measurements is
elucidated by the predictions of the masses of heavy fundamental
particles, such as W boson, top quark and Higgs boson, which are
contrasted with the direct measurements of these quantities.

\clearpage

\section{Fermion-Pair production and the Z resonance}
\label{sec:ew:2f}

\subsection{Introduction}

\begin{figure}[htbp]
\begin{center}
\includegraphics[width=0.9\textwidth]{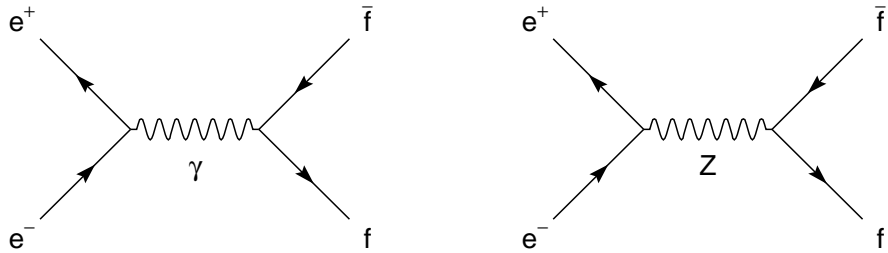}
\caption[]{The lowest-order $s$-channel Feynman diagrams for the
reaction $\ee \rightarrow \ff$. For $\ee$ and $\nu_e\bar\nu_e$ final
states, additional $t$-channel diagrams with photon, Z- or W-boson
exchange contribute as well. The contribution of Higgs boson exchange
diagrams is negligible.}
\label{fig:intro_eeff}
\end{center}
\end{figure}

Fermion-pair production is the dominant interaction in
electron-positron collisions.  The two important Feynman diagrams in
fermion-antifermion production, shown in Figure~\ref{fig:intro_eeff},
proceed through $s$-channel exchange of the neutral electroweak gauge
bosons, namely photon $(\gamma)$ and Z boson.

\begin{figure}[htbp]
\begin{center}
  \includegraphics[width=0.495\textwidth]{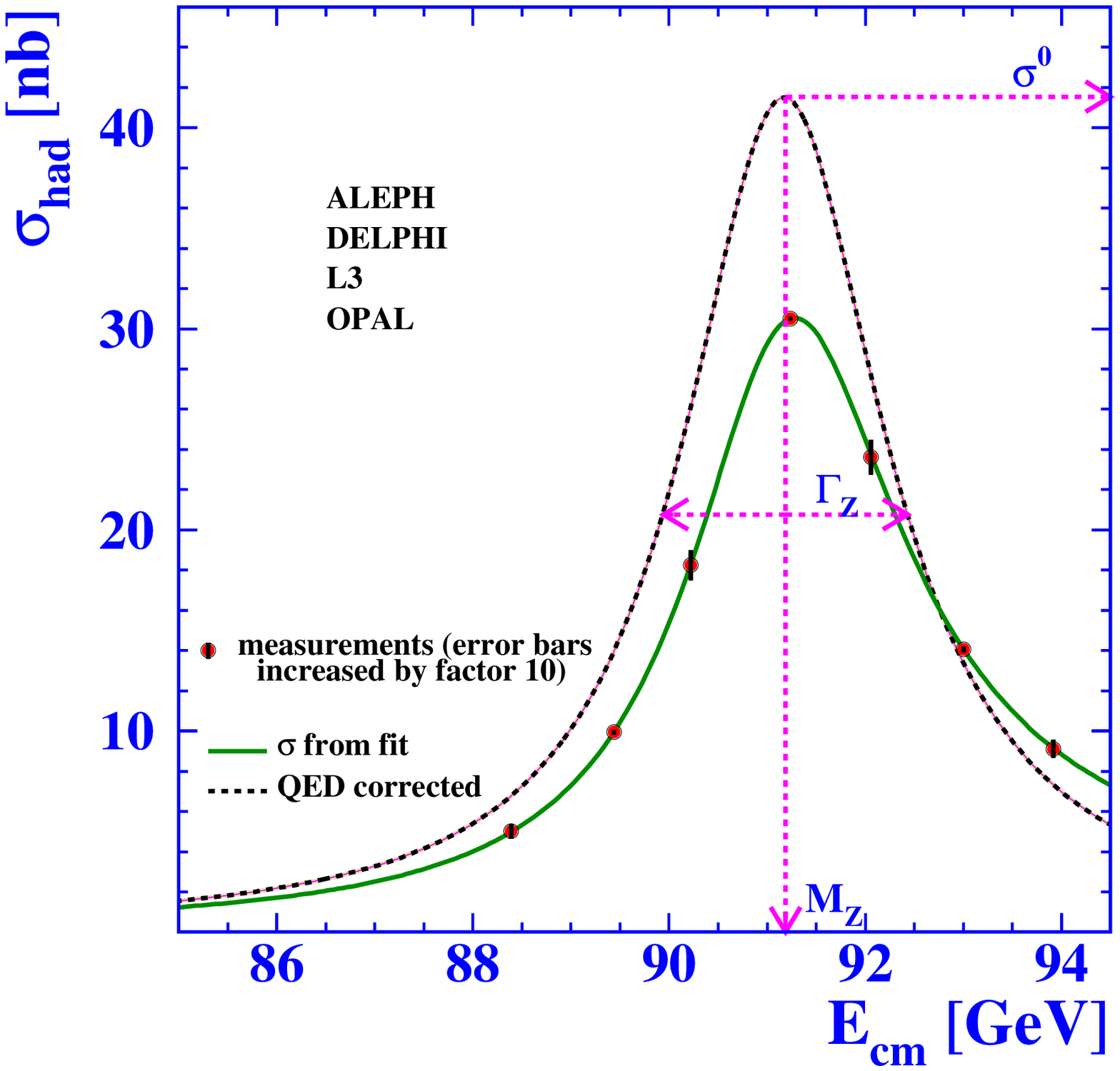}
  \includegraphics[width=0.495\textwidth]{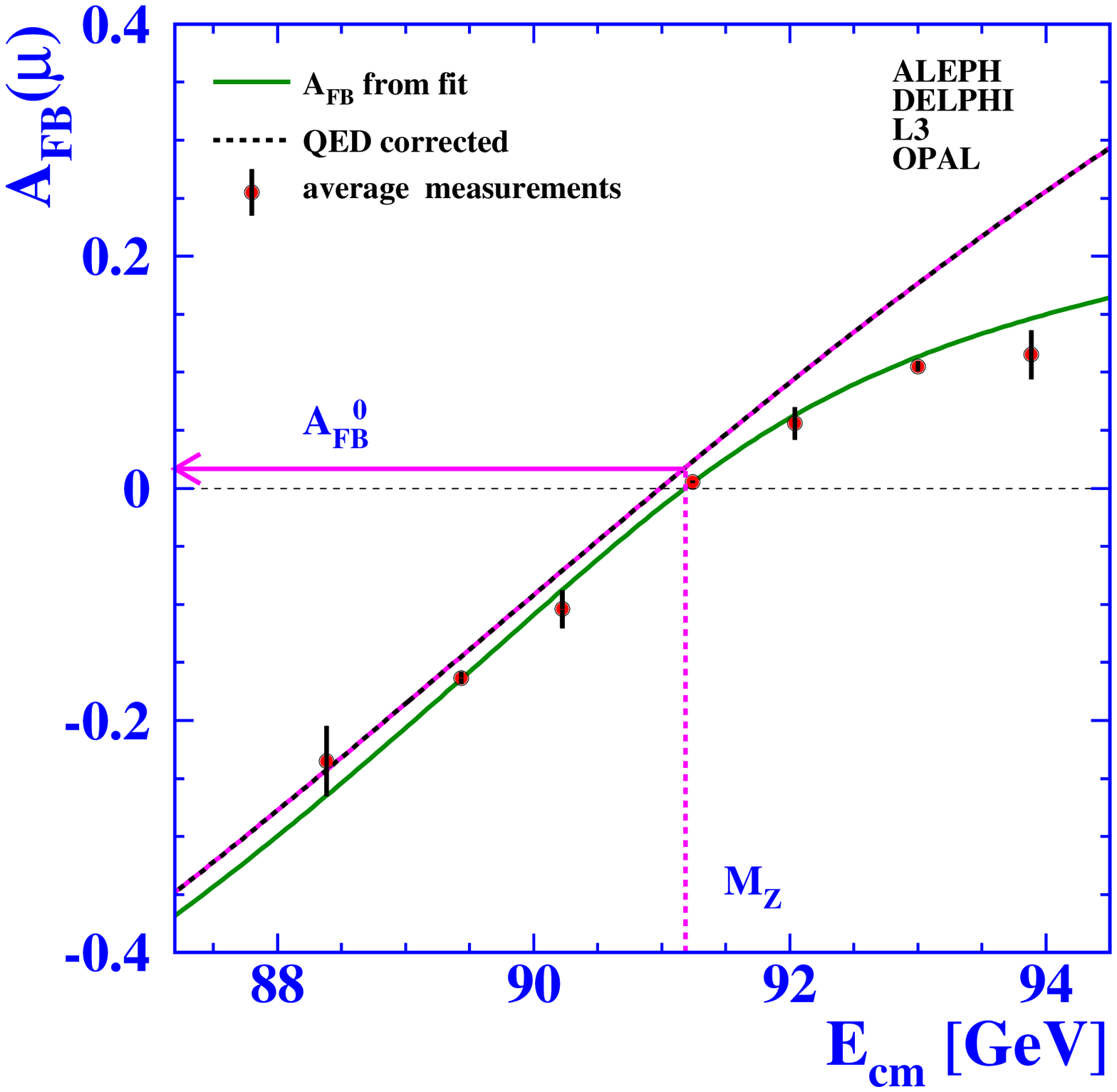}
\end{center}
\caption[Cross Sections and Asymmetries at LEP]{\label{fig:xshafb}
  Left: average over measurements of the total inclusive hadronic
  cross-section of the process $\ee\to\qq$,
  Equation~\ref{eq:Z-pole-xsec}.  Right: average over measurements of
  the forward-backward asymmetry $\Afbm$ in muon production,
  $\ee\to\mumu$, Equation~\ref{eq:Z-pole-afb}.  The measurements are
  shown as a function of the centre-of-mass energy. The full line
  represents the result of a fit including QED radiative corrections
  to the measurements as described in the text.  Unfolding for QED
  radiative effects yields the dashed curves, corresponding to the
  $\Zzero$ parameters shown in Equations~\ref{eq:Z-pole-xsec}
  and~\ref{eq:Z-pole-afb}.  }
\end{figure}

The Z-exchange diagram causes a resonant enhancement of the total
cross section by nearly three orders of magnitude when the
centre-of-mass energy, $\sqrt{s}$, is close to the mass of the Z
boson, $\MZ$. In the vicinity of the Z pole, the total and
differential cross sections for fermion-pair production, $\ee\to\ff$,
are given by:
\begin{eqnarray}
\label{eq:Z-pole-xsec}
\sigma^0(\ee\to\ff;s) & ~=~ & \frac{12\pi}{\MZ^2}
                \frac{\Gee\Gff}{\GZ^2}
\frac{s\GZ^2}{(s-\MZ^2)^2+s^2\GZ^2/\MZ^2}     \nonumber \\
          &     & ~+~ \hbox{$\gamma$/Z interference} 
          ~ + ~ \hbox{photon exchange}\\
\frac{d\sigma^0(s)}{d\cos\theta} & ~=~ & \sigma^0(s)\left[
\frac{3}{8}(1+\cos^2\theta) + \Afbzf(s)\cos\theta\right]
\label{eq:Z-pole-afb}
\end{eqnarray}
where the additional terms not further specified are small.  The angle
$\theta$ denotes the polar scattering angle of the outgoing fermion
with respect to the direction of the incoming electron.  The
dependence of the total cross section and the forward-backward
asymmetry on the centre-of-mass energy is shown in
Figure~\ref{fig:xshafb}.  The effect of radiation of photons in the
initial state on cross section and forward-backward asymmetry, very
large as shown in Figure~\ref{fig:xshafb}, is incorporated in the
calculations by a convoluting the above expressions with QED radiator
functions to yield realistic predictions, which are compared to the
experimental measurements.  These realistic, QED-convoluted
observables are indicated by dropping the superscript 0 from the above
pole-like quantities.  The QED-deconvoluted quantities, indicated by
the superscript 0, are sometimes also called pseudo observables.  The
numerical caluclations are performed with Monte Carlo programs as well
as the semi-analytical programs TOPAZ0~\cite{\TOPAZref} and
ZFITTER~\cite{\ZFITTERref}, which incorporate state-of-the-art
higher-order radiative corrections.

The total and partial decays widths of the Z boson, $\GZ$ and $\Gff$
for fermion species f ($\mathrm{Z}\to\ff$), are:
\begin{eqnarray}
\GZ & = & \sum_{f\ne t}\Gff ~=~ \Ghad+\Gee+\Gmumu+\Gtautau+3\Gnn\\
\Gff & = & N^f_C \frac{\GF\MZ^3}{6\pi\sqrt{2}}\left(\gvf^2+\gaf^2\right)\,
\end{eqnarray}
where $\Ghad$ is the inclusive hadronic decay width summed over the
five light quark flavours, u, d, s, c, and b (top is too heavy),
$N^f_C$ is the QCD colour factor (3 for quarks and 1 for leptons), and
$\gvf$ and $\gaf$ are the effective vector and axial-vector coupling
constants of the neutral weak current, i.e., the coupling of the Z
boson to fermion species f.  The QED-deconvoluted forward-backward
asymmetry $\Afbzf$ at $\sqrt{s}=\MZ$, excluding photon exchange and
$\gamma$Z interference, decomposes into factors of coupling constants
for the inital state and for the final state as follows:
\begin{eqnarray}
\Afbzf  & = & \frac{3}{4}\cAe\cAf\\
\cAf          & = & 2\frac{\gvf\gaf}{\gvf^2+\gaf^2}
              ~ = ~ 2\frac{\gvf/\gaf}{1+(\gvf^2/\gaf^2)}\,,
\end{eqnarray}
where $\cAf$ is called the asymmetry parameter.  Hence, measurements
of both cross-sections and forward-backward asymmetries allow us to
disentangle and determine the effective vector and axial-vector
coupling constants $\gvf$ and $\gaf$.  The asymmetry parameters and
thus forward-backward asymmetries alone determine directly the
effective electroweak mixing angle, because:
\begin{eqnarray}
\frac{\gvf}{\gaf} & = & 1-4|\Qf|\swsqefff\,,
\end{eqnarray}
where $\Qf$ is the electric charge of fermion f.

\subsection{Z Lineshape and Forward-Backward Asymmetries}
\label{sec:Lineshape}

At LEP, data samples were collected at various centre-of-mass energy
points around the Z resonance.  Measurements are routinely performed
separating four final states: the three charged lepton species, $\ee$,
$\mumu$, and $\tautau$ and the inclusive hadronic final state arising
from $\qq$ production.\footnote{For a separation of quark flavours,
see Section~\ref{sec:Heavy}.}  Total cross sections are measured for
all four final states.  Differential cross sections and
forward-backward asymmetries are measured for the lepton-pair final
states.\footnote{For a measurement of an inclusive hadronic charge
asymmetry, see Section~\ref{sec:InclusiveQfb}.}  Each LEP experiment
collected a total integrated luminosity of about $200~\pb$, with the
datasets of the four LEP experiments combined consisting of about 15.5
million hadronic events and 1.7 million lepton-pair events.

\begin{figure}[p]
\begin{center}
\vskip -0.5cm
\includegraphics[width=0.6\linewidth]{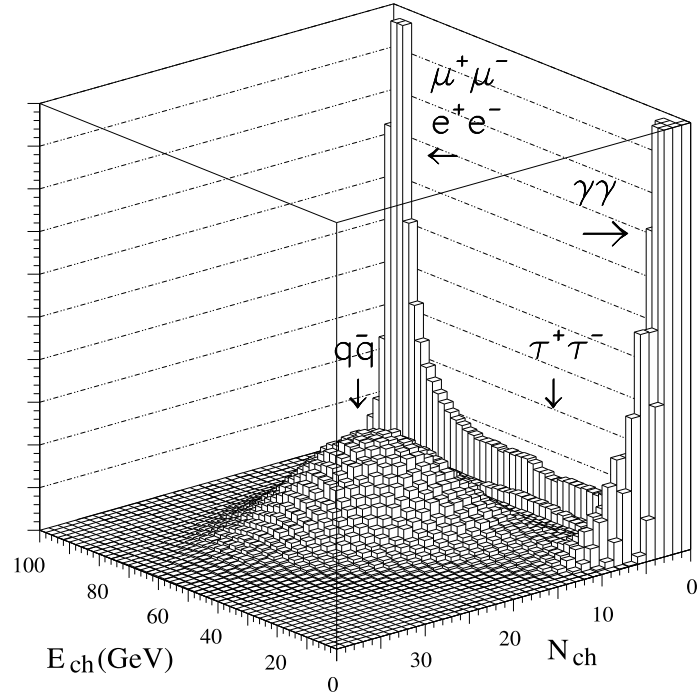}
\end{center}
\vskip -0.5cm
\caption[Separation of final states]{\label{fig:EvsN} Experimental
  separation of the final states using only two variables, the sum of
  the track momenta, $E_{ch}$, corresponding to the energy carried by
  charged hadrons, and the track multiplicity, $N_{ch}$, in the
  central detector of the ALEPH experiment. }
\end{figure}

The different final states are separated based on multiplicity
(high-multiplicity hadronic jets arising from quark fragmentation,
versus low-multiplicity charged leptons) and the characteristic
signatures of the different charged lepton species as illustrated in
Figure~\ref{fig:EvsN}: electrons deposit all their energy in the
electromagnetic calorimeter while muons traverse the complete detector
as minimal ionising particles with their momentum measured by the
inner and outer tracking systems. Tau leptons decay before they are
measured by the detectors and are thus identified by their
low-multiplicity decay signature different of that of di-electrons and
di-muons and the high-multiplicity quark-jets, combined with missing
energy due to neutrinos.

\begin{figure}[p]
\begin{center}
\vskip -0.5cm
\includegraphics[width=0.8\linewidth]{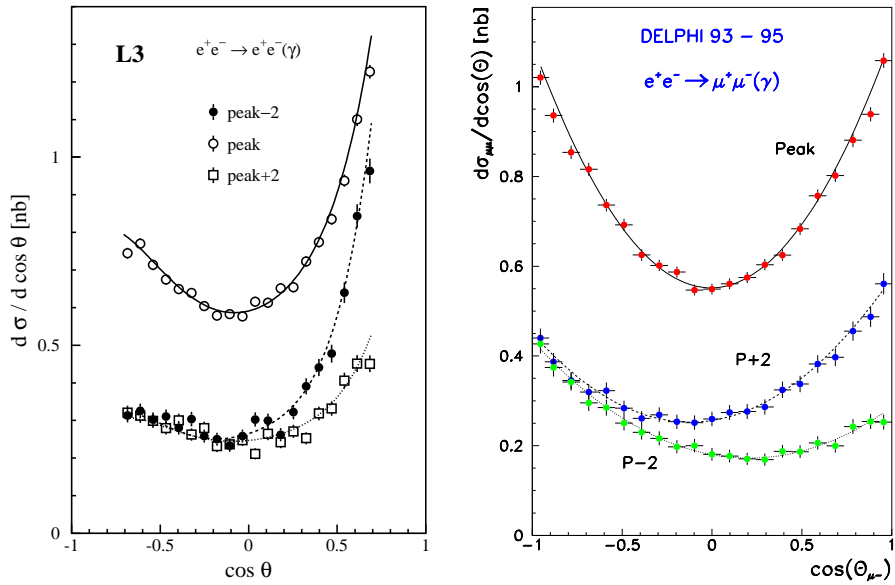}
\end{center}
\vskip -0.5cm
\caption[$\mumu$ differential cross-section]{\label{fig:xsdiff}
  Distribution of the production polar angle, $\cos\theta$, for $\ee$
  and $\mumu$ events at the three principal energies during the years
  1993--1995, measured in the L3 (left) and DELPHI (right) detectors,
  respectively. The curves show the $\SM$ prediction from
  ALIBABA~\cite{ALIBABA} for $\ee$ and a fit to the data for $\mumu$
  assuming the parabolic form of the differential cross-section given
  in the text. The labels ``peak'' and ``peak$\pm2$'' correspond to
  the centre-of-mass energy of the peak, and energy points about
  $\pm2~\GeV$ below and above the cross section peak. }
\end{figure}

Owing to the highly advanced detectors with near $4\pi$ coverage,
backgrounds in the selected samples are typically at the 1\% level
only. Event selection efficiencies within the acceptance range from
70\% for $\tautau$ to more than 99\% for hadrons. Systematic errors
are smallest for the inclusive hadronic final states, typically 0.04\%
to 0.1\%, and somewhat larger for the leptonic finals states, up to
0.7\% especially for tau-pairs. In addition, the luminosity
uncertainty of 0.07\% to 0.24\%, depending on data taking period, is
correlated between all cross section measurements. Detailed
informations on the analyses are given in the
References~\cite{\ALEPHls,\DELPHIls,\Lls,\OPALls}.

The large data samples enable to perform precision cross section and
asymmetry measurements as shown in Figures~\ref{fig:xshafb}
and~\ref{fig:xsdiff}.  The effect of initial-state photon radiation is
clearly visible in Figure~\ref{fig:xshafb}.  It must be controlled to
per-mill precision and better to allow the determination of the
underlying Z resonance parameters such as mass, total and partial
decay widths, pole cross sections and pole asymmetries, with unspoiled
precision.  Differential cross sections measured in $\ee$ and $\mumu$
production are shown in Figure~\ref{fig:xsdiff}.  The $t$-channel
exchange contribution in $\ee$ production changes the differential and
also total cross section compared to the $s$-channel only production
of the other charged fermions.  This effect is of course taken into
account in the analysis of $\ee$ production.

\begin{figure}[p]
\begin{center}
\includegraphics[width=0.495\textwidth]{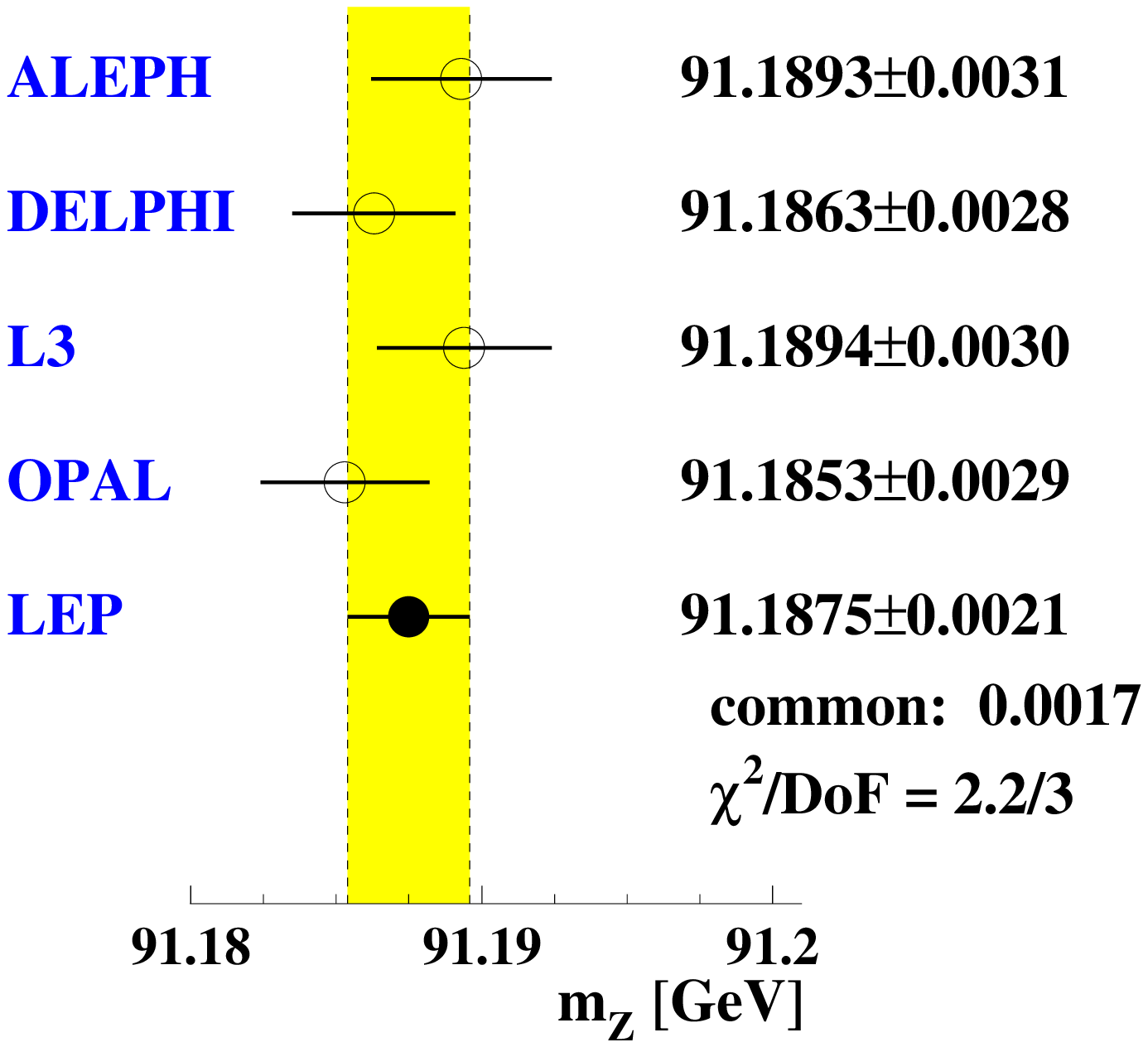}
\includegraphics[width=0.495\textwidth]{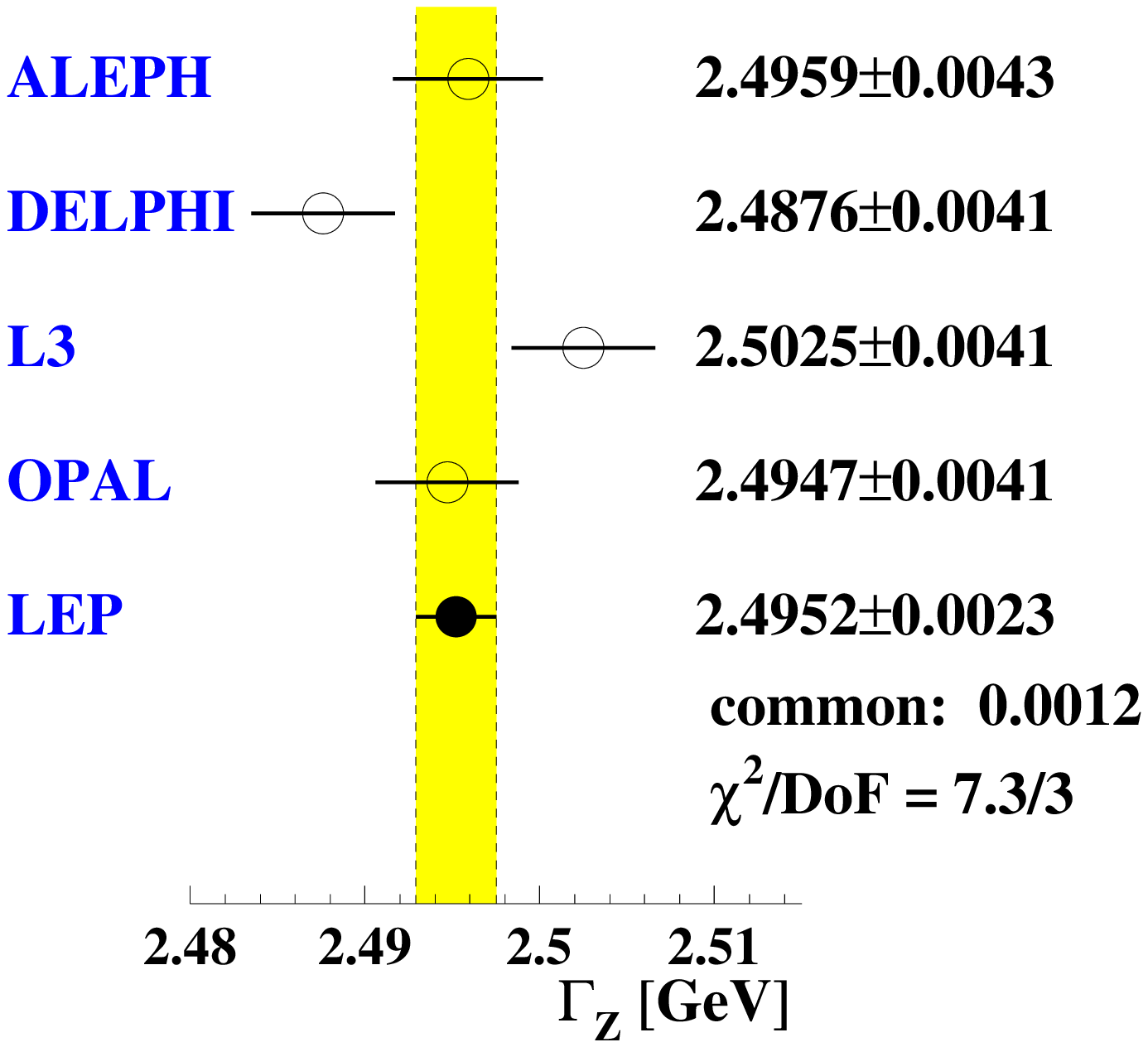}
\includegraphics[width=0.495\textwidth]{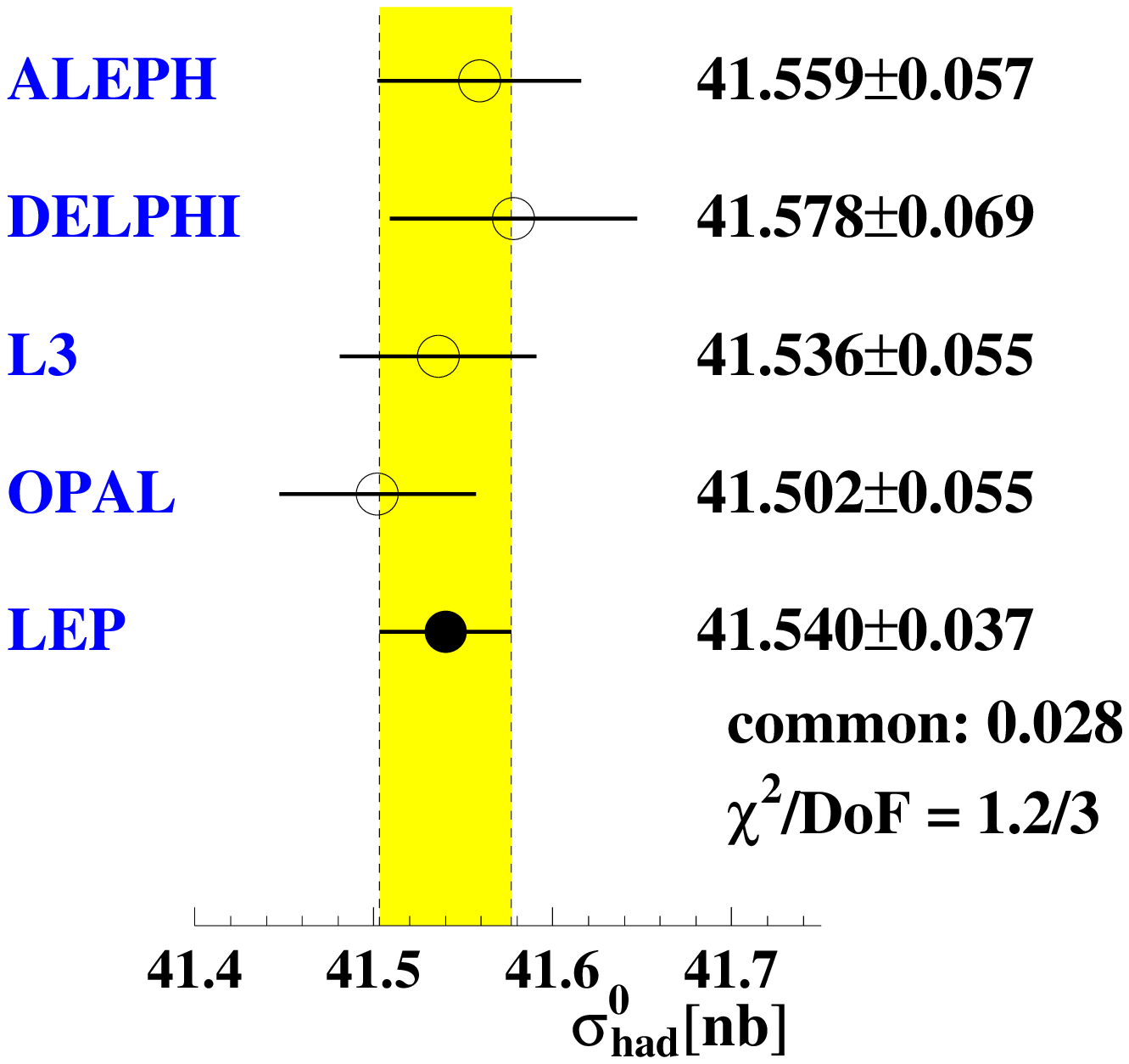}
\includegraphics[width=0.495\textwidth]{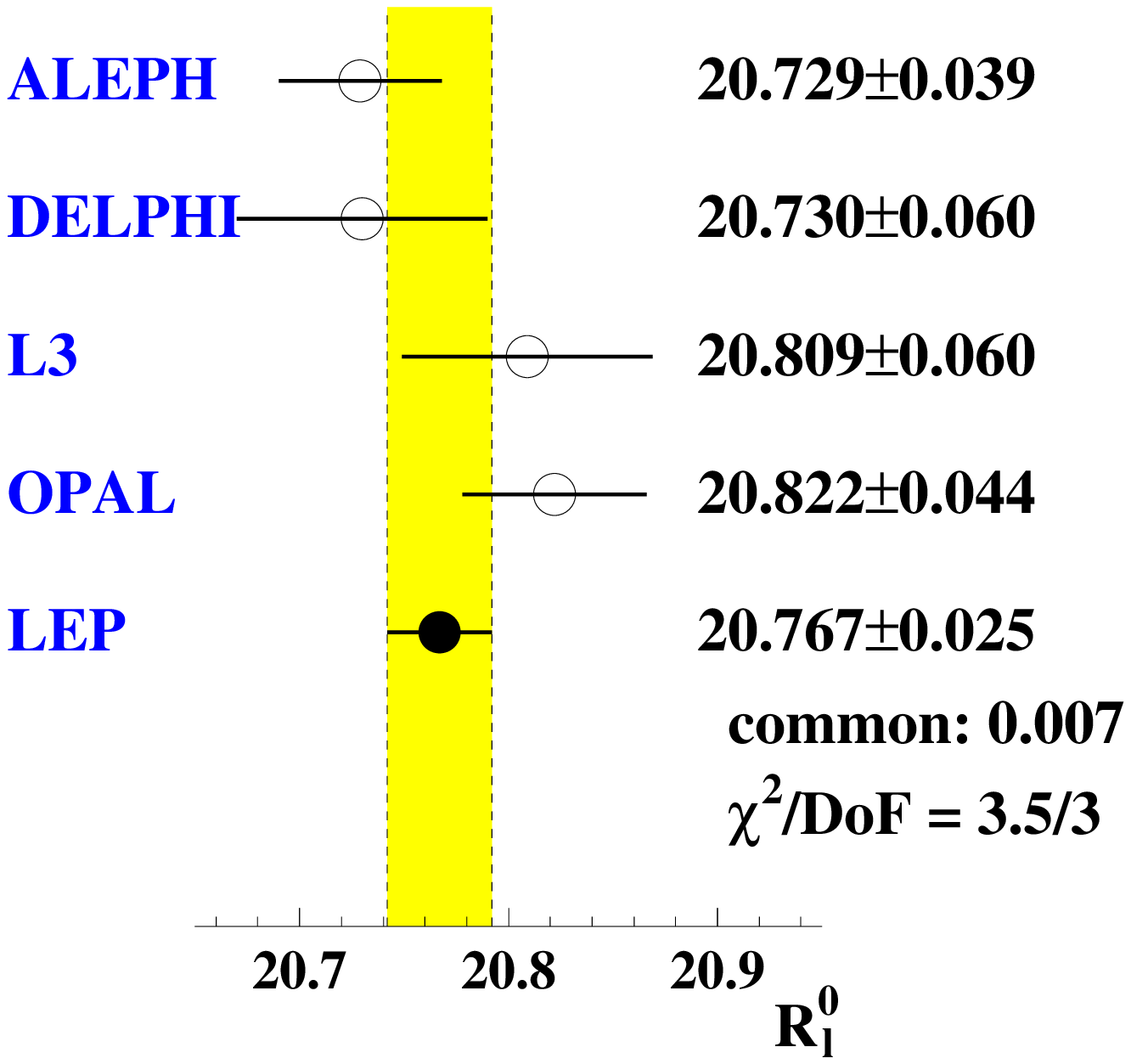}
\includegraphics[width=0.495\textwidth]{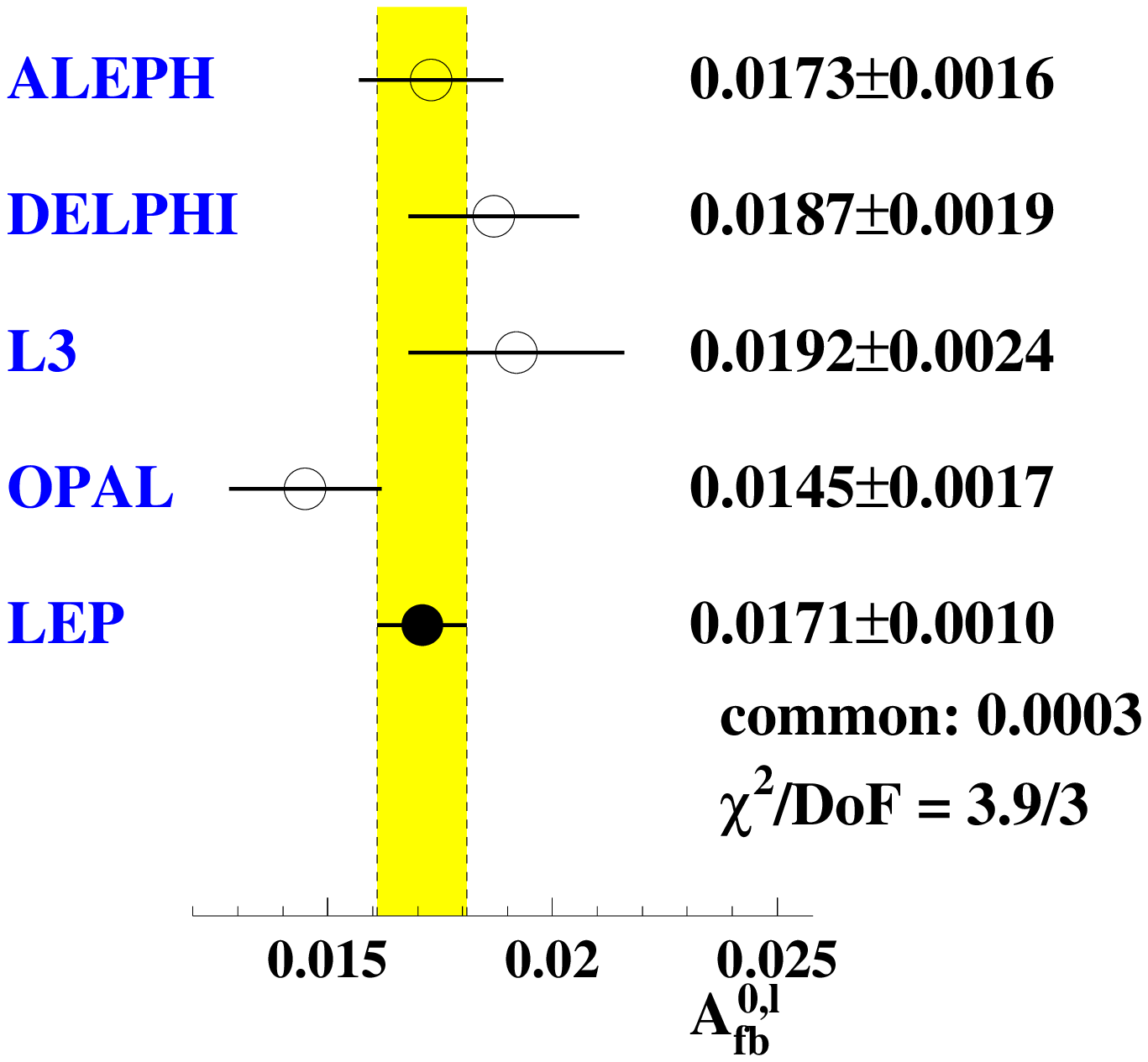}
\caption[Measurements of of $\MZ$,$\GZ$, $\shad$, $\Rl$ and $\Afbzl$]
{\label{fig:lsafb} Measurements of $\MZ$, $\GZ$, $\shad$, $\Rl$ and
$\Afbzl$. The averages indicated were obtained taking correlated
uncertainties into account.  The values of $\chi^2$ per degree of
freedom were calculated considering error correlations between
measurements of the same parameter, but not error correlations between
different parameters.}
\end{center}
\end{figure}

\begin{figure}[p]
\begin{center}
\vskip -1cm
\includegraphics[width=0.6\linewidth]{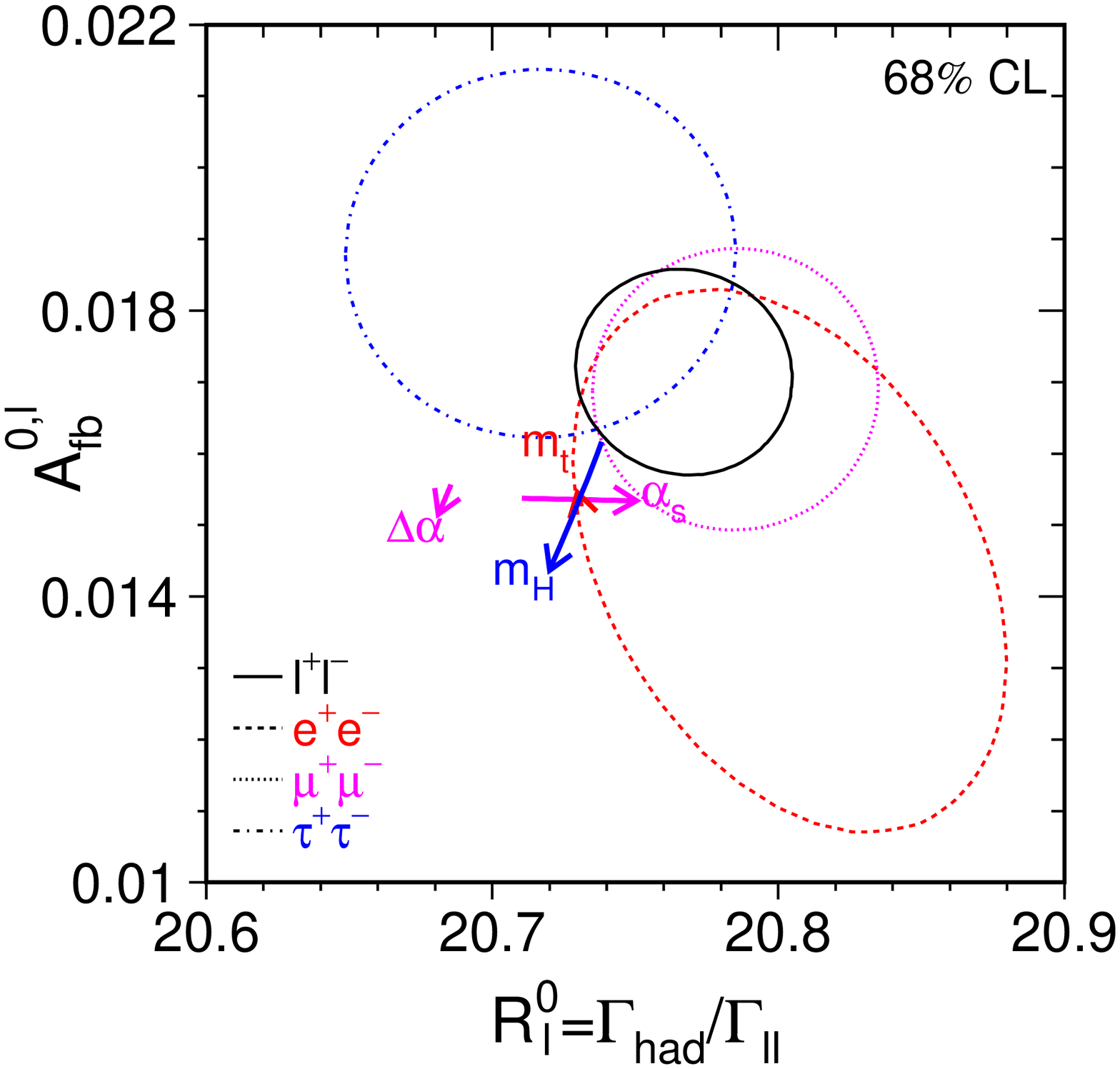}
\end{center}
\vskip -1cm
\caption[$\Rl$ vs. $\Afbzl$] {\label{fig:rlafb} Contour lines (68\,\%
CL) in the $\Rl$--$\Afbzl$ plane for $\ee$, $\mumu$ and $\tautau$
pairs, without and with the assumption of lepton universality,
$\leptlept$. The results for the $\tau$ lepton are corrected to
correspond to the massless case.  The $\SM$ prediction is shown as
arrows corresponding to variations of $\MZ=91.1875~\GeV$,
$\Mt=170.9\pm1.8~\GeV$, $\MH=300^{+700}_{-186}~\GeV$,
$\alfmz=0.118\pm0.003$.  The arrow showing the dependence on the
hadronic vacuum polarisation $\dalhad=0.02758\pm0.00035$ is displaced
for clarity.}
\end{figure}

For reduced correlations between the observables, the measurements of
Z-boson parameters are presented using the parameters~\cite{Z-Pole}:
$\MZ$, $\GZ$, $\shad=\sigma^0(\ee\to\hbox{hadrons})$ at
$\sqrt{s}=\MZ$, $\Rl=\shad/\slept=\Ghad/\Gll$ and $\Afbzl=3\cAe\cAl/4$
for $\ell=e,\,u,\tau$.  These nine pseudo observables reduce to five
under the assumption of neutral-current charged-lepton universality.
The results of the four LEP experiments are compared in
Figure~\ref{fig:lsafb}; combined numerical results~\cite{Z-Pole} are
also reported in Table~\ref{tab:lsafbresult}.  The $\chidf$ of the
combination indicates that the model describes the data well.  The
mass of the Z boson is determined with a precision of 23~ppm, and its
total width with per-mill precision.  Uncertainties due to the
absolute and relative LEP beam-energy calibration~\cite{bib-ECAL95}
affect the averaged results on $\MZ$ and $\GZ$ at the level of
$1.7~\MeV$ and $1.2~\MeV$, respectively, showing how precise a LEP
beam energy calibration was achieved.  The measurement of luminosity
affects $\shad$ only.  Both statistical and systematic uncertainties
on efficiency and acceptance contribute to the total uncertainty on
$\Rl$, while the forward-backward asymmetries are still statistics
dominated.  Neutral current lepton universality is strongly favoured
by the results for $\Rl$ and $\Afbzl$ as shown in
Figure~\ref{fig:rlafb}, where the contours for electrons, muons and
taus largely overlap.  Numerical results imposing lepton universality
are also reported in Table~\ref{tab:lsafbresult}.

\begin{table}[p]\begin{center}
\caption[]{\label{tab:lsafbresult} Combined results for the $\Zzero$
  parameters, without (top) and with (bottom) the assumption of lepton
  universality.}
\begin {tabular} {|lr||r@{\,}r@{\,}r@{\,}r@{\,}r@{\,}r@{\,}r@{\,}r@{\,}r|}
\hline 
\multicolumn{2}{|c||} {Without lepton universality} & 
                                    \multicolumn{9}{l|}{~~~Correlations} \\
\hline 
\hline 
\multicolumn{2}{|c||}{$\pzz \chidf\,=\,32.6/27 $} &
   $\MZ$ & $\GZ$ & $\shad$ &
     $\Ree$ &$\Rmu$ & $\Rtau$ & $\Afbze$ & $\Afbzm$ & $\Afbzt$ \\
\hline 
 $\MZ$ [\GeV{}]  & 91.1876$\pm$ 0.0021 &
 ~1.000 & \multicolumn{8}{c|}{} \\
 $\GZ$ [\GeV]  & 2.4952 $\pm$ 0.0023 &
 $-$0.024 & ~1.000 & \multicolumn{7}{c|}{}\\ 
 $\shad$ [nb]  & 41.541 $\pm$ 0.037$\pz$ &
 $-$0.044 & $-$0.297 & ~1.000 & \multicolumn{6}{c|}{}\\ 
 $\Ree$        & 20.804 $\pm$ 0.050$\pz$ &
 ~0.078 & $-$0.011 & ~0.105 & ~1.000 & \multicolumn{5}{c|}{}\\ 
 $\Rmu$        & 20.785 $\pm$ 0.033$\pz$ & 
 ~0.000 & ~0.008 & ~0.131 & ~0.069 & ~1.000 & \multicolumn{4}{c|}{}\\ 
 $\Rtau$       & 20.764 $\pm$ 0.045$\pz$ &  
 ~0.002 & ~0.006 & ~0.092 & ~0.046 & ~0.069 & ~1.000 & \multicolumn{3}{c|}{}\\ 
 $\Afbze$      & 0.0145 $\pm$ 0.0025 &
 $-$0.014 & ~0.007 & ~0.001 & $-$0.371 & ~0.001 & ~0.003 & ~1.000 & \multicolumn{2}{c|}{}\\ 
 $\Afbzm$      & 0.0169 $\pm$ 0.0013 &
 ~0.046 & ~0.002 & ~0.003 & ~0.020 & ~0.012 & ~0.001 & $-$0.024 & ~1.000 & \multicolumn{1}{c|}{}\\ 
 $\Afbzt$      & 0.0188 $\pm$ 0.0017 &
 ~0.035 & ~0.001 & ~0.002 & ~0.013 & $-$0.003 & ~0.009 & $-$0.020 & ~0.046 & ~1.000 \\ 
\hline 
\multicolumn{3}{c}{~}\\
\end{tabular} \\
\begin {tabular} {|lr||r@{\,}r@{\,}r@{\,}r@{\,}r|}
\hline 
\multicolumn{2}{|c||} {With lepton universality} & \multicolumn{5}{c|}{Correlations} \\
\hline 
\hline 
\multicolumn{2}{|c||}{$\pzz \chidf\,=\,36.5/31 $}  & 
   $\MZ$ & $\GZ$ & $\shad$ & $\Rl$ &$\Afbzl$ \\
\hline 
 $\MZ$ [\GeV{}]  & 91.1875$\pm$ 0.0021$\pz$    &
 ~1.000 & \multicolumn{4}{c|}{}\\ 
 $\GZ$ [\GeV]  & 2.4952 $\pm$ 0.0023$\pz$    &
 $-$0.023  & ~1.000 & \multicolumn{3}{c|}{}\\ 
 $\shad$ [nb]  & 41.540 $\pm$ 0.037$\pzz$ &
 $-$0.045 & $-$0.297 &  ~1.000 & \multicolumn{2}{c|}{}\\ 
 $\Rl$         & 20.767 $\pm$ 0.025$\pzz$ &
 ~0.033 & ~0.004 & ~0.183 & ~1.000 & \multicolumn{1}{c|}{}\\ 
 $\Afbzl$     & 0.0171 $\pm$ 0.0010$\pz$   & 
 ~0.055 & ~0.003 & ~0.006 & $-$0.056 &  ~1.000 \\ 
\hline 
\end{tabular} 
\end{center}
\end{table}

\clearpage

\subsection{Polarised Asymmetries at SLC}

\begin{figure}[htbp]
\begin{center}
\includegraphics[angle=-90,width=0.8\textwidth]{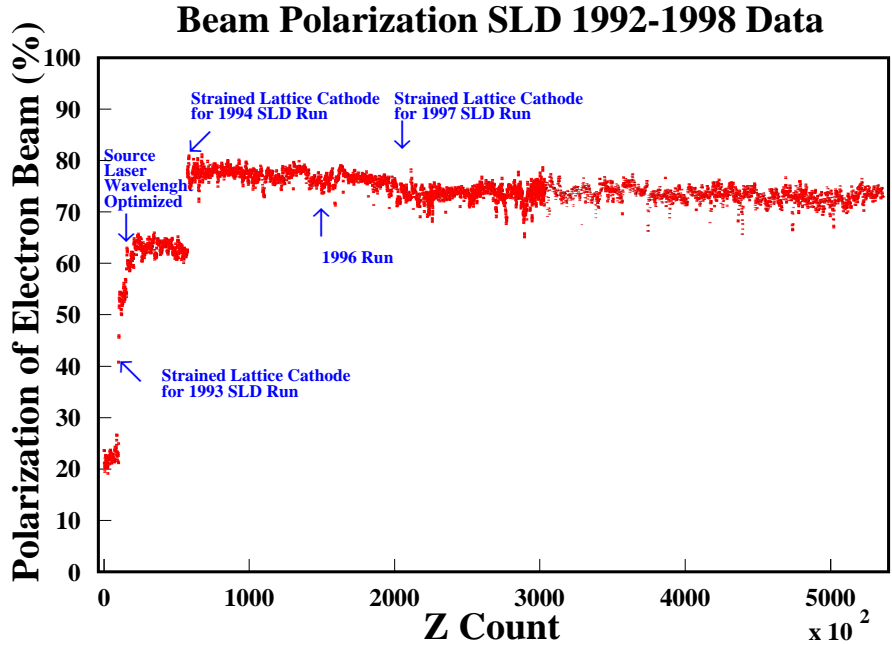}
\caption[Longitudinal polarisation at SLC]{The amount of
  longitudinal electron polarisation as a function of the number of
  recorded Z decays at SLD.}
\label{fig:slc_polar}
\end{center}
\end{figure}

In contrast to LEP where the beams were unpolarised, SLC provided
electron-positron collisions with longitudinally polarised electrons.
The beam polarisation as a function of Z event count is shown in
Figure~\ref{fig:slc_polar}, indicating that polarisation in excess of
70\% was routinely achieved for the bulk of the data.  In order to
measure polarised asymmetries such as left-right and left-right
forward-backward asymmetries with highest possible statistical
precision, SLC maximised the event count by providing collisions at a
single fixed centre-of-mass energy only, corresponding to the maximum
of the annihilation cross section.

\subsubsection{Left-Right Asymmetry}

The left-right asymmetry is a experimentally robust quantity,
depending only on event counting:
\begin{eqnarray}
\ALR & = & \frac{N_L-N_R}{N_L+N_R}\frac{1}{\pole}\,,
\end{eqnarray}
where $N_L$ and $N_R$ are the event yields per luminosity unit
produced with left- and right-handed polarised electron beams, with
the degree of longitudinal polarisation denoted by $\pole$.  Many
systematic effects, such as those affecting acceptance or efficiency,
cancel in the ratio.  QED radiative corrections $\Delta\ALR$ are small
and calculated to high precison. The corresponding pole quantity
$\ALRz$ is then simply:
\begin{eqnarray}
\ALR+\Delta\ALR ~=~ \ALRz & = & \cAe ~=~ \frac{2\gve/\gae}{1+(\gve/\gae)^2}\,.
\end{eqnarray}
The left-right asymmetry measurement determines directly the asymmetry
parameter $\cAf$ for the inital state fermions.  Any final state can
be used for the measurement.

\begin{figure}[htbp]
\begin{center}
\includegraphics[width=0.8\linewidth]{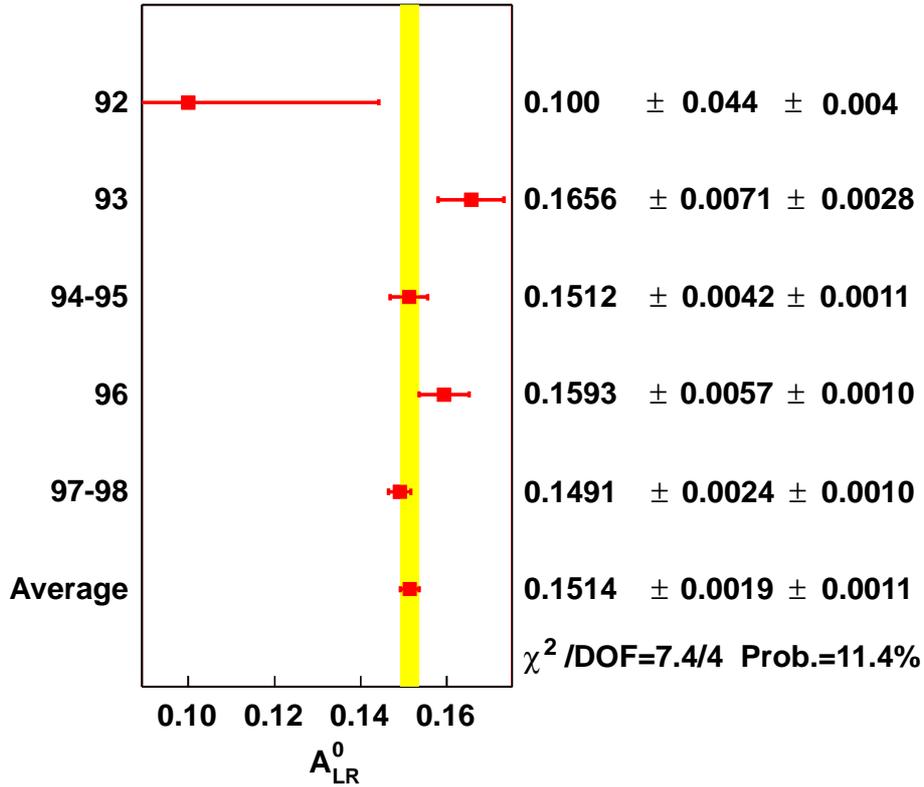}
\caption[History of the SLD $\ALRz$ measurements.]  {A compilation of
the published SLD $\ALRz$ results, ordered by year the data was taken.
The final average is calculated including correlations of systematic
uncertainties.}
\label{fig:alrhistory}
\end{center}
\end{figure}

The SLD collaboration performed the measurement of $\ALR$ with the
high-statistics inclusive hadronic channel in order to obtain a
high-precision measurement of $\ALRz$~\cite{\SLDalr}.  Leptonic final
states are separated out and used in the independent determination of
the left-right forward-backward asymmetry discussed below.  The
results on $\ALRz$ for the different data taking periods with their
different degree of beam polarisation, compared in
Figure~\ref{fig:alrhistory}, show reasonable overall consistency.

The average over the complete SLD data sample, taking correlated
uncertainties into account, results in~\cite{\SLDalr}:
\begin{eqnarray}
\ALRz     & = & 0.1514  \pm 0.0022
\end{eqnarray}
or equivalently:
\begin{eqnarray}
\swsqeffl & = & 0.23097 \pm 0.00027 \,.
\end{eqnarray}
This measurement is the single most precise determination of
$\swsqeffl$. It is still statistics dominated; the systematic errors
included in the total errors listed above are $\pm0.0011$ and
$\pm0.00013$, respectively.

\subsubsection{Left-Right Forward-Backward Asymmetry}

In contrast to the left-right asymmetry, the left-right
forward-backward asymmetry determines the asymmetry parameter $\cAf$
specific to the analysed final state:
\begin{eqnarray}
\AFBLRf & = & \frac{1}{\pole}
              \frac{(N_{F,L}-N_{B,L}) - (N_{F,R}-N_{B,R})}
                   {(N_{F,L}+N_{B,L}) + (N_{F,R}+N_{B,R})}
~\to~ \AFBLRz ~=~ \frac{3}{4}\cAf\,,
\end{eqnarray}
where the subscripts indicate forward and backward events ($F,B$)
accumulated with left- or right-handed electron beams ($L,R$).  Owing
to $|\pole|=75\%$, the left-right forward-backward asymmetries yield a
statistical precision equivalent to measurements of the unpolarised
forward-backward asymmetry, $\Afbzf=(3/4)\cAe\cAf$, using a 25 times
larger event sample.

\begin{figure}[htbp]
\begin{center}
\includegraphics[width=0.7\linewidth]{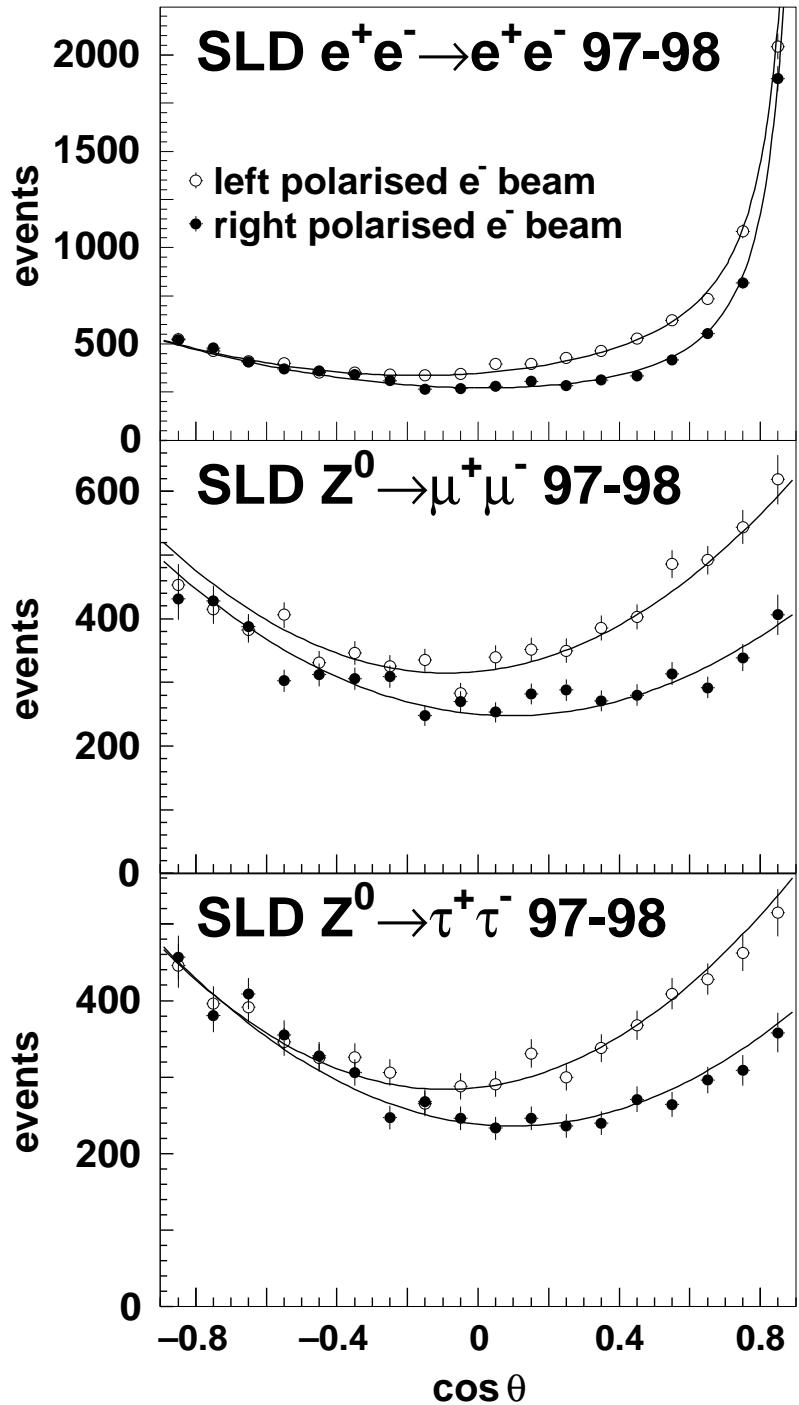}
\caption[Polarisation dependent angular distributions for leptons]
{Polar-angle distributions for observed for $\ee$, $\mumu$ and
$\tautau$ events selected in the 1997-1998 SLD run.  The solid line
represents the fit, while the points with error bars show the data in
bins of 0.1 in $\cos\theta$.  For $|\cos\theta|>0.7$, the data are
corrected for a decrease in the detection efficiency with increasing
$|\cos\theta|$.  }
\label{fig:alr:distr}
\end{center}
\end{figure}

The final states $\ee$, $\mumu$ and $\tautau$ were used to determine
the asymmetry parameter $\cAl$ for the three charged lepton
species~\cite{ref:sld-al2000,ref:sld-al1997}.  For increased
statistical precision, the differential distributions in terms of
$\cos\theta$, the polar scattering angle, for the left- and
right-handed data samples, shown in Figure~\ref{fig:alr:distr}, are
analysed, allowing a combined determination of both $\cAe$ and $\cAf$.
In case of $\ee$ production, the additional contributions due to
$t$-channel scattering, skewing the differential distribution, are
accounted for in the analysis.

\subsubsection{Summary}

The combined results~\cite{Z-Pole} derived from the measurements of
the left-right and left-right forward-backward asymmetries by SLD at
SLC are listed in Table~\ref{tab:alr:result}.  These results are in
good agreement as expected from neutral-current lepton
universality. The measurement of $\cAe$, fully dominated by the
$\ALRz$ result, is by far the most precise determination of any of the
$\cAl$ asymmetry parameters.  The combined result, assuming lepton
universality and accounting for small correlated systematic
uncertainties, is:
\begin{equation}
 \cAl = 0.1513 \pm 0.0021,
\label{eq:al:result}
\end{equation}
where the total error includes a systematic error of $\pm 0.0011$.
This measurement is equivalent to a determination of:
\begin{equation}
 \swsqeffl = 0.23098 \pm 0.00026,
\end{equation}
where the total error includes a systematic error of $\pm 0.00013$.

\begin{table}[htbp]
\begin{center}
\caption[Results on the leptonic asymmetry parameters $\cAl$ from SLD]
{Results on the leptonic asymmetry parameters $\cAl$ not assuming
  neutral-current lepton universality obtained at SLD. The result on
  $\cAe$ includes the result on $\ALRz$. }
\label{tab:alr:result}
\renewcommand{\arraystretch}{1.25}
\begin{tabular}{|c||r@{$\pm$}l||rrr|}
\hline
Parameter & \multicolumn{2}{|c||}{Average} 
          & \multicolumn{3}{|c| }{Correlations}    \\
          & \multicolumn{2}{|c||}{ }
          & {$\cAe$} & {$\cAm$}& {$\cAt$}          \\
\hline
\hline
$\cAe$ &$0.1516$&$0.0021$ & $ 1.000$&$      $&$      $ \\
$\cAm$ &$0.142 $&$0.015 $ & $ 0.038$&$ 1.000$&$      $ \\
$\cAt$ &$0.136 $&$0.015 $ & $ 0.033$&$ 0.007$&$ 1.000$ \\
\hline
\end{tabular}
\end{center}
\end{table}

\clearpage

\subsection{Tau Polarisation at LEP}

Even in case of unpolarised beams, the final-state fermions in
$\ee\to\ff$ exhibit a non-zero polarisation (average helicity) arising
from the parity-violating coupling of the exchanged Z boson to the
initial- and final-state fermions.  For final-state fermions
undergoing charged weak decay, the parity-violating nature of the
charged weak decay serves as a polarisation analyser and thus allows
to measure the polarisation~\cite{Eberhard:1989ve}, here in case of
$\tautau$ production with subsequent V--A tau decay.

At $\sqrt{s}=\MZ$ and considering only Z-boson exchange, the
dependence of the polarisation on the polar scattering angle reads:
\begin{equation}
\label{eq-ptcos}
\ptau(\cos\theta_{\tau^-})= - \frac
{\cAt(1+\cos^2\theta_{\tau^-}) + 2\cAe \cos\theta_{\tau^-}}
{(1+\cos^2\theta_{\tau^-}) + \frac{8}{3}\Afb^{\tau}\cos\theta_{\tau^-}}\,,
\end{equation}
where $\ptau\equiv \ptaum=-\ptaup$ since the $\tau^-$ and $\tau^+$
have opposite helicities in Z decays.  Analysis of the differential
distributions allows the separate determination of the asymmetry
parameters $\cAe$ and $\cAt$. The average polarisation and the
forward-backward asymmetry of the polarisation are related to the
asymmetry parameters $\cAl$ as:
\begin{eqnarray}
  \label{eq:ptau}
\langle \ptau^{0} \rangle & = & -\phantom{\frac{3}{4}} \cAt  \\
  \label{eq:afbpol}
  \AFBpolz & = & - \frac{3}{4}\cAe \, .
\end{eqnarray}
The analysis of the differential distribution Equation~\ref{eq-ptcos}
in terms of both $\cAe$ and $\cAt$ yields statistically somewhat more
precise results on the asymmetry parameters than using just the
average polarisation and its forward-backward asymmetry.

\begin{figure}[htbp]
\includegraphics[width=0.465\linewidth]{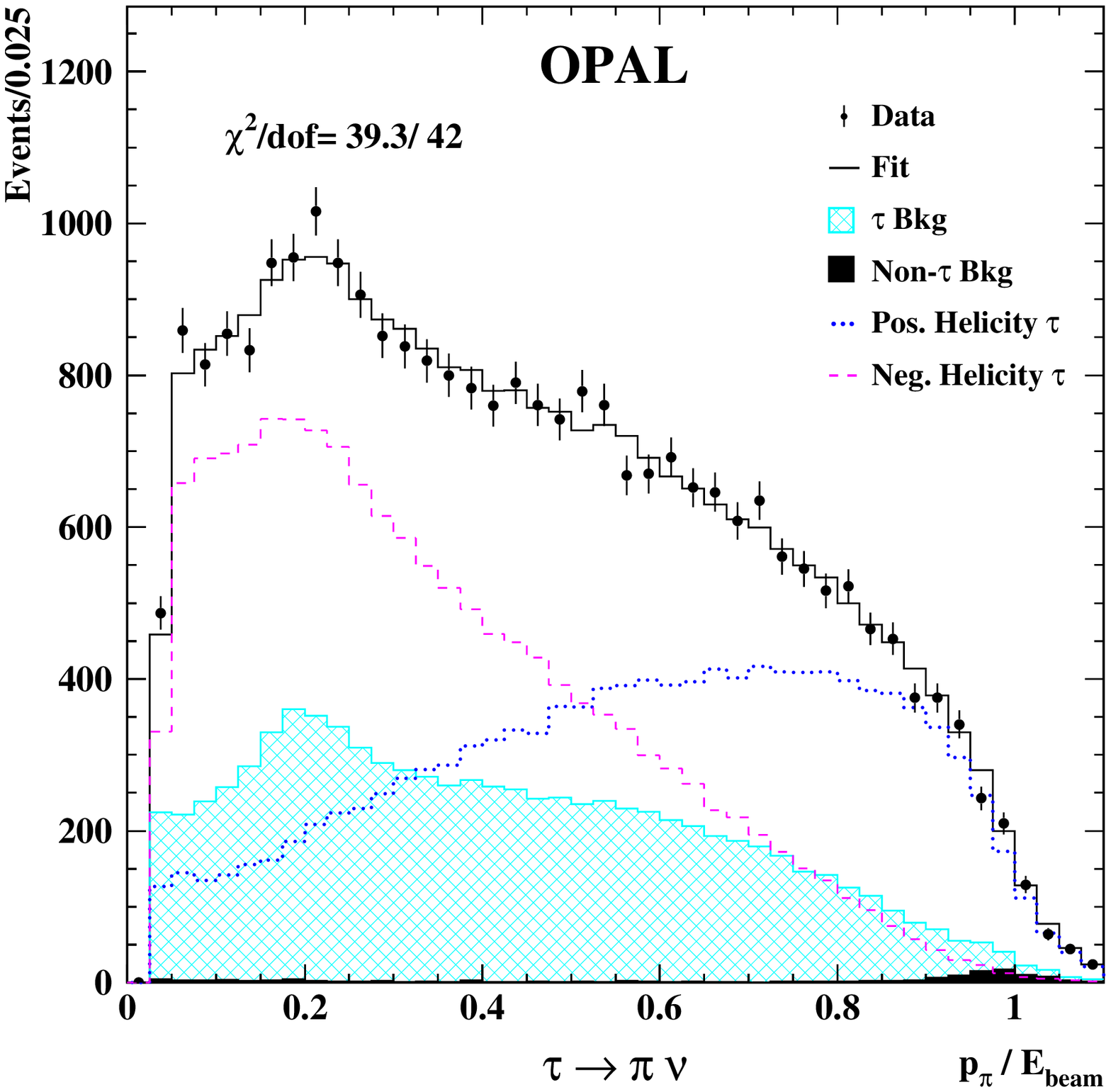}
\hfill
\includegraphics[width=0.52\linewidth]{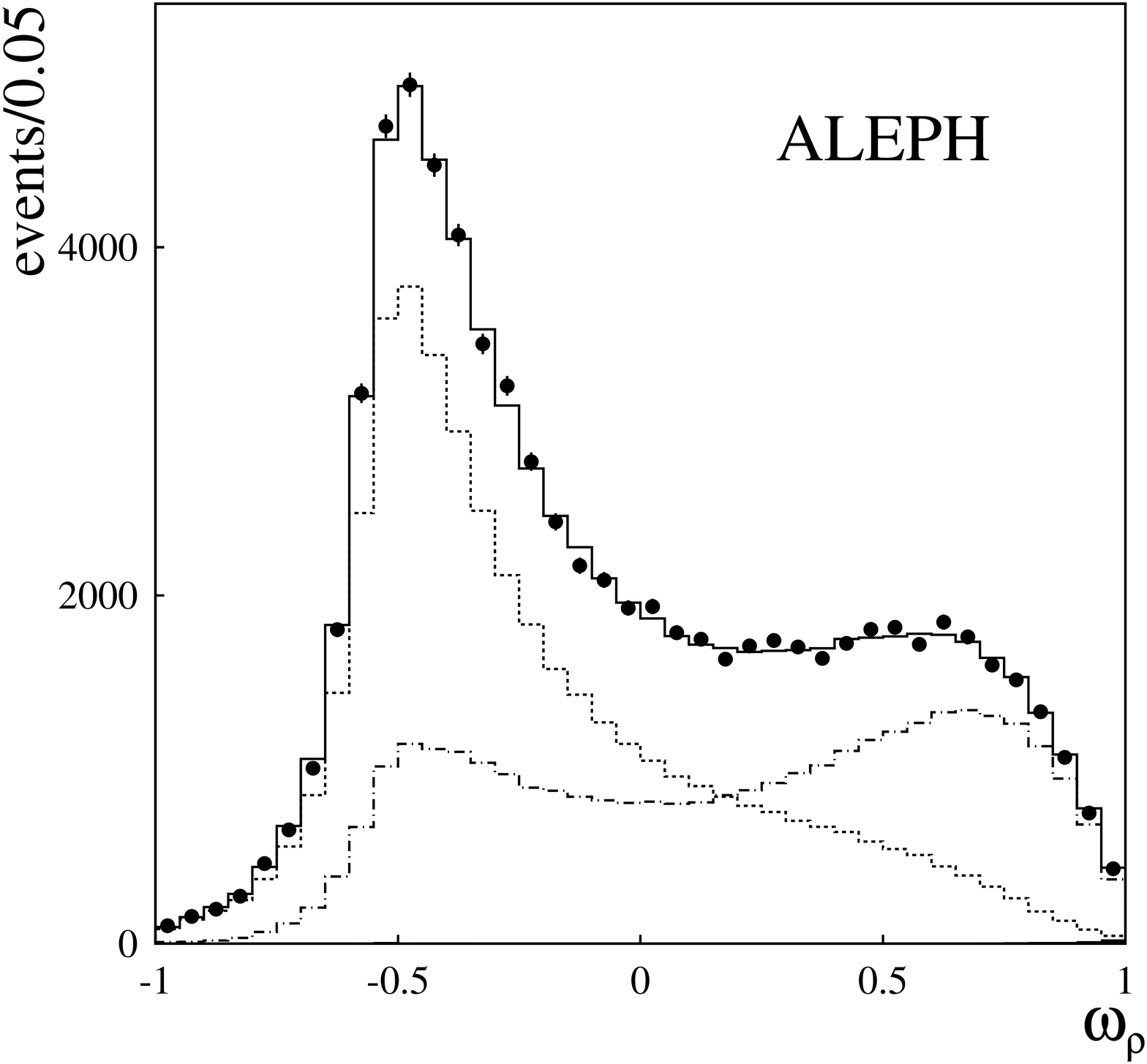}\\
\includegraphics[width=0.51\linewidth]{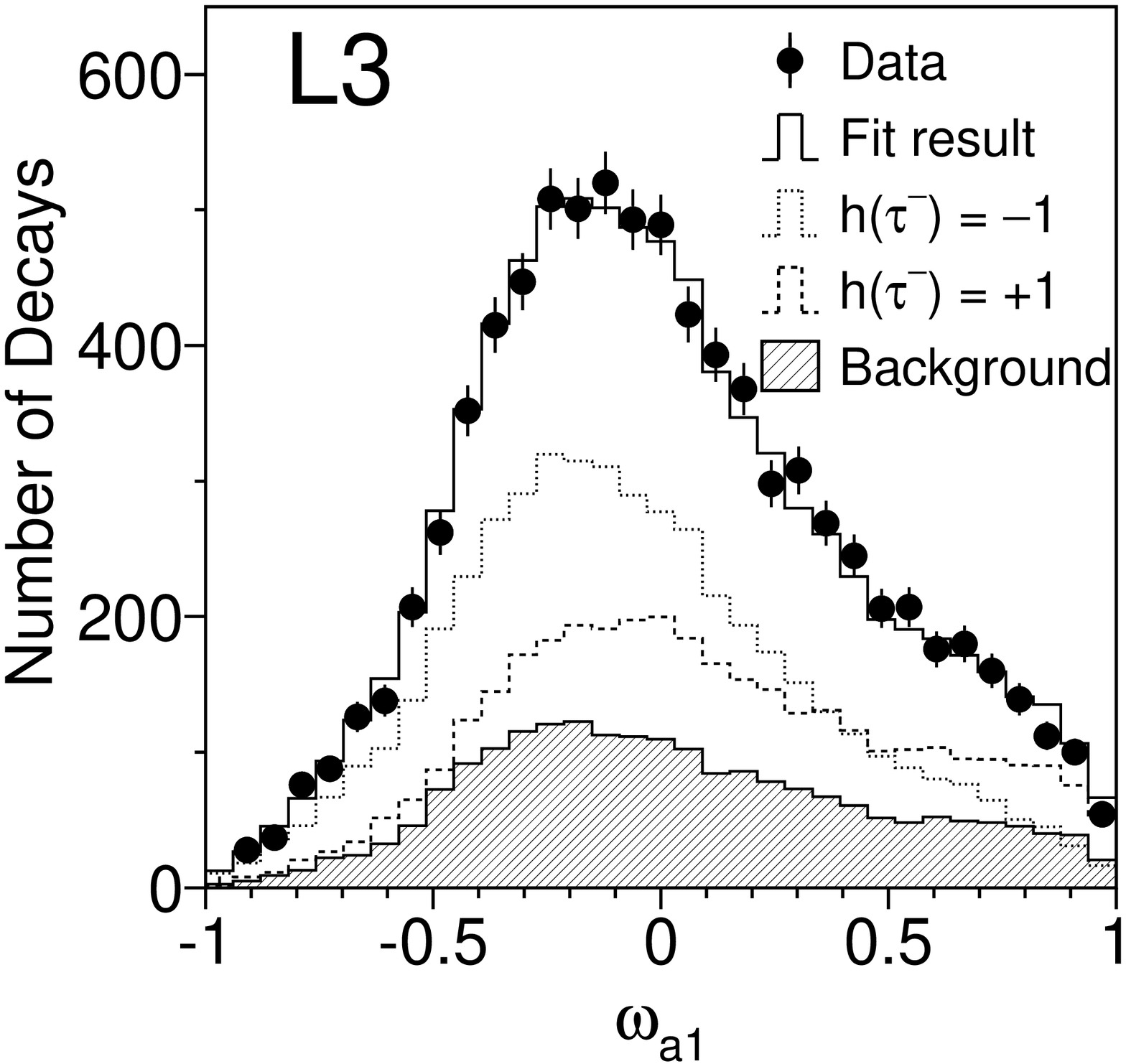}
\hfill
\includegraphics[width=0.51\linewidth]{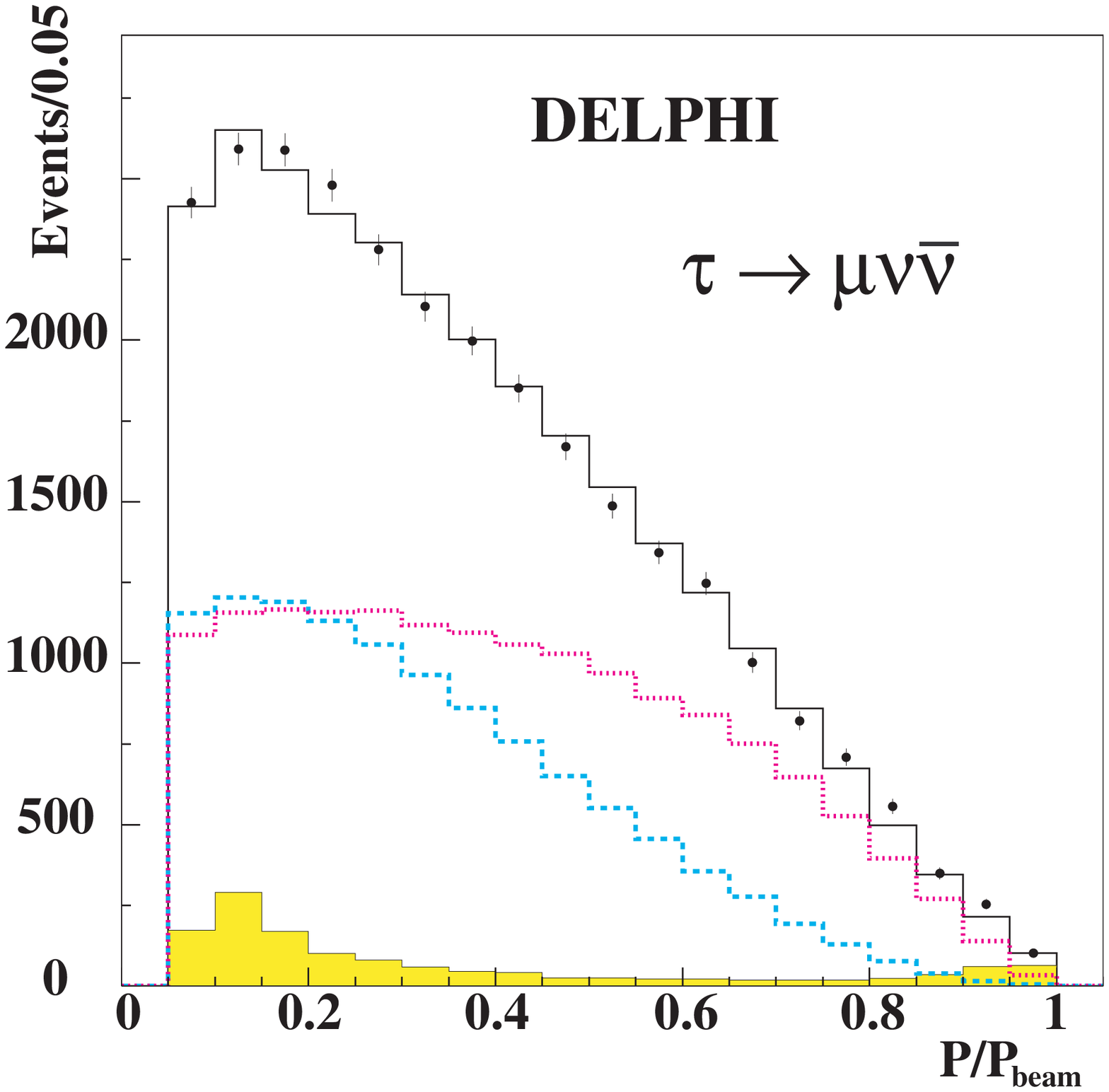}\\
\caption[$\tau$ polarisation spectra] { \label{fig:taupol} The
measured distributions in the polarisation-sensitive variable for the
$\tpinu$ decays (OPAL), $\trhonu$ decays (ALEPH), $\tanu$ decays (L3)
and $\tmununu$ decays (DELPHI), in each case integrated over the whole
$\cos\theta_{\tau^-}$ ~range, with $\omega$ denoting the relevant
optimal observable as discussed in the test.  Overlaying this
distribution are Monte Carlo distributions for the two helicity
states, and for their sum including background, assuming a value for
$\pta$ equal to the fitted polarisation.}
\end{figure}

\begin{figure}[htbp]
\begin{center}
\includegraphics[width=0.7\linewidth]{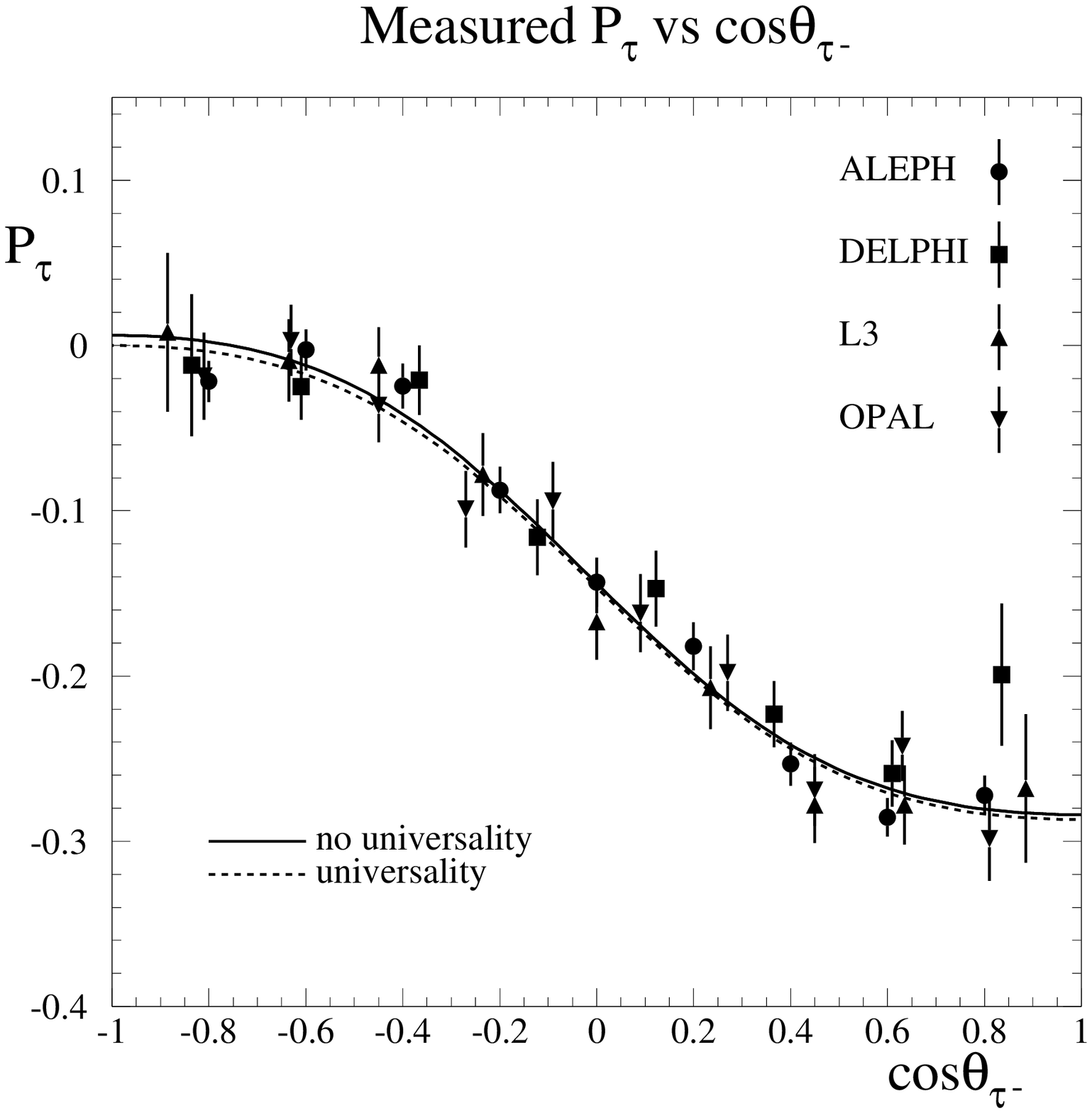}
\end{center}
\caption[Measured $\tau$ polarisation {\em vs.}
$\cos\theta_{\tau}$ for all LEP experiments] {\label{fig:taupol_7} The
  values of \ptau ~as a function of $\cos\theta_{\tau^-}$ as measured
  by each of the LEP experiments.  Only the statistical errors are
  shown.  The values are not corrected for radiation, interference or
  pure photon exchange. The solid curve overlays
  Equation~\ref{eq-ptcos} for the LEP values of $\cAt$ and $\cAe$. The
  dashed curve overlays Equation~\ref{eq-ptcos} under the assumption
  of lepton universality for the LEP value of $\cAl$.}
\end{figure}

Each LEP experiment analysed up to five of the most important $\tau$
decay modes with high branching
fractions~\cite{\ALEPHTAU,\DELPHITAU,\LTAU,\OPALTAU}: $\tpinu$,
$\trhonu$, $\tanu$, $\tenunu$ and $\tmununu$.  The tau polarisation
affects the distribution of decay angles in the rest frame of the
decaying tau lepton, which translates into a polarisation dependent
energy spectrum of the decay products in the laboratory or detector
system; for example for $\tpinu$ and $\tau\to\ell\nu_\ell\nu_\tau$
decays:
\begin{eqnarray}
\frac{1}{\Gamma}\frac{d\Gamma}{dx_\pi  } & = & 1+\ptau(2x_\pi-1) \\
\frac{1}{\Gamma}\frac{d\Gamma}{d x_\ell} & = & 
\frac{1}{3}\left[ (5 - 9 x_{\ell}^2 + 4 x_{\ell}^3) + 
           \ptau (1 - 9 x_{\ell}^2 + 8 x_{\ell}^3 )\right]\,,
\end{eqnarray}
where $x$ denotes the fractional energy $E/E_\tau$ of the visible
decay product.  Distributions of kinematic observables of the visible
$\tau$ decay products are hence used to measure the polarisation as a
function of $\cos\theta$.  For multi-hadronic $\tau$ decays through
intermediate mesons, such as $\rho$ or $a_1$, more complicated,
so-called optimal observables, $\omega$, are used, which are designed
for optimal statistical sensitivity to the tau
polarisation~\cite{Davier:1993nw}.  Example distributions are shown in
Figure~\ref{fig:taupol}.  The tau polarisation is extracted by fitting
a linear combination of positive and negative helicity contributions,
obtained from Monte-Carlo simulations~\cite{KORALZ,PHOTOS}, to the
observed data distribution. The measured values of $\ptau$ as a
function of $\cos\theta_{\tau^-}$ is shown in
Figure~\ref{fig:taupol_7}.

\begin{figure}[p]
\begin{center}
\vskip -1cm
\includegraphics[width=0.65\linewidth]{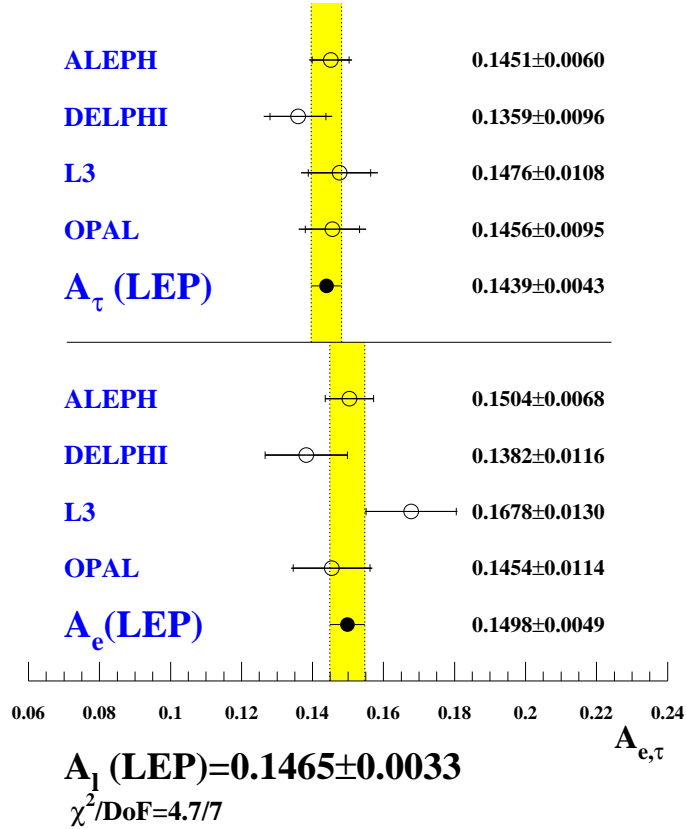}
\end{center}
\vskip -0.75cm
\caption[Comparison of $\tau$ polarisation measurements of $\cAt$ and
$\cAe$] {\label{fig:taupol_8} Measurements of $\cAt$ and $\cAe$ from
the four LEP experiments and the combined values. The error bars
indicate the total errors.  The magnitude of the statistical error
alone is indicated by the small tick marks on each error bar. }
\end{figure}

For the LEP combination~\cite{Z-Pole}, each LEP experiment provides a
result for $\cAe$ and $\cAt$, which are already averaged over the
experiment's analyses~\cite{\ALEPHTAU,\DELPHITAU,\LTAU,\OPALTAU}.
These results are reported in Table~\ref{tab-taupol} and compared in
Figure~\ref{fig:taupol_8}, showing that the results from the different
experiments agree well.

Systematic errors are summarised in Table~\ref{table-systematics};
they are generally much smaller for $\cAe$ than for $\cAt$, as $\cAe$
is equivalent to a forward-backward asymmetry of the $\cos\theta$
dependent tau polarisation, whereas $\cAt$ is given by the average tau
polarisation.  The systematic uncertainty labelled ZFITTER includes
decay-independent systematic uncertainties arising from corrections of
the experimental distributions to the pole quantities $\cAe$ and
$\cAt$, such as QED radiative corrections, energy dependence of the
tau polarisation, and mass effects leading to helicity flips. These
uncertainties are calculated with ZFITTER and are treated as fully
correlated between all decay channels and all experiments in the
combination.  The remaining systematic errors are dependent on the
specific tau decay mode analysed and are thus treated as uncorrelated
between different decay modes but as correlated between experiments.
These errors decribe the dependence on tau branching fractions taken
from References~\cite{PDG98,PDG2000}, affecting the purity of the data
sample, the contamination of the selected samples by background
processes, mainly arising from two-photon collision processes, and the
modelling of the specific decay in hadronic tau
decays~\cite{Decker,Finkemeier}.  In all cases, the systematic errors
are much smaller than the statistical errors.

The combined LEP results for $\cAe$ and $\cAt$ are:
\begin{eqnarray} 
  \cAt & = & 0.1439 \pm 0.0043 \label{eq:ptau:At}\\
  \cAe & = & 0.1498 \pm 0.0049 \label{eq:ptau:Ae}\,,
\end{eqnarray}
where the average has a $\chi^2$ of 3.9 for six degrees of freedom.
The correlation between the results is +0.012.  The uncertainties of
both averages are still dominated by statistics, in particular for
$\cAe$, being related to the forward-backward asymmetry of the tau
polarisation.  The results are in good agreement with each other, as
expected from neutral-current lepton universality.  Assuming $e-\tau$
universality, a combined value of:
\begin{eqnarray}
  \cAl & = & 0.1465 \pm 0.0033 \label{eq:ptau:Al} \,,
\end{eqnarray}
is obtained, where the total error contains a systematic error of
0.0015.  The $\chi^2$ of this average is 4.7 for seven degrees of
freedom. The above value of $\cAl$ corresponds to:
\begin{eqnarray}
\swsqeffl = 0.23159 \pm   0.00041 \,.
\end{eqnarray}

\begin{table}[p]
\renewcommand{\arraystretch}{1.15}
\begin{center}
\caption[LEP results for $\cAt$ and $\cAe$]{  
  LEP results for $\cAt$ and $\cAe$.  The first error is statistical
  and the second systematic.  }
\label{tab-taupol}
\begin{tabular}{|l||c|c|}
\hline
Experiment  & $\cAt$ & $\cAe$ \\
\hline
\hline
ALEPH   & $0.1451\pm0.0052\pm0.0029$  & $0.1504\pm0.0068\pm0.0008$  \\
DELPHI  & $0.1359\pm0.0079\pm0.0055$  & $0.1382\pm0.0116\pm0.0005$  \\
L3      & $0.1476\pm0.0088\pm0.0062$  & $0.1678\pm0.0127\pm0.0030$  \\
OPAL    & $0.1456\pm0.0076\pm0.0057$  & $0.1454\pm0.0108\pm0.0036$  \\
\hline
LEP     & $0.1439\pm0.0035\pm0.0026$  & $0.1498\pm0.0048\pm0.0009$  \\
\hline
\end{tabular}
\end{center}
\end{table}

\begin{table}[p]
\begin{center}
\renewcommand{\arraystretch}{1.2}
\caption[Common systematic errors in $\tau$ polarisation measurements]
{The magnitude of the major common systematic errors on $\cAt$ and
  $\cAe$ by category for each of the LEP experiments.}
\label{table-systematics}
\begin{tabular}{|l||cc|cc|cc|cc|} \hline
                      & \multicolumn{2}{|c|}{ALEPH} 
                      & \multicolumn{2}{|c|}{DELPHI} 
                      & \multicolumn{2}{|c|}{L3} 
                      & \multicolumn{2}{|c|}{OPAL} \\ 
                      & $\delta\cAt$ & $\delta\cAe$ 
                      & $\delta\cAt$ & $\delta\cAe$ 
                      & $\delta\cAt$ & $\delta\cAe$ 
                      & $\delta\cAt$ & $\delta\cAe$ \\ \hline \hline
ZFITTER  &0.0002&0.0002&0.0002&0.0002&0.0002&0.0002&0.0002&0.0002 \\
$\tau$ branching fractions &0.0003&0.0000&0.0016&0.0000&0.0007&0.0012&    0.0011    & 0.0003   \\
two-photon bg              &0.0000&0.0000&0.0005&0.0000&0.0007&0.0000&    0.0000    & 0.0000   \\
had. decay model           &0.0012&0.0008&0.0010&0.0000&0.0010&0.0001&    0.0025    & 0.0005   \\
\hline 
\end{tabular}
\end{center}
\end{table}

\clearpage

\subsection{Heavy Quark Flavours b and c}
\label{sec:Heavy}

\subsubsection{Introduction}

The Z lineshape analyses in the inclusive hadronic channel as
discussed in Section~\ref{sec:Lineshape} do not separate the different
quark flavours, hence do not allow the determination of individual
quark couplings.  In contrast, this section describes the complex
analyses required in order to isolate samples of specific quark
flavor, namely containing the heavy hadrons containing b and c quarks
arising in $\bb$ and $\cc$ production~\cite{Z-Pole}. These data
samples allow the determination of electroweak observables similar to
those discussed in the previous sections, namely:
\begin{eqnarray}
\Rqz     & = & \frac{\Gqq}{\Ghad}  \\
\Afbzq   & = & \frac{3}{4}\cAe\cAq \\
\AFBLRq  & = & \frac{3}{4}\cAq     \,,
\end{eqnarray}
for the quark flavours q = b, c. The quantities $\Rqz$ are measured by
SLD and at LEP, $\Afbzq$ is measured at LEP, and $\AFBLRq$ is measured
by SLD.

\subsubsection{Tagging Methods}

\begin{figure}[htb]
\begin{center} 
\includegraphics[width=0.475\linewidth]{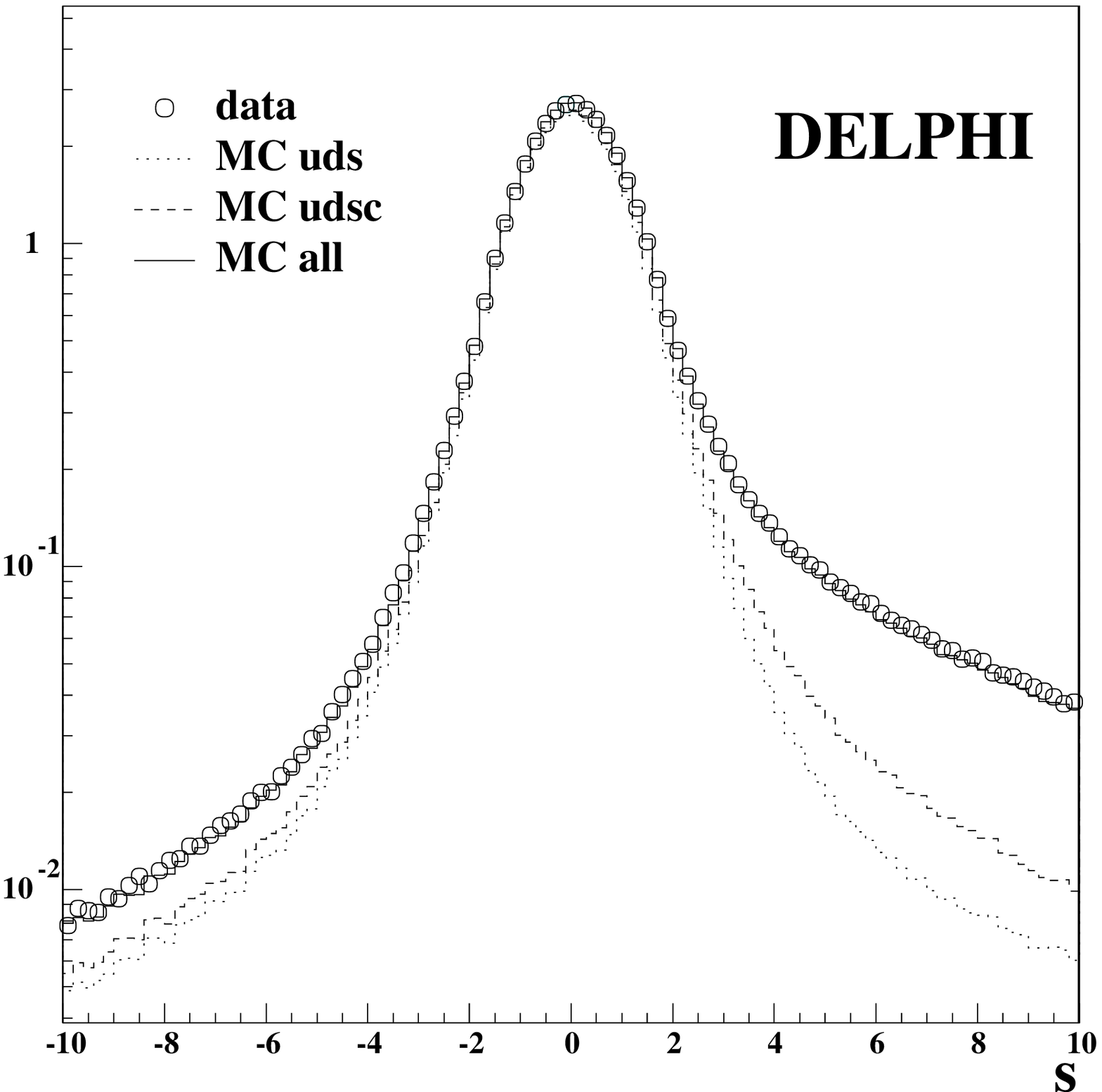}
\hfill
\includegraphics[width=0.515\linewidth,bb=0 20 514 514]{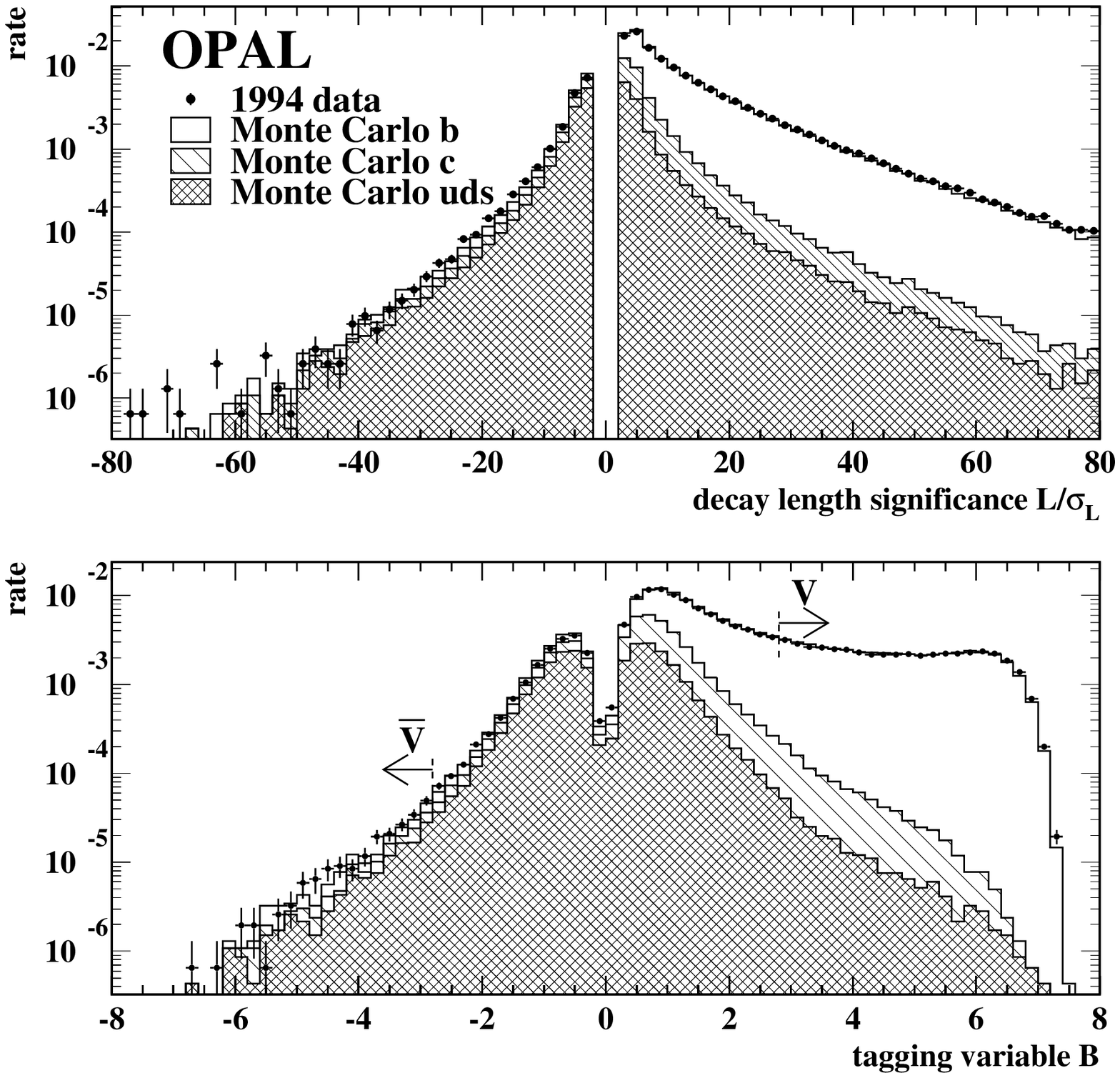}\\
\includegraphics[width=0.515\linewidth]{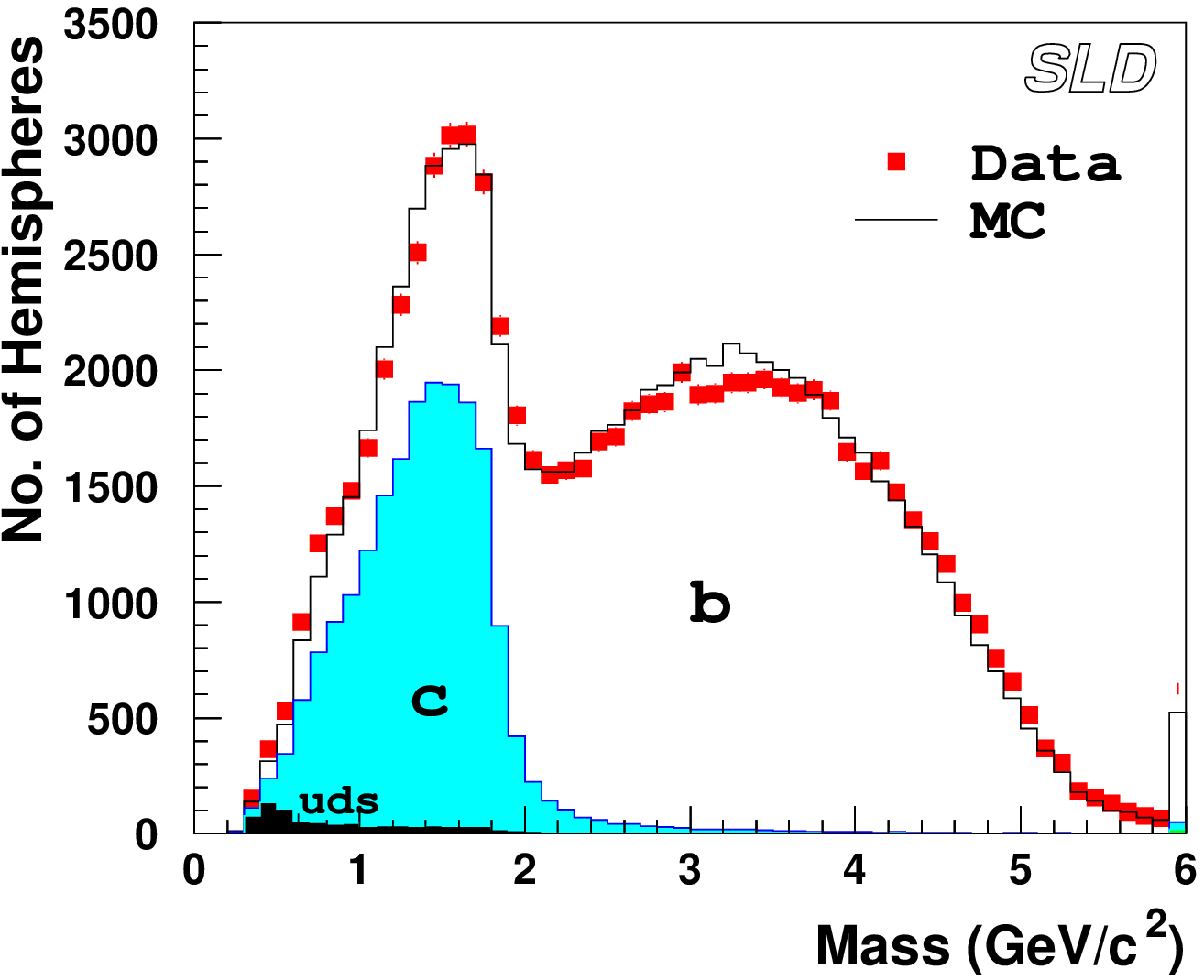}
\hfill
\includegraphics[width=0.475\linewidth,bb=20 175 535 655]{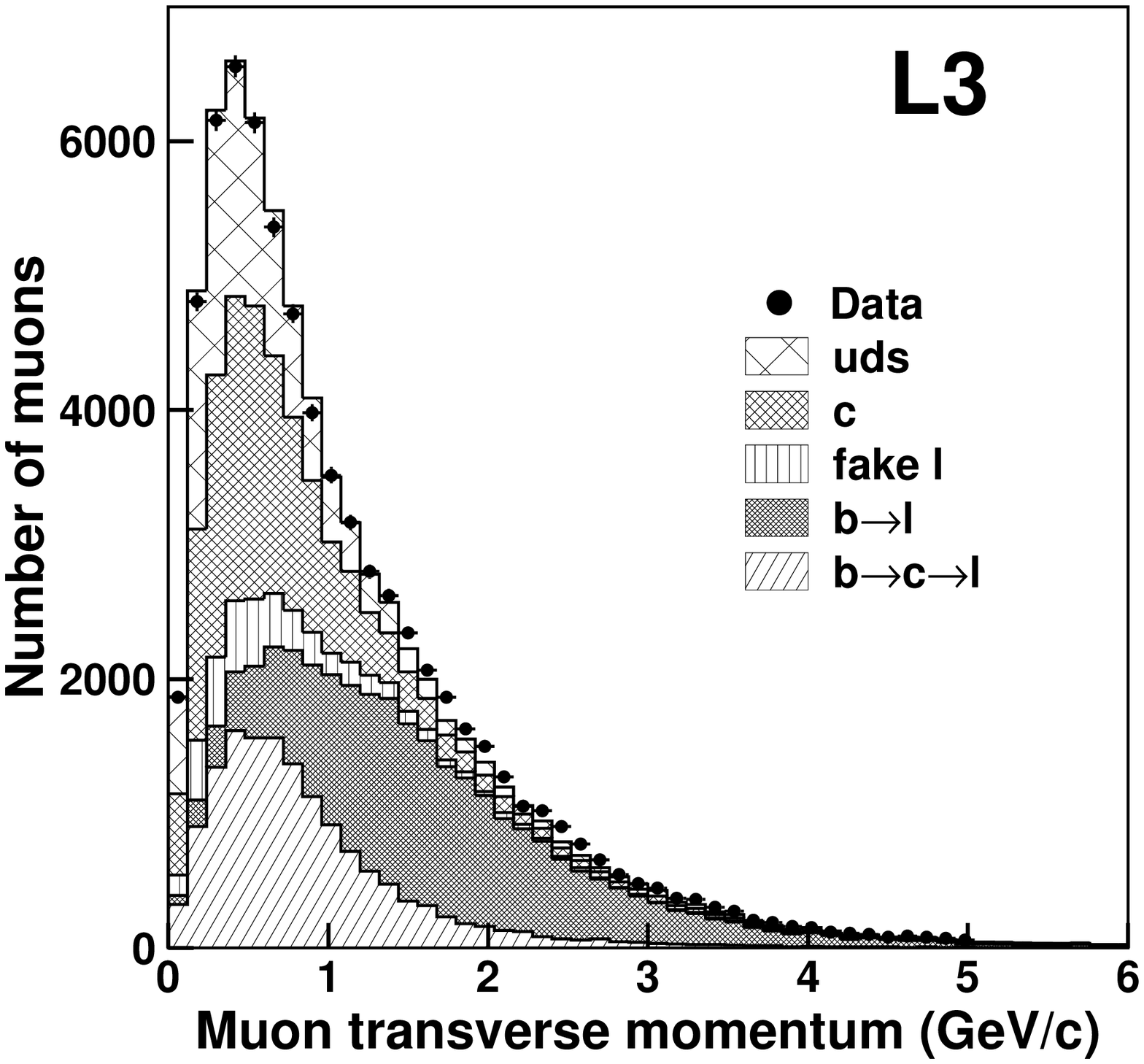}
\end{center}
\caption[Distributions of tagging variables]{ Distribution of tagging
variables. Top left: impact parameter significance; top right: decay
length significance and neural-net tagging output; bottom left: vertex
mass; bottom right: transverse momentum of inclusive muon arising from
the decay of b- and c-flavourd hadrons transverse to the containing
jet. Monte Carlo contributions are separated according to different
quark flavours. }
\label{fig:HF-tag} 
\end{figure}

Hadrons containing heavy quarks are identified experimentally by their
high mass, semileptonic decay modes, fully or partially reconstructed
decays, or long lifetime leading to secondary decay vertices and high
impact parameter tracks.  Distributions of several tagging
observables~\cite{ref:atag, ref:aimp, ref:drb, ref:dbtag,
ref:lrbmixed, ref:omixed, SLD_ZVTOP}, comparing heavy quarks samples
with light quark background, are shown in Figure~\ref{fig:HF-tag}.
The impact parameter is the distance of closest approach of a track to
the reconstructed interaction point or primary vertex, and is thus
large for tracks originating from the decay of long-lived hadrons.  In
case the secondary decay vertex is reconstructed, the siginificance of
its distance from the primary production vertex is again large for
long-lived hadrons containing heavy quarks.  In case the decay vertx
is measured, the mass of the decaying object is reconstructed using a
pseudo-mass technique, and used to check whether it is compatible with
a heavy objects.  Semileptonic decays of heavy hadrons produce leptons
with high total momentum as well as high momentum transverse to the
jet containing the decaying hadron.  As an example, the performace of
the combined tagging algorithms used in the $\Rb$ analyses is reported
in Table~\ref{tab:hq_iptag} for the five experiments.  The small beam
pipe and low repetition rate of SLC allows the SLD experiment to
employ detector systems reaching superior tagging performance.

\begin{table}[htbp]
\begin{center}
\renewcommand{\arraystretch}{1.1}
\caption[Tagging performance of the different experiments.]  {Tagging
performance of the different experiments as used for the $\Rb$
analyses.}
\label{tab:hq_iptag}
\begin{tabular} {| l || c | c  | c | c | c |} \hline
                  &~ALEPH~   &~DELPHI~ & ~~~L3~~~&~~OPAL~~&~~SLD~~\\
\hline
\hline
b Purity     [\%] &  97.8    & 98.6    &    84.3 & 96.7   & 98.3 \\
b Efficiency [\%] &  22.7    & 29.6    &    23.7 & 25.5   & 61.8 \\
\hline
\end{tabular}
\end{center}
\end{table}

\subsubsection{Partial Width Measurements}

Each tagging method, applied to a hadronic jet or event hemisphere,
identifies the underlying quark flavor with a certain efficiency,
tuned to be high for b and c quarks but low for uds quarks.  Since
there are two jets or hemispheres in each event, the fraction of
events with a single tag and a double tag are expressed as:
\begin{eqnarray}
f_s & = & \epsb \Rb + \epsc \Rc + \epsl(1-\Rb-\Rc) \label{eq:rb}\\
f_d & = & \epsb^{(d)} \Rb + \epsc^{(d)} \Rc + \epsl^{(d)} (1-\Rb-\Rc) \,,
\nonumber
\end{eqnarray}
where $\epsf$ is the tagging efficiency for quark flavour f.  The
double-tagging efficiency $\epsf^{(d)} $ is written as
\begin{equation}
\epsf^{(d)} = (1+\calCf) \epsf^2 
\label{eq:hq_hcor}
\end{equation}
where the correction factor $\calCf \not= 0$ accounts for the the two
hemispheres in an event being somewhat correlated.  Neglecting
correlations and backgrounds, one finds $\Rb=f_s^2/f_d$ independent of
$\epsb$, ie, a robust measurement.  However, the neglected terms are
not small and must be determined from Monte Carlo simulations, leading
to model dependencies and thus correlated systematic uncertainties
affecting the measurements.  In multi-tag methods, several different
tagging algorithms are used simultaneously, leading to a system of
linear equations similar to the above which is overconstrained and
allows the extraction of rates as well as several efficiencies from
the data, resulting in reduced model dependence. Further detailes on
the various measurements of the five experiments are given
in~\cite{ref:alife, ref:drb, ref:lrbmixed, ref:omixed, Abe:2005nq} for
$\Rb$ and in~\cite{ref:arcd, ref:arcc, ref:arcd, ref:drcc, ref:drcd,
ref:drcc, ref:orcc, ref:orcd, Abe:2005nq} for $\Rc$.

\subsubsection{Asymmetry Measurements}

The asymmetry measurements are performed by analysing the event rate
as a function of the polar angle of the event thrust axis.  This in
turn requires the assignment of a charge to a quark jet or event
hemisphere.  The measured asymmetry is given by:
\begin{equation}
A_{\rm FB}^{\rm meas} ~=~ 
\sum_{\rm q} (2 \omega_{\rm q} -1) \eta_{\rm q} \Aqq\,,
\label{eq:afbsum}
\end{equation}
where $\eta_{\rm q}$ is the fraction of $\qq$ events in the sample,
$\omega_{\rm q}$ is the probability to tag the quark charge correctly,
and the sum is taken over all quark flavours.

Various charge tagging methods are employed by the five experiments.
For heavy quarks decaying semileptonically, the charge of the lepton
is correlated with the charge of the decaying heavy quark. Independent
of decay mode, the jet or vertex charge is used, constructed as the
momentum weighted sum of charges of tracks belonging to the jet or the
identified secondary decay vertex:
\begin{equation}
  Q_h ~ = ~ \frac{ \sum_i q_i p_{\| i}^\kappa }{ \sum_i p_{\| i}^\kappa }\,,
\label{eq:jetch}
\end{equation}
and where $p_\|$ is the momentum of the track along the thrust axis,
and $\kappa$ is a tunable parameter with typical values between 0.3
and 1.

\begin{figure}[htb]
\begin{center} 
\includegraphics[width=0.67\linewidth,bb=30 12 515 345]{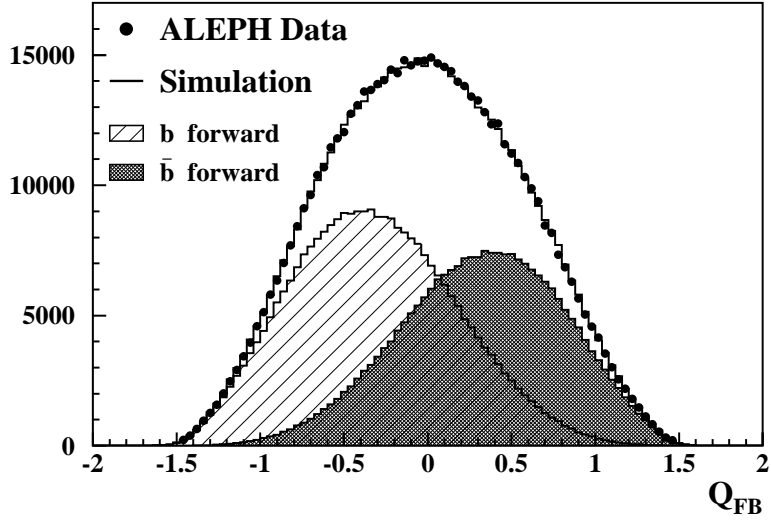}
\end{center}
\caption[Charge separation of the ALEPH neural net tag] {Charge
separation of the ALEPH neural net tag using jet charge, vertex charge
and charged kaons~\cite{ref:ajet}.  The asymmetry reflects $\Abb$
diluted by the non-perfect charge tagging.  }
\label{fig:hq_vtxnn}
\end{figure}

A precise measurement of an asymmetry requires good charge separation
(for example, semi-leptonic decays, which however are limited by their
branching fractions) and/or high statistics (jet/vertex charge which,
however, has a worse charge separation).  As an example, the charge
separation $\delta_{\rm q}$ for $\bb$ events achieved by ALEPH is
shown in Figure~\ref{fig:hq_vtxnn}, where:
\begin{eqnarray}
  \avQfb & = & \langle Q_{{\rm F}} - Q_{{\rm B}} \rangle \label{eq:hf_qfb}
         ~ = ~ \delta_{\rm q} \Aqq \nonumber \\
  \delta_{\rm q} & = & \langle Q_{\rm q} - Q_{\bar{{\rm q}}} \rangle\,, 
  \nonumber
\end{eqnarray}
which is largely measured from data, for example:
\begin{equation}
\left( \frac{\delta_{\rm q}}{2} \right)^2 ~ = ~
\frac {\langle Q_{\rm{F}} \cdot Q_{\rm{B}} \rangle + 
\rho_{\qq} \sigma(Q)^2 + \mu(Q)^2}{1+\rho_{\qq}},
\label{eq:hf:deltaq}
\end{equation}
where $\mu(Q)$ is the mean value of $Q$ for all hemispheres and
$\sigma(Q)$ is its variance.  Note that $\mu(Q)$ is slightly positive
due to an excess of positive hadrons in secondary hadronic
interactions.  Only the hemisphere correlations, $\rho_{\qq}$, which
arise from charge conservation, hard gluon radiation and other small
effects, must be taken from simulation.

\begin{figure}[htbp]
\begin{center} 
\includegraphics[width=\linewidth,bb=8 47 555 285]{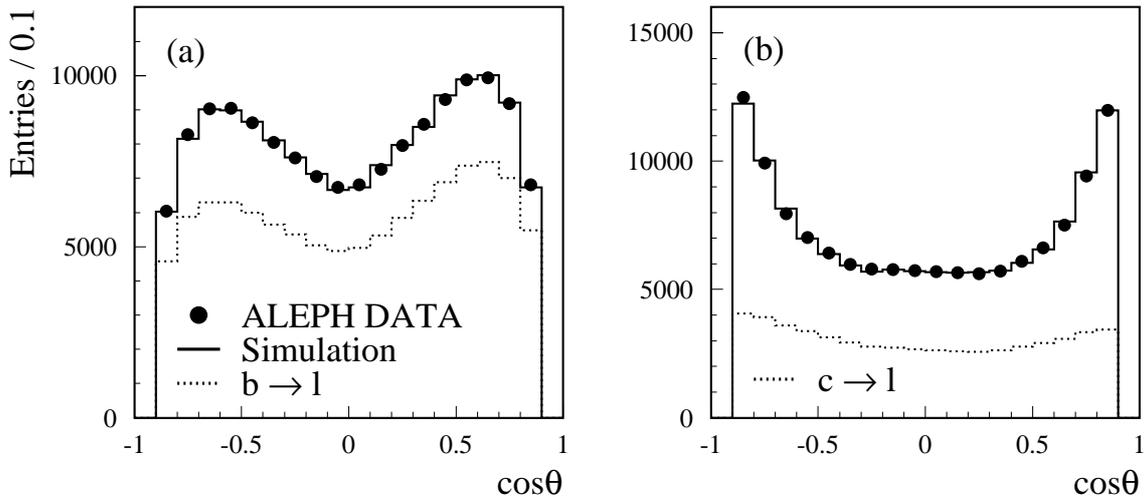}
\end{center}
\caption[$\cos \theta$ distribution from the ALEPH b-asymmetry
measurement with leptons] {Reconstructed $\cos \theta$ distribution
from the ALEPH asymmetry measurements with leptons for a) the
b-enriched and b) the c-enriched sample~\cite{\alasy}.  The full
histogram shows the expected raw angular distribution in the
simulation. The dashed histogram show the signal component.}
\label{fig:hq_asyplot}
\end{figure}

\begin{figure}[htbp]
\begin{center} 
\includegraphics[width=\linewidth]{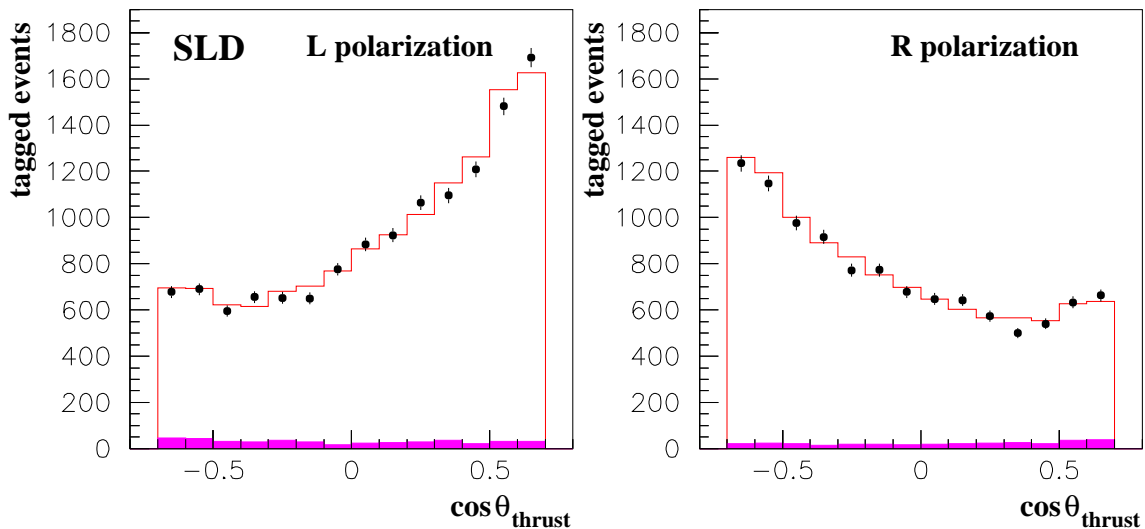}
\end{center}
\caption[Reconstructed $\cos\theta$ distributions from the SLD vertex
charge $\cAb$ analysis.]  {%
Reconstructed $\cos\theta$ distributions
from the SLD vertex charge $\cAb$ analysis for events with left-handed
and right-handed electron beam polarisations. The shaded region
corresponds to udsc background in the sample estimated from Monte
Carlo.  }
\label{fig:hq_asyplot_pol}
\end{figure}

The resulting differential distributions in polar angle are shown in
Figures~\ref{fig:hq_asyplot} and~\ref{fig:hq_asyplot_pol} for the
forward-backward and left-right forward-backward asymmetry in $\bb$
production, respectively.  These distributions, evaluated in terms of
the expected asymmetries between bins at $\pm\cos\theta$:
\begin{eqnarray}
  \Aqq(\cos \theta) &=& \frac{8}{3} \Aqq \frac{\cos \theta}
  {1 + \cos^2 \theta} 
  \label{eq:hq_difasy} \\
  \Aqqlr(\cos \theta) &=& |\Pe|\cAq \frac{2\cos \theta}{1 + \cos^2 \theta}\,,
\label{eq:hq_asylr_difasy}
\end{eqnarray}
allow to determine $\Afbzq$ and $\cAq$, respectively. The analyses of
the four LEP experiments are described in detail in~\cite{\alasy,
ref:ajet, \dlasy, ref:ddasy, ref:dnnasy, \llasy, ref:ljet, ref:ojet,
ref:olasy, ref:odsac} for b quarks, and in~\cite{\alasy, ref:adsac,
\dlasy, ref:ddasy, \llasy, ref:olasy, ref:odsac} for c quarks, while
the SLD results for the polarised asymmetries are decscirbed
in~\cite{\SLDacbl, \SLDabj, \SLDabk, ref:SLD_vtxasy} and
in~\cite{\SLDacbl, ref:SLD_ACD, ref:SLD_vtxasy}, respectively.

\subsubsection{Corrections and Systematic Uncertainties}

The various flavour and charge tagging analyses neccessarily depend on
assumptions about the fragmentation of heavy quarks and decays of
hadrons containg these quarks. This leads to non-trivial correlations
between the results of the LEP and SLD experiments, which have to be
taken into account in a combination, as well as to a dependence on
additional model parameters. In order to derive consistent and correct
averages in a fit to the experimental results, the five experiments
have agreed on a common set of values for external parameters and
their associated uncertainties. The associated uncertainties are
propagated to all results using parametrised dependencies.

Additional observables besides the six electroweak observables are
included in the combination procedure as well in case they are also
measured by the five experiments and correlated to the main analyses.
These additional fit parameters are:
\begin{itemize}
\item the probability that a c-quark fragments into a $\Dstarp$ that
  decays into $\Dzero \pi^+$: $P\mathrm{( c \rightarrow D^{*+})}
  \times \BR\mathrm{( D^{*+} \rightarrow \pi^+ D^0 )}$, denoted
  $\PcDst$ in the following,
\item the fraction of charm hemispheres fragmenting into a specified
  weakly-decaying charmed hadron: $\fDp$, $\fDs$, $\fcb$,\footnote{ 
  The quantity $\fDz$ is calculated from the constraint $\fDz + \fDp +
  \fDs + \fcb = 1.$ }
\item the prompt and cascade semileptonic branching fraction of the
  b-hadrons $\Brbl$\footnote{ Unless otherwise stated, charge
  conjugate modes are always included.} and $\Brbclp$ and the prompt
  semileptonic branching fraction of the c-hadrons $\Brcl$,
\item the $\BB$ effective mixing parameter $\chiM$, which is the
  probability that a semileptonically decaying b-quark has been
  produced as an anti-b-quark.
\end{itemize}

Small corrections have to be applied to the raw experimental results.
The $\Rb$ and $\Rc$ analyses measure ratios of production rates,
$R_{\rm{q}} = \sigma_{\rm{q\bar{q}}}/\sigma_{\rm{had}}$.  To obtain
the ratios of partial widths $R_{\rm{q}}^0 =
\Gamma_{\rm{q\bar{q}}}/\Gamma_{\rm{had}}$, corrections for photon
exchange and $\gammaZ$ interference must be applied. These corrections
are small, typically $+0.0002$ for \Rb{} and $-0.0002$ for \Rc{}
depending slightly on the invariant mass cutoff of the
$\rm{q\bar{q}}$-system imposed in the analysis.  Corrections of
similar origin have to be applied to the asymmetries, as even the data
denoted ``peak'', taken at the maximum annihilation cross section, is
at a $\sqrt{s}$ value slightly different from $\MZ$ while the
asymmetries are strongly varying as a function of energy.  These
corrections on the asymmetries are summarised in
Table~\ref{tab:aqqcor}.

\begin{table}[h]
\begin{center}
\renewcommand{\arraystretch}{1.1}
\caption[Corrections to be applied to the quark asymmetries.]
{Corrections to be applied to the quark asymmetries as $\Afbzq =
A_{\mathrm{FB}}^{\qq}({\rm pk}) + \delta A_{\mathrm{FB}}$.  The row
labelled ``other'' denotes corrections due to $\gamma$ exchange,
$\gammaZ$ interference, quark-mass effects and imaginary parts of the
couplings.  The uncertainties of the corrections are negligible. }
\label{tab:aqqcor}
\begin{tabular}{|l||l|l|}
\hline
Source   & $\delta \Abb$
         & $\delta \Acc$ \\
\hline
\hline
$\sqrt{s} = \MZ $       & $ -0.0014 $  & $ -0.0035$  \\
QED corrections         & $ +0.0039 $  & $ +0.0107$  \\
other                   & $ -0.0006 $  & $ -0.0008$  \\
\hline
Total                   & $ +0.0019 $  & $ +0.0064$  \\
\hline
\end{tabular}
\end{center}
\end{table}

However, for the asymmetries, the largest corrections are those for
QCD effects, which are, in contrast to the case of partial widths, by
construction not absorbed in the definition of the electroweak
asymmetry parameters. The dominant QCD corrections arise from gluon
radiation and depend on whether the jet axis or thrust axis is used to
determine the polar angle in the analysis. The QCD corrections on the
asymmetries are written as follows~\cite{ourpap}:
\begin{eqnarray}
  \left(\Aqq \right)_{\rm meas} & = &
(1-C_{\mathrm{QCD}}) \left(\Aqq \right)_{\rm no \, QCD} \\
 & = &
  \left( 1 - \frac{\alpha_s(\MZ^2)}{\pi} c_1 -   
    \left( \frac{\alpha_s(\MZ^2)}{\pi} \right)^2 c_2 \right)
  \left(\Aqq \right)_{\rm no \, QCD}  \,. \nonumber 
\end{eqnarray}
The first-order corrections are known including mass
effects~\cite{lampe}.  Taking the thrust axis as the direction and
using the pole mass, they are $c_1=0.77$ for $\Abb$ and $c_1=0.86$ for
$\Acc$. The second-order corrections have been recalculated
in~\cite{neerv} and~\cite{sey}; the second order coefficients used are
$c_2 = 5.93$ for $\Abb$ and $c_2 = 8.5$ for $\Acc$. The final QCD
correction coefficients, including further corrections due to
fragmentation effects and using the thrust axis as reference direction
($C_{\mathrm{QCD}}^{\mathrm{had,T}}$), are
$C_{\mathrm{QCD}}^{\mathrm{had,T}} = 0.0354 \pm 0.0063$ for $\Abb$ and
$C_{\mathrm{QCD}}^{\mathrm{had,T}} = 0.0413 \pm 0.0063$ for $\Acc$.
The breakdown of the errors relating to the QCD corrections is given
in Table~\ref{tab:qcderr}.

\begin{table}[htbp]
\begin{center}
\renewcommand{\arraystretch}{1.2}
\caption{Error sources for the QCD corrections to the forward-backward
  asymmetries.}
\label{tab:qcderr}
\begin{tabular}{|lc||l|l|}
\hline
\multicolumn{2}{|l||}{Error on $C_{\mathrm{QCD}}^{\mathrm{had,T}}$
 \rule{0pt}{12pt} }
& \multicolumn{1}{c|}{$\bb$} & \multicolumn{1}{c|}{$\cc$} \\ 
\hline
\hline
Higher orders              &\cite{sey}    & 0.0025& 0.0046\\
Mass effects               &\cite{ourpap} & 0.0015& 0.0008\\
Higher order mass          &\cite{sey}    & 0.005 & 0.002\\
$\alpha_s = 0.119\pm 0.003$&              & 0.0012& 0.0015\\
Hadronisation              &\cite{ourpap} & 0.0023& 0.0035\\
\hline Total               &              & 0.0063& 0.0063\\
\hline
\end{tabular}
\end{center}
\end{table}

The dominant sources of systematic errors in the determination of the
six electroweak observables are reported in Table~\ref{tab:hferrbk}.
For the two partial width observables, the systematic errors are of
similar size to the statistical errors. In contrast, the four
asymmetry measurements are by far statistically dominated.

\begin{table}[tbp]
\begin{center}
\renewcommand{\arraystretch}{1.1}
\caption[Dominant error sources for the heavy-flavour electroweak
parameters]{ The dominant error sources for the heavy-flavour
electroweak parameters from the 14-parameter fit, see text for
details. The total systematic error is separated into a contribution
uncorrelated between the measurements, and in several components which
are correlated between measurements, as resulting from the combination
procedure. }
\label{tab:hferrbk}
\begin{tabular}{|c||c|c|c|c|c|c|}
\hline
Source
&\makebox[1.2cm]{\Rbz}
&\makebox[1.2cm]{\Rcz}
&\makebox[1.2cm]{$\Afbzb$}
&\makebox[1.2cm]{$\Afbzc$}
&\makebox[0.9cm]{\cAb}
&\makebox[0.9cm]{\cAc}\\
 & $[10^{-3}]$ & $[10^{-3}]$ & $[10^{-3}]$ & $[10^{-3}]$ 
 & $[10^{-2}]$ & $[10^{-2}]$ \\
\hline
\hline
statistics & 
$0.44$ & $2.4$ & $1.5$ & $3.0$ & $1.5$ & $2.2$ \\
\hline
internal systematics &
$0.28$ & $1.2$ & $0.6$ & $1.4$ & $1.2$ & $1.5$ \\
\hline
QCD effects &
$0.18$ & $0  $ & $0.4$ & $0.1$ & $0.3$ & $0.2$ \\
$\BR$(D $\rightarrow$ neut.)&
$0.14$ & $0.3$ & $0$   & $0$ & $0$ & $0$ \\
D decay multiplicity &
$0.13$ & $0.6$ & $0  $ & $0.2$ & $0$ & $0$ \\
B decay multiplicity &
$0.11$ & $0.1$ & $0  $ & $0.2$ & $0$ & $0  $ \\
$\BR$(D$^+ \rightarrow$ K$^- \pi^+ \pi^+) $&
$0.09$ & $0.2$ & $0  $ & $0.1$ & $0$ & $0  $ \\
$\BR$($\Ds \rightarrow \phi \pi^+) $&
$0.02$ & $0.5$ & $0  $ & $0.1$ & $0$ & $0$ \\
$\BR$($\Lambda_{\mathrm{c}} \rightarrow $p K$^- \pi^+) $&
$0.05$ & $0.5$ & $0  $ & $0.1$ & $0$ & $0  $ \\
D lifetimes&
$0.07$ & $0.6$ & $0  $ & $0.2$ & $0$ & $0$ \\
B decays&
$0$ & $0$ & $0.1$ & $0.4$ & $0$ & $0.1$  \\
decay models&
$0$ & $0.1$ & $0.1$ & $0.5$ & $0.1$ & $0.1$ \\
non incl. mixing&
$0$ & $0.1$ & $0.1$ & $0.4$ & $0$ & $0$ \\
gluon splitting &
$0.23$ & $0.9$ &$0.1$& $0.2$ & $0.1$ & $0.1$ \\
c fragmentation &
$0.11$ & $0.3$ & $0.1$ & $0.1$ & $0.1$ & $0.1$ \\
light quarks&
$0.07$ & $0.1$ & $0  $ & $0  $ & $0$ & $0  $ \\
beam polarisation&
$0$ & $0$ & $0$ & $0$ & $0.5$ & $0.3$ \\
\hline
total correlated&
$0.42$ & $1.5$ & $0.4$ & $0.9$ & $0.6$ & $0.4$ \\
\hline
total error&
$0.66$ & $3.0$ & $1.6$ & $3.5$ & $2.0$ & $2.7$ \\
\hline
\end{tabular}
\end{center}
\end{table}

\subsubsection{Combined Results}

\begin{figure}[htbp]
\begin{center}
 \includegraphics[width=0.495\linewidth,bb=10 20 483 455]{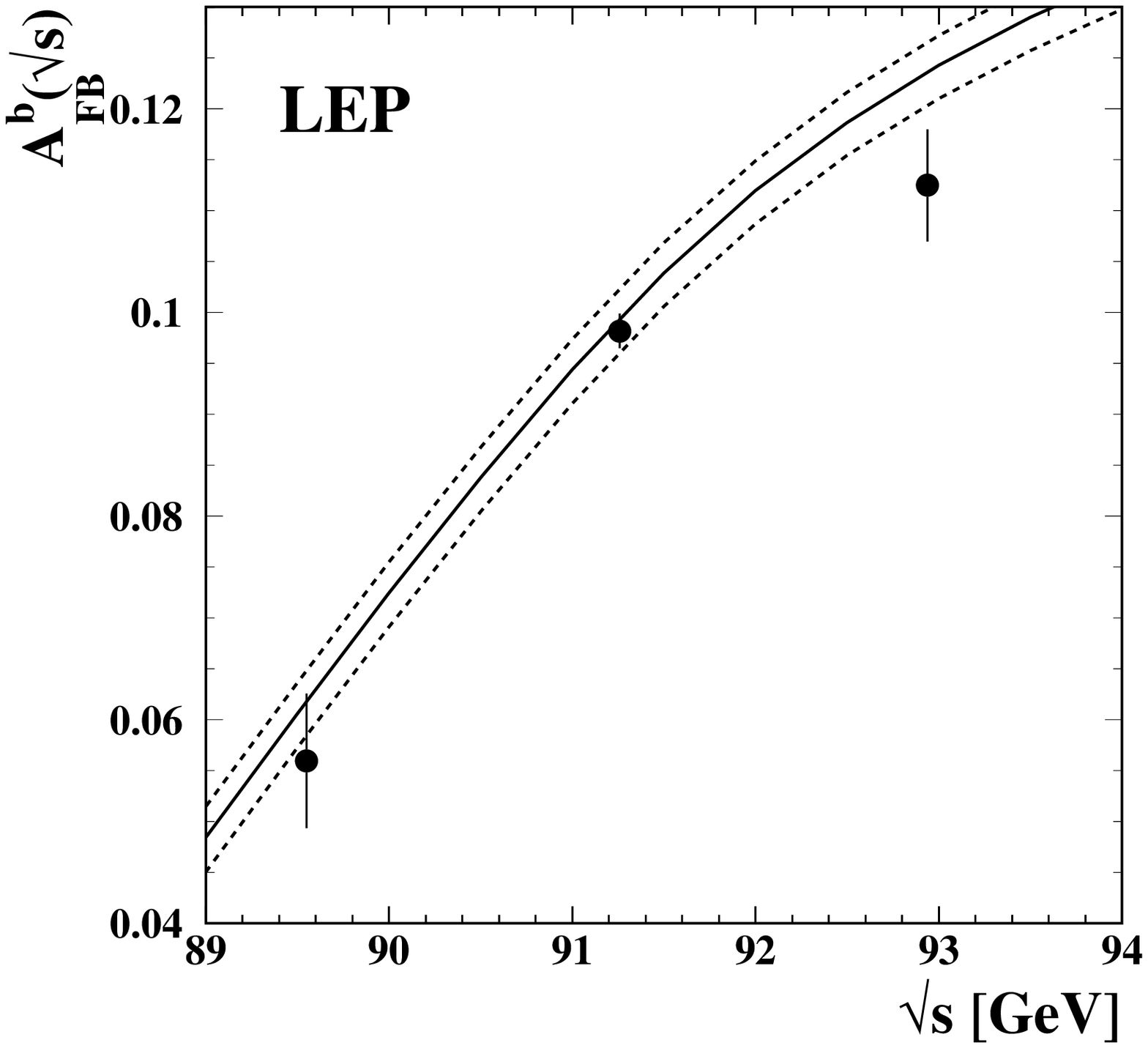}
\hfill
 \includegraphics[width=0.495\linewidth,bb=10 20 483 455]{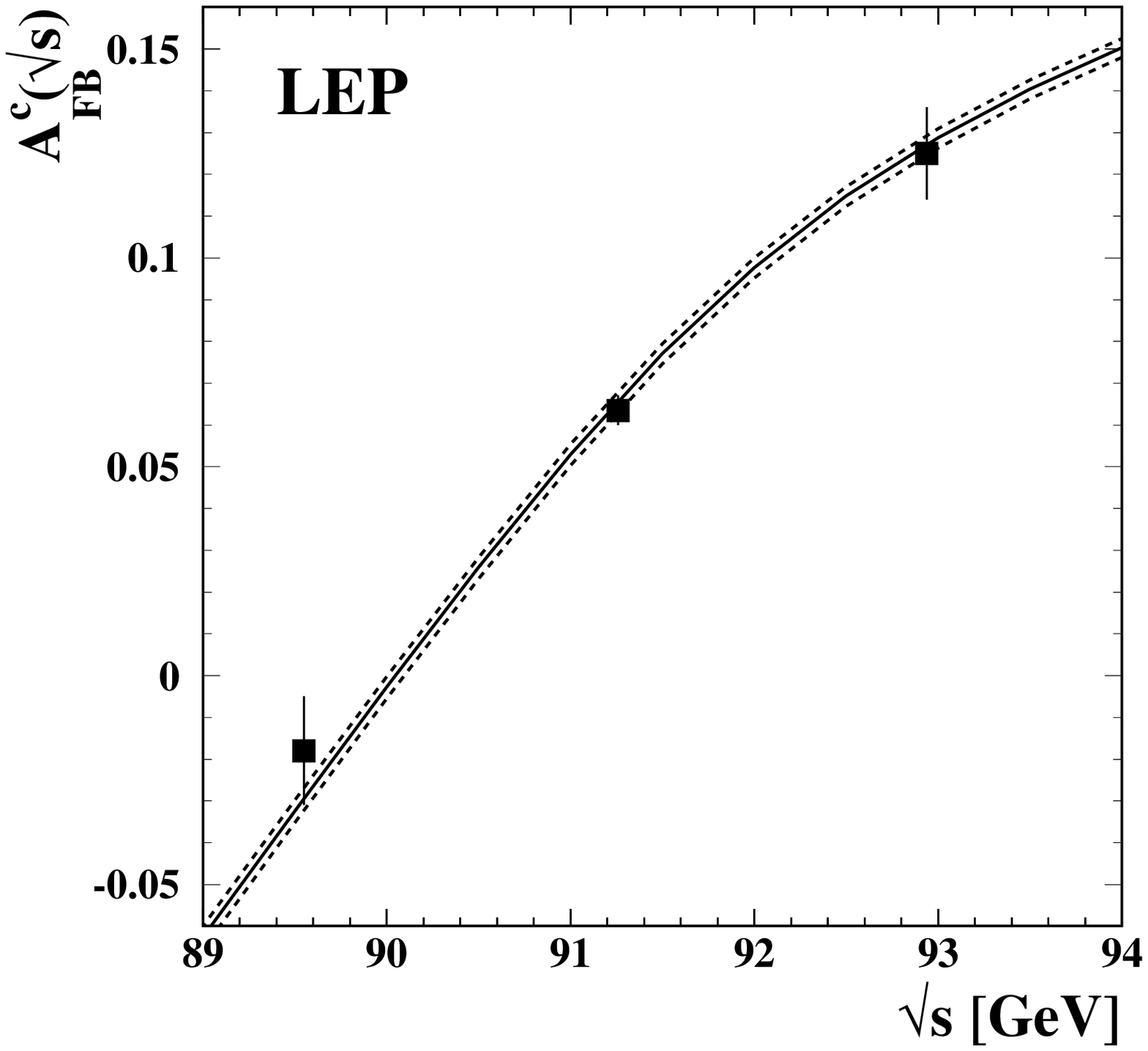}
\end{center}
\caption[Energy dependence of $\Afb^b$ and $\Afb^c$. ]{Dependence of
 of $\Afb^b$ and $\Afb^c$ on the centre-of-mass energy $\sqrt{s}$.
 The solid line represents the $\SM$ prediction for $\Mt=178 \,
 \GeV,\, \MH = 300 \,\GeV$, the upper (lower) dashed line is the
 prediction for $\MH = 100 \, (1000) \, \GeV$. The measurements are
 shown for the three main energies. }
\label{fig:hfafbvsene} 
\end{figure}

The heavy flavour results are combined using a $\chi^2$
minimisation~\cite{ref:lephf}.  In a first analysis to the set of
experimental results, the forward-backward asymmetry measurements at
LEP are corrected to three common centre-of-mass energy points,
corresponding to the ``peak'' of the Z resonance curve, and
``peak$\pm2~\GeV$''~\cite{Z-Pole}.  The asymmetry measurements are
compared to the expectation as a function of $\sqrt{s}$ in
Figure~\ref{fig:hfafbvsene}.  Good agreement with the expected
$\sqrt{s}$ dependence is observed.

In a second analysis, the off-peak forward-backward asymmetry
measurements are all transported to the ``peak'' centre-of-mass
energy~\cite{Z-Pole}. The peak asymmetries in $\bb$ and $\cc$
production are then corrected~\cite{hfasycor} to the corresponding
pole quantity $\Afbzq$ using ZFITTER, recall Table~\ref{tab:aqqcor}.
The results of the corresponding 14-parameter fit are reported in
Tables~\ref{tab:14parres} and~\ref{tab:14parcor}.

\begin{figure}[p]
\begin{center}
 \includegraphics[width=0.495\linewidth,bb=30 190 495 505]{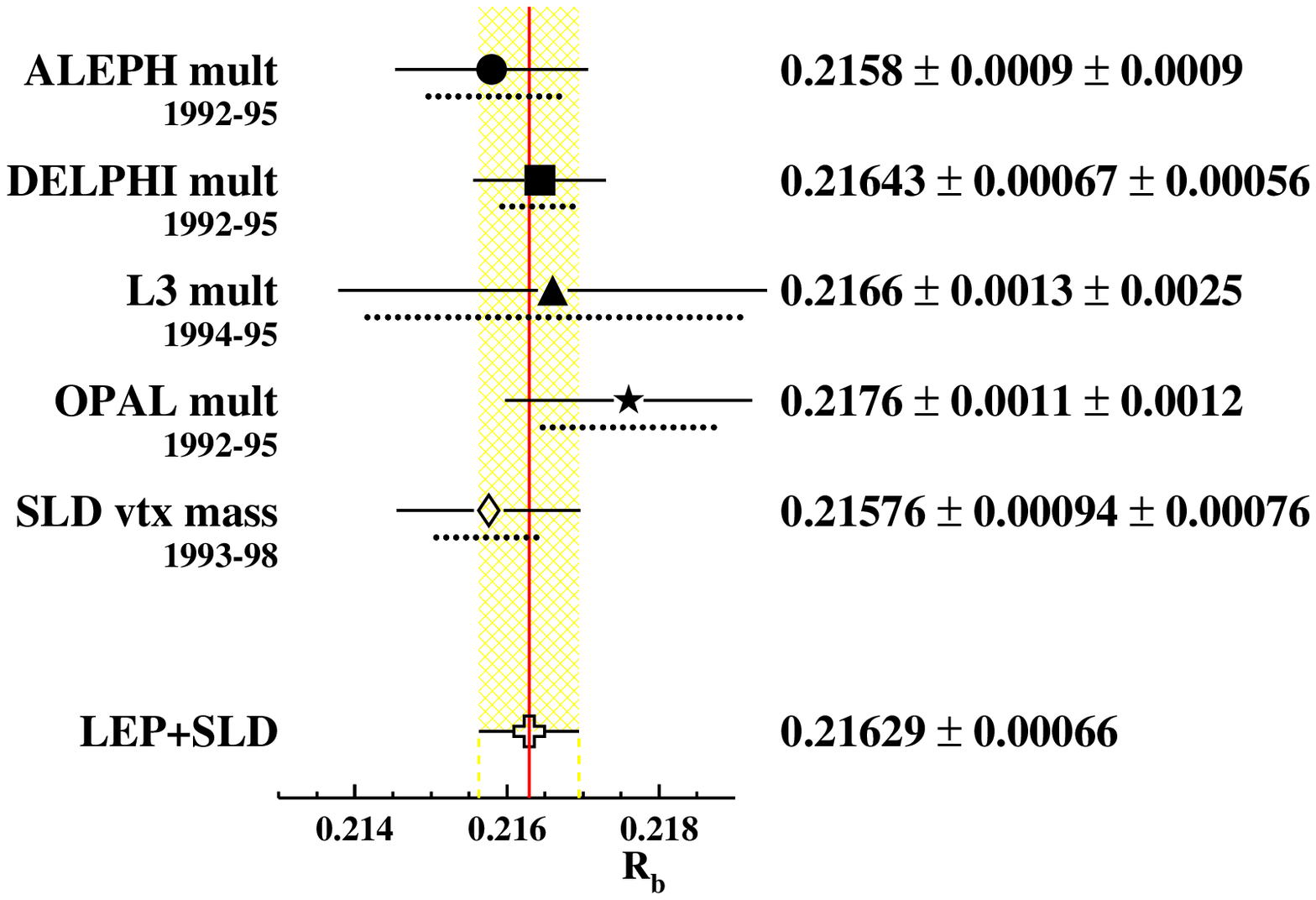}
 \includegraphics[width=0.495\linewidth,bb=10  80 475 480]{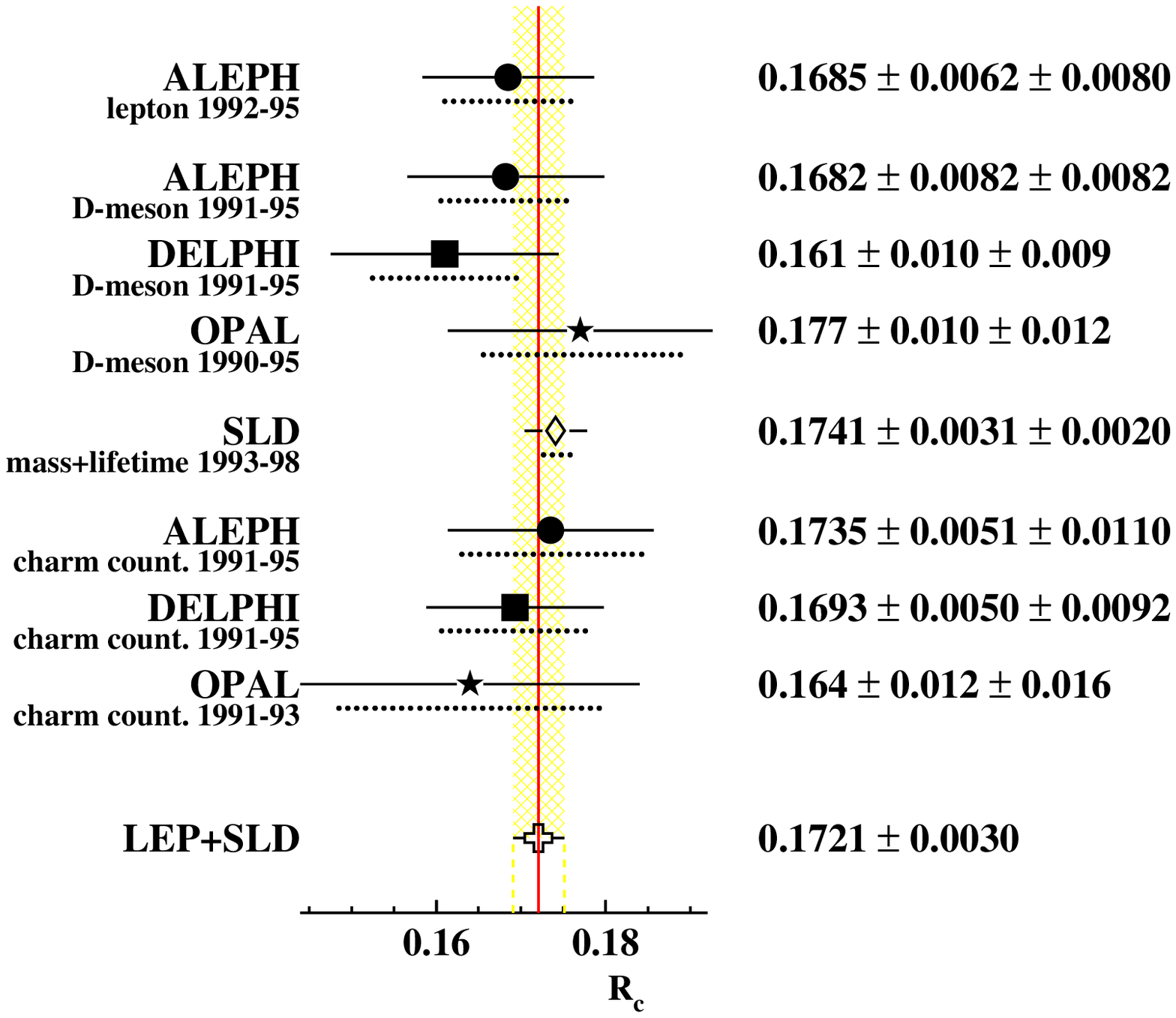}
\end{center}
\caption[$\Rbz$ and $\Rcz$ measurements used in the heavy flavour
  combination.]{$\Rbz$ and $\Rcz$ measurements used in the heavy
  flavour combination, corrected for their dependence on parameters
  evaluated in the multi-parameter fit described in the text.  The
  dotted lines indicate the size of the systematic error.  }
\label{fig:hqbar_rbc}
\end{figure}

\begin{figure}[p]
\begin{center}
 \includegraphics[width=0.495\linewidth,bb=24 76  472 485]{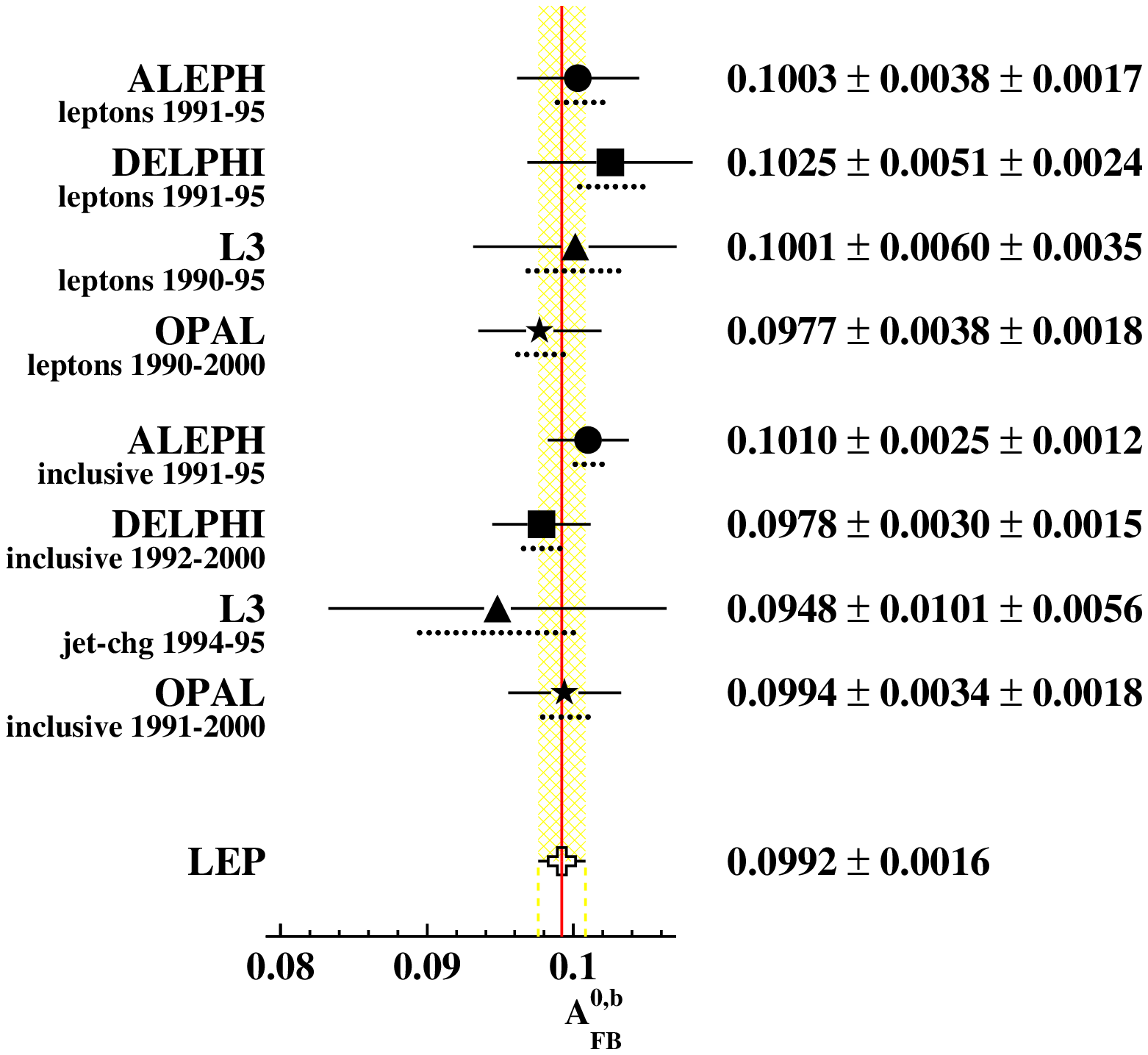}
 \includegraphics[width=0.495\linewidth,bb=24 76  472 485]{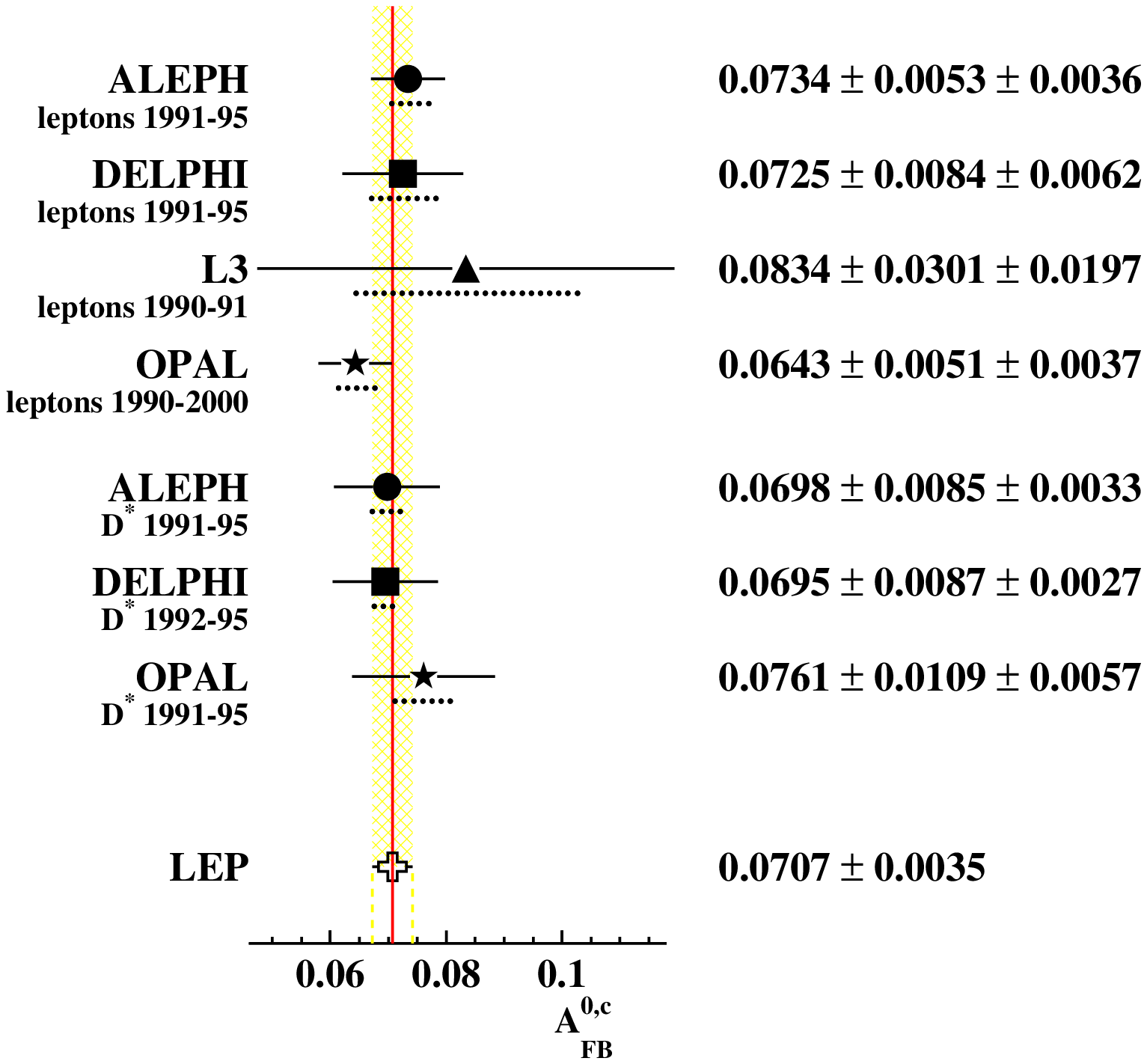}
\end{center}
\caption[$\Afbzb$ and $\Afbzc$ measurements used in the heavy flavour
  combination.]{$\Afbzb$ and $\Afbzc$ measurements used in the heavy
  flavour combination, corrected for their dependence on parameters
  evaluated in the multi-parameter fit described in the text. The
  $\Afbzb$ measurements with D-mesons do not contribute significantly
  to the average and are not shown in the plots.  The experimental
  results combine the different centre of mass energies. The dotted
  lines indicate the size of the systematic error. }
\label{fig:hqbar_asy}
\end{figure}

\begin{figure}[p]
\begin{center}
 \includegraphics[width=0.495\linewidth,bb=33 60 511 313]{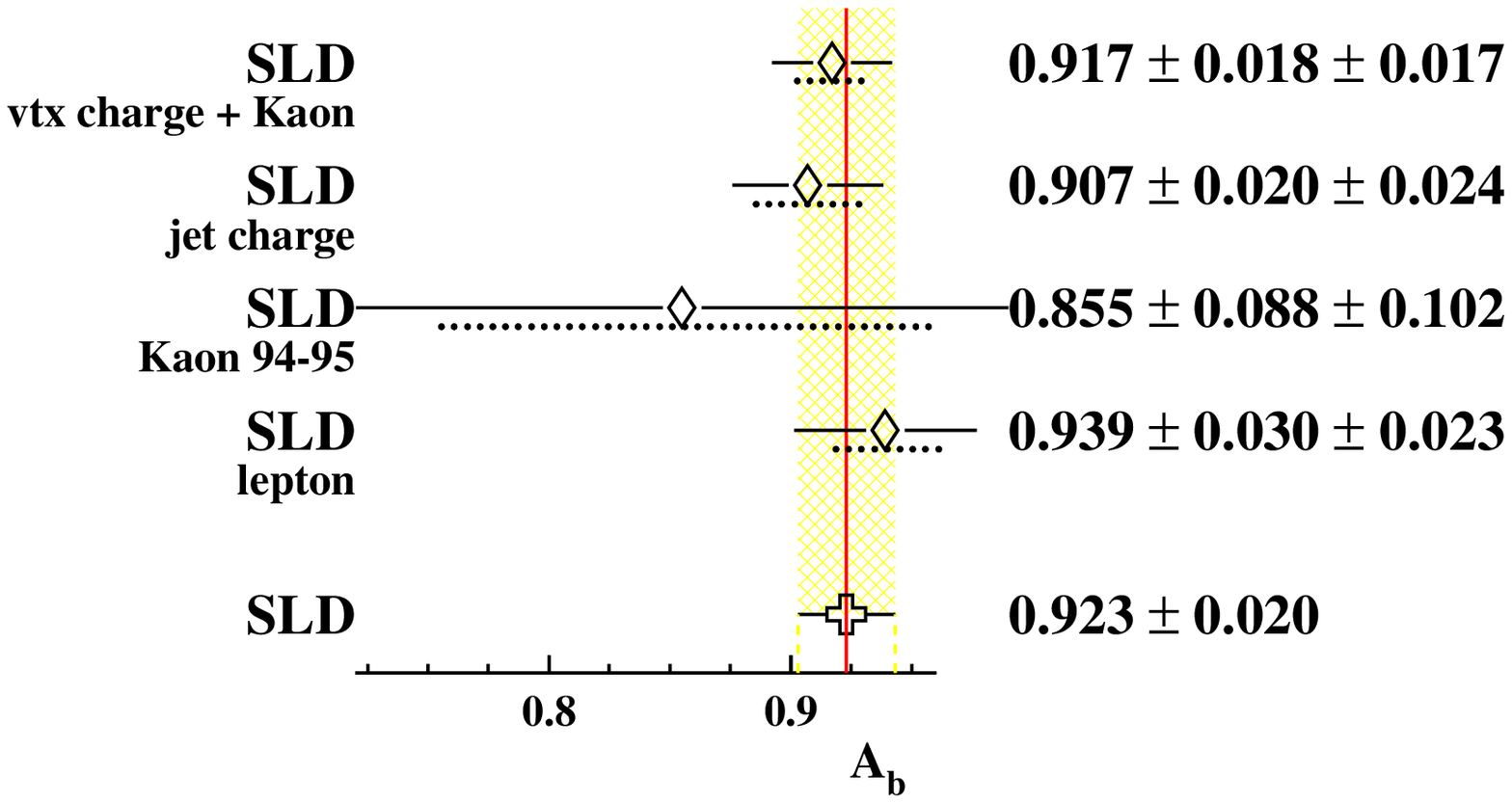}
 \includegraphics[width=0.495\linewidth,bb=33 60 511 313]{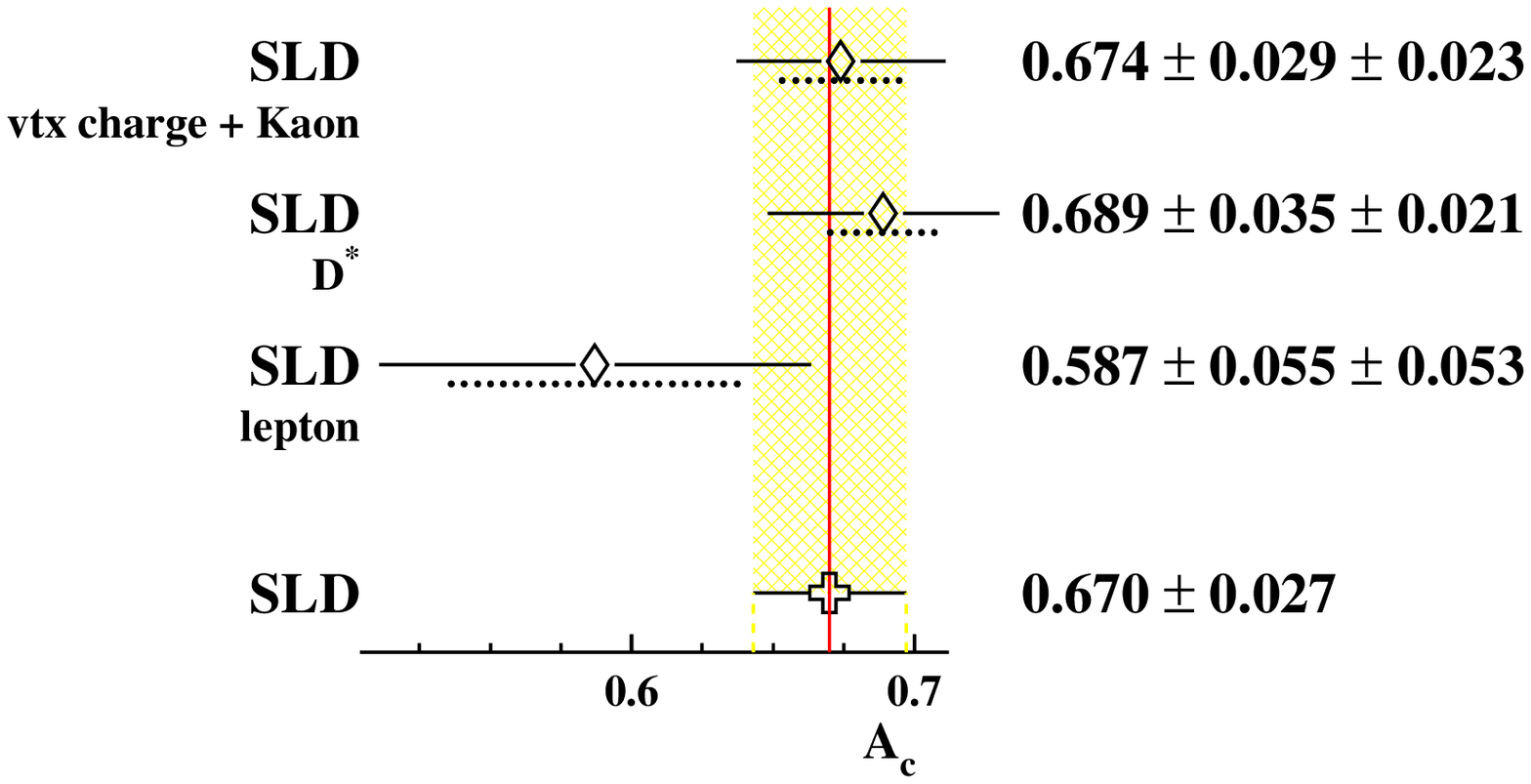}
\end{center}
\caption[ $\cAb$ and $\cAc$ measurements used in the heavy flavour
  combination.]{$\cAb$ and $\cAc$ measurements used in the heavy
  flavour combination, corrected for their dependence on parameters
  evaluated in the multi-parameter fit described in the text. The
  dotted lines indicate the size of the systematic error. }
\label{fig:hqbar_abc}
\end{figure}

The main measurements of the experiments used in the determination of
the six electroweak observables are compared in
Figures~\ref{fig:hqbar_rbc} to~\ref{fig:hqbar_abc}.  For each
electroweak observable, the measurements of the five experiments are
very consistent. This is also reflected in the overall $\chi^2$ of the
fit, which is $53$ for $105-14$ degrees of freedom. The small $\chidf$
is largely a result of statistical effects and systematic errros
evaluated conservatively. For example, the forward-backward asymmetry
measurements, all statisticaly dominated, show very consistent central
values. For the partial width measurements, several systematic
uncertainties are evaluated by comparing test quantities between data
as weall as between model simulations. Even if no effect is found or
expected, the statistical errors of these tests are taken as
systematic uncertainties.

\begin{figure}[p]
\begin{center}
\includegraphics[width=0.495\linewidth,bb=16 16 488 457]{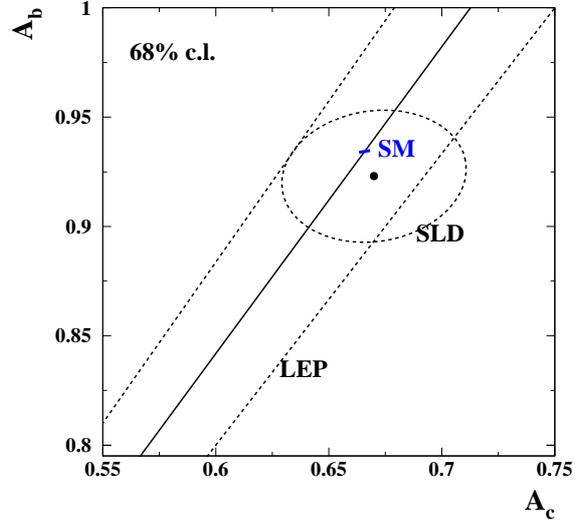}
\end{center}
\caption[Contours in the $\cAc$-$\cAb$ plane from the Z-pole data] {
Contours in the $\cAc$-$\cAb$ plane and ratios of forward-backward
asymmetries from the SLD and LEP, corresponding to 68~\% confidence
levels assuming Gaussian systematic errors. The SM prediction is shown
for the same parameters as used for Figure~\ref{fig:hfsmcomp2}. }
\label{fig:hfsmcomp1}
\end{figure}

\begin{figure}[p]
\begin{center}
\includegraphics[width=0.495\linewidth]{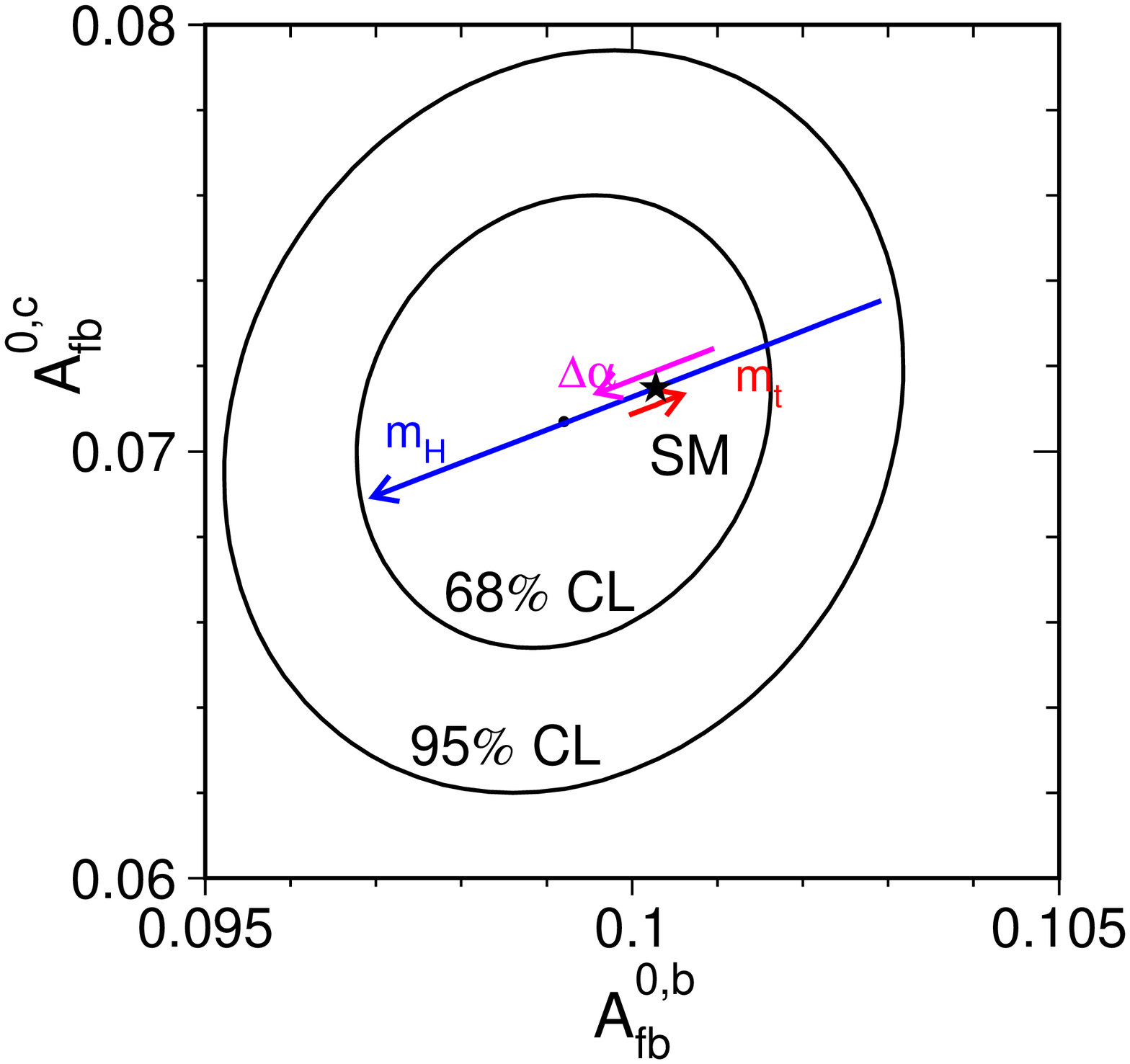}
\hfill
\includegraphics[width=0.495\linewidth]{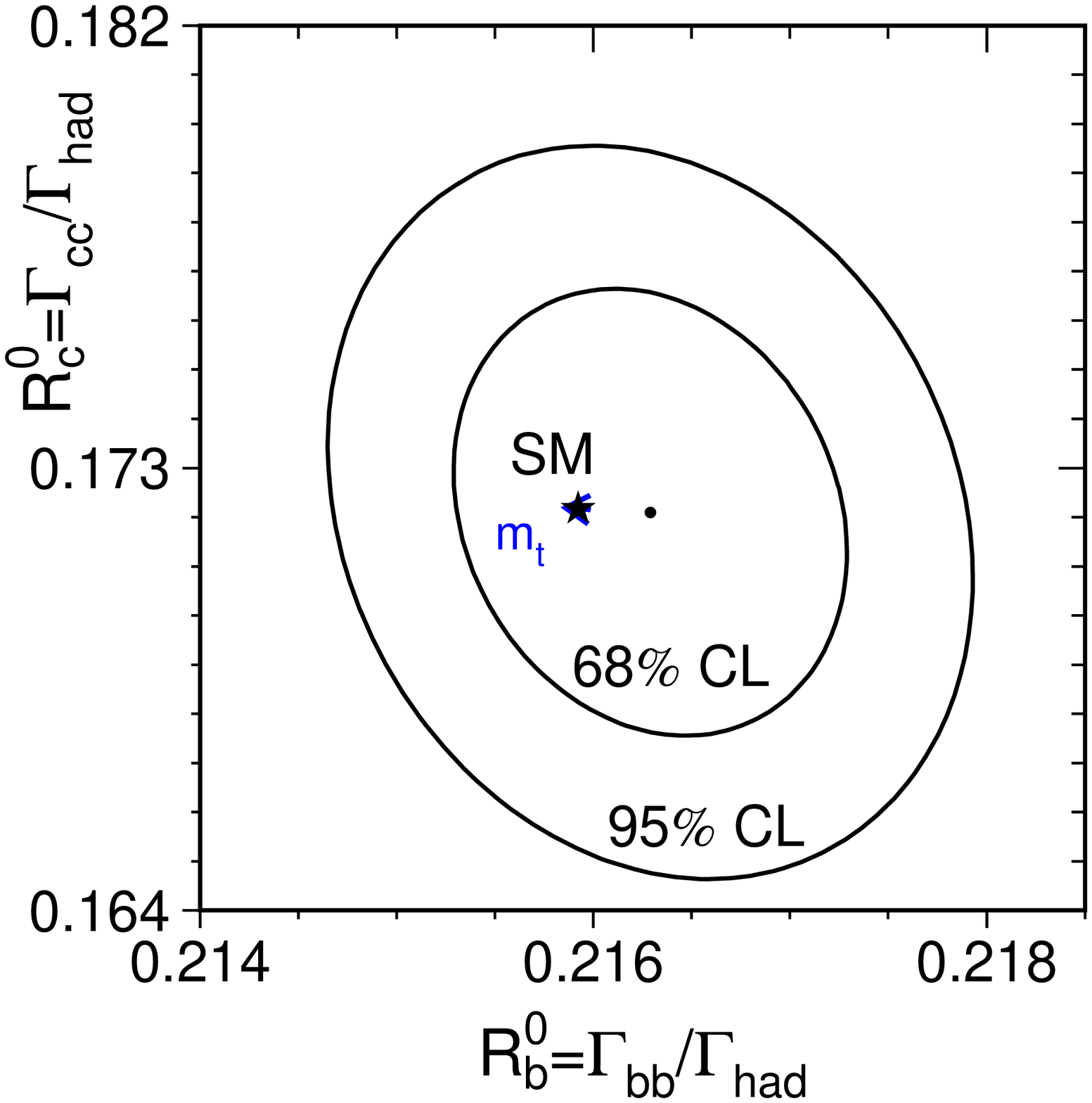}
\end{center}
\caption[Contours in the $\Afbzc$-$\Afbzb$ and $\Rcz$-$\Rbz$ planes
from the Z-pole data]{ Contours in the $\Afbzc$-$\Afbzb$ and
$\Rcz$-$\Rbz$ planes from the LEP and SLD data, corresponding to 68~\%
confidence levels assuming Gaussian systematic errors. The $\SM$
prediction for $\Mt=170.9 \pm 1.8\,\GeV$, $\MH =
300^{+700}_{-186}\,\GeV$ and $\dalhad=0.02758\pm0.00035$ is also
shown.  } \label{fig:hfsmcomp2}
\end{figure}

The consistency of the combined results with the expectations
calculated within the $\SM$ is shown in Figures~\ref{fig:hfsmcomp1}
and~\ref{fig:hfsmcomp2}, including the consistency between the
forward-backward and forward-backward left-right asymmetry results.
Note that the asymmetries are in perfect agreement with the $\SM$
expectation for high masses of the Higgs boson, in contrast to the
leptonic quantities discussed in the previous sections which prefer a
much lower Higgs-boson mass.  This different behaviour has
consequences for joint analyses of all results as presented in
Sections~\ref{chap:Z+coup} and~\ref{chap:MSM}.

\begin{table}[htbp]
\begin{center}
\renewcommand{\arraystretch}{1.1}
\caption[The results of the 14-parameter fit to the LEP/SLD heavy
flavour data]{The results of the 14-parameter fit to the LEP/SLD heavy
flavour data. 
}
\label{tab:14parres}
\begin{tabular}{|l||c|}
\hline
Observable & Result \\
\hline
\hline
 $ \Rbz    $&$ 0.21629  \pm 0.00066 $\\
 $ \Rcz    $&$ 0.1721   \pm 0.0030  $\\
 $ \Afbzb  $&$ 0.0992   \pm 0.0016  $\\
 $ \Afbzc  $&$ 0.0707   \pm 0.0035  $\\
 $ \cAb    $&$ 0.923    \pm 0.020   $\\
 $ \cAc    $&$ 0.670    \pm 0.027   $\\
 $ \Brbl   $&$ 0.1071    \pm 0.0022 $\\
 $ \Brbclp $&$ 0.0801    \pm 0.0018 $\\
 $ \Brcl   $&$ 0.0969    \pm 0.0031 $\\
 $ \chiM   $&$ 0.1250    \pm 0.0039 $\\
 $ \fDp    $&$ 0.235     \pm 0.016  $\\
 $ \fDs    $&$ 0.126     \pm 0.026  $\\
 $ \fcb    $&$ 0.093     \pm 0.022  $\\
 $ \PcDst  $&$ 0.1622    \pm 0.0048 $\\
\hline
\end{tabular}
\end{center}
\end{table}

\clearpage

\begin{table}[p]
\begin{center}
\renewcommand{\arraystretch}{1.1}
\caption[The correlation matrix for the set of the 14 heavy flavour
parameters.]{ The correlation matrix for the set of the 14 heavy
flavour parameters. $\BR(1)$, $\BR(2)$ and $\BR(3)$ denote $\Brbl$,
$\Brbclp$ and $\Brcl$ respectively, $P$ denotes $\PcDst$.  }
\label{tab:14parcor}
\begin{center}
\tiny 
\begin{tabular}{|l||rrrrrrrrrrrrrr|}
\hline
%
&\makebox[0.45cm]{\Rb}
&\makebox[0.45cm]{\Rc}
&\makebox[0.45cm]{$\Afbzb$}
&\makebox[0.45cm]{$\Afbzc$}
&\makebox[0.45cm]{\cAb}
&\makebox[0.45cm]{\cAc}
&\makebox[0.45cm]{$\BR(1)$}
&\makebox[0.45cm]{$\BR(2)$}
&\makebox[0.45cm]{$\BR(3)$}
&\makebox[0.45cm]{\chiM}
&\makebox[0.45cm]{$\fDp$}
&\makebox[0.45cm]{$\fDs$}
&\makebox[0.45cm]{$f(c_{bar.})$}
&\makebox[0.55cm]{$P$}\\
\hline
\hline
 {\Rb}          & $ 1.00$ &         &         &         &         &         
&         &         &         &         &         &         &         &        \\ 
 {\Rc}          & $-0.18$ & $ 1.00$ &         &         &         &         
&         &         &         &         &         &         &         &        \\ 
 {$\Afbzb$}     & $-0.10$ & $ 0.04$ & $ 1.00$ &         &         &         
&         &         &         &         &         &         &         &        \\ 
 {$\Afbzc$}     & $ 0.07$ & $-0.06$ & $ 0.15$ & $ 1.00$ &         &         
&         &         &         &         &         &         &         &        \\ 
 {\cAb}         & $-0.08$ & $ 0.04$ & $ 0.06$ & $-0.02$ & $ 1.00$ &         
&         &         &         &         &         &         &         &        \\ 
 {\cAc}         & $ 0.04$ & $-0.06$ & $ 0.01$ & $ 0.04$ & $ 0.11$ & $ 1.00$ 
&         &         &         &         &         &         &         &        \\ 
 {$\BR(1)$}     & $-0.08$ & $ 0.05$ & $-0.01$ & $ 0.18$ & $-0.02$ & $ 0.02$
& $ 1.00$ &         &         &         &         &         &         &        \\ 
 {$\BR(2)$}     & $-0.03$ & $-0.01$ & $-0.06$ & $-0.23$ & $ 0.02$ & $-0.04$ 
& $-0.24$ & $ 1.00$ &         &         &         &         &         &        \\ 
 {$\BR(3)$}     & $ 0.00$ & $-0.30$ & $ 0.00$ & $-0.21$ & $ 0.03$ & $-0.02$ 
& $ 0.01$ & $ 0.10$ & $ 1.00$ &         &         &         &         &        \\ 
 {\chiM}        & $ 0.00$ & $ 0.02$ & $ 0.11$ & $ 0.08$ & $ 0.06$ & $ 0.00$ 
& $ 0.29$ & $-0.23$ & $ 0.16$ & $ 1.00$ &         &         &         &        \\ 
 {$\fDp$}       & $-0.15$ & $-0.10$ & $ 0.01$ & $-0.04$ & $ 0.00$ & $ 0.00$ 
& $ 0.04$ & $ 0.02$ & $ 0.00$ & $ 0.02$ & $ 1.00$ &         &         &        \\ 
 {$\fDs$}       & $-0.03$ & $ 0.13$ & $ 0.00$ & $-0.02$ & $ 0.00$ & $ 0.00$ 
& $ 0.01$ & $ 0.00$ & $-0.01$ & $-0.01$ & $-0.40$ & $ 1.00$ &         &        \\ 
 {$f(c_{bar.})$}& $ 0.11$ & $ 0.18$ & $-0.01$ & $ 0.04$ & $ 0.00$ & $ 0.00$ 
& $-0.02$ & $-0.01$ & $-0.02$ & $ 0.00$ & $-0.24$ & $-0.49$ & $ 1.00$ &        \\ 
 {$P$}          & $ 0.13$ & $-0.43$ & $-0.02$ & $ 0.04$ & $-0.02$ & $ 0.02$ 
& $-0.01$ & $ 0.01$ & $ 0.13$ & $ 0.00$ & $ 0.08$ & $-0.06$ & $-0.14$ & $ 1.00$\\ 
 \hline
\end{tabular}
\normalsize
\end{center}
\end{center}
\end{table}


\clearpage

\subsection{Inclusive Hadronic Charge Asymmetry}
\label{sec:InclusiveQfb}

\begin{figure}[htb]
\begin{center} 
\includegraphics[width=0.6\linewidth]{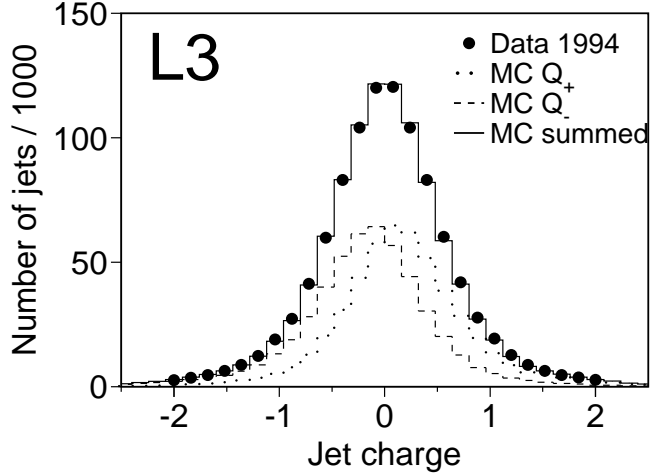}
\caption[The $Q_{+}$ and $Q_{-}$ distributions obtained from 
Monte Carlo simulation by L3] {The $Q_{+}$ and $Q_{-}$ distributions
  obtained from Monte Carlo simulation by L3. Their sum is compared to
  the sum of the $ \QF + \QB \equiv Q_+ + Q_- $ distributions for 1994
  data.}
\label{fig:Qfbpic} 
\end{center}
\end{figure}

The hadronic final state occurs in Z boson production with a fraction
of about 70\%.  The determination of the inclusive hadronic partial
decay width of the Z boson described in Section~\ref{sec:Lineshape}
does not attempt to separate the quark flavours.  In a similar manner,
a high-statistics inclusive forward-backward charge asymmetry can be
defined, avoiding heavy-quark tagging and hence based on charge flow
rather than flavour flow. The inclusive hadronic forward-backward
charge asymmetry, averaged over all quark flavours u, d, s, c and b,
is measured through the charges of jets or hemispheres associated to
the primary quark-antiquark pair:
\begin{eqnarray}
\avQfb & = & \sum_{\mathrm{q \ne t}} \Rq \Aqq ~sign(Q_q)\,,
\end{eqnarray}
where $Q_q$ is the electric charge of quark flavour $q$.  Since $\Aqq$
denotes the flavour forward-backward asymmetry, up-type quarks and
down-type quarks contribute with opposite sign to the forward-backward
charge asymmetry. The charge separation, ie, the difference of the
measured charges of the two hadronic jets, indicates how well the
detector can separate charges in order to achieve a significant
asymmetry measurement. As an example, the distribution of jet charges
in the inclusive analysis of the L3 experiment is shown in
Figure~\ref{fig:Qfbpic}.

Using the $\SM$ dependence of $\Rq$ and $\Aqq$ in terms of the
effective electroweak mixing angle, the
analyses~\cite{ALEPHcharge1996,DELPHIcharge,ref:ljet,OPALcharge} of
the forward-backward charge asymmetry are directly expressed in terms
of $\swsqeffl$. The results listed in Table~\ref{tab:avQfb} agree very
well.  Taking correlated uncertainties into account, the combined
result is~\cite{Z-Pole}:
\begin{eqnarray}
\swsqeffl & = & 0.2324 \pm 0.0012\,,
\label{eq:avQfb}
\end{eqnarray}
The total error is dominated by systematic effects amounting to
0.0010.  The largest systematic uncertainties arise from the model
input required to describe the properties of jets created in the
fragmentation of light quarks and thus the description of the charge
separation achievable.  Note that the correlation of this measurement
with the asymmetry measurements of identified heavy-flavour samples
presented in Section~\ref{sec:Heavy} is negligible.  Since the
inclusive charge separation is worse compared to that of identified
$\qq$ flavours, and due to the larger model dependence, the resulting
measurement of $\swsqeffl$ is of limited precision compared to the
measurements with leptonic or heavy-quark final states.

\begin{table}[htbp]
\begin{center}
\renewcommand{\arraystretch}{1.2}
\caption[$\swsqeffl$ from inclusive hadronic charge asymmetry]{
  Determination of $\swsqeffl$ from inclusive hadronic charge
  asymmetries at LEP. For each experiment, the first error is
  statistical and the second systematic.  Also listed is the
  correlation matrix for the total errors, summing statistical and
  systematic uncertainties in quadrature, used in the average of
  $\Qfbhad$ results.}
\label{tab:avQfb}
\begin{tabular}{|ll||c|cccc|}
\hline
Experiment & & $\swsqeffl$ & \multicolumn{4}{|c|}{Correlations} \\
\hline
\hline
ALEPH &(1990-94)&$0.2322\pm0.0008\pm0.0011$& $1.00$ & $    $ & $    $ & $    $\\
DELPHI&(1990-91)&$0.2345\pm0.0030\pm0.0027$& $0.12$ & $1.00$ & $    $ & $    $\\
L3    &(1991-95)&$0.2327\pm0.0012\pm0.0013$& $0.27$ & $0.13$ & $1.00$ & $    $\\
OPAL  &(1990-91)&$0.2321\pm0.0017\pm0.0029$& $0.14$ & $0.37$ & $0.15$ & $1.00$\\
\hline
LEP Average & &$0.2324\pm0.0007\pm0.0010$& \multicolumn{4}{|c|}{ } \\
\hline
\end{tabular}
\end{center}
\end{table}

\clearpage

\subsection{Z Boson Properties and Effective Couplings}
\label{chap:Z+coup}

The results presented above are derived in largely model independent
analyses and are quoted in terms of experimentally motivated pseudo
observables defined such that correlations between them are reduced.
Other familiar pseudo observables, such as partial decay widths,
effective vector and axial-vector coupling constants, or the effective
electroweak mixing angle, can easily be derived through parameter
transformations~\cite{Z-Pole} and are presented in the following.

\subsubsection{Z-Boson Decay Widths and Number of Neutrinos}

A first example is given by the partial decay widths of the Z boson.
The results, derived from the ratios of partial widths determined
above, are reported in Table~\ref{tab:width}. The invisible decay
width, i.e., the decay width for Z decays into invisible particles
such as neutrinos, is included and calculated as the difference
between the total decay width and the sum of all visible decay width
caused by decays into hadrons and charged leptons. Assuming only $\SM$
particles, the invisible decay width constrains the number of light
neutrino species to:
\begin{eqnarray}
N_{\nu} & = & \frac{\Ginv}{\Gll} \left(\frac{\Gll}{\Gnn}\right)_{\SM}
        ~ = ~ 2.9840 \pm 0.0082\,,
\end{eqnarray}
in agreement with the three known neutrino species, excluding the
existence of additional families or generations of fermions with light
neutrinos.  Using ratios in the above equation is advantageous because
the first ratio is determined with higher precision experimentally,
and in the $\SM$ calculation of the second ratio, many parametric
uncertainties cancel.

\begin{table}[ht] \begin{center} 
\renewcommand{\arraystretch}{1.15}
\caption[Partial $\Zzero$ widths] {\label{tab:width} Partial $\Zzero$
decay widths, derived from the results of
Tables~\ref{tab:lsafbresult}, \ref{tab:14parres}
and~\ref{tab:14parcor}. The width denoted as $\leptlept$ is that of a
single charged massless lepton species.  The width to invisible
particles is calculated as the difference of total and all other
partial widths.  }
\begin{tabular} {|l||r||rrrrrrr|}
\hline
Parameter & Average &  \multicolumn{7}{|c|}{Correlations}\\
$\Gff$    & [\MeV]~ &  \multicolumn{7}{|c|}{            }\\
\hline
\hline
\multicolumn{9}{|c|}{Without Lepton Universality} \\
\hline 
 & & $\Ghad$ & $\Gee$ & $\Gmumu$ & $\Gtautau$ & $\Gbb$ & $\Gcc$ & $\Ginv$ \\
\hline 
$\Ghad$    & 1745.8 $\pzz\pm$ 2.7$\pzz$ &    1.00 & \multicolumn{6}{c|}{} \\
$\Gee$     &   83.92 $\pz\pm$ 0.12$\pz$ & $-$0.29&1.00& \multicolumn{5}{c|}{} \\ 
$\Gmumu$   &   83.99 $\pz\pm$ 0.18$\pz$ &   ~0.66& $-$0.20&1.00&\multicolumn{4}{c|}{} \\
$\Gtautau$ &   84.08 $\pz\pm$ 0.22$\pz$ &    0.54& $-$0.17&  ~0.39&1.00& \multicolumn{3}{c|}{}\\  
$\Gbb$     &  377.6 $\pzz\pm$ 1.3$\pzz$ &    0.45& $-$0.13&   0.29&  ~0.24&1.00&\multicolumn{2}{c|}{}\\
$\Gcc$     &  300.5 $\pzz\pm$ 5.3$\pzz$ &    0.09& $-$0.02&   0.06&   0.05&$-$0.12&1.00&\multicolumn{1}{c|}{}\\
$\Ginv$    &  497.4 $\pzz\pm$ 2.5$\pzz$ & $-$0.67&    0.78&$-$0.45&$-$0.40&$-$0.30&$-$0.06&1.00\\
\hline 
\hline 
\multicolumn{9}{|c|}{With Lepton Universality} \\
\hline 
 & & $\Ghad$ & $\Gll$ & $\Gbb$ & $\Gcc$ & $\Ginv$ & & \\
\hline 
$\Ghad$ & 1744.4 $\pzz\pm$ 2.0$\pzz$   &  1.00&\multicolumn{6}{c|}{} \\
$\Gll$  &   83.985 $\pm$   0.086       &   ~0.39&1.00&\multicolumn{5}{c|}{} \\ 
$\Gbb$  &  377.3 $\pzz\pm$ 1.2$\pzz$   &   ~0.35&~0.13&1.00&\multicolumn{4}{c|}{} \\
$\Gcc$  &  300.2 $\pzz\pm$ 5.2$\pzz$   &   ~0.06&~0.03&$-$0.15&1.00& \multicolumn{3}{c|}{}\\  
$\Ginv$ &  499.0 $\pzz\pm$ 1.5$\pzz$   & $-$0.29&~0.49&$-$0.10&$-$0.02&1.00&\multicolumn{2}{c|}{}\\
\hline 
\end{tabular} 
\end{center} 
\end{table}

\subsubsection{The Asymmetry Parameters}

The forward-backward asymmetries determine products of the asymmetry
parameters, $\Afbzf=(3/4)\cAe\cAf$, while the polarised asymmetries
measure $\cAf$ directly. Using all measurements determining any
(combination of) leptonic asymmetry parameters, combined values are
calculated and reported in Table~\ref{tab:coup:al}. The results for
the different charged lepton species agree well as expected from
lepton universality. With this assumption, a combined value of
\begin{eqnarray}
\cAl & = & 0.1501\pm0.0016\,
\label{eq:coup:al}
\end{eqnarray}
is obtained, where this average has a $\chidf$ of 7.8/9, corresponding
to a probability of 56\%.

\begin{table}[htbp]
\begin{center}
\renewcommand{\arraystretch}{1.25}
\caption[Results on the leptonic asymmetry parameters $\cAl$] {Results
on the leptonic asymmetry parameters $\cAl$ using the 14 electroweak
measurements of Tables~\ref{tab:lsafbresult} and~\ref{tab:alr:result},
and Equations~\ref{eq:ptau:At} and~\ref{eq:ptau:Ae}.  The combination
has a $\chidf$ of 3.6/5, corresponding to a probability of 61\%.}
\label{tab:coup:al}
\begin{tabular}{|c||r@{$\pm$}l||rrr|}
\hline
Parameter & \multicolumn{2}{|c||}{Average} 
          & \multicolumn{3}{|c| }{Correlations}    \\
          & \multicolumn{2}{|c||}{ }
          & {$\cAe$} & {$\cAm$}& {$\cAt$}          \\
\hline
\hline
$\cAe$ &$0.1514$&$0.0019$ & $ 1.00$&$     $&$     $ \\
$\cAm$ &$0.1456$&$0.0091$ & $-0.10$&$ 1.00$&$     $ \\
$\cAt$ &$0.1449$&$0.0040$ & $-0.02$&$ 0.01$&$ 1.00$ \\
\hline
\end{tabular}
\end{center}
\end{table}

\begin{figure}[htbp]
\begin{center}
\includegraphics[width=0.495\linewidth]{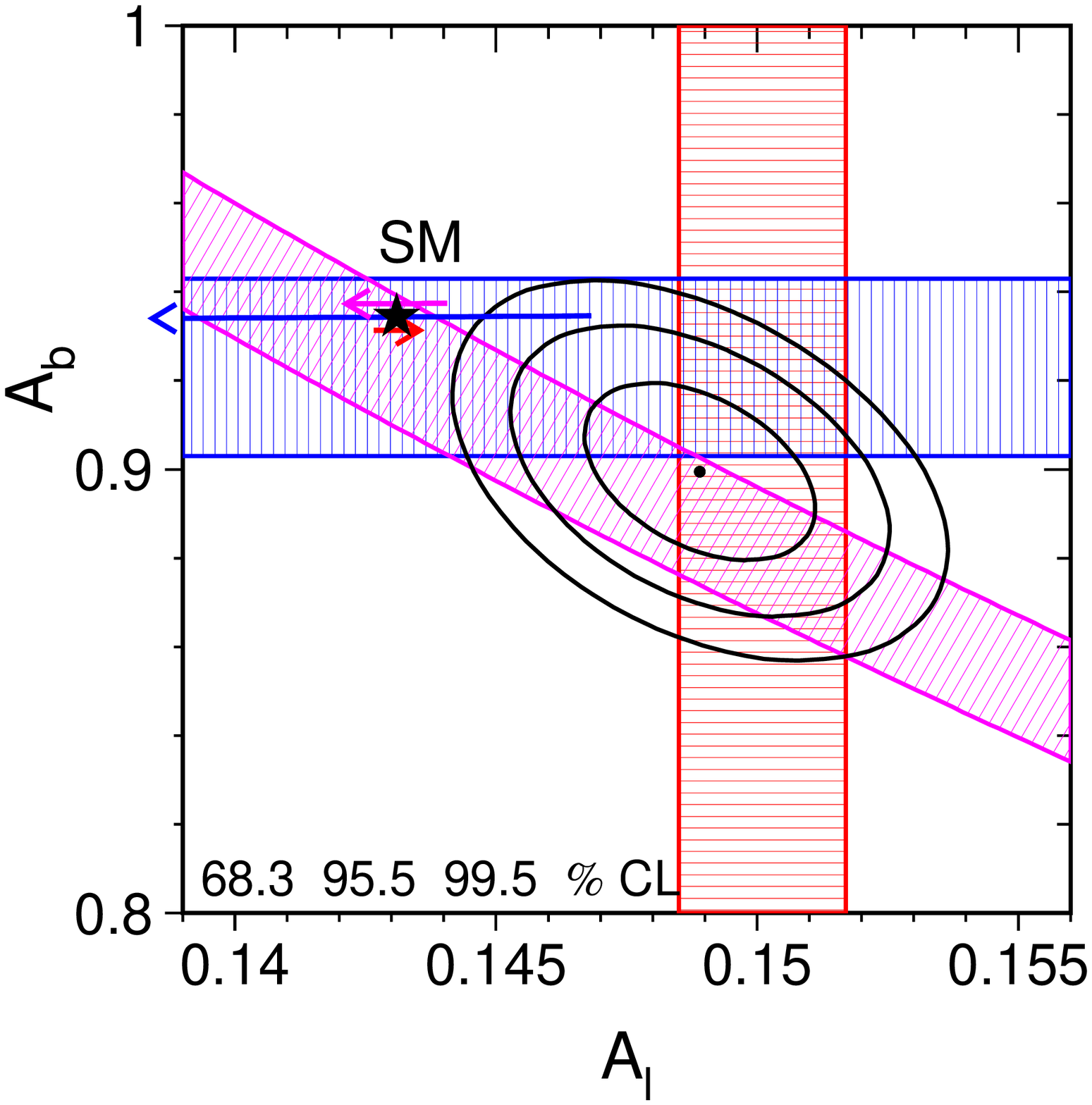} 
\hfill
\includegraphics[width=0.495\linewidth]{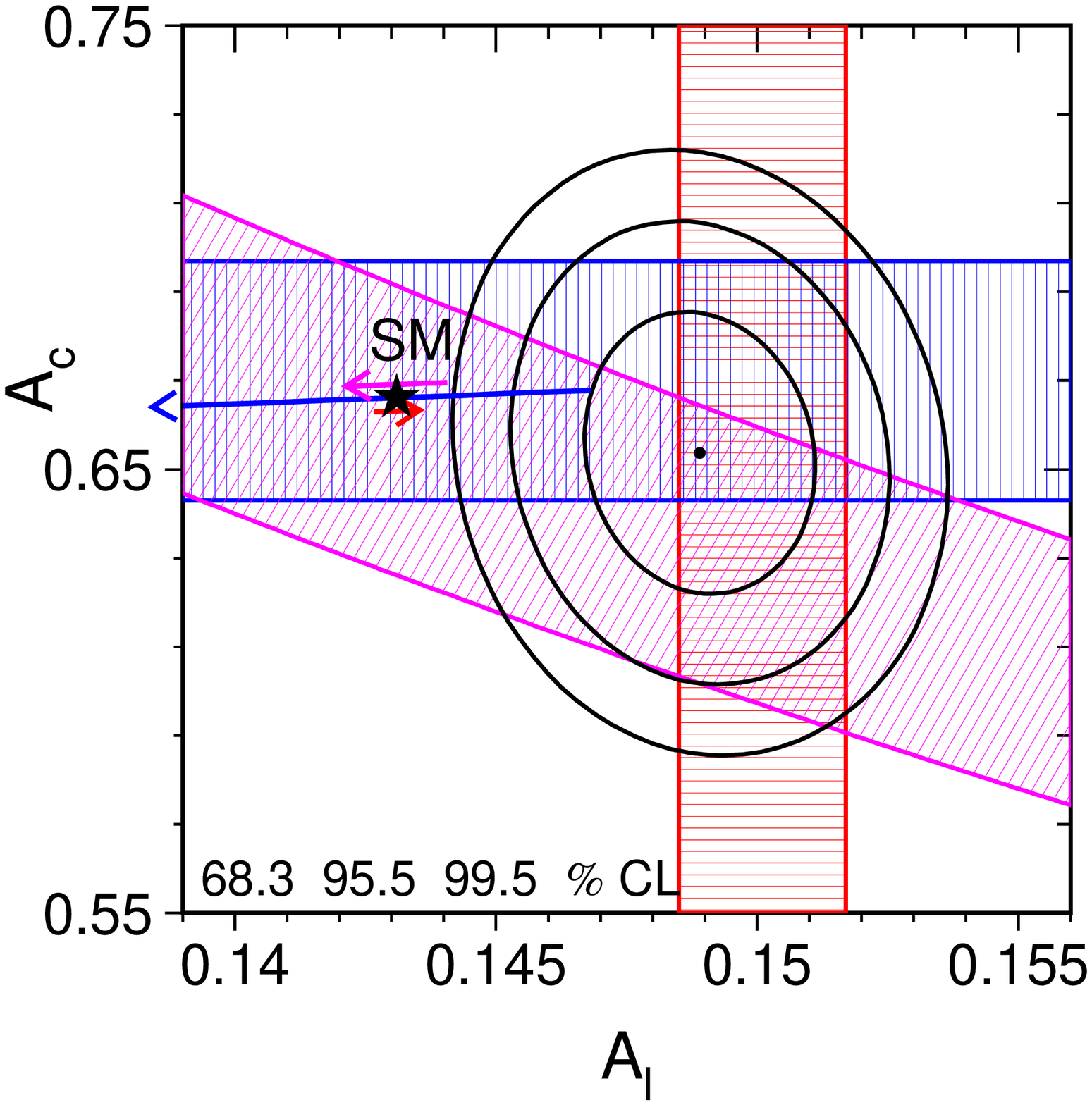} 
\caption[Comparison of the asymmetry parameters $\cAf$] { Comparison
of the measurements of $\cAl$, $\cAq$ and $\Afbzq$ for (top) b-quarks,
and (bottom) c-quarks, assuming lepton universality.  Bands of $\pm1$
standard deviation width in the $(\cAl,\cAq)$ plane are shown for the
measurements of $\cAl$ (vertical band), $\cAq$ (horizontal band), and
$\Afbzq=(3/4)\cAe\cAq$ (diagonal band).  Also shown are the 68\%, 95\%
and 99.5\% confidence level contours for the two asymmetry parameters
resulting from the joint analysis (Table~\ref{tab:coup:aq}).  The
arrows pointing to the right and to the left show the variation in the
$\SM$ prediction for varying $\dalhad$ in the range
$0.02758\pm0.00035$ (arrow displaced vertically upwards), $\MH$ in the
range of $300^{+700}_{-186}~\GeV$, and $\Mt$ in the range
$170.9\pm1.8~\GeV$ (arrow displaced vertically downwards).  All arrows
point in the direction of increasing values of these parameters.  }
\label{fig:coup:aq} 
\end{center}
\end{figure}

Using this combined value of the leptonic asymmetry parameter $\cAl$,
the mutual consistency of the heavy-flavour measurements
$\Afbzq=(3/4)\cAl\cAq$ and $\cAq$ is shown in
Figure~\ref{fig:coup:aq}.  While the bands show good overlap for c
quarks, the b-quark case exhibits some tension between the
measurements. This is made strikingly obvious by extracting the
asymmetry parameter $\cAb$ from the forward-backward asymmetry
measurement using the above value of $\cAl$: $\cAb=(4/3)\Afbzb/\cAl$.
All $\cAq$ determinations are compared in Table~\ref{tab:coup:test},
together with the $\SM$ expectation.  While there is good agreement
for c quarks, the ratio $\cAb=(4/3)\Afbzb/\cAl$ shows a 3.2 standard
deviation discrepancy from the $\SM$ expectation; a consequence of the
same effect as discussed towards the end of Section~\ref{sec:Heavy}.
The numerical results of the joint analysis in terms of asymmetry
parameters are in reported in Table~\ref{tab:coup:aq}.

\begin{table}[htbp]
\begin{center}
\renewcommand{\arraystretch}{1.25}
\caption[Determination of the quark asymmetry parameters $\cAq$]
{Determination of the quark asymmetry parameters $\cAq$, based on the
ratio $\Afbzq/\cAl$ and the direct measurement $\AFBLRq$.  Lepton
universality for $\cAl$ is assumed. The correlation between
$4\Afbzb/3\cAl$ and $4\Afbzc/3\cAl$ is $+0.24$, while it is $+0.11$
between the direct measurements $\cAb$ and $\cAc$.  The expectation of
$\cAq$ in the $\SM$ is listed in the last column.}
\label{tab:coup:test}
\begin{tabular}{|c||r@{$\pm$}l|r@{$\pm$}l||r@{$\pm$}l|}
\hline
Flavour $q$  & \multicolumn{2}{|c|}{$\cAq=\frac{4}{3}\frac{\Afbzq}{\cAl}$} 
             & \multicolumn{2}{|c||}{Direct $\cAq$}
             & \multicolumn{2}{|c|}{SM} \\
\hline
\hline
b  &$0.881 $&$0.017 $& $0.923$&$0.020$ & $0.935$&$0.001$ \\
c  &$0.628 $&$0.032 $& $0.670$&$0.027$ & $0.668$&$0.002$ \\
\hline
\end{tabular}
\end{center}
\end{table}

\begin{table}[htbp]
\begin{center}
\renewcommand{\arraystretch}{1.25}
\caption[Results on the asymmetry parameters $\cAf$]
{Results on the quark asymmetry parameters $\cAq$ and the leptonic
  asymmetry parameter $\cAl$ assuming neutral-current lepton
  universality using the 13 electroweak measurements of
  Tables~\ref{tab:lsafbresult}, \ref{tab:14parres}
  and~\ref{tab:14parcor}, and Equations~\ref{eq:al:result}
  and~\ref{eq:ptau:Al}.  The combination has a $\chidf$ of 4.5/4,
  corresponding to a probability of 34\%. }
\label{tab:coup:aq}
\begin{tabular}{|c||r@{$\pm$}l||rrr|}
\hline
Parameter & \multicolumn{2}{|c||}{Average} 
          & \multicolumn{3}{|c|}{Correlations} \\
          & \multicolumn{2}{|c||}{ }
          & {$\cAl$} & {$\cAb$}& {$\cAc$} \\
\hline
\hline
$\cAl$ &$0.1489$&$0.0015$& $ 1.00$& $     $&$     $ \\
$\cAb$ &$0.899 $&$0.013 $& $-0.42$& $ 1.00$&$     $ \\
$\cAc$ &$0.654 $&$0.021 $& $-0.10$& $ 0.15$&$ 1.00$ \\
\hline
\end{tabular}
\end{center}
\end{table}

\subsubsection{Effective Couplings of the Neutral Weak Current}

The electroweak measurements presented above determine asymmetry
parameters $\cAf$, their products such as $\Afbzf=(3/4)\cAe\cAf$, as
well as ratios of partial decay widths. In terms of effective vector
and axial-vector coupling constants of the neutral weak current,
$\gvf$ and $\gaf$, the asymmetry parameters $\cAf$ depend only on the
ratio $\gvf/\gaf$, while partial widths are proportional to
$\gvf^2+\gaf^2$. Hence, the combined set of measurements allows to
disentangle the effective coupling constants $\gvf$ and $\gaf$.

\begin{figure}[htbp]
\begin{center}
\vskip -0.5cm
\includegraphics[width=0.67\linewidth]{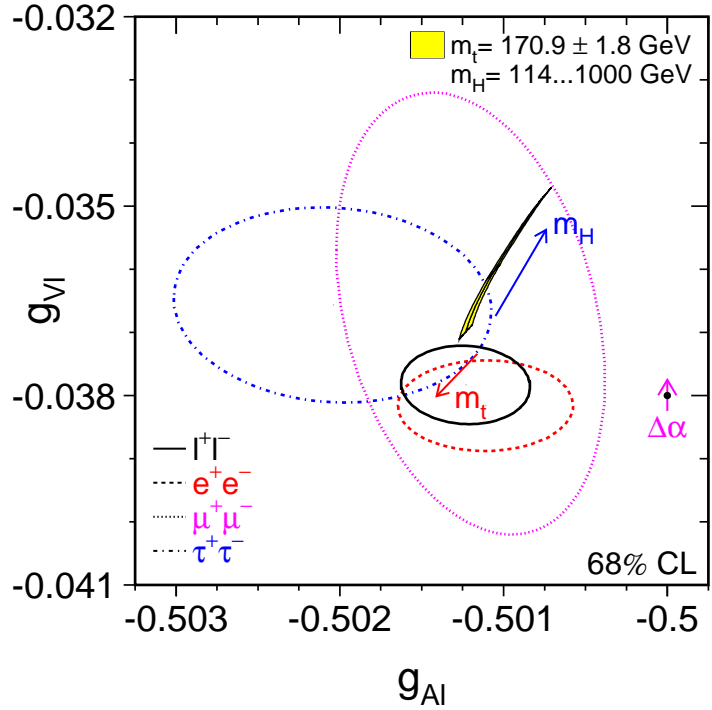}
\vskip -0.5cm
\caption[Comparison of the effective coupling constants for leptons] {
Comparison of the effective vector and axial-vector coupling constants
for leptons (Tables~\ref{tab:coup:gemt} and~\ref{tab:coup:gl}).  The
shaded region in the lepton plot shows the predictions within the
$\SM$ for $\Mt=170.9\pm1.8~\GeV$ and $\MH=300^{+700}_{-186}~\GeV$;
varying the hadronic vacuum polarisation by
$\dalhad=0.02758\pm0.00035$ yields an additional uncertainty on the
$\SM$ prediction shown by the arrow labeled $\Delta\alpha$. }
\label{fig:coup:gl} 
\end{center}
\end{figure}

\begin{figure}[htbp]
\begin{center}
\vskip -0.5cm
\includegraphics[width=0.495\linewidth]{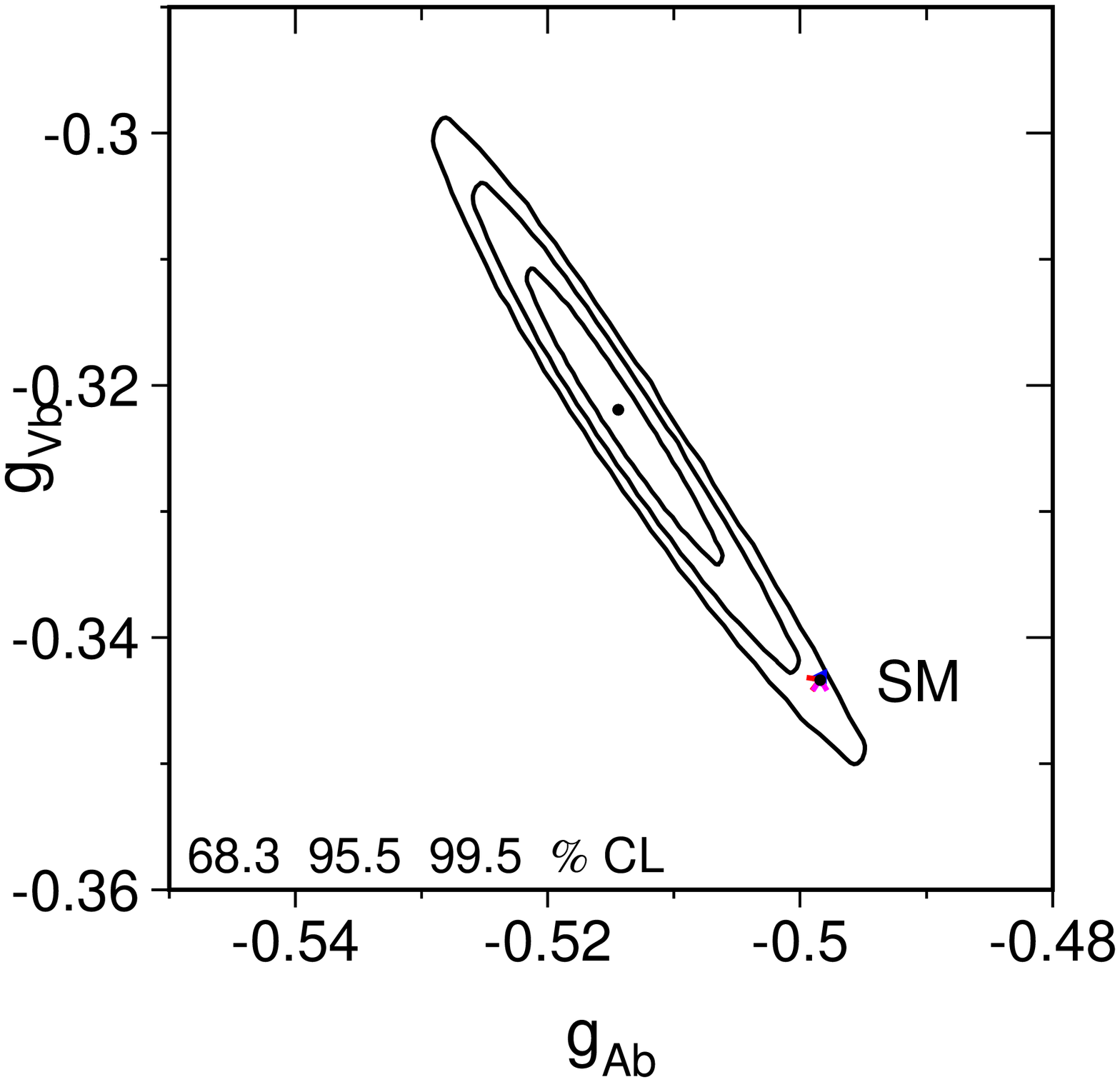}
\hfill
\includegraphics[width=0.495\linewidth]{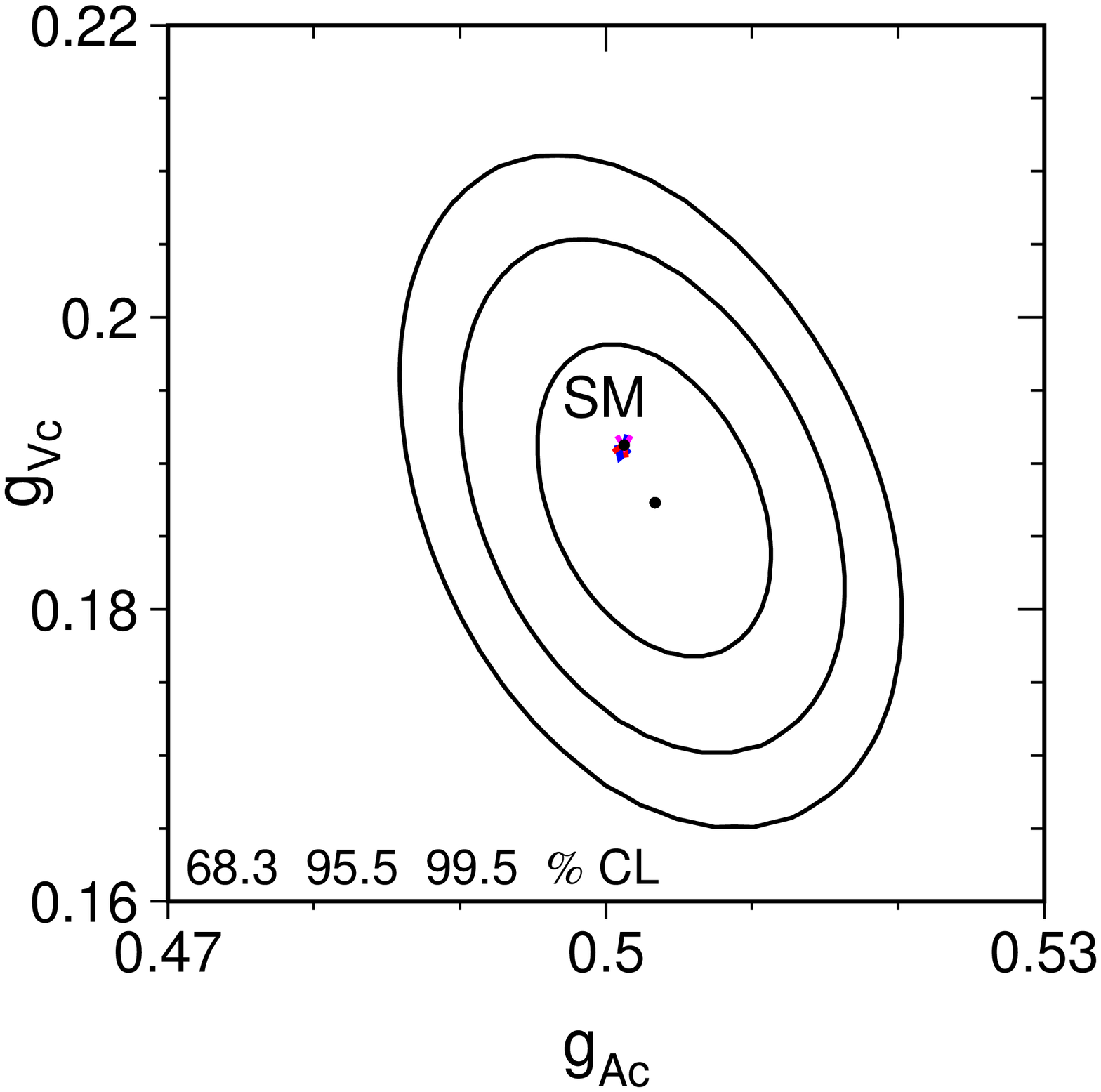}
\vskip -0.5cm
\caption[Comparison of the effective coupling constants for heavy
quarks] { Comparison of the effective vector and axial-vector coupling
constants for heavy quarks, using results on leptons and assuming
lepton universality (Table~\ref{tab:coup:gq}). Top: b quarks; bottom:
c quarks.  Compared to the experimental uncertainties, the $\SM$
predictions for the heavy quarks b and c have negligible dependence on
the $\SM$ input parameters.}
\label{fig:coup:gq} 
\end{center}
\end{figure}

Results for the leptonic couplings are reported in
Tables~\ref{tab:coup:gemt}.  The values for the three charged leptons
agree well as expected from neutral-curent lepton universality. This
is also shown in Figure~\ref{fig:coup:gl}.  Imposing lepton
universality results with increased precision are obtained and
reported in Table~\ref{tab:coup:gl}.  Including also the results on
heavy quarks, the results are listed in Table~\ref{tab:coup:gq}.  The
value of the leptonic axial coupling $\gal$ is different from the
corresponding Born-level value of $\Tl=-1/2$ by 4.8 standard
deviations, indicating the presence of non-QED electroweak radiative
corrections. The contour curves corresponding to the quark couplings
are presented in Figure~\ref{fig:coup:gq}.  The difference between
b-quark couplings and the $\SM$ expectation is caused by the b-quark
asymmetry parameter extracted from the forward-backward asymmetry as
discussed above.

\begin{table}[htbp]
\begin{center}
\renewcommand{\arraystretch}{1.25}
\caption[Results on the effective coupling constants for leptons]
{Results on the effective coupling constants for leptons, using the 14
electroweak measurements of Tables~\ref{tab:lsafbresult}
and~\ref{tab:alr:result}, and Equations~\ref{eq:ptau:At}
and~\ref{eq:ptau:Ae}.  The combination has a $\chidf$ of 3.6/5,
corresponding to a probability of 61\%.}
\label{tab:coup:gemt}
\begin{tabular}{|c||r@{$\pm$}l||rrrrrrr|}
\hline
Parameter & \multicolumn{2}{|c||}{Average} 
          & \multicolumn{7}{|c| }{Correlations} \\
          & \multicolumn{2}{|c||}{ }
          & {$\gan$}
          & {$\gae$} & {$\gamu$} & {$\gatau$} 
          & {$\gve$} & {$\gvmu$} & {$\gvtau$}    \\
\hline
\hline
$\gan\equiv\gvn$ 
        &$+0.5003 $&$0.0012 $&$ 1.00$&$     $&$     $&$     $&$     $&$     $&$     $\\
$\gae$  &$-0.50111$&$0.00035$&$-0.75$&$ 1.00$&$     $&$     $&$     $&$     $&$     $\\
$\gamu$ &$-0.50120$&$0.00054$&$ 0.39$&$-0.13$&$ 1.00$&$     $&$     $&$     $&$     $\\
$\gatau$&$-0.50204$&$0.00064$&$ 0.37$&$-0.12$&$ 0.35$&$ 1.00$&$     $&$     $&$     $\\
$\gve$  &$-0.03816$&$0.00047$&$-0.10$&$ 0.01$&$-0.01$&$-0.03$&$ 1.00$&$     $&$     $\\
$\gvmu$ &$-0.0367 $&$0.0023 $&$ 0.02$&$ 0.00$&$-0.30$&$ 0.01$&$-0.10$&$ 1.00$&$     $\\
$\gvtau$&$-0.0366 $&$0.0010 $&$ 0.02$&$-0.01$&$ 0.01$&$-0.07$&$-0.02$&$ 0.01$&$ 1.00$\\
\hline
\end{tabular}
\end{center}
\end{table}

\begin{table}[htbp]
\begin{center}
\renewcommand{\arraystretch}{1.25}
\caption[Results on the effective coupling constants for leptons]
{Results on the effective coupling constants for leptons, using the 14
electroweak measurements of Tables~\ref{tab:lsafbresult}
and~\ref{tab:alr:result}, and Equations~\ref{eq:ptau:At}
and~\ref{eq:ptau:Ae}.  Lepton universality is imposed. The combination
has a $\chidf$ of 7.8/9, corresponding to a probability of 56\%.}
\label{tab:coup:gl}
\begin{tabular}{|c||r@{$\pm$}l||rrr|}
\hline
Parameter & \multicolumn{2}{|c||}{Average} 
          & \multicolumn{3}{|c|}{Correlations} \\
          & \multicolumn{2}{|c||}{ }
          & {$\gn $} & {$\gal$}& {$\gvl$} \\
\hline
\hline
$\gan\equiv
 \gvn$ &$+0.50076$&$0.00076$& $ 1.00$& $     $&$     $ \\
$\gal$ &$-0.50123$&$0.00026$& $-0.48$& $ 1.00$&$     $ \\
$\gvl$ &$-0.03783$&$0.00041$& $-0.03$& $-0.06$&$ 1.00$ \\
\hline
\end{tabular}
\end{center}
\end{table}

\begin{table}[htbp]
\begin{center}
\renewcommand{\arraystretch}{1.25}
\caption[Results on the effective coupling constants] {Results on the
effective coupling constants for leptons and quarks assuming
neutral-current lepton universality, using the 13 electroweak
measurements of Tables~\ref{tab:lsafbresult}, \ref{tab:14parres}
and~\ref{tab:14parcor}, and Equations~\ref{eq:al:result}
and~\ref{eq:ptau:Al}.  The combination has a $\chidf$ of 4.5/4,
corresponding to a probability of 34\%.}
\label{tab:coup:gq}
\begin{tabular}{|c||r@{$\pm$}l||rrrrrrr|}
\hline
Parameter & \multicolumn{2}{|c||}{Average} 
          & \multicolumn{7}{|c| }{Correlations} \\
          & \multicolumn{2}{|c||}{ }
          & {$\gan$} 
          & {$\gal$} & {$\gab$} & {$\gac$} 
          & {$\gvl$} & {$\gvb$} & {$\gvc$}          \\
\hline
\hline
$\gan\equiv\gvn$ 
      &$+0.50075$&$0.00077$&$ 1.00$&$     $&$     $&$     $&$     $&$     $&$     $\\
$\gal$&$-0.50125$&$0.00026$&$-0.49$&$ 1.00$&$     $&$     $&$     $&$     $&$     $\\
$\gab$&$-0.5144 $&$0.0051 $&$ 0.01$&$-0.02$&$ 1.00$&$     $&$     $&$     $&$     $\\
$\gac$&$+0.5034 $&$0.0053 $&$-0.02$&$-0.02$&$ 0.00$&$ 1.00$&$     $&$     $&$     $\\
$\gvl$&$-0.03753$&$0.00037$&$-0.04$&$-0.04$&$ 0.41$&$-0.05$&$ 1.00$&$     $&$     $\\
$\gvb$&$-0.3220 $&$0.0077 $&$ 0.01$&$ 0.05$&$-0.97$&$ 0.04$&$-0.42$&$ 1.00$&$     $\\
$\gvc$&$+0.1873 $&$0.0070 $&$-0.01$&$-0.02$&$ 0.15$&$-0.35$&$ 0.10$&$-0.17$&$ 1.00$\\
\hline
\end{tabular}
\end{center}
\end{table}

\subsubsection{The Effective Electroweak Mixing Angle}

The effective electroweak mixing angle is in one to one correspondence
with the ratio $\gvf/\gaf$ and thus the asymmetry parameters $\cAf$.
Owing to the values of weak isospin and electric charge, the leptonic
asymmetry parameter $\cAl$ shows the largest sensitivity to the mixing
angle, and is thus determined with highest precision even in case of
heavy-quark forward-backward asymmetry measurements.  

\begin{figure}[htbp]
\begin{center}
\includegraphics[width=0.7\linewidth]{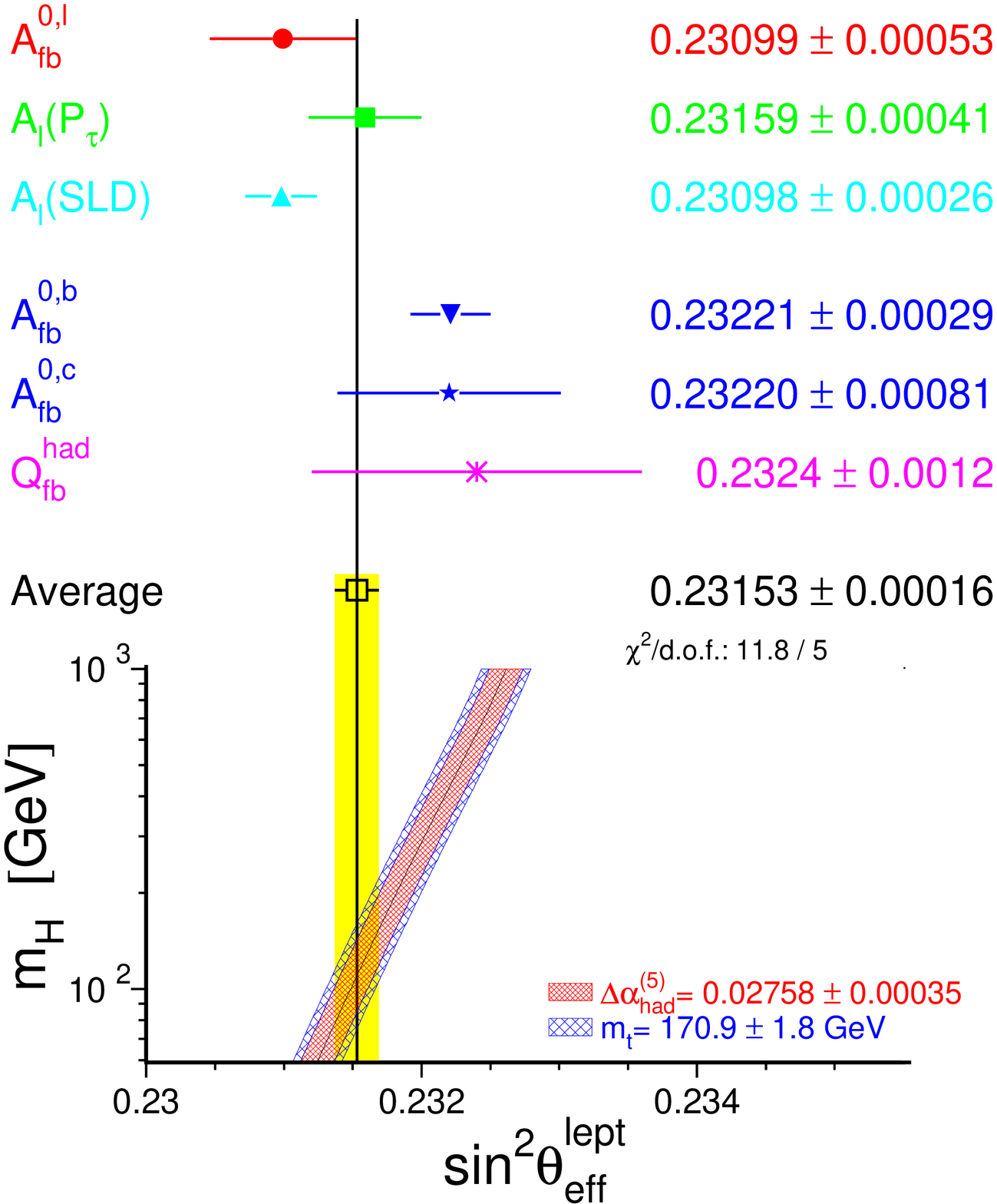}
\caption[Comparison of the effective electroweak mixing angle
$\swsqeffl$] { Comparison of the effective electroweak mixing angle
$\swsqeffl$ derived from measurements depending on lepton couplings
only (top) and also quark couplings (bottom).  Also shown is the $\SM$
prediction for $\swsqeffl$ as a function of $\MH$.  The additional
uncertainty of the $\SM$ prediction is parametric and dominated by the
uncertainties in $\dalhad$ and $\Mt$, shown as the bands.  The total
width of the band is the linear sum of these effects. }
\label{fig:coup:sef2} 
\end{center}
\end{figure}

The comparison of all $\swsqeffl$ values derived from asymmetry
measurements is shown in Figure~\ref{fig:coup:sef2}.  The average of
these six $\swsqeffl$ determinations is:
\begin{eqnarray}
\swsqeffl & = & 0.23153\pm0.00016\,,
\label{eq:coup:swsqeffl}
\end{eqnarray}
where the average has a $\chidf$ of 11.8/5, corresponding to a
probability of only 3.7\%. This large $\chidf$ value is caused by the
two most precise determinations of $\swsqeffl$, the measurements of
$\cAl$ by SLD, dominated by the $\ALRz$ result, and of $\Afbzb$ at
LEP. These two measurements fall on opposite sides of the average and
differ by 3.2 standard deviations. This is a consequence of the same
effect as discussed above: the deviation in b-quark couplings is
reflected in the value of $\swsqeffl$ extracted from $\Afbzb$.

\clearpage

\section{The W Boson}
\label{sec:ew:4f}

\subsection{W Bosons at Hadron Colliders}

\subsubsection{Production of W Bosons}

\begin{figure}[htbp]
\begin{center}
\includegraphics[width=0.8\linewidth]{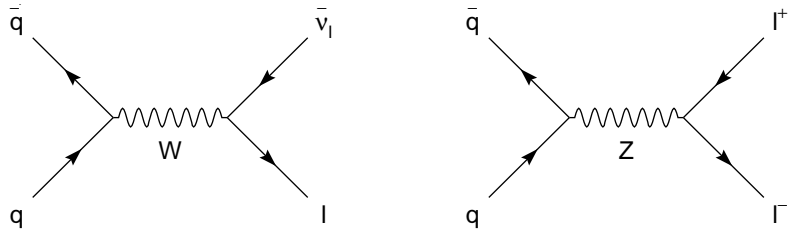}
\caption[Feynman diagrams of W and Z production and decay
in $\pp$ collisions] {Feynman diagrams of W and Z production
  and decay in $\pp$ collisions on parton level. }
\label{fig:feyn-ppsff}
\end{center}
\end{figure}

At hadron colliders, W and Z bosons are produced by quark-antiquark
fusion, $\pp\to\mathrm{W+X_W}$ and $\pp\to\mathrm{Z+X_Z}$, where
$\mathrm{X_V},~V=W,Z$, denotes the $\pp$ remnant recoiling against the
heavy intermediate vector boson V.  Owing to the hadronic event
environment and overwhelming background of QCD multi-jet production,
only events where the Z or W decay into leptons can be selected:
$\Wtolnu$ and $\mathrm{Z}\to\leptlept$.  The lowest-order Feynman
diagrams on parton level for W and Z production with subsequent
leptonic decay are shown in Figure~\ref{fig:feyn-ppsff}.

In 1983, the W and Z bosons were discovered~\cite{\Wdisco,\Zdisco} in
$\pp$ collisions by the experiments UA1 and UA2 taking data at CERN's
SPS collider at a centre-of-mass energy between $0.5~\TeV$ and
$0.6~\TeV$.  In total, a few hundred events were eventually observed.
The experiments CDF and D\O, taking data at the Tevatron accelerator
providing $\pp$ collisions with a centre-of-mass energy of $1.8~\TeV$
in Run-I, have observed several 10,000 leptonic W and Z decays using
the complete Run-I data of about $100\pb$ for each experiment.

\subsubsection{Determination of W-Boson Mass and Width}

Owing to the neutrino in W decays, and loss of remnant particles along
the beam axis, the kinematics of the W boson can not be fully
reconstructed.  However, the kinematics is closed when observed in the
plane transverse to the beam axis and in case the transverse part of
the hadronic recoil X is measured.  For this reason, the W-boson mass
and width analyses exploit kinematic variables measured in the
transverse plane only: (i) $E^{\ell}_T$, the transverse energy of the
charged lepton, (ii) $E^{\nu}_T$, the missing transverse energy -
solely given by the transverse energy of the neutrino in the ideal
case, and (iii) $m^2_T$, the transverse mass of the reconstructed W
boson calculated as:
\begin{eqnarray}
m^2_T(\ell,\nu)
     &  =  & 2 E^{\ell}_T E^{\nu}_T\left(1-\cos\phi_{\ell\nu}\right) 
     ~ \le ~ m^2_{\inv}(\ell,\nu) ~ = ~ \MW^2\,,
\end{eqnarray}
where the angle $\phi_{\ell\nu}$ denotes the azimuthal opening angle
between the direction of the lepton momentum and the missing-energy
vector (neutrino) in the transverse plane.  The distributions of all
three quantities are affected by the mass and total decay width of the
W boson, allowing to measure these quantities.

The transverse mass is always smaller than the invariant mass of the
decaying W boson.  It approaches the invariant mass when the W boson
decay products lie in the transverse plane. This cutoff or Jacobian
peak allows the W mass to be extracted from the position of the upper
edge of the transverse mass distribution. The result is more precise
than that derived from the other distributions.  The transverse mass
distribution also facilitates the measurement of the total width of
the W boson.  A non-zero total decay width effectively leads to a
smearing, since the decaying W bosons have an invariant mass
distributed according to a Breit-Wigner functional form.  Since a
Breit-Wigner falls off more slowly than Gaussian resolution effects,
the number of W events with measured transverse mass above the nominal
cut-off of $\MW$ depends strongly on the total width of the W boson.

\begin{figure}[p]
\begin{center}
\includegraphics[width=0.51\linewidth]{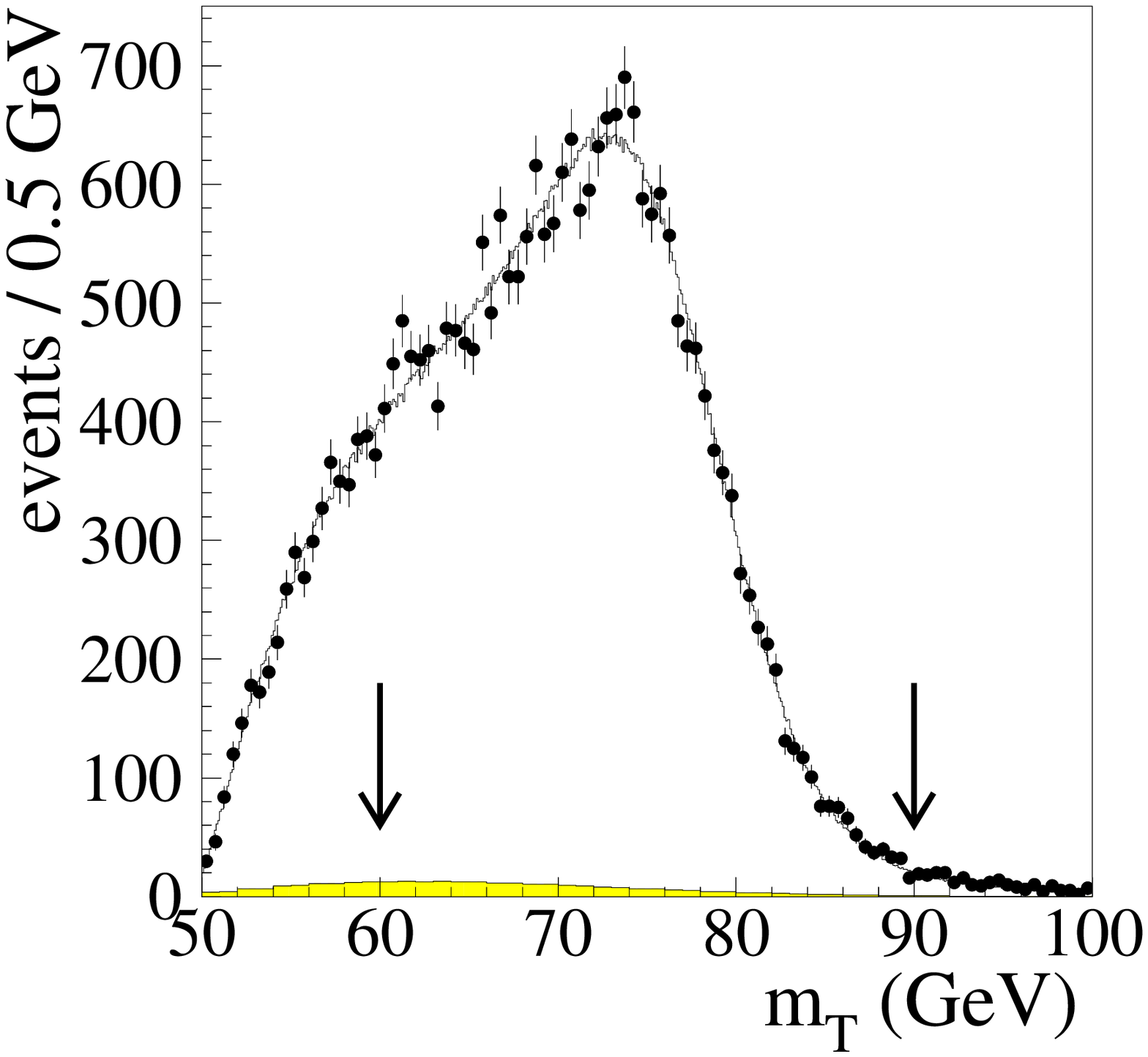}
\hfill
\includegraphics[width=0.48\linewidth]{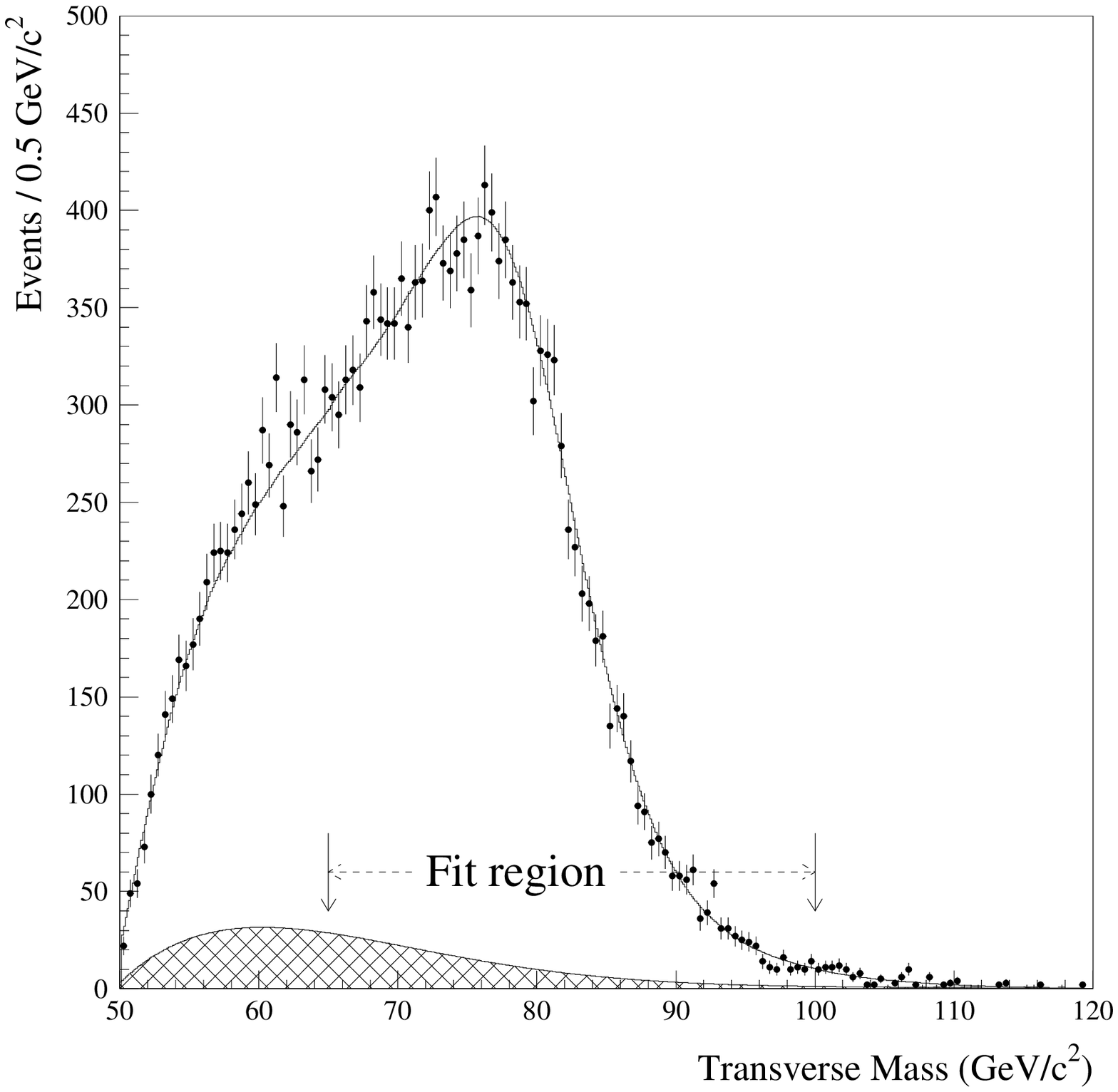}
\caption[Distributions of the transverse mass] {Distributions of the
reconstructed transverse mass from events collected in Run-I.  Left:
$\Wtoenu$ events from D\O; right: $\Wtomunu$ events from CDF.  The
solid line shows the fit result including background.  The arrows
indicate the fit region.  The shaded part denotes the background. }
\label{fig:exp-tev-mtw}
\end{center}
\end{figure}

The distributions of the reconstructed transverse mass for $\Wtoenu$
events from D\O~\cite{Abbott:1999uw,Abazov:2002bu} and $\Wtomunu$
events from CDF~\cite{Affolder:2000bp} are shown in
Figure~\ref{fig:exp-tev-mtw}.  D\O\ selects electrons in a range of
pseudo rapidity $|\eta|\le2.5$, while CDF selects electrons and muons
up to $|\eta|\le1$.  The region in transverse mass analysed for $\MW$
is extended around the Jacobian peak to include normalisation
information.  The results on $\MW$ obtained by CDF and D\O\ based on
all Run-I data are in good agreement: $\MW =
80.433\pm0.079~\GeV$~\cite{Affolder:2000bp} and $\MW=
80.483\pm0.084~\GeV$~\cite{Abbott:1999uw,Abazov:2002bu}, respectively.
The contributions to the total uncertainty of each measurement are
reported in Table~\ref{tab:mw-tev-syst}.  The dominant systematic
error arises from the calibration of the lepton energy scale.  Since
this calibration is derived from data analysing $\pi^0\to\gamma\gamma$
and $J/\Psi/\Upsilon/$Z$\to\leptlept$ decays, it will decrease along
with the statistical error of the measurements with the increased data
samples collected in Run-II, as shown by the very recent CDF result
obtained from 200/pb of Run-II data:
$\MW=80.413\pm0.048$~\cite{CDF2MW}.

\begin{figure}[p]
\begin{center}
\includegraphics[width=0.51\linewidth]{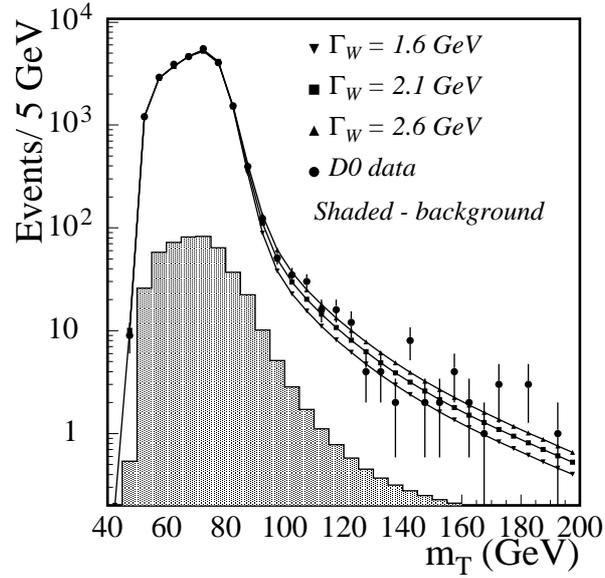}
\caption[Distribution of the transverse mass with tail] { Distribution
of the reconstructed transverse mass in $\Wtoenu$ events collected by
D\O\ in Run-II, showing predicted curves for various values of
$\GW$. The shaded part denotes the background. }
\label{fig:exp-tev-gtw}
\end{center}
\end{figure}

The principle of the width measurement is shown in
Figure~\ref{fig:exp-tev-gtw}. The integral of the high-end tail in the
transverse mass distribution depends on $\GW$.  The width results from
Run-I data are $\GW = 2.05\pm0.13~\GeV$ (CDF~\cite{GW-CDF-Ia,
Affolder:2000mt}) and $\GW=2.231^{+0.175}_{-0.170}~\GeV$
(D\O~\cite{Abazov:2002xj}).  The preliminary analysis from the D\O\
collaboration based on Run-II data and shown in
Figure~\ref{fig:exp-tev-gtw} yields $\GW =
2.011\pm0.142~\GeV$~\cite{GW-D0-II-0408}.  The contributions of
systematic uncertainties to the total uncertainty of each measurement
are listed in Table~\ref{tab:gw-tev-syst}.

The combination~\cite{PP-MW-GW:combination,PP-GW:2005-GW} of the mass
and width results from Run-I and Run-II yields the values:
\begin{eqnarray}
\MW = 80.429 \pm 0.039~\GeV\\
\GW =  2.078 \pm 0.087~\GeV\,.
\end{eqnarray}
Both results are dominated by statistics, and many systematic
uncertainties, for example energy calibration effects, will also
decrease with increased luminosity collected in Run-II.

\clearpage

\begin{table}[p]
\begin{center}
\caption[Systematic errors in the W mass measurements at the
Tevatron]{Systematic uncertainties in the measurements of the mass of
the W boson by the Tevatron experiments CDF and D\O in Run-I. All
numbers are in $\MeV$.  The first group denotes uncorrelated
uncertainties, the second group denotes uncertainties correlated
between measurements.  }
\label{tab:mw-tev-syst}
\renewcommand{\arraystretch}{1.25}
\begin{tabular}{|l||ccc|}
\hline
Source                    & D\O\ e & CDF e & CDF $\mu$ \\
\hline
\hline
W statistics              & 60 & 65 & 100 \\
Lepton energy scale       & 56 & 75 &  85 \\
Lepton energy resolution  & 19 & 25 &  20 \\
W transverse momentum     & 15 & 15 &  20 \\
Recoil model              & 35 & 37 &  35 \\
Selection                 & 12 & -- &  18 \\
Background                &  9 &  5 &  25 \\
\hline
PDF and parton luminosity & 7$\oplus$4 & \multicolumn{2}{c|}{15}\\
Radiative corrections     &         12 & \multicolumn{2}{c|}{11}\\
$\GW$                     &         10 & \multicolumn{2}{c|}{10}\\
\hline
\end{tabular}
\end{center}
\end{table}

\begin{table}[p]
\begin{center}
\renewcommand{\arraystretch}{1.30}
\caption[Input measurements]{Summary of the five measurements of $\GW$
performed by CDF and D\O. All numbers are in $\MeV$. For the
systematic uncertainties, the first group denotes uncorrelated
uncertainties and the second group denotes uncertainties correlated
between measurements. }
\label{tab:gw-tev-syst}
\begin{tabular}{|l||rrr|r||r|}
\hline       
       & \multicolumn{4}{|c||}{{Run-I}} & \multicolumn{1}{|c|}{{Run-II}} \\
\hline
       & \multicolumn{3}{|c|}{ CDF } & \multicolumn{1}{|c||}{ D\O\ } & \multicolumn{1}{|c|}{ D\O\ } \\
\hline       
       &  $e$ (Ia) &  $e$ (Ib) & $\mu$&  $e$ & $e$ \\
\hline 			       
\hline 			       
Result & 2110 & 2175 & 1780 & 2231 & 2011 \\
\hline                         
\hline                         
Stat.  &  280 &  125 &  195 &  142 &   93 \\
\hline                         
\hline                         
E-scale&   42 &   20 &   15 &   42 &   23 \\
Non-lin&    - &   60 &    5 &    - &    - \\
Recoil &  103 &   60 &   90 &   59 &   80 \\
$p_T(W)$& 127 &   55 &   70 &   12 &   29 \\
BG     &   17 &   30 &   50 &   42 &    3 \\
DM, $\ell$-ID  
       &    - &   30 &   40 &   10 &   16 \\
Resol. &   13 &   10 &   20 &   27 &   51 \\
\hline
PDF    &   15 &   15 &   15 &   39 &   27 \\
QED RC &   28 &   10 &   10 &   10 &    3 \\
$\MW$  &   10 &   10 &   10 &   15 &   15 \\
\hline                         
Syst.  &  173 &  114 &  135 &   99 &  108 \\
\hline                         
\hline                         
Total  &  329 &  169 &  237 &  173 &  142 \\
\hline
\end{tabular}
\end{center}
\end{table}

\clearpage

\subsection{W Bosons at LEP-II}

\subsubsection{Production of W Bosons}

\begin{figure}[htbp]
\begin{center}
\includegraphics[width=0.8\linewidth]{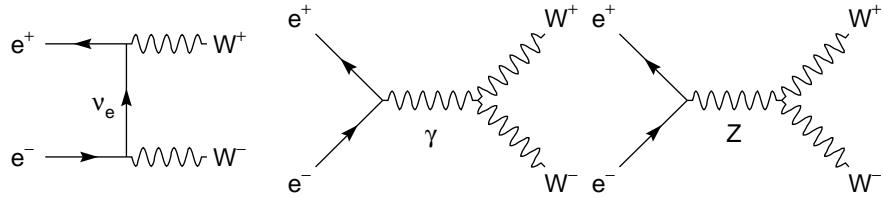}
\caption[Feynman diagrams of W-pair production in $\ee$
collisions]{Feynman diagrams of W-pair production in $\ee$
collisions. The contribution of Higgs boson exchange diagrams is
negligible.}
\label{fig:feyn-cc03}
\end{center}
\end{figure}

The dominant production mode for W bosons in the \LEPII\ centre of
mass energy range of $130~\GeV$ to $209~\GeV$ is W-pair production,
$\eeWW$.  The phase space for the production of on-shell W-boson pairs
opens up at the kinematic threshold of $\sqrt{s}=2\MW$.  The
lowest-order Feynman diagrams contributing to this process are shown
in Figure~\ref{fig:feyn-cc03}. Besides the $t$-channel neutrino
exchange diagram, two other interesting diagrams contribute, namely
the $s$-channel diagrams involving the triple-gauge-boson vertex
between the neutral and charged gauge bosons of the electroweak
interaction. The W boson decays via the charged weak current into a
fermion-antifermion pair, 3 leptonic modes ($\enu$, $\munu$, $\taunu$)
and 6 hadronic modes ($\qq'$ except top), so that W-pair production
leads to four-fermion final states.

\begin{figure}[htbp]
\includegraphics[width=\linewidth]{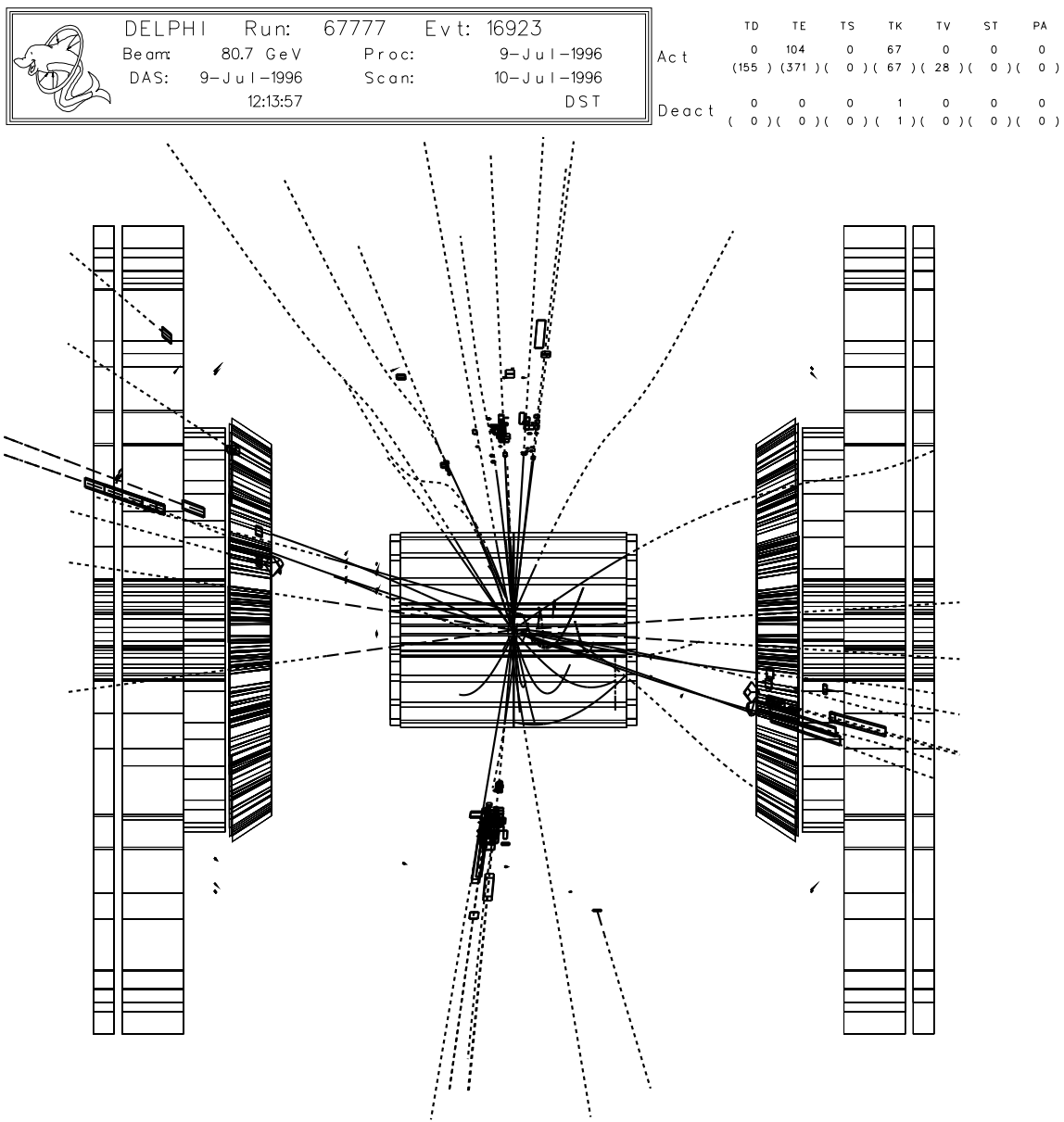}
\caption[Selected W-pair events]{W-pair events selected at \LEPII:
$\WWtoqqqq$ event observed in the DELPHI detector, showing
four well separated jets.}
\label{fig:exp-lep2-ww-events-1}
\end{figure}

\begin{figure}[htbp]
\includegraphics[width=\linewidth]{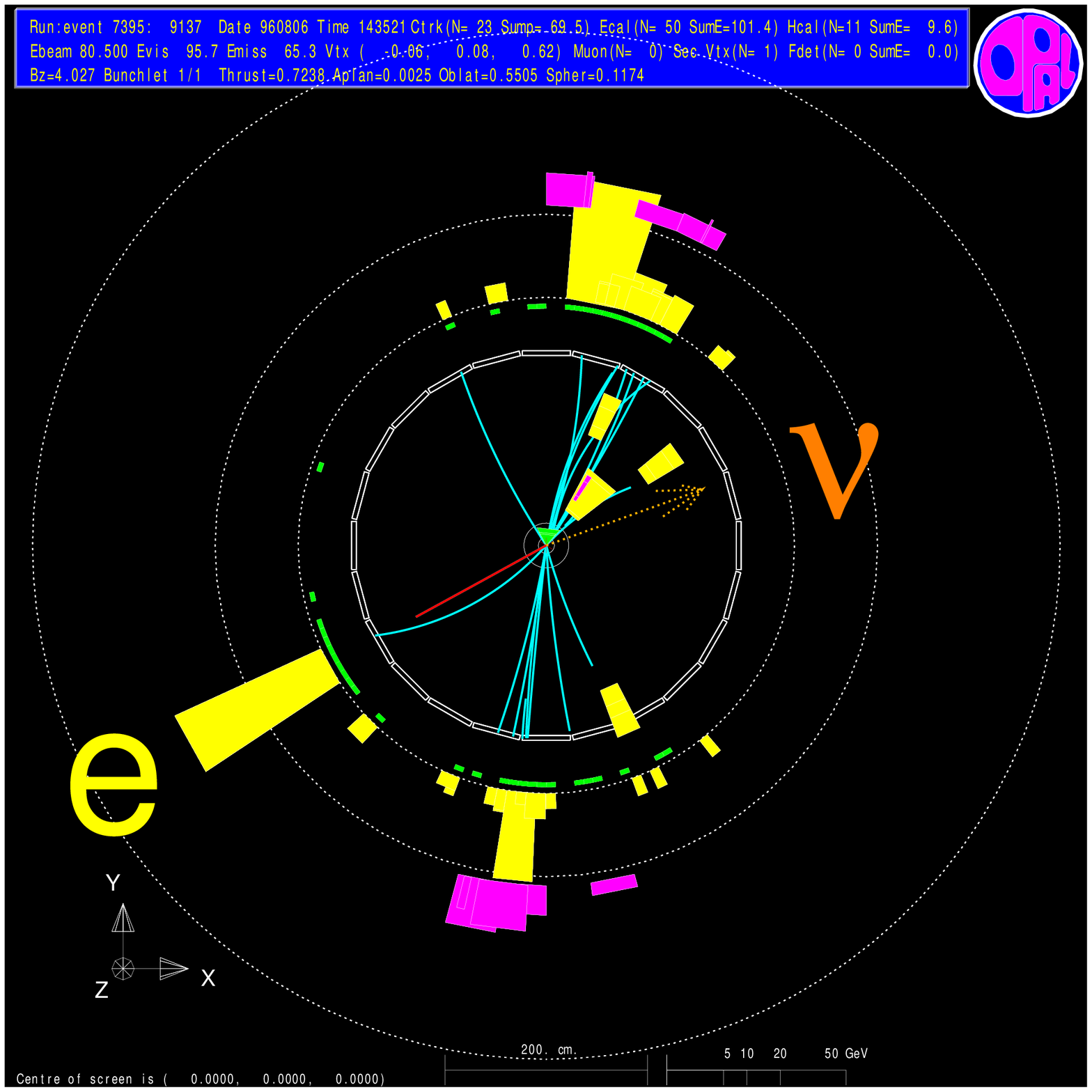}
\caption[Selected W-pair events]{W-pair events selected at \LEPII:
$\WW\to\qq\enu$ event observed in
the OPAL detector, showing two hadronic jets and an electron.}
\label{fig:exp-lep2-ww-events-2}
\end{figure}

\begin{figure}[htbp]
\includegraphics[height=\linewidth,angle=270,clip=true]{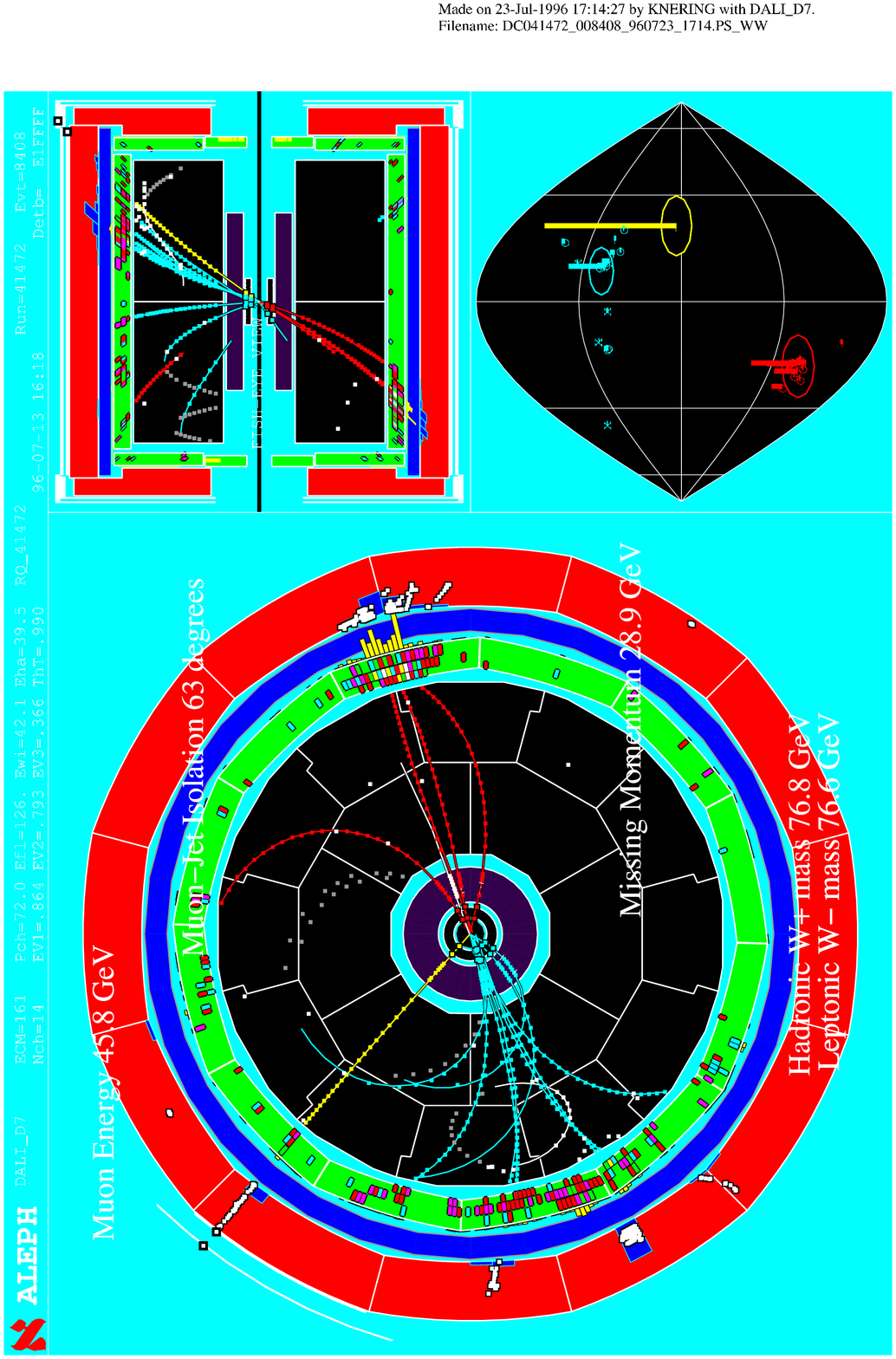}
\caption[Selected W-pair events]{W-pair events selected at \LEPII:
$\WW\to\qq\munu$ event observed in the ALEPH detector. }
\label{fig:exp-lep2-ww-events-3}
\end{figure}

\begin{figure}[htbp]
\includegraphics[width=\linewidth]{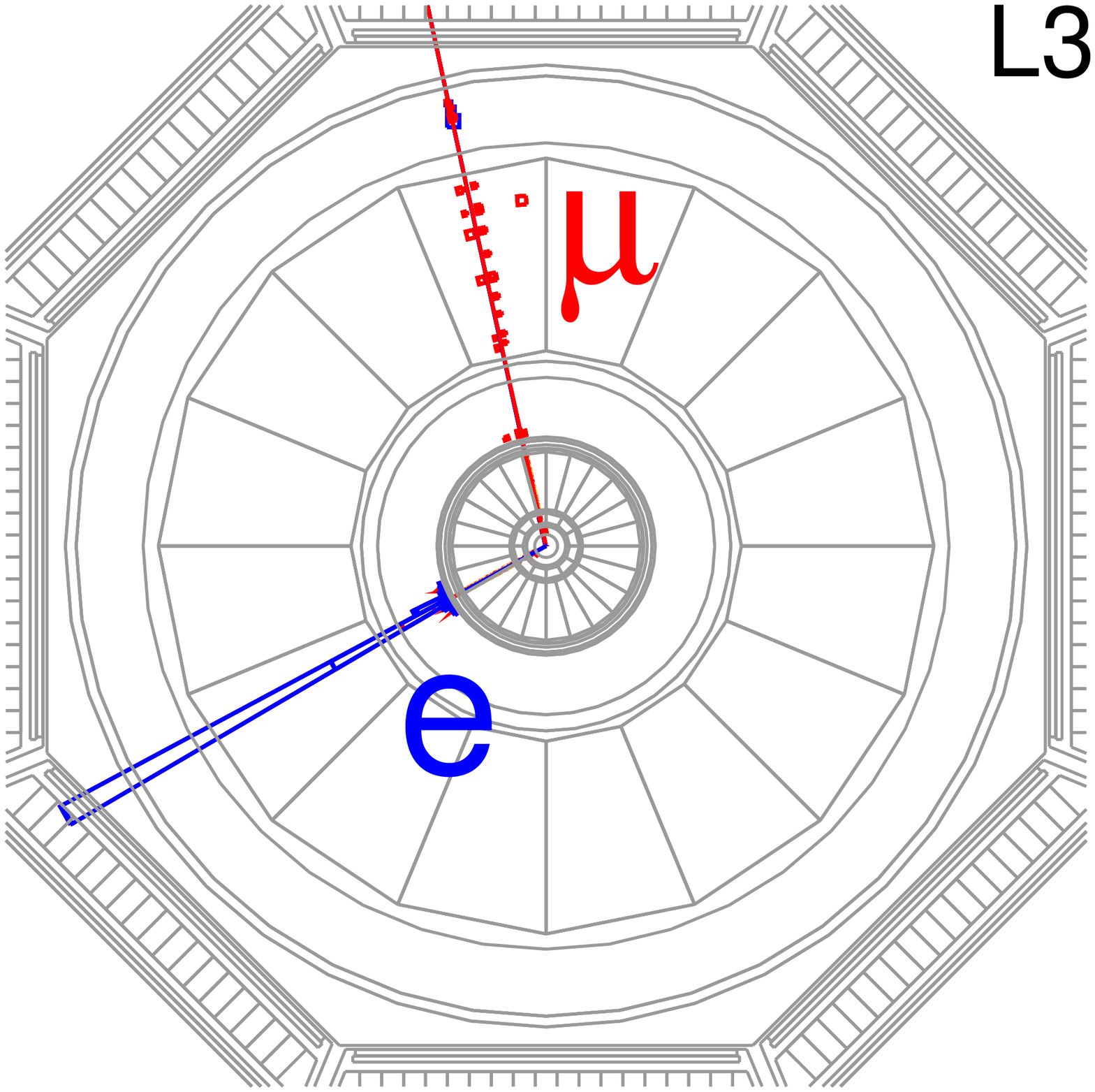}
\caption[Selected W-pair events]{W-pair events selected at \LEPII:
$\WW\to\enu\munu$ event observed in the L3 detector.}
\label{fig:exp-lep2-ww-events-4}
\end{figure}

Pair production of W bosons at \LEPII\ exhibits a clean signature in
the detectors since the events contain only the decay products of the
two W bosons.  Typical events of each type are shown in
Figures~\ref{fig:exp-lep2-ww-events-1}
to~\ref{fig:exp-lep2-ww-events-4}.  The LEP experiments analyse all
possible decay modes and event signatures. Hadronic events,
$\WWtoqqqq$ are selected with typical efficiencies of 85\% and
purities of 80\%. Events must contain at least four well separated
jets and no large missing energy. The dominant background arises from
QCD multi-jet production. Leptonic events, $\WWtolvlv$ are selected by
requiring two acoplanar charged leptons, which rejects the main
background arising from dilepton production, $\ee\to\leptlept$.
Efficiencies ranging from 30\% to 70\% and purities ranging from 75\%
to 90\% are achieved, where the lower values are obtained if both
leptons are $\tau$ leptons and the higher values if both leptons are
electrons or muons.  Semi-leptonic events, $\WWtoqqlv$, are tagged by
the presence of a high-energy charged lepton. In addition, events must
contain two hadronic jets and missing energy due to the neutrino.  The
main background arises through inclusive lepton production in $\qq$
events, where the missing energy is given by initial-state radiative
photons lost in the beam pipe.  Semi-leptonic events are selected with
an efficiency between 30\% and 90\% and a purity between 70\% and
95\%, where the lower values are obtained for $\qq\taunu$ events and
the higher values for $\qq\enu$ and $\qq\munu$ events.

\begin{figure}[htbp]
\begin{center}
\includegraphics[width=0.495\linewidth]{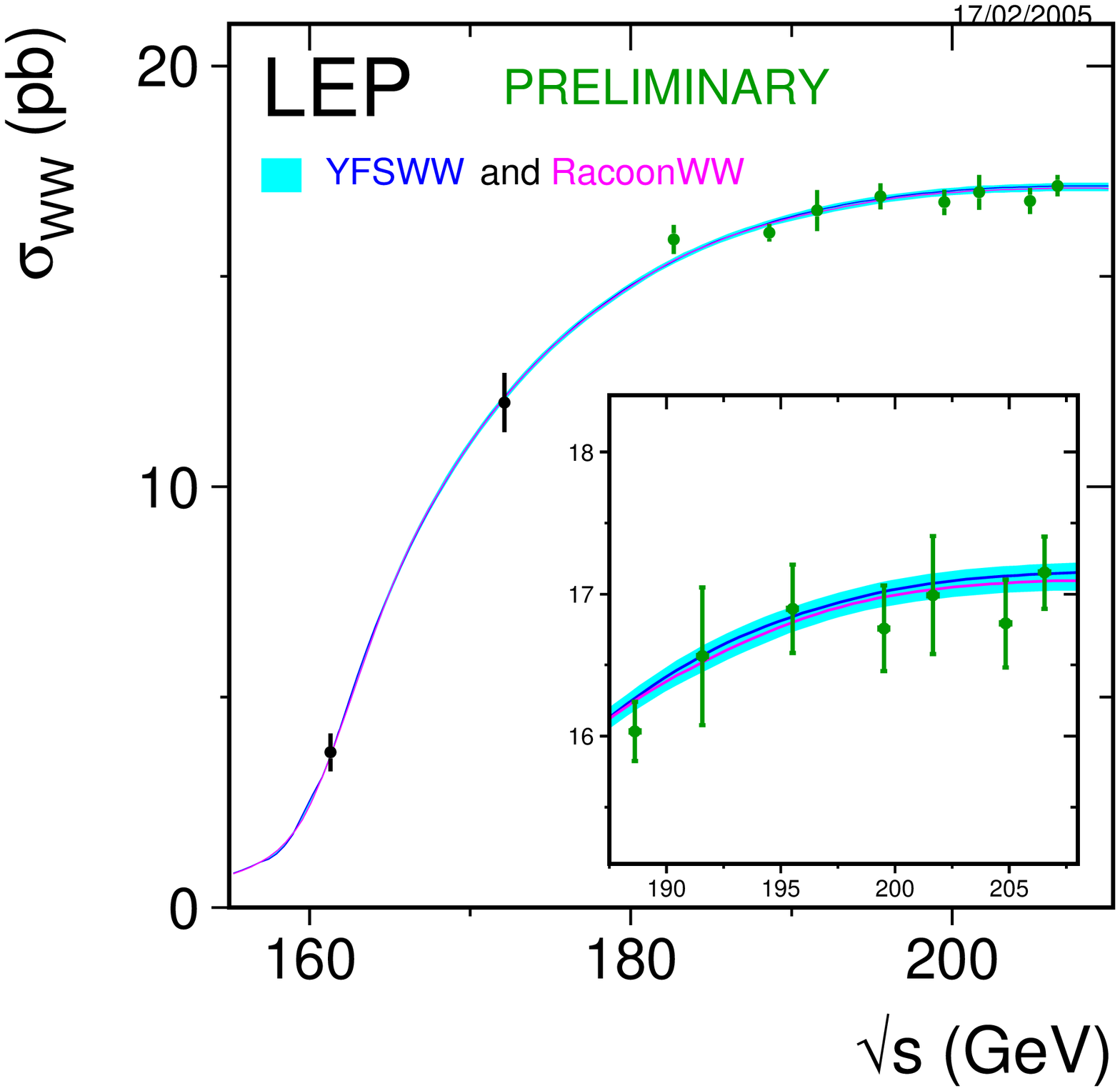}
\includegraphics[width=0.495\linewidth]{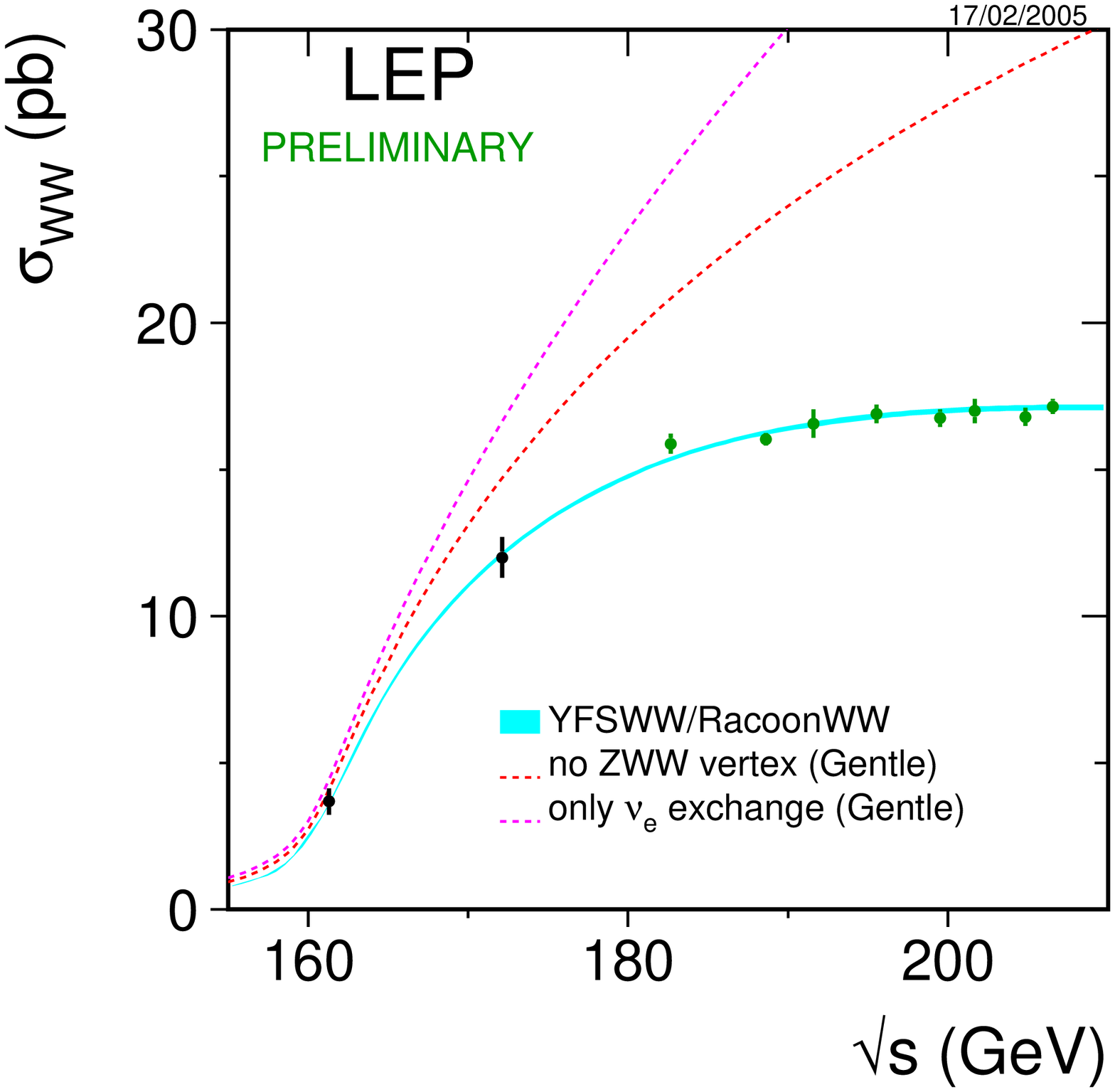}
\caption[W-pair cross sections measured at LEP]{Combined W-pair cross
sections, shown as dots with error bars, as measured by the four LEP
experiments as a function of the centre-of-mass energy, $\sqrt{s}$.
The results are preliminary.  Left: Measured cross sections
compared to the $\SM$ calculation. The theoretical uncertainty is
shown as the width of the curve.  Right: Cross sectionc expected for
theories without the $\gamma$WW and ZWW gauge vertices. The
measurements clearly require the presence of these vertices with their
expected properties.}
\label{fig:exp-lep-ww-xsec}
\end{center}
\end{figure}

The measurement of production cross sections for each final state
allows the simultaneous determination of the total W-pair production
cross section and the branching fractions of W decay. The total W-pair
cross section measured as a function of the centre-of-mass energy,
$\sqrt{s}$, is shown in Figure~\ref{fig:exp-lep-ww-xsec}, having
combined the results of the four LEP experiments at each $\sqrt{s}$
point~\cite{bib-EWEP-07}.  Especially at higher energies, the
influence of the $s$-channel Feynman diagrams is clearly visible,
establishing qualitatively as well as quantitativley the $\gamma$WW
and ZWW triple-gauge-boson couplings of the electroweak interaction.
Combining all energy points, the ratio of measurement to theory is
determined to be $0.995\pm0.009$, a successfull precision test of the
theory~\cite{bib-EWEP-07}.


\clearpage

\begin{figure}[ht]
\begin{center}
\includegraphics[width=0.495\linewidth]{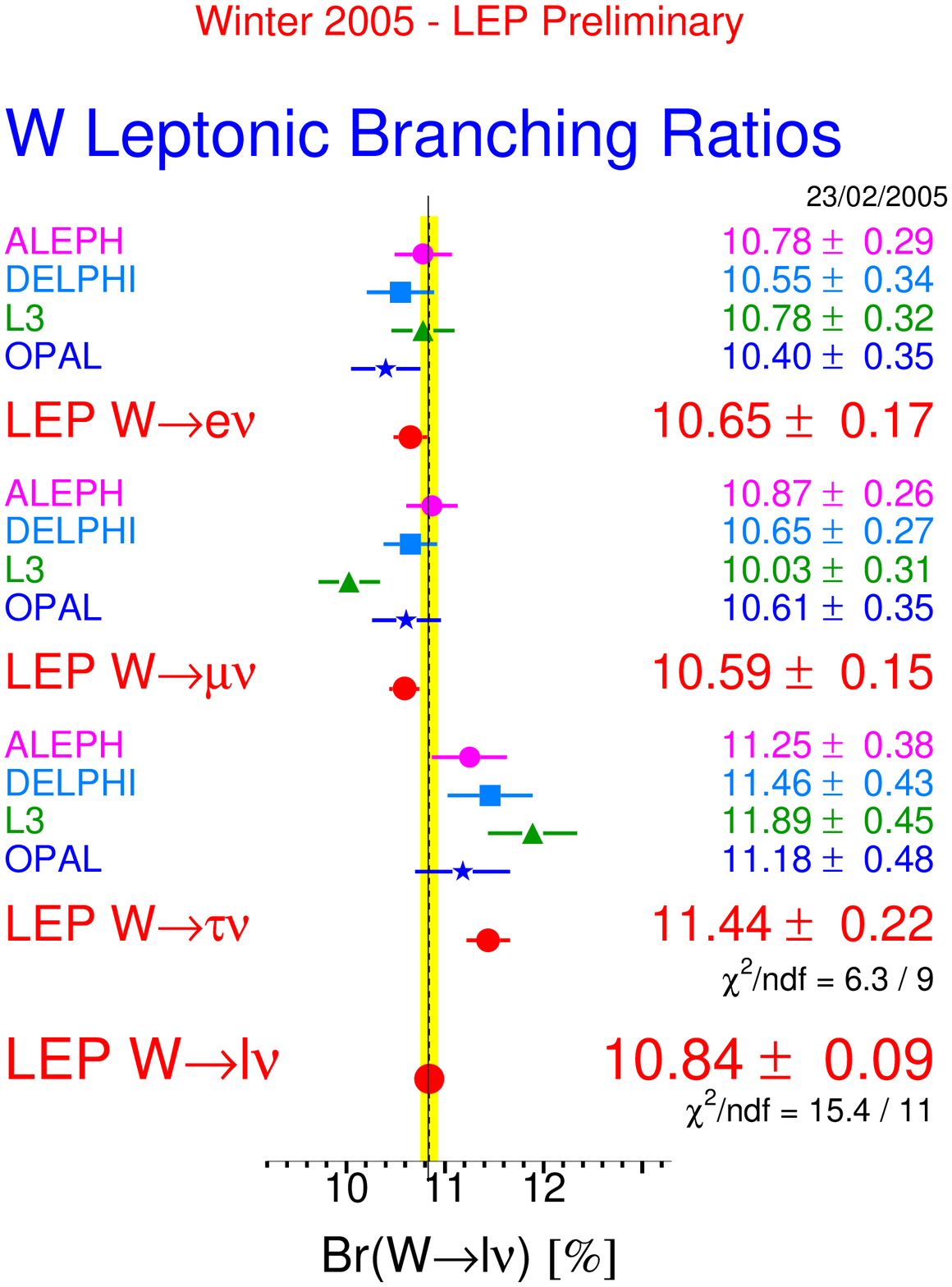}
\includegraphics[width=0.495\linewidth]{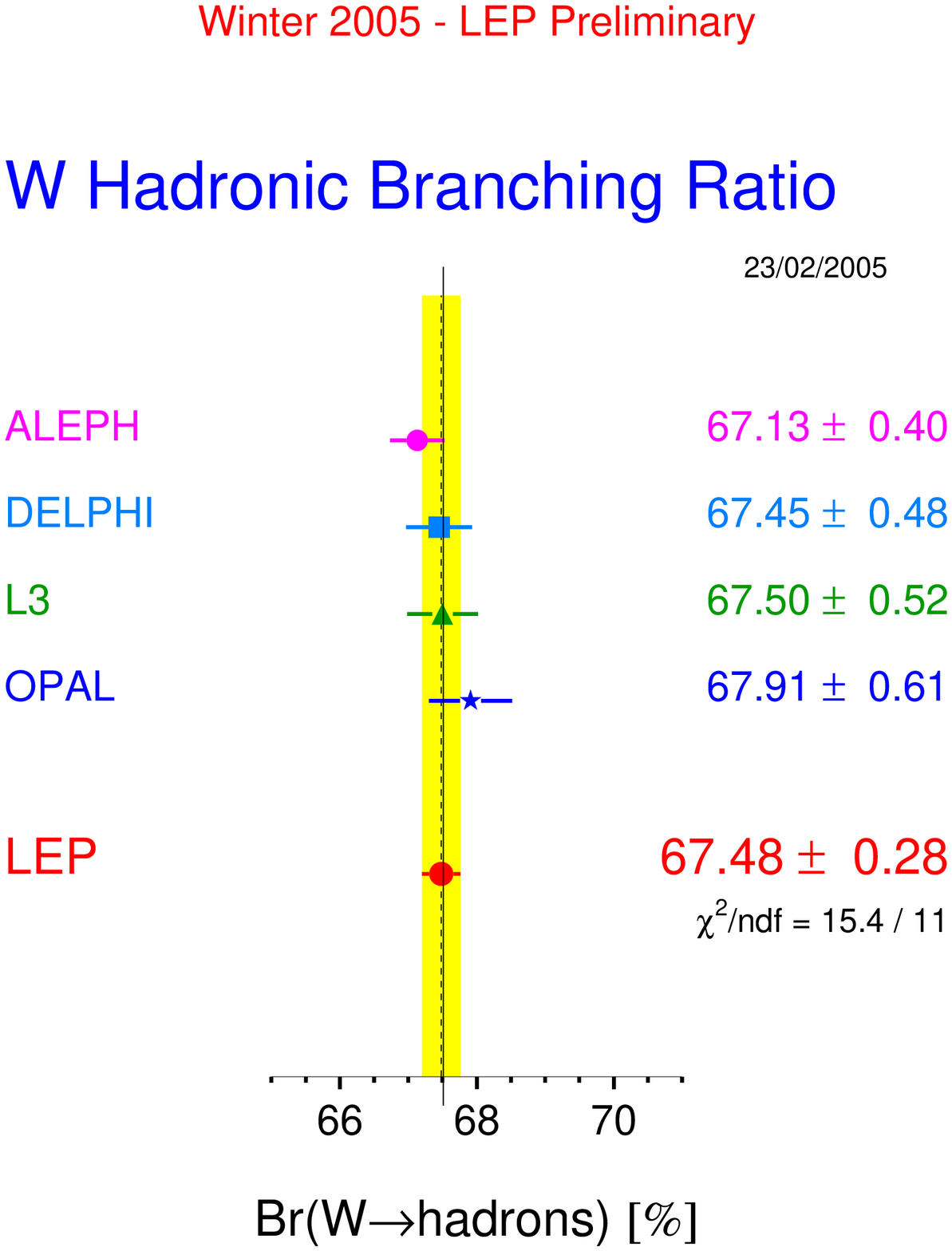}
\caption[W-decay branching fractions measured at LEP]{ W-decay
branching fractions measured by the four LEP experiments and their
combination. The results are preliminary. The sum of the branching
fractions is constrained to be unity. Left: branching fractions of
leptonic W decays. Right: branching fractions of the inclusive
hadronic W decay.}
\label{fig:exp-lep-ww-bf}
\end{center}
\end{figure}

The comparison of the branching fractions of W decay, determined for
the three leptonic modes and the inclusive hadronic mode, is shown in
Figure~\ref{fig:exp-lep-ww-bf}. In general, good agreement is observed
between the results of the four experiments. The combined
results~\cite{bib-EWEP-07} agree well with the $\SM$ expectation
except in the case of $\Wtotaunu$ decays, where all experiments
measured a branching fraction higher than expected. The combination
under assumption of lepton universality also agrees with the
expectation, because the electron and muon decay modes are measured to
be slightly smaller than expected. Comparing the tau decay mode to the
average of the electron and muon mode results, which maximises the
significance of the effect a posteriori, results in the two values
being different at the level of 2.8 standard deviations.

\subsubsection{Determination of W-Boson Mass and Width}

At the W-pair production threshold of $\sqrt{s}\simeq 161~\GeV$, the
cross section is sensitive to the mass of the W boson, used for the
first measurement of $\MW$ by the four LEP
experiments~\cite{mw:bib:A-mw161, mw:bib:D-mw161, mw:bib:L-mw161,
mw:bib:O-mw161}.  However, at higher centre-of-mass energies, the
cross section flattens out and becomes independent of $\MW$.  At these
energies, where the bulk of the \LEPII\ luminosity was collected, the
method of direct reconstruction is most sensitive in determining the
mass and also the total width of the W boson.

The four fermions arising in W-pair decay are reconstructed in the
detector, in particluar energy and direction, and paired to form two W
bosons. The invariant mass of each paired fermion system corresponds
to the invariant mass of the decaying W boson, and follows a
Breit-Wigner distribution convoluted with the detector resolution.  A
kinematic fit of the measured energies and angles, enforcing a
4-fermion final state, overall energy-momentum conservation as well as
equal mass of the two W bosons in an event, is performed.  Such a
procedure allows the determination of the unmeasured neutrino
kinematics in $\qq\lv$ events and improves the resolutions in the
kinematics of the measured fermions by a factor of up to four.

\begin{figure}[htbp]
\begin{center}
\includegraphics[width=0.495\linewidth]{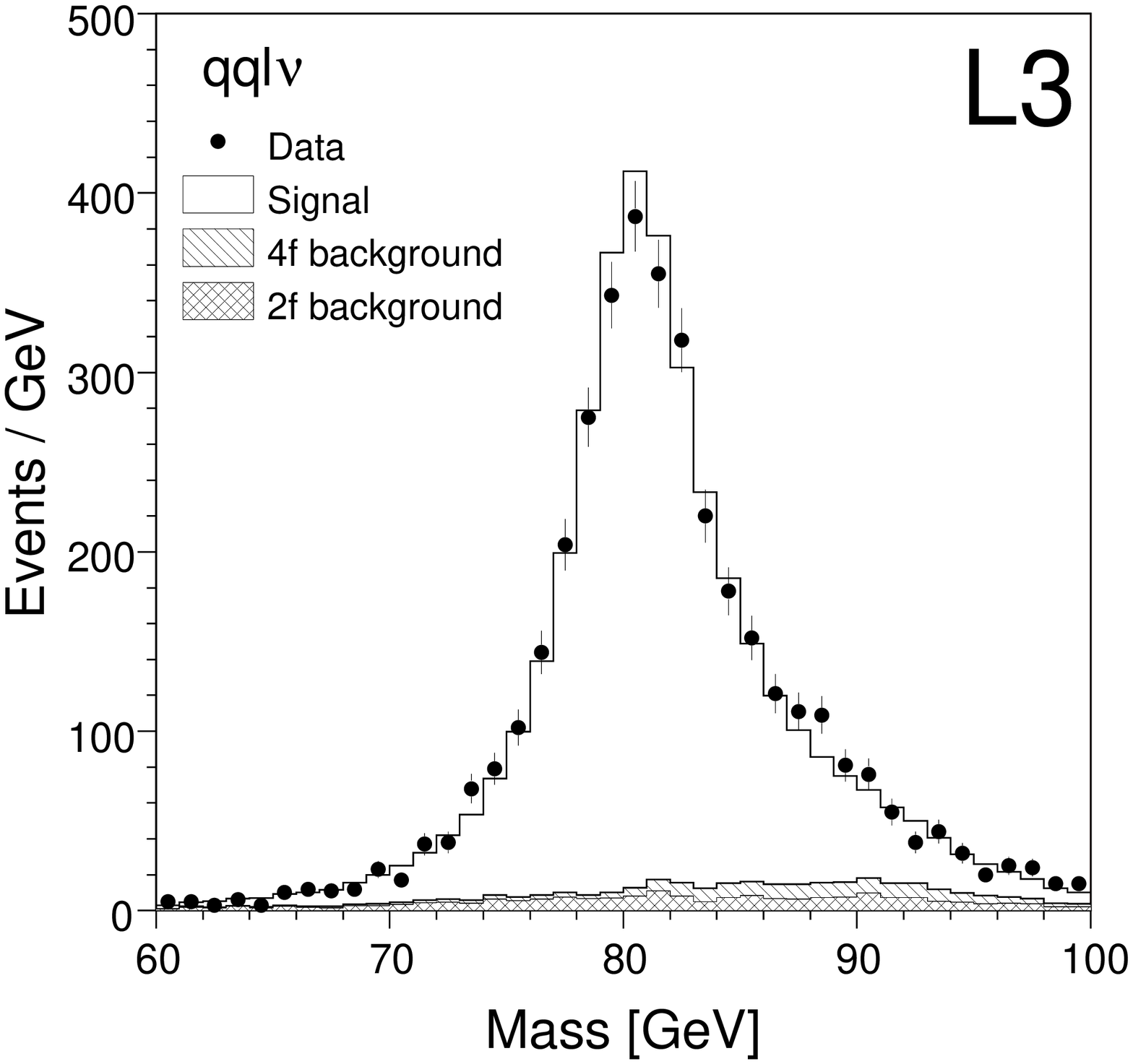}
\includegraphics[width=0.495\linewidth]{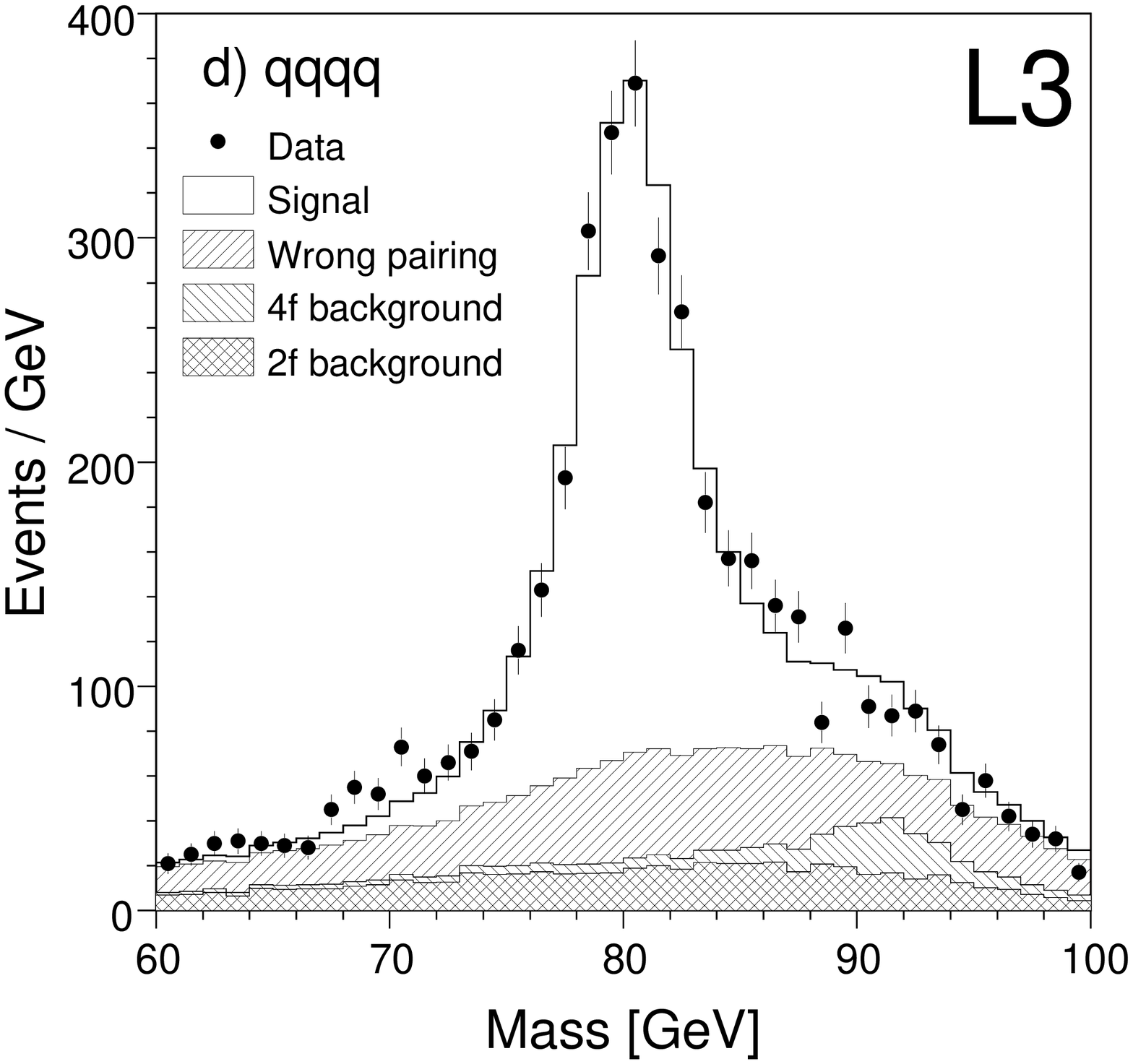}
\caption[Reconstructed invariant mass spectra]{Distribution of
reconstructed invariant masses from the L3 experiment. Left: $\qq\lv$
final states. Right: $\qq\qq$ final states.}
\label{fig:exp-lep-w-mass-rec}
\end{center}
\end{figure}

\begin{figure}[htbp]
\begin{center}
\includegraphics[width=0.495\textwidth]{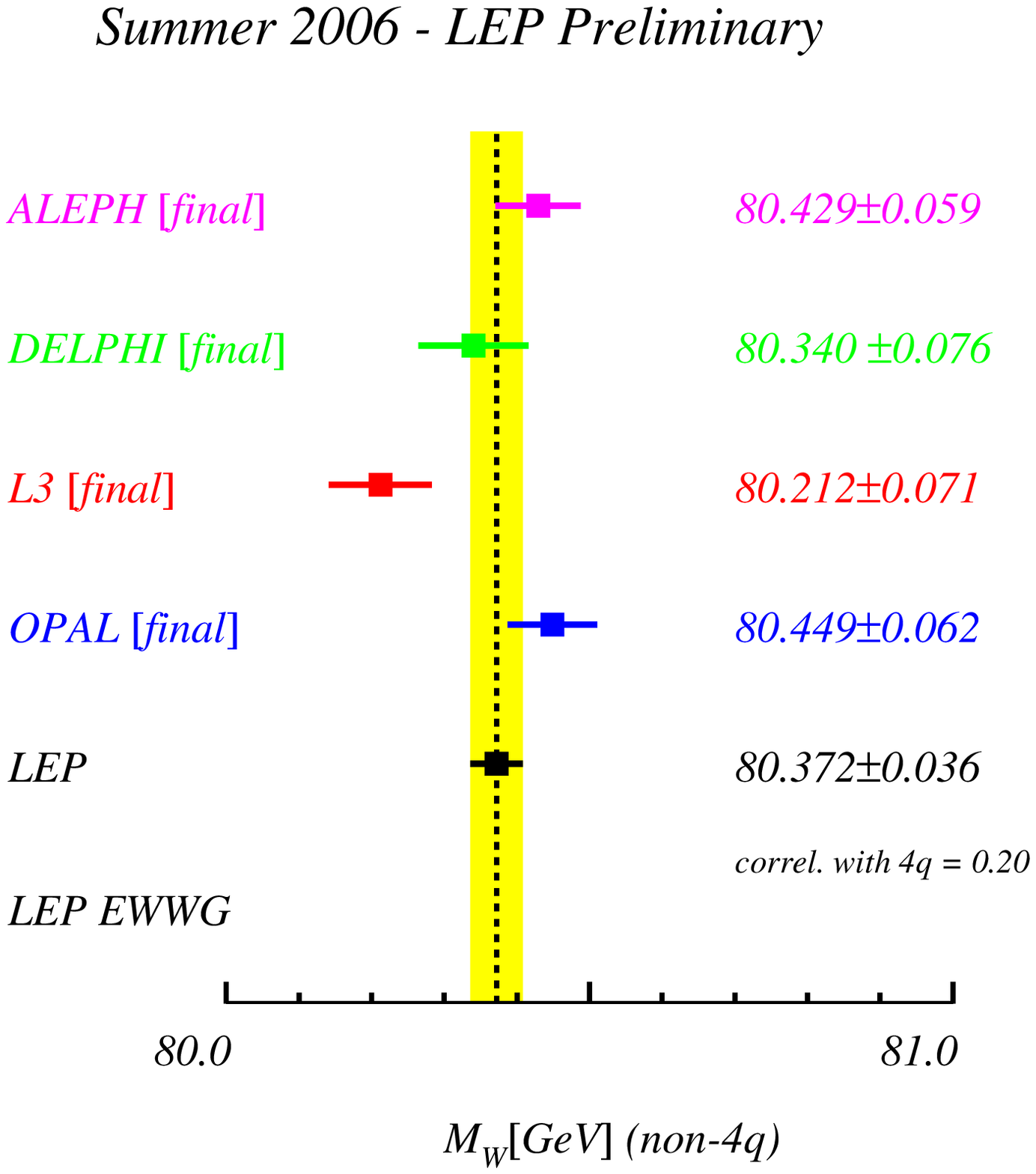}
\hfill
\includegraphics[width=0.495\textwidth]{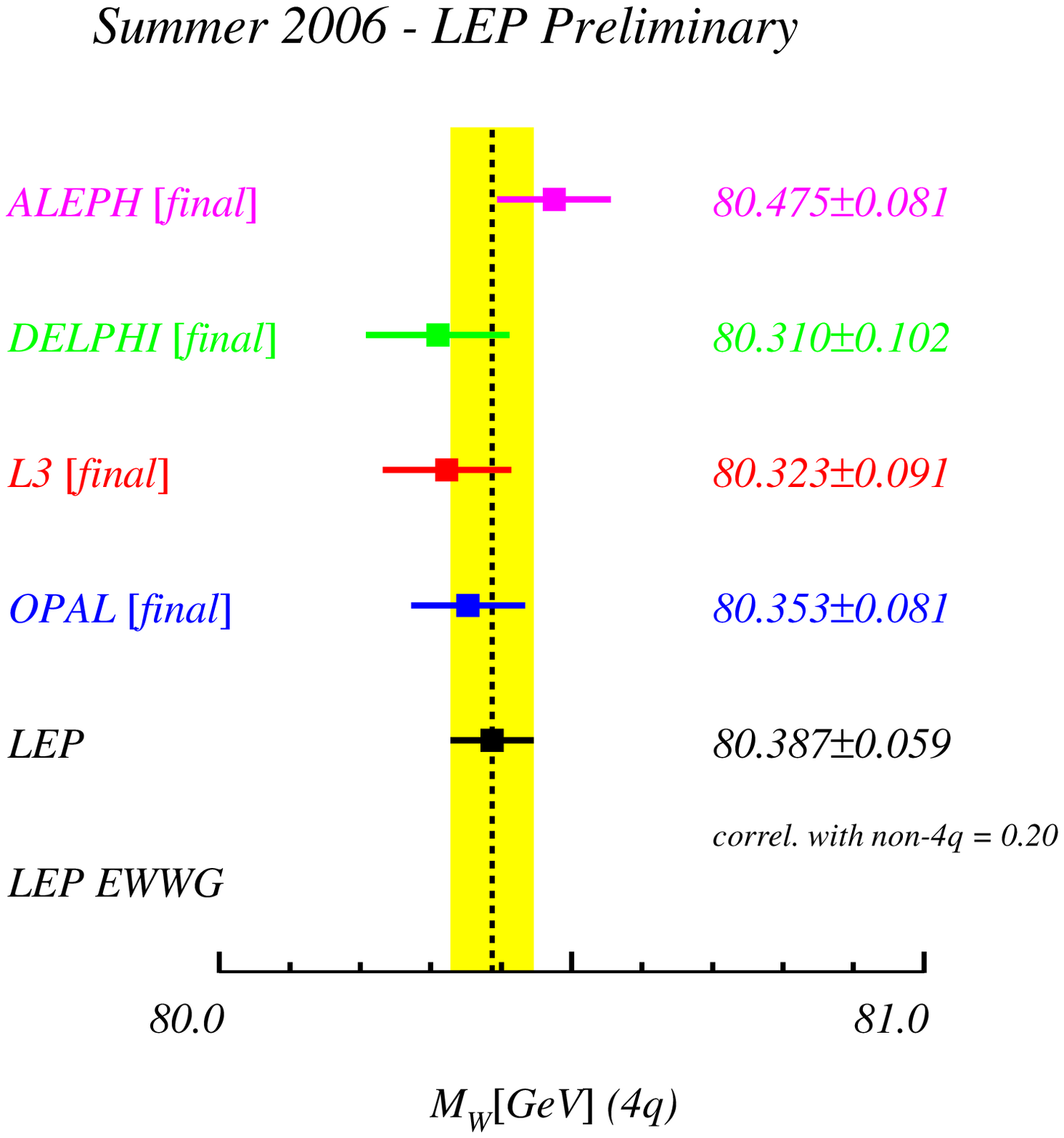}
\vskip -1cm
\caption{\label{mw:fig-qqlnqqqq-com} The W mass measurements from the
          $\WWtoqqlv$ (left) and $\WWtoqqqq$ (right) channels obtained
          by the four LEP collaborations, propagating the common LEP
          estimates of FSI effects to the mass. The combined values
          take into account correlations between experiments, years
          and the two channels.  The $\qq\lv$ and $\qq\qq$ results are
          correlated since they are obtained from a fit to both
          channels taking into account inter-channel correlations. The
          LEP combined values and common estimates for CR
          uncertainties are still preliminary.}
\end{center}
\end{figure}

\begin{figure}[htbp]
\begin{center}
\includegraphics[width=0.495\textwidth]{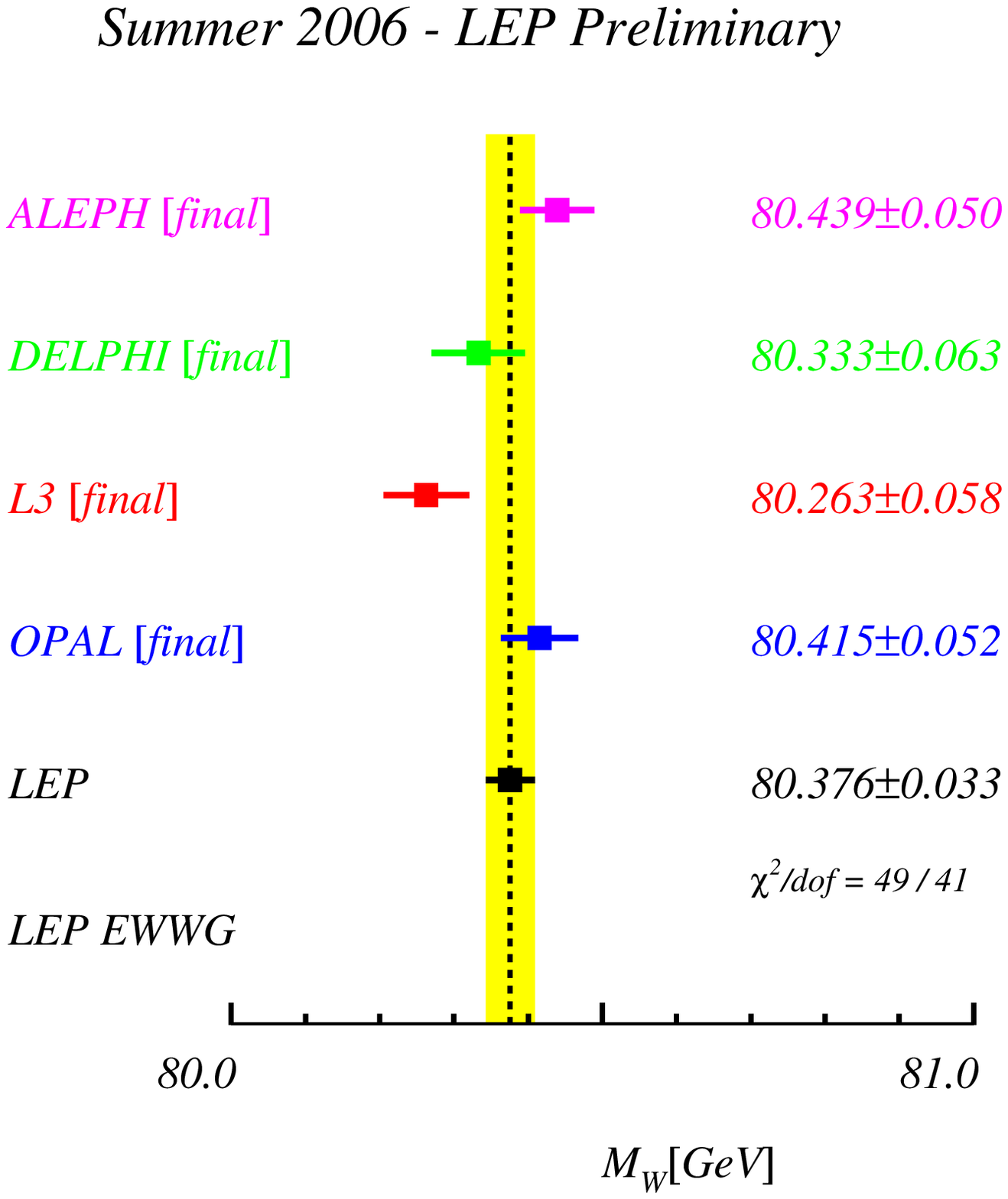}
\hfill
\includegraphics[width=0.495\textwidth]{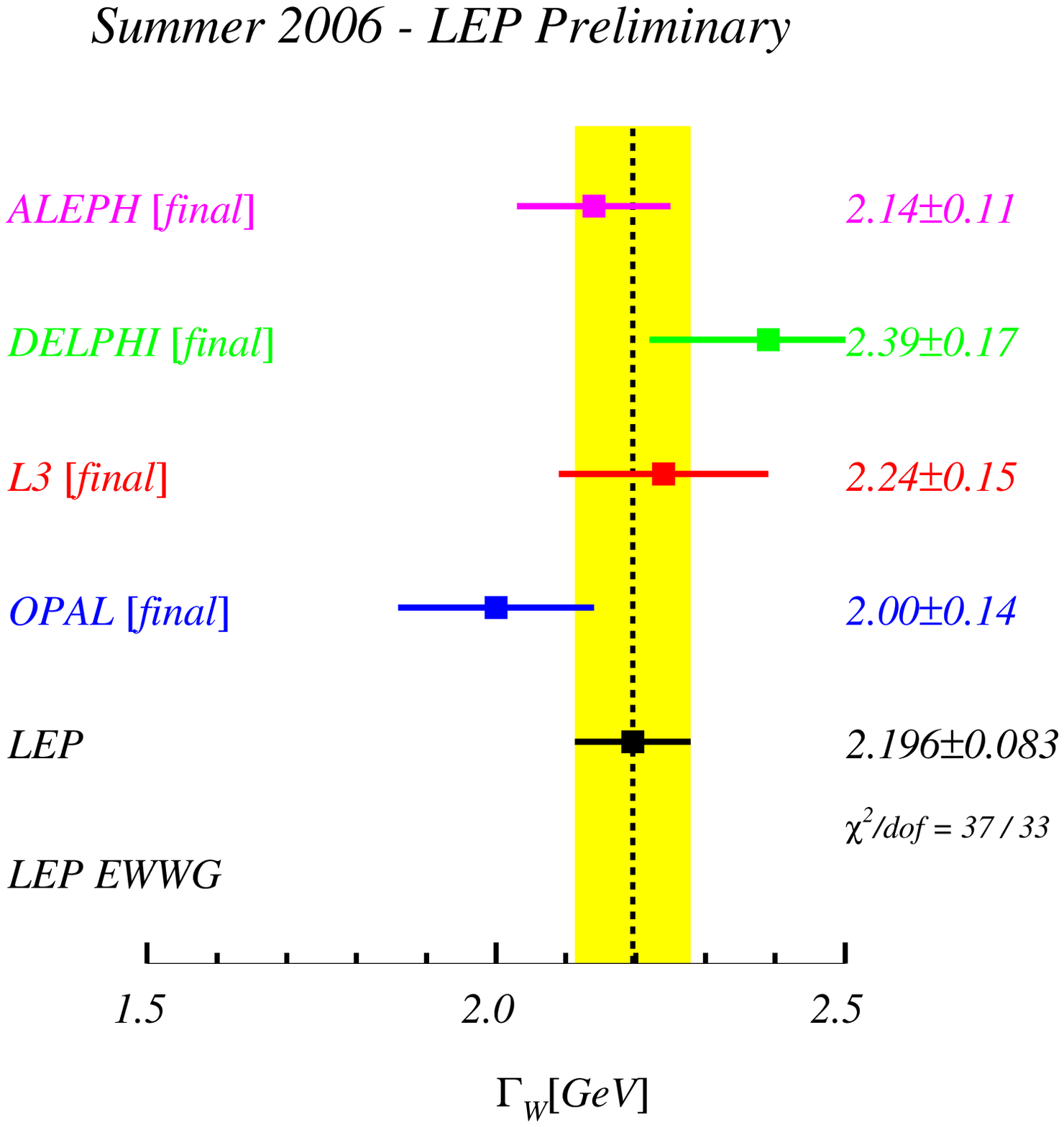}
\vskip -1cm
\caption{\label{mw:fig:mwgw-com} The combined results for the
          measurements of the W mass (left) and total W width (right)
          compared to the results obtained by the four LEP
          collaborations, propagating the common LEP estimates of FSI
          effects to mass and width (see text). The combined values
          take into account correlations between experiments and years
          and hence, in general, do not give the same central value as
          a simple average.  The individual and combined $\Mw$ results
          include the measurements from the threshold cross
          section. The LEP combined values and common estimates for CR
          uncertainties are still preliminary.}
\end{center}
\end{figure}

Spectra of reconstructed invariant masses are shown in
Figure~\ref{fig:exp-lep-w-mass-rec}.  The W-mass peak around $80~\GeV$
is clearly visible. Its width is given by the total decay width of the
W boson and the detector resolution. These distributions are analysed
by the four LEP experiments in order to extract the pole mass and
total decay width of the W boson~\cite{mw:bib:ALEPH, mw:bib:delphi,
mw:bib:mwl3,mw:bib:O-final}, where a definition of mass and width
using a Breit-Wigner with $s$-dependent width is used.  The results of
the experiments for the different channels and their
combinations~\cite{bib-EWEP-07} are compared in
Figures~\ref{mw:fig-qqlnqqqq-com} and~\ref{mw:fig:mwgw-com}.  The
combined W-mass values obtained in the $\qq\lv$ and $\qq\qq$ channels
are~\cite{bib-EWEP-07}:
\begin{eqnarray}
\MW(\qq\lv) & = & 80.372\pm0.030(stat.)\pm0.020(syst.)~\GeV\\
\MW(\qq\qq) & = & 80.387\pm0.040(stat.)\pm0.044(syst.)~\GeV\,,
\end{eqnarray}
where the two results show a correlation of 20\%.

\begin{table}[htbp]
\begin{center}
 \caption{Error decomposition for the combined LEP W mass results.
          Detector systematics include uncertainties in the jet and
          lepton energy scales and resolution. The `Other' category
          refers to errors, all of which are uncorrelated between
          experiments, arising from: simulation statistics, background
          estimation, four-fermion treatment, fitting method and event
          selection. }
 \label{mw:tab:errors}
\begin{tabular}{|l|r|r||r|}\hline
       Source  &  \multicolumn{3}{|c|}{Systematic Error on \Mw\ ($\MeV$)}  \\  
                             &  \qq\lv & \qq\qq  & Combined  \\ \hline   
 Initial-/final-state radiation         &  8 &  5 &  7 \\
 Hadronisation                          & 13 & 19 & 14 \\
 Detector Systematics             & 10 &  8 & 10\\
 LEP Beam Energy                 & 9 & 9 & 9 \\
 Colour Reconnection            & $-$& 35 & 8 \\
 Bose-Einstein Correlations  & $-$& 7 &  2 \\
 Other                                        &  3 &  11 & 4 \\ \hline
 Total Systematic                    & 21 & 44 & 22 \\ \hline
 Statistical                                & 30 & 40 & 25 \\ \hline\hline
 Total                                        & 36& 59 & 33 \\ \hline
  & & & \\
 Statistical in absence of Systematics  & 30 & 27 & 20 \\ \hline

\end{tabular}
\end{center}
\end{table}

Systematic uncertainties affecting the mass measurements are compared
in Table~\ref{mw:tab:errors}.  The largest systematic uncertainty
affecting both channels arises from the uncertainty in the modeling of
quark fragmentation and hadronisation.  In addition, the hadronic
channel suffers from additional sources of uncertainty, which are
related to the fact that the two hadronic W-boson decays strongly
overlap in space-time.  Thus, cross talk between the two hadronic
system arising through the strong interaction may lead to
four-momentum exchange between them so that the invariant masses of
the visible decay products measured in the detector can not be
attributed uniquely to one of the two decaying W bosons.  Such FSI
effects can arise due to (i) Bose-Einstein correlations (BE) between
identical neutral hadrons, mainly pions, or (ii) colour-reconnection
effects (CR) in the perturbative or non-perturbative evolution of the
$\qq$ systems originating from the two W bosons.  Dedicated studies
are performed to analyse in detail particle correlations and the
particle and energy flow in $\qq\qq$ events between jets to search for
BE and CR effects.

In case of BE, the effects within a jet are measured to be at the same
strength as observed between jets from Z boson decay, excluding b
quarks as these are suppressed in W decays. Of relevance for the W
mass, however, is the strength of inter-W BE. These are measured by
comparing genuine $\WWtoqqqq$ with two combined $\WWtoqqlv$ events
where the leptons are removed. The mixed events properly contain the
intra-W correlations while inter-W correlations are absent. Hence any
difference to $\WWtoqqqq$ events must be attributed to inter-W
correlations. The analyses of the LEP experiments result in a combined
limit of 30\% in the data on the strength of a model describing such
BE correlations in the Monte Carlo simulation~\cite{bib-EWEP-07}.
This limit is turned into a systematic uncertainty on the mass of the
W boson using the same model.

A similar strategy is followed in limiting possible CR effects.
Again, for the W mass reconstruction, only inter-W effects are
relevant since intra-W effects do not change the invariant mass of a
hadronic system. To limit CR, the particle flow in $\WWtoqqqq$ between
jets from the same W boson is compared to that between jets from
different W bosons.  Interpreting the particle flow with the help of
CR models allows to constrain model parameters, for example limiting
the model parameter $k_i$ describing CR in the SK-I model to a value
of at most 2.13 combining the LEP analyses~\cite{bib-EWEP-07}.  The
consequences for the mass are then evaluated using the same SK-I
model, comparing $k_i=0$ with $k_i=2.13$.

The difference of the W-boson masses obtained in the two channels may
also indicate the presence of FSI effects. Excluding FSI related
uncertainties in its calculation, the mass difference is evaluated to
be $-12\pm45~\MeV$, showing no indication of any significant FSI
effects.

The combined results for mass and total decay width of the W boson
obtained by the four LEP experiments, including also the threshold
measurements~\cite{mw:bib:A-mw161, mw:bib:D-mw161, mw:bib:L-mw161,
mw:bib:O-mw161}, are~\cite{bib-EWEP-07}:
\begin{eqnarray}
\MW & = & 80.376\pm0.033~\GeV\\
\GW & = &  2.196\pm0.083~\GeV\,,
\end{eqnarray}
where the contributions of the statistical and systematic errors to
the total error are also reported in Table~\ref{mw:tab:errors}.  Owing
to the additional FSI errors in the $\qq\qq$ channel, the average is
dominated by the $\qq\lv$ channel, having a weight of 78\% in the
average.

\clearpage

\clearpage

\section{The Top Quark}
\label{sec:ew:6f}

\subsection{Top Quark Production}

\begin{figure}[htbp]
\begin{center}
\includegraphics[width=0.8\linewidth]{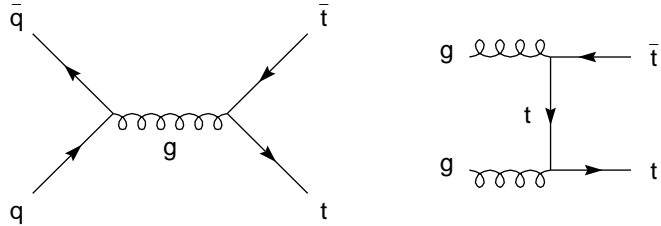}
\caption[Feynman diagrams of $\TT$ production in $\pp$ collisions]
{Feynman diagrams of $\TT$ production in $\pp$ collisions on parton
  level. }
\label{fig:feyn-pp2tt}
\end{center}
\end{figure}

Currently the Tevatron collider is the only accelerator in the world
powerful enought to produce top quarks.  Proton-antiproton collisions
produce $\TT$ pairs in the reaction $\pp\to\TT+X$ where $X$ denotes
the $\pp$ remnant recoiling against the $\TT$ system.  The
lowest-order Feynman diagrams on parton level contributing to this
process are QCD mediated processes as shown in
Figure~\ref{fig:feyn-pp2tt}.

For a $\pp$ centre-of-mass energy of $1.8~\TeV$ ($1.96~\TeV$), the
total $\TT$ production cross section is expected to be about 5.2~pb
(6.7~pb), with a theoretical uncertainty of about
10\%~\cite{Cacciari:2003fi,Kidonakis:2003qe}.  About 90\% (85\%) of
the cross section is due to the $\qq$ diagram.  The top quark
immediately decays via the charged-current reaction:
\begin{eqnarray}
\mathrm{t} & \rightarrow & \mathrm{b} \mathrm{W} \quad,\quad
\mathrm{W} ~ \rightarrow ~ \ff'~(\mathrm{f,f'}\ne\mathrm{t})\,,
\end{eqnarray}
since the Cabibbo-Kobayashi-Maskawa matrix element $\Vtb$ dominates.
Since $\Mt>\MW+\Mb$, the daughter W boson is on-shell. The total decay
width of the top quark is then $\Gamma_{\mathrm{t}}\propto\GF\Mt^3$,
for example $\Gamma_{\mathrm{t}}=1.4~\GeV$ for $\Mt=175~\GeV$. Thus
the top quark decays before there is enough time for it to combine
with other quarks to form top-flavoured bound states of hadrons.

The decay modes of a $\TT$ system are those of a W-pair accompanied by
a $\bb$ system. Hence, in analogy to W-pair production discussed in
electron-positron collisions, fully hadronic modes, semileptonic modes
and dilepton modes occur. The branching fractions for these final
states are given by the corresponding products of W branching
fractions:
\begin{eqnarray}
\TT & \rightarrow & \bb\,\WW 
    ~ \rightarrow ~ \bb\,(\ff')_1\,(\ff')_2 \\
     & \rightarrow &
\left\{
\begin{array}{ll}
  \bb \, \qq' \, \qq'                      & 
  ~~45.6\%                                  \\
  \bb \, \qq' \, \lv                       & 
  ~~14.6\%~\mathrm{each~for}~\ell=e,\mu,\tau       \\
  \bb \, \lv  \, \lv & 
  ~~10.6\%~\mathrm{for}~\ell=e,\mu,\tau~\mathrm{combined}
\end{array}
\right. \,.
\end{eqnarray}
Like in W and Z physics at the Tevatron, the main leptonic W decay
modes analysed are $\Wtoenu$ and $\Wtomunu$, including $\Wtotaunu$
where the tau decays to leptons. These leptons are easy to identify in
the hadronic event environment and constitute an important selection
tool; the more leptons, the higher the efficiency and purity.  The
common signature of all $\TT$ events consists of the $\bb$ system;
hence in several $\TT$ analyses b-tagging is explicitely used in the
event selection.  On the balance of branching fraction, efficiency and
purity of the selected sample, the semileptonic channel offers best
prospects for high precision measurements of top-quark properties.

Electroweak single-top production, a charged weak current process
mediated by W-boson echange, has an expected cross section only
slightly smaller than $\TT$ production, a strong process.  However,
single-top production is much harder to observe experimentally, as
only one heavy top quark is produced.  Only very recently, first
evidence for single-top production has been reported by D\O\ and also
found by CDF~\cite{Abazov:2006gd, CDF-Single-Top}.  For a robust and
precise mass measurement, large samples top quarks must be selected
with high efficiency and purity.  Hence, for the study of top-quark
properties such as its mass, $\TT$ events are used.

\subsection{Mass of the Top Quark}

In $\TT\to\bb\WW$ events where both W bosons decay leptonically,
$\TT\to\bb\lv\lv$, the two unmeasured neutrinos in the event prevent a
complete reconstruction of the decayed top quarks and hence a direct
determination of their invariant mass. However, the mass of the top
quark affects the kinematic properties of the measured decay products,
such as lepton energies, quark jet energies and jet-lepton angular
separations.  Analysing the distribution of these observables yields a
determination of $\Mt$ though with comparably large errors. While the
purity of the selected event sample is high, the low branching
fractions cause large statistical uncertainties.

In hadronic events, $\TT\to\bb\qq\qq$, all decay products are quarks,
leading to an event signature with six jets or more due to gluon
radiation and no unobserved partons. Hence the kinematics of the top
quarks can be reconstructed completely, up to combinatorial
ambiguities in associating light quark jets to the correct W boson,
and combining W bosons with b-quark jets to top quarks. The number of
combinations can be reduced by using b-tagging to identify the two b
jets among the six jets, and by a matrix-element based probability
assignment to find the most likely combination representing a $\TT$
decay. While the branching fraction is high, the lack of leptons to
reduce the QCD multi-jet background causes low efficiencies and
purities and hence limited precision in the mass determination.

In principle, semileptonic events, $\TT\to\bb\qq\lv$ also allow a
complete reconstruction of the invariant masses of the decaying top
quarks. However, it is necessary to consider the complete production
and decay chain including the hadronic system recoiling against the
$\TT$ system:
\begin{eqnarray}
\pp  & \rightarrow & \mathrm{ t_1 t_2 + X_t }   \nonumber\\
\mathrm{t_1} & \rightarrow & \mathrm{b_1 + W_1} \qquad
\mathrm{t_2} ~ \rightarrow ~ \mathrm{b_2 + W_2} \\
\mathrm{W_1} & \rightarrow & \ell + \nu         \qquad\quad
\mathrm{W_2} ~ \rightarrow ~ \mathrm{q_1 + q_2} \nonumber\,.
\end{eqnarray}
These five four-vector equations contain seven unknown four-momenta,
namely those of q$_1$, q$_2$, $\ell$, $\nu$, b$_1$, b$_2$, and
X$_{\mathrm{t}}$, corresponding to 28 unknowns. There are 17
measurements, namely the three-momenta of the five fermions
$\mathrm{q_1,q_2,\ell,b_1,b_2}$ and the two transverse momentum
components of the underlying event $X_t$. Overall four-momentum
conservation, using the known masses of
$\mathrm{\ell,\nu,q_1,q_2,b_1,b_2,W_1,W_2}$ and enforcing
$m_{\inv}(t_1)=m_{\inv}(t_2)$ introduce 13 constraints, leading to a
2C kinematic fit in the determination of all four-momenta.
Nevertheless, also here combinatorial ambiguities arise in associating
decay fermions to W bosons and top quarks. This channel combines high
branching fractions with a lepton in the final state to allow for
selection of event sampels with reasonable efficiency and purity.

\begin{figure}[tp]
\begin{center}
\includegraphics[width=0.495\textwidth]{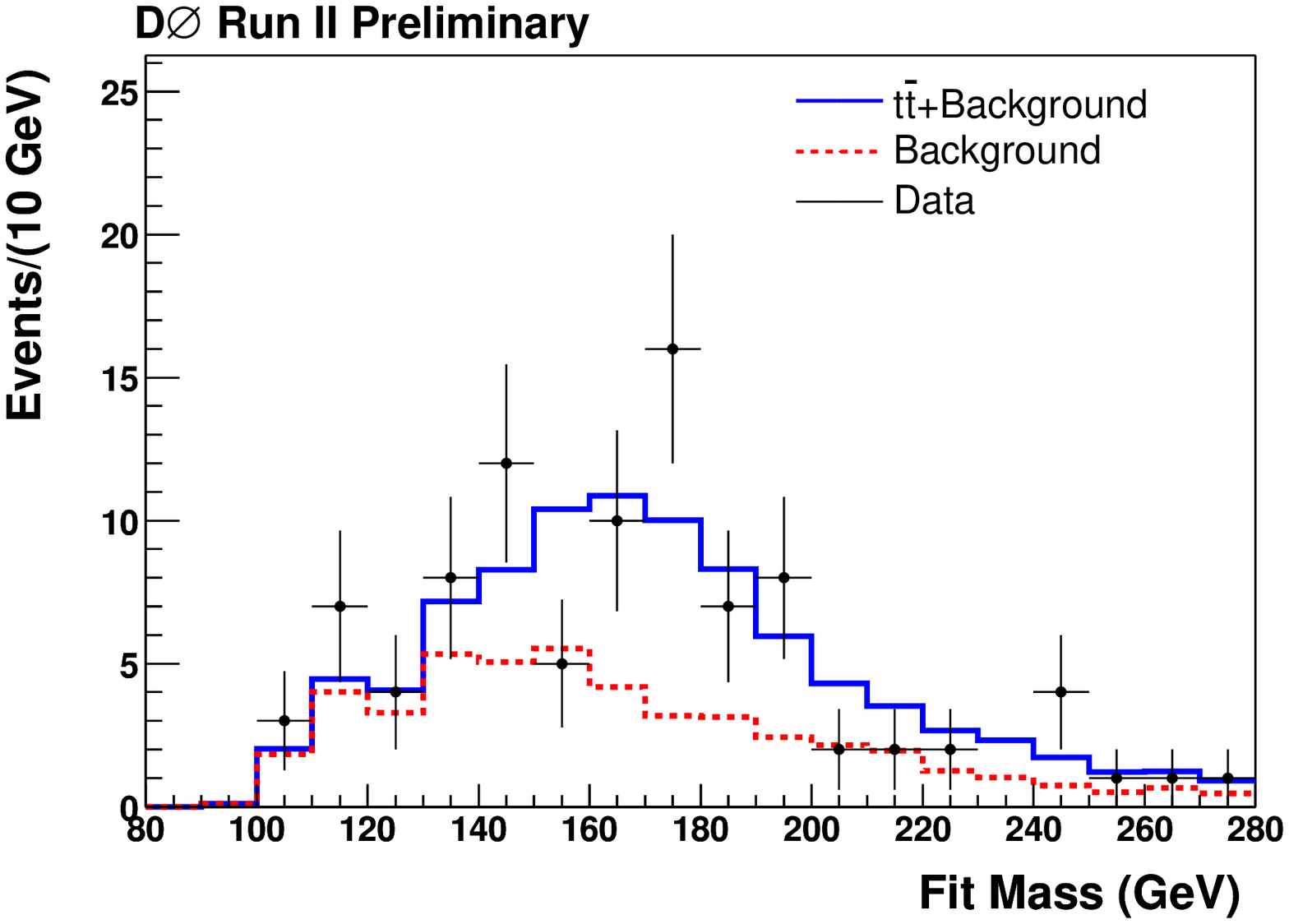}
\hfill
\includegraphics[width=0.495\textwidth]{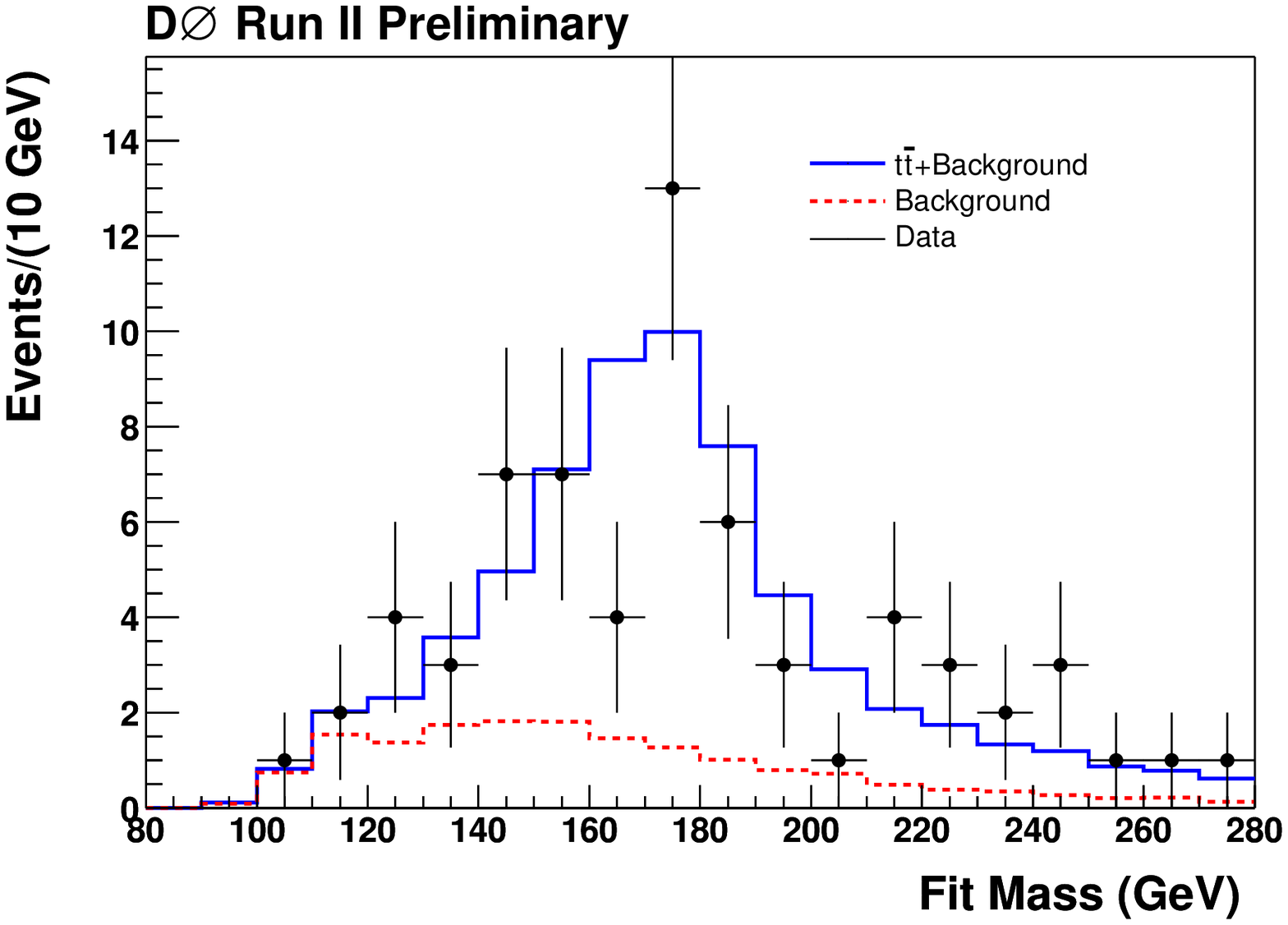}
\caption{\label{mw:fig:mtop-reco-d0} Distribution of top-quark masses
reconstructed in semi-leptonic $\TT$ events selected by D\O\ in a data
sample of 230/pb, without (left) and with (right) requiring at least
one jet tagged as a b-quark jet. The data is shown as dots with error
bars and is compared to the MC expectations shown as lines. }
\end{center}
\end{figure}

\begin{figure}[h]
\begin{center}
\vskip -0.5cm
\includegraphics[width=0.6\textwidth]{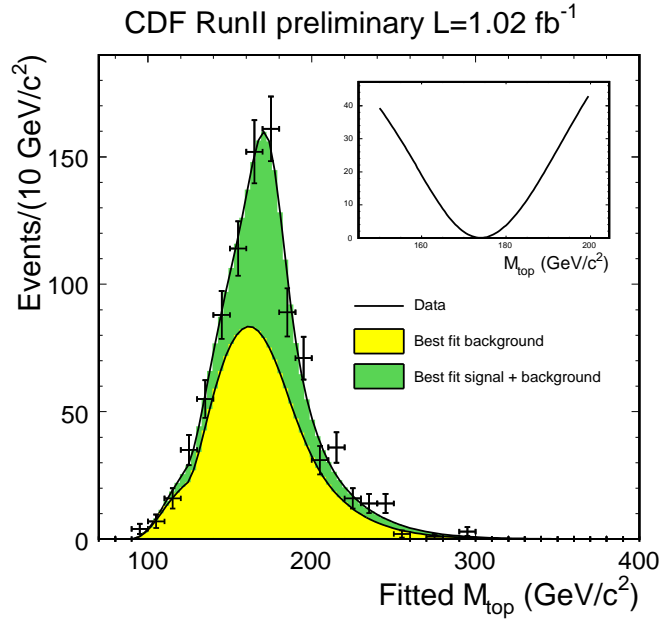}
\vskip -0.5cm
\caption{\label{mw:fig:mtop-reco-cdf} Distribution of top-quark masses
reconstructed in hadronic $\TT$ events selected by CDF in a data
sample of 1.02/fb, where at least one jet is tagged as b-quark
jet. The data is shown as dots with error bars and is compared to the
MC expectations shown as lines.}
\end{center}
\end{figure}

Example invariant mass distributions are shown in
Figure~\ref{mw:fig:mtop-reco-d0} from D\O\ in the semileptonic
channel~\cite{D0-Mtop-qqlv}, and in~Figure~\ref{mw:fig:mtop-reco-cdf}
from CDF in the hadronic channel~\cite{CDF-Mtop-qqqq}.
Figure~\ref{mw:fig:mtop-reco-d0} compares topologically selected $\TT$
events with a sample requiring in addition at least one jet tagged as
a b-jet.  Figure~\ref{mw:fig:mtop-reco-cdf} shows the invariant mass
distribution reconstructed in a sample of hadronic $\TT$ events,
having required a b-tag in the event.  In both cases, the b-tagging
requirements lead to a significantly higher purity and a more precise
mass measurement.

The mass analyses in the various final states are augmented by using
neural networks on pre-selected samples, and/or the $\TT$ matrix
element to assign probabilities, to resolve combinatorial ambiguities,
and to increase the sensitivity to $\Mt$ by exploiting the dependence
of the matrix element on $\Mt$.  Essentially, the matrix element is
evaluated using the reconstructed kinematics of the visible fermions
and the missing transverse energy as estimates for the fermion
kinematics; convolutions are used to account for resolution effects,
and unmeasured parameters are integrated out. The resulting
differential cross section is used as a relative probability estimate
to estimate correct pairings and the mass dependence.  The D\O\
collaboration pioneered the use of this computing-intensive approach
in the determination of the mass of the top quark in the semileptonic
channel~\cite{Mtop1-D0-l+j-new1}.

\begin{figure}[htbp]
\begin{center}
\includegraphics[width=0.56\textwidth]{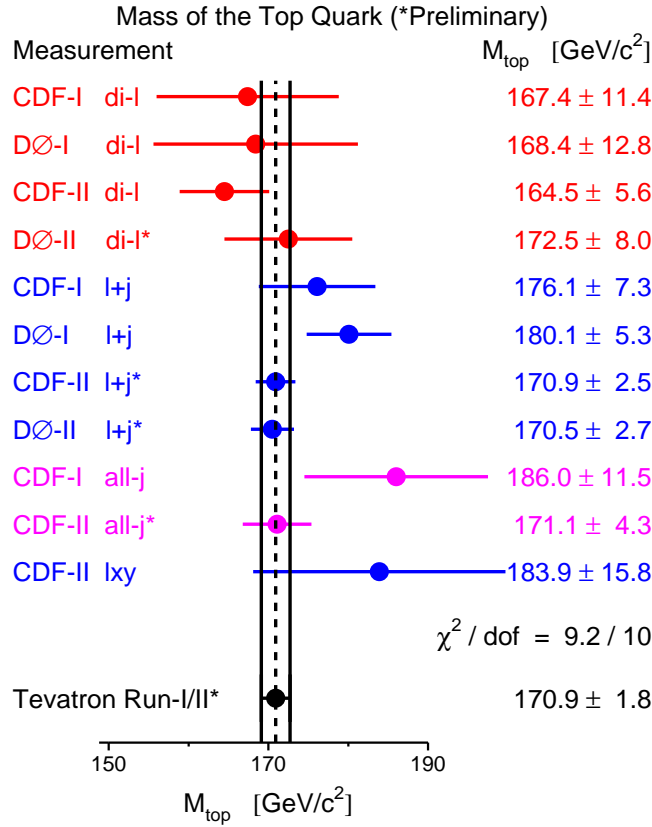}
\end{center}
\vskip -0.5cm
\caption[Summary plot for the world average top-quark mass]
  {Comparison of measurements of the top-quark mass from CDF and D\O\
  and the resulting world average mass of the top quark.}
\label{fig:mtop-results}
\end{figure}

The mass of the top quark has been measured by the Tevatron
experiments CDF and D\O\ in all channels, with various techniques, and
in data collected during Run-I and Run-II of the Tevatron.  For the
combination, only the most precise results of each experiment in the
various channels are used, namely
References~\cite{Mtop1-CDF-di-l-PRLb, Mtop1-CDF-di-l-PRLb-E,
Mtop1-D0-di-l-PRD, Mtop1-CDF-l+j-PRD, Mtop1-D0-l+j-new1,
Mtop1-CDF-all-j-PRL} for the final and published Run-I results, 
two published Run-II CDF 
results~\cite{Mtop2-CDF-di-l-1fbPRD, Mtop2-CDF-lxy-new}, two preliminary 
Run-II CDF results~\cite{Mtop2-CDF-l+j-new, Mtop2-CDF-all-j-new} and two
preliminary Run-II D\O\ results~\cite{Mtop2-D0-l+j-new,Mtop2-D0-di-l-new}.
A comparison of these results, as used in the most recent
combination, is shown in Figure~\ref{fig:mtop-results}.  Taking
correlated uncertainties into account, the combined result
is~\cite{Mtop-tevewwgWin07}:
\begin{eqnarray}
  \Mt & = & 170.9 \pm 1.8~\GeV\,.
\end{eqnarray}
The combination has a $\chi^2$ of 9.2 for 10 degrees of freedom,
which corresponds to a probability of 51\%, indicating good agreement
between the various measurements as also visible in
Figure~\ref{fig:mtop-results}.  

The total uncertainty contains a statistical part of $1.1~\GeV$ and a
systematic part of $1.5~\GeV$.  The dominant part of the systematic
error of each measurement is given by the jet-energy scale, i.e., the
energy calibration for hadronic jets.  This uncertainty alone
contributes $1.1~\GeV$ to the combined result.  Recent $\Mt$ analyses
use a W-mass constraint to fix the jet energy scale in situ, adjusting
the jet energies such that the mass of the hadronically decaying W
boson averages to the known W mass.  This technique fixes the jet
energy scale for light-quark jets occuring in W decays.  Only energy
scale of b-quark jets then remains to be calibrated relative to
light-quark jets.  Other important systematic uncertainties include
the modelling dependence of the simulated event samples for signal and
background used in the mass extraction procedure.  

With the advent of new mass analysis techniques and the larger
luminosity collected in Run-II, the mass of the top quark is now known
with an accuracy of 1.1\%, much improved compared to just a few years
ago. The prediction of $\Mt$ within the $\SM$ based on the analysis of
electroweak radiative corrections measured at the Z-pole is discussed
in Section~\ref{sec:msm:add:W}.

\clearpage

\section{Low-Energy Measurements}
\label {sec:lowQ2}

Effective coupling constants of the weak current are not only measured
in high-$Q^2$ interactions such as at the Z-pole ($Q^2\approx\MZ^2$),
but also in low-$Q^2$ processes, $Q^2\ll\MZ^2$.  Because of the
running of effective coupling constants with $Q^2$, the couplings
measured in these reactions are different from those measured at the Z
pole.  This running has to be accounted for before comparisons can be
made.

\subsection{Parity Violation in Atoms}

A parity violating effect occurs in atomic transitions due to the
parity-violating $t$-channel $\gamma$Z exchange between the shell
electron and the quarks in the atomic nucleus. Its strength is given
by the weak charge of the atomic nucleus as probed by the shell
electron, $\QW(Z,N) = -2[(2Z+N)C_{1\mathrm{u}}+(Z+2N)C_{1\mathrm{d}}]$
for a nucleus with $Z$ protons and $N$ neutrons. The weak charges
$C_{1\mathrm{q}}$ of up and down quarks can be expressed in terms of
effective vector and axial-vector coupling constants introduced
earlier, $C_{1\mathrm{q}}=2\gae\gvq$.

The most precise results have been obtained from measurements with
cesium~\cite{QWCs:exp:1,QWCs:exp:2}. Over the last years, the
corrections due to nuclear many-body perturbation theory and QED
radiative corrections needed in the experimental analyses have bee
revised~\cite{QWCs:theo:2003:new}.  The newly corrected experimental
results for cesium is:
$\QWCs=-72.74\pm0.46$~\cite{QWCs:theo:2003:new}.  This result is now
in good agreement with the $\SM$ expectation.

\subsection{Parity Violation in M\o ller Scattering}

The weak charge of the electron, $\QW(\mathrm{e})=-4\gae\gve$, is
measured by M\o ller scattering, $\emem$, using polarised beams. The
experiment was performed at an average momentum transfer of
$Q^2=0.026~\GeV^2$ by the E-158 collaboration at SLAC.  Expressed in
terms of the electroweak mixing angle, the result
is~\cite{E158RunI,E158RunI+II+III}: $\swsqeff(Q^2)=0.2397\pm0.0013$ or
$\swsqMSb = 0.2330\pm0.0015$ using the $\SM$ running of the
electroweak mixing angle with $Q^2$.  The effective electroweak mixing
angle, $\swsqeffl$, is obtained by adding 0.00029~\cite{PDG2004} to
$\swsqMSb$.

\subsection{Neutrino-Nucleon Scattering}

\begin{figure}[htbp]
\begin{center}
\includegraphics[width=0.8\linewidth]{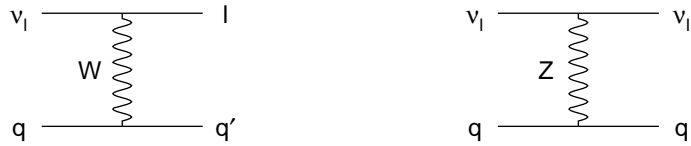}
\caption[Feynman diagrams in neutrino-nucleon scattering]
{Feynman diagrams in neutrino-nucleon scattering on parton level.
  Left: charged-current reaction. Right: neutral-current reaction.}
\label{fig:feyn-nntff}
\end{center}
\end{figure}

Neutrino-nucleon scattering allows to measure both charged weak
current and neutral weak current interactions.  As shown in
Figure~\ref{fig:feyn-nntff}, the interactions proceed via the
$t$-channel exchange of a W or Z boson, connecting the incoming
neutrino or anti-neutrino to a quark in the nucleons of the target
material.

Using both a neutrino and an anti-neutrino beam, as done for the first
time by the NuTeV experiment, it is possible to exploit the
Paschos-Wolfenstein relation~\cite{Paschos-Wolfenstein:1973}:
\begin{eqnarray}
R_- & = & \frac{\sigma_{NC}(\nu)-\sigma_{NC}(\bar\nu)}
               {\sigma_{CC}(\nu)-\sigma_{CC}(\bar\nu)} 
~ = ~ 4 g^2_{L\nu}\sum_{u,d}\left[g^2_{Lq} - g^2_{Rq}\right] 
~ = ~ \rho_\nu\rho_{ud}\left[1/2 -
               \sin^2\theta_W^{on-shell}\right]\,,
\end{eqnarray}
where the sum runs over the valence quarks u and d. This relation
holds for iso-scalar targets and up to small electroweak radiative
corrections. Thus $R_-$ is a measurement of the on-shell electroweak
mixing angle.

\begin{figure}[htbp]
\begin{center}
\vskip -0.75cm
\includegraphics[width=0.8\linewidth]{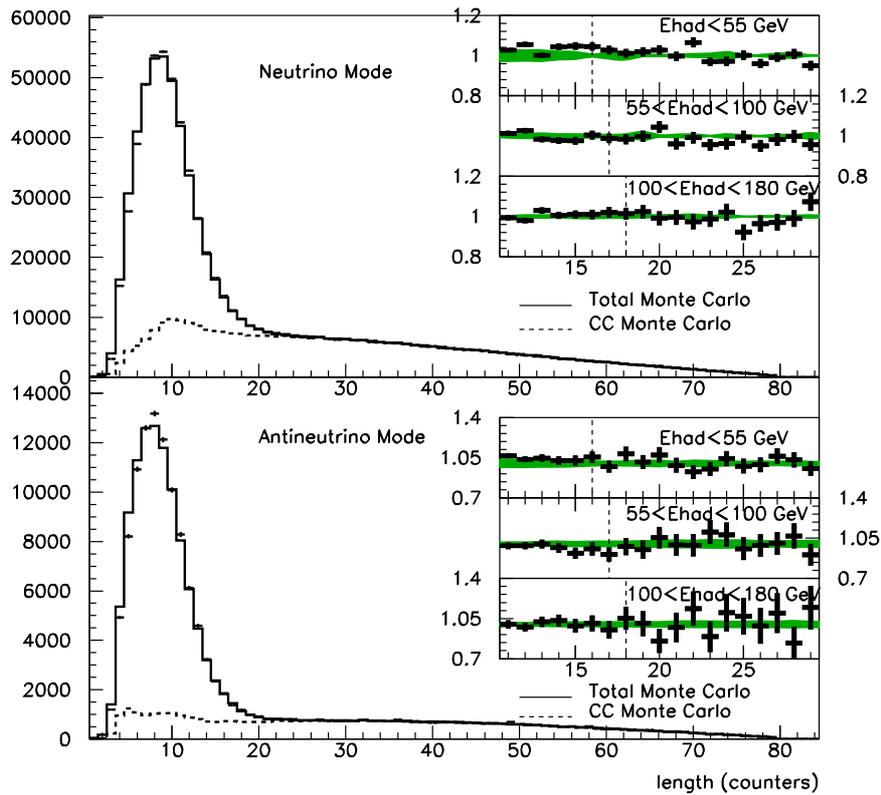}
\caption{Distribution of event lengths.}
\label{fig:nutev-length}
\end{center}
\end{figure}

Using a muon (anti-) neutrino beam, CC reactions contain a primary
muon in the final state, while NC reactions do not.  The muon as a
minimum ionising particle traverses the complete detector, while the
hadronic shower alone is confined in a small target volume. The length
of the event thus discriminates between CC and NC events.  The
distributions of event lengths as observed for neutrino and
anti-neutrino beams are shown in Figure~\ref{fig:nutev-length}. In
total, close to 2 million events were recorded by the NuTeV
collaboration, 1167K CC and 457K NC events with neutrino beams, and
250K CC and 101K NC events with anti-neutrino beams. The separation
between CC and NC events is dependent on the energy of the hadronic
shower and ranges from 16 to 18 in units of counters (equivalent to
10~cm of steel) as indicated in the inserts of
Figure~\ref{fig:nutev-length}.

\clearpage

In order to extract $R_-$ from the measured distributions, a Monte
Carlo simulation of the spectra of the \hbox{(anti-)} neutrino beams,
radiative corrections and detector response is used. In terms of the
on-shell electroweak mixing angle, NuTeV's final results
reads~\cite{bib-NuTeV-final}:
\begin{eqnarray}
\sin^2\theta_W^{on-shell} & \equiv & 1-\MW^2/\MZ^2 \\
& = & 0.2277 \pm 0.0013  \pm 0.0009  \nonumber \\
&   & - 0.00022 \frac{\Mt^2-(175~\GeV)^2}{(50~\GeV)^2} 
      + 0.00032 \ln(\MH/150~\GeV)\,, 
\end{eqnarray}
where the first error is statistical and the second is systematic.
Here $\rho=\rho_{SM}$ is assumed. This result is a factor of two more
precise than the average of all previous neutrino-nucleon
measurements.

The two main contributions to the total systematic uncertainty of
0.0009 are about equal, 0.0006 each for experimental systematics and
for modelling. The experimental systematics are dominated by the
uncertainty on the (anti-)electron neutrino flux, as for such beams
both CC and NC interactions lead to final states without primary
muons.  The model systematics are dominated by charm production and
the strange-quark sea, effects which are much reduced compared to
previous single-beam experiments.  With a statistical error of 0.0013
and a total systematic error of 0.0009, NuTeV's final result is
statistics limited.

When presented this result caused a great deal of excitement, as the
global SM analysis of all electroweak measurements, presented later in
this paper, predicts a value of $0.2232\pm0.0004$ for the on-shell
electroweak mixing angle, showing a deviation from the NuTeV result at
the level of 2.8 standard deviations.

\begin{figure}[htbp]
\begin{center}
\includegraphics[width=0.495\linewidth]{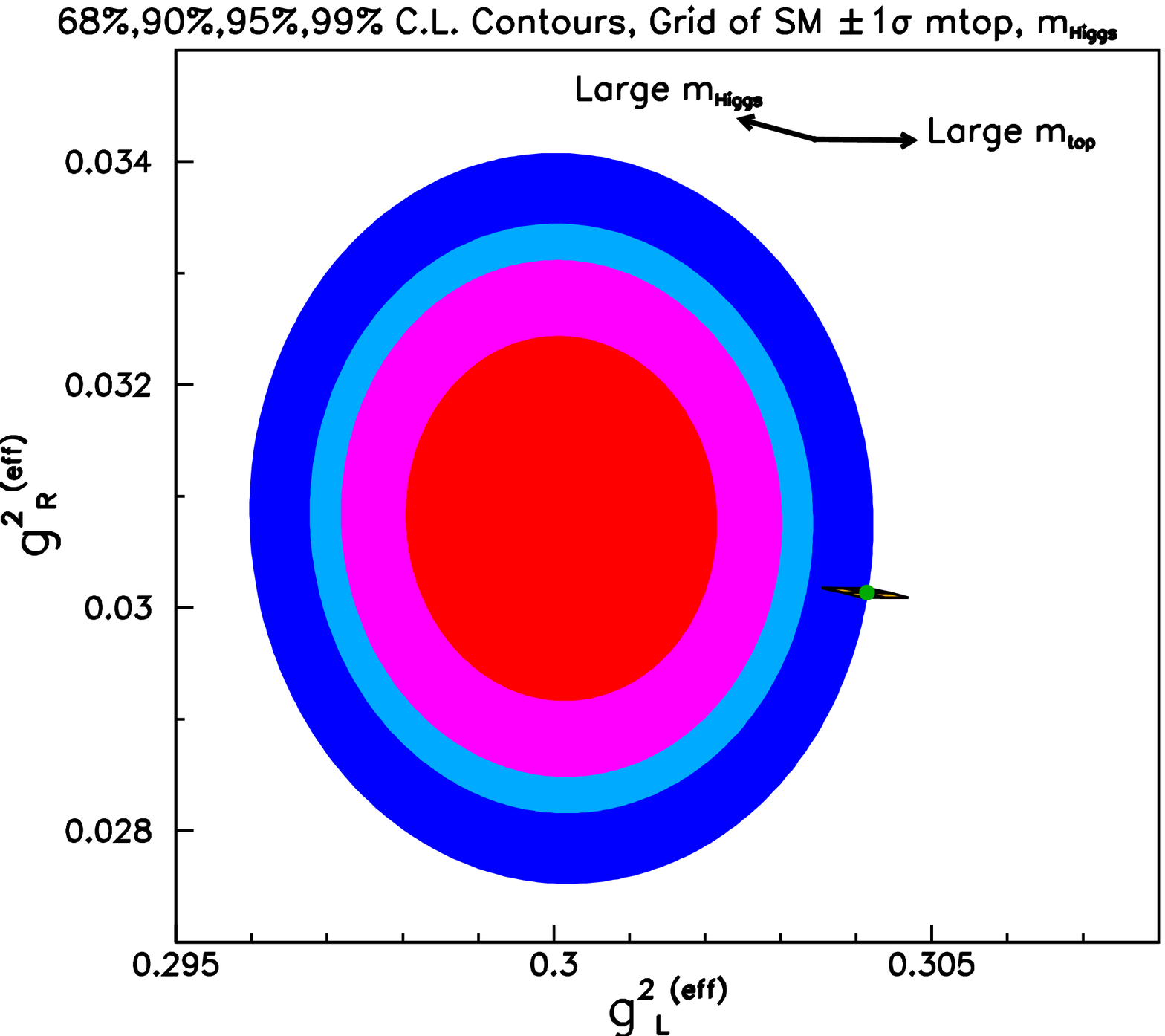}
\includegraphics[width=0.495\linewidth]{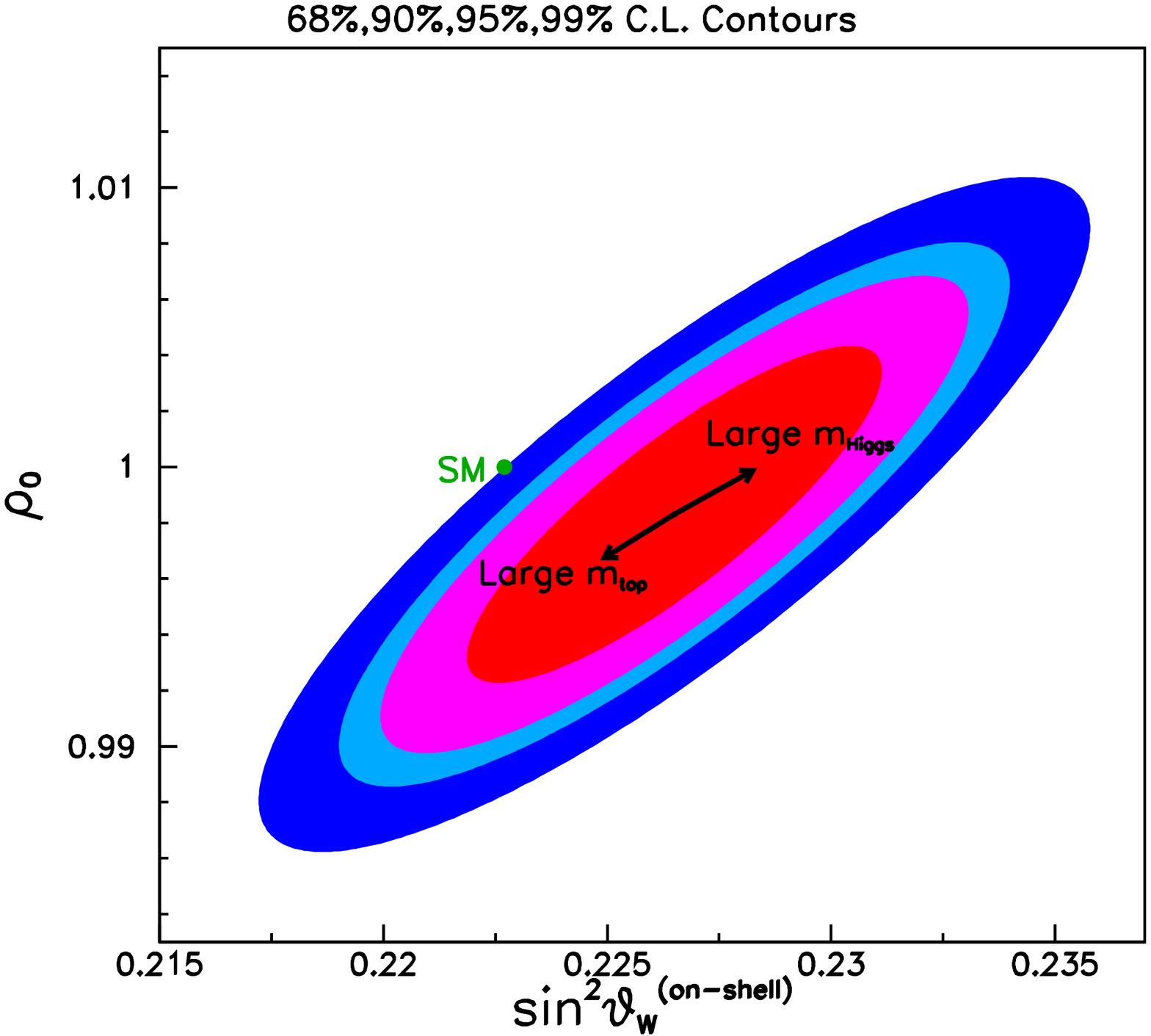}
\caption{NuTeV result in the plane of (left:) effective right- and
  left-handed couplings, and (right:) on-shell angle versus
  $\rho$-scale factor $\rho_0$. In both cases, the SM expectation is
  shown as the small dot located at the outer-most contour edge.}
\label{fig:nutev-2d}
\end{center}
\end{figure}

In a more model-independent analysis, the NuTeV result is also
interpreted in terms of effective left- and right-handed couplings,
shown in Figure~\ref{fig:nutev-2d}, defined as: $g^2_X(eff) =
4g^2_{L\nu}\sum_q g^2_{Xq} $ for $X=L,R$.  Here the deviation is
confined to the effective left-handed coupling product.  Modifying all
$\rho$ parameters by a scale factor $\rho_0$, also shown in
Figure~\ref{fig:nutev-2d}, shows that either $\rho_0$ or the mixing
angle, but not both, could be in agreement with the SM.  Assuming the
electroweak mixing angle to have it's expected value, the change in
the $\rho$ factors can be absorbed in $\rho_\nu$, \ie, interpreted as
a change in the coupling strength of neutrinos, then lower than
expected by about $(1.2\pm0.4)\%$.  A similar trend is observed with
the neutrino coupling as measured by the invisible width of the Z
boson at LEP-1, yielding a much less significant deficit of
$(0.5\pm0.3)\%$ in $\rho_\nu$.

To date, various explanations ranging from old and new physics effects
have been put forward.  Some old physics effects are: theoretical
uncertainties in PDFs, iso-spin violating PDFs, quark-antiquark
asymmetries for sea quarks, nuclear shadowing asymmetries between W
and Z interactions, etc. Some new physics effects are: a new heavy Z
boson, contact interactions, lepto-quarks, new fermions, neutrinos
oscillations, etc. Most of the old and new physics effects are,
however, severely constrained by NuTeV itself or other precision
electroweak measurements, thus cannot explain the full effect.  It
seems, however, that PDF uncertainties should be investigated, and in
particular the partly leading-order analysis employed by NuTeV should
be assessed and eventually improved to next-to-leading order.

\subsection{Anomalous Magnetic Moment of the Muon}

The Pauli and Dirac equations of quantum mechanics describe pointlike
spin-1/2 particles of mass $m$ with a magnetic moment $\vec{\mu} =
(g/2)(e/m)\vec{S}$ where $g$ is predicted to be 2.  In quantum field
theory, small corrections to $g$ arise, leading to a non-zero
so-called anomalous magnetic moment, $a=(g-2)/2\ne0$.  Recently, a
precision measurement of the anomalous magnetic moment for muons was
performed by the experiment E821 in Brookhaven, which measured the
precession of the muon spin relative to its direction of flight,
$\omega_a$, when moving on a circular orbit in a homogeneous magnetic
field:
\begin{eqnarray}
\omega_a & = & \omega_s-\omega_c ~ = ~ a_\mu\frac{e B}{m_\mu}\,,
\end{eqnarray}
where $\omega_s$ and $\omega_c$ are the spin precession and cyclotron
frequencies, respectively.  In a storage ring, electric fields are
also present which are felt my the moving muon as additional magnetic
fields due to the Lorentz transformation from laboratory system to
muon system, modifying the above equation:
\begin{eqnarray}
\omega_a & = & \frac{e }{m_\mu}\left[ a_\mu B - \left(
a_\mu - \frac{1}{\gamma^2-1}\right)\frac{\beta E}{c}\right]\,.
\end{eqnarray}
The additional term vanishes for $\gamma=29.30$ corresponding to the
``magic'' muon momentum of $3.094~\GeV$, establishing the working
point used by all $a_\mu$ measurements to suppress the effect of
electric fields.

\begin{figure}[t]
\begin{center}
\includegraphics[height=0.8\linewidth,angle=-90]{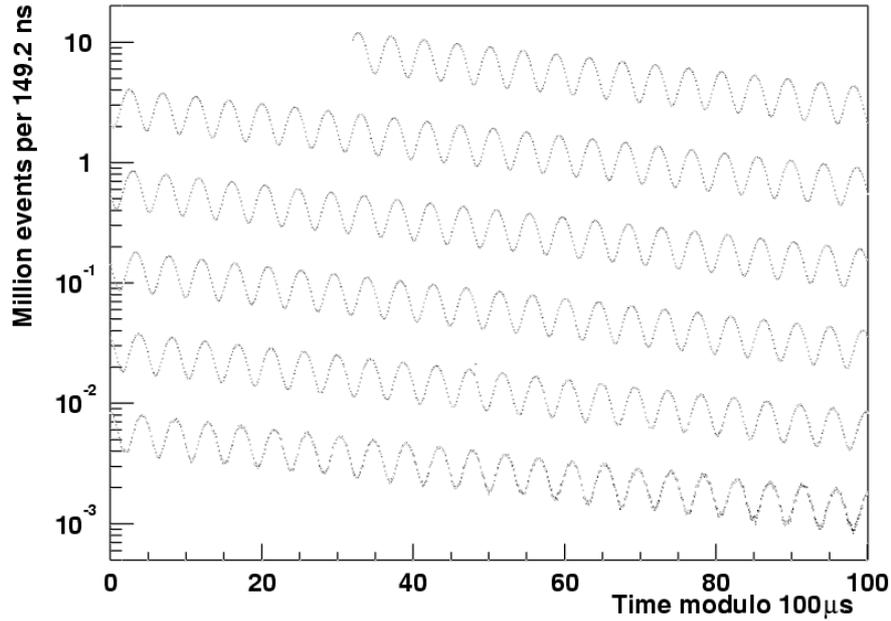}
\caption{Distribution of electron counts as a function of time.  The
data is wrapped every 100$\mu$s.}
\label{fig:g-2_wiggles}
\end{center}
\end{figure}

Longitudinally polarised muons are injected into the storage ring with
the magic momentum, leading to a time-dilated lifetime of $64.4\mu$s.
The muon decays through the charged weak current into an electron and
two neutrinos. Because of the parity violation of the charged weak
current, the muon spin is correlated with the direction of flight of
the electron.  This correlation allows the muon spin to be measured by
observing the electron signal at a fixed direction from the muon beam.
The E821 experiment used 24 electromagnetic calorimeters, placed
symmetrically along the inside of the storage ring, to measure the
time of the muon decay and the decay electron energy. The distribution
of events as a function of time is shown in
Figure~\ref{fig:g-2_wiggles}, showing the exponential decay following
the time-dilated muon lifetime, modulated by the relative precession
frequency $\omega_a$:
\begin{eqnarray}
N(t) & = & N_0 \exp(-t/\gamma\tau_\mu)[1-A\cos(\omega_a t -\phi)]\,,
\end{eqnarray}
where $N_0$, $A$ and $\phi$ implicitly depend on the electron energy
threshold, $1.8~\GeV$ for E821, used to select events.

Measurements of $\omega_a$ and hence $a_\mu$ were made by E821 for
both positive and negative muons, and found to be in good agreement as
expected from CPT invariance, with an average value
of~\cite{Bennett:2006fi}:
\begin{eqnarray}
a_\mu & = & (116592080\pm63)\cdot10^{-11}\,,
\end{eqnarray}
a 0.54~ppm measurement.  The total uncertainty combines a statistical
error of 54 units and a systematic error of 33 units in quadrature.

\begin{figure}[b]
\includegraphics[width=0.245\linewidth]{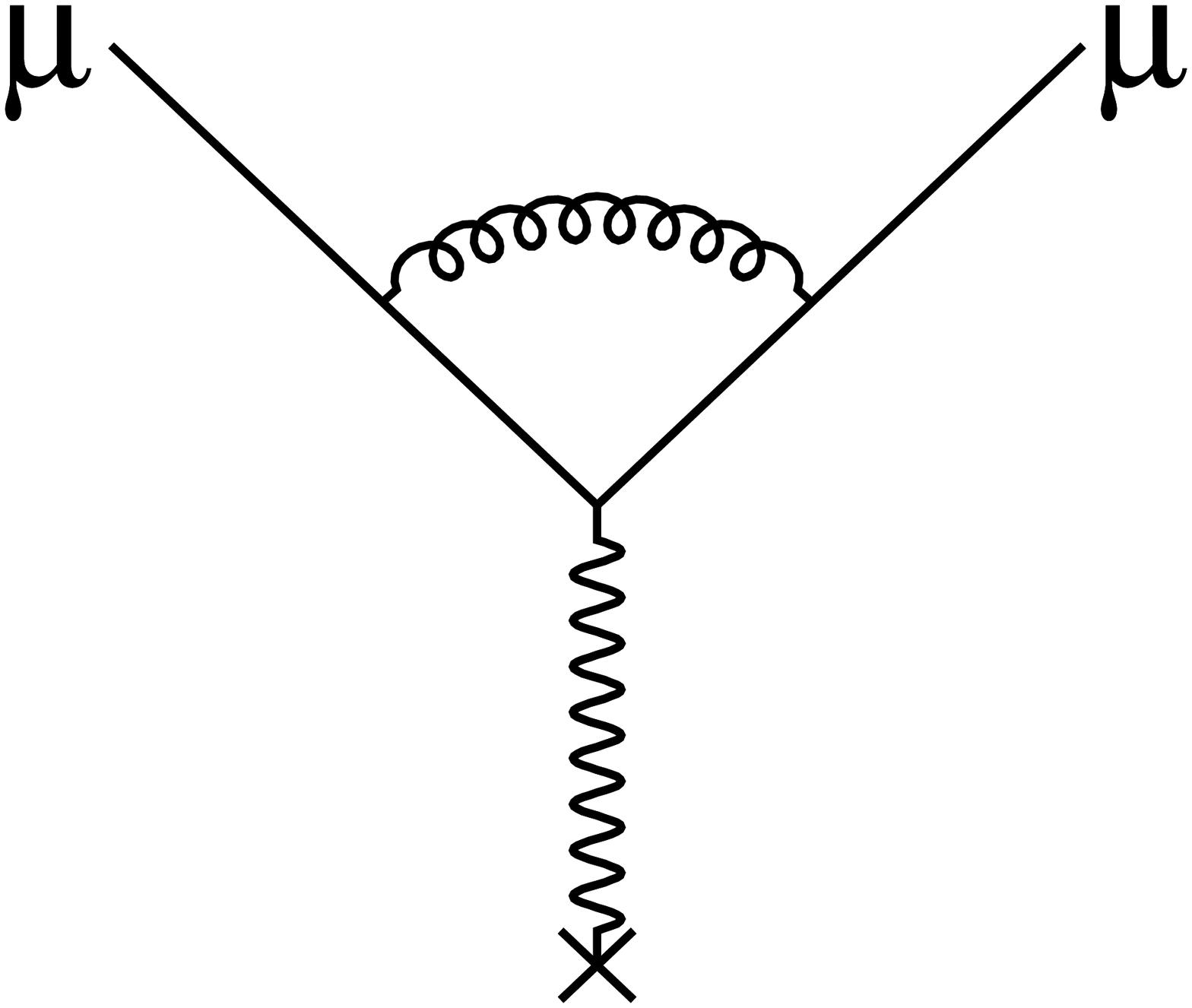}
\includegraphics[width=0.245\linewidth]{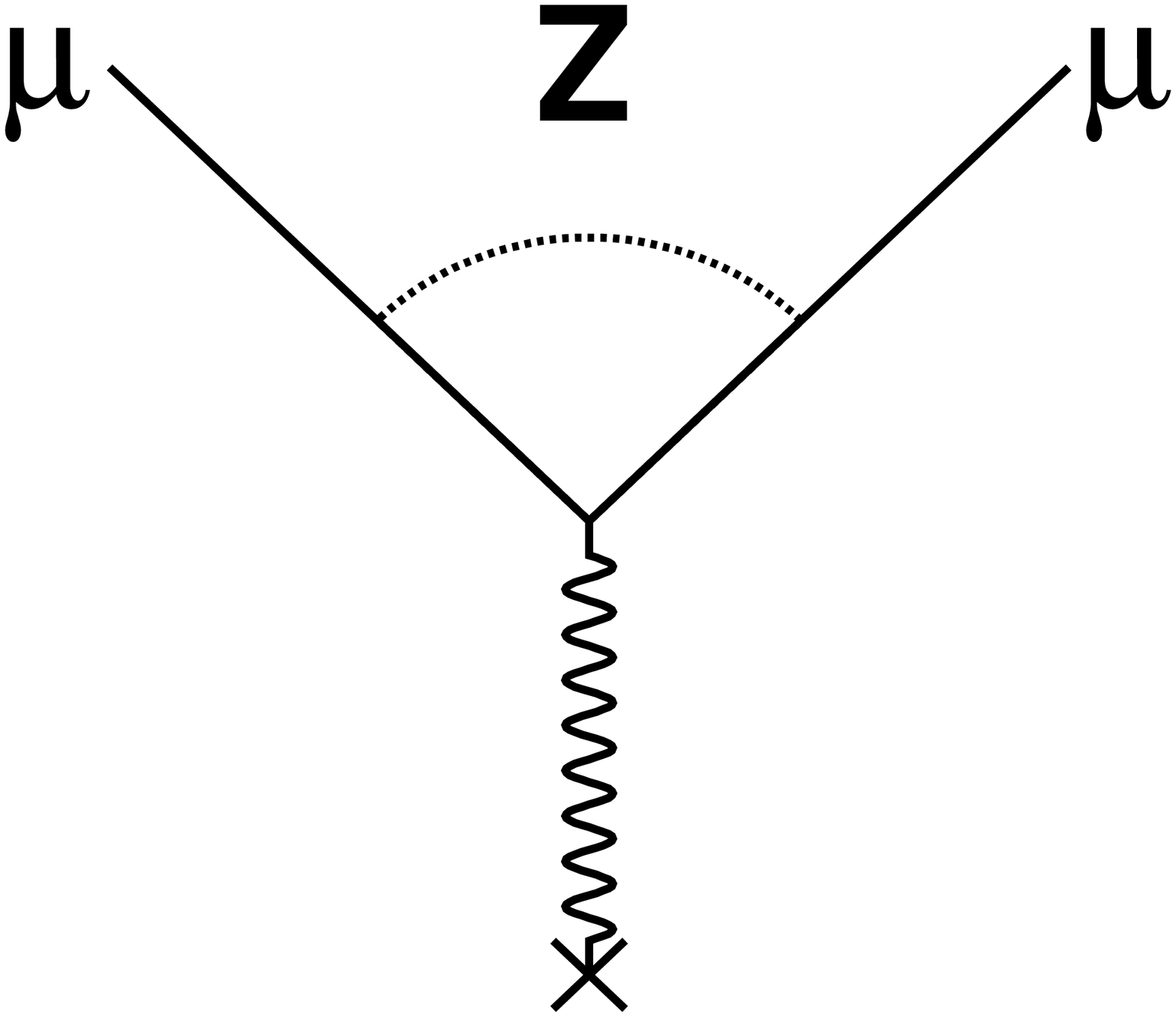}
\includegraphics[width=0.245\linewidth]{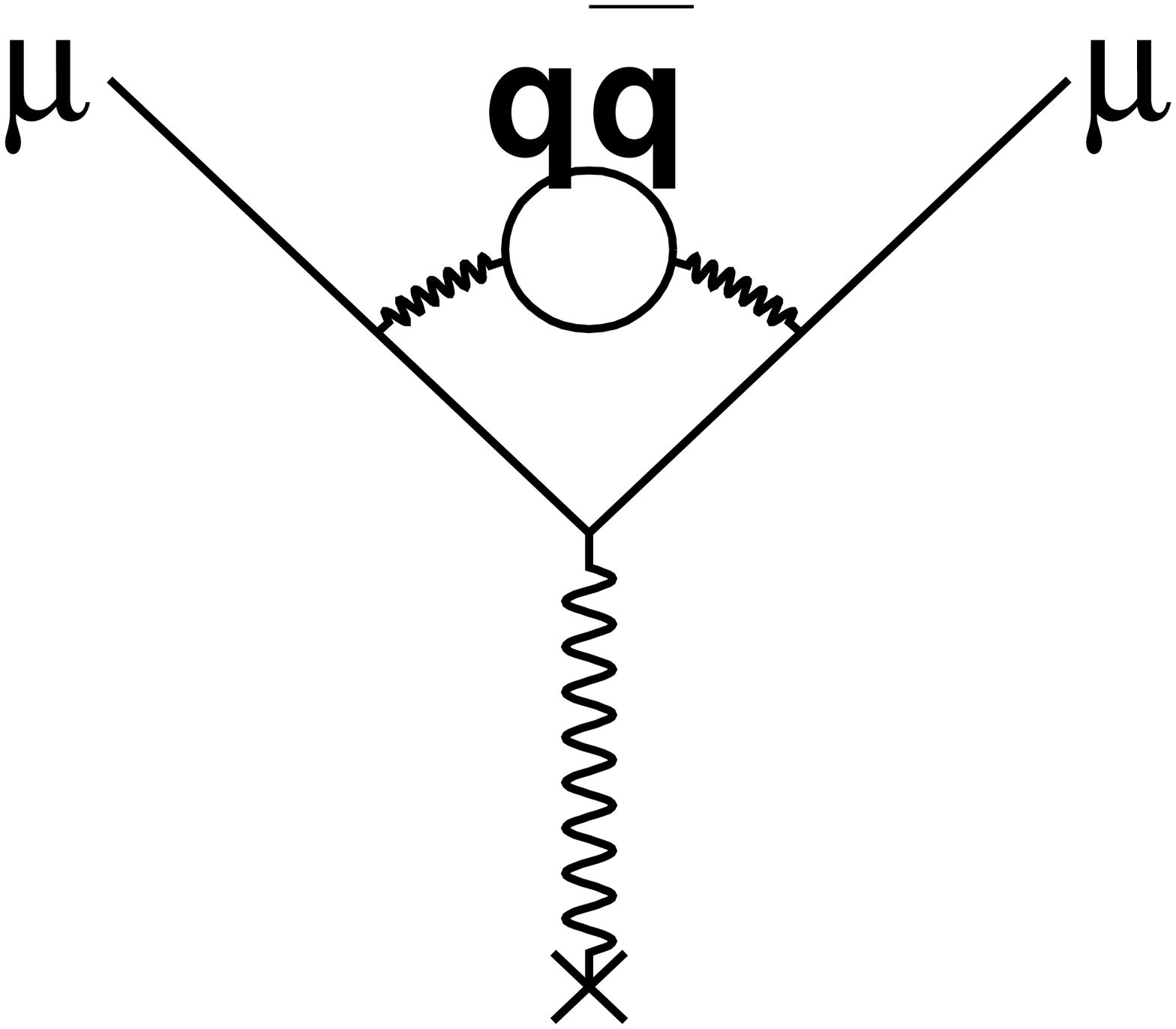}
\includegraphics[width=0.245\linewidth]{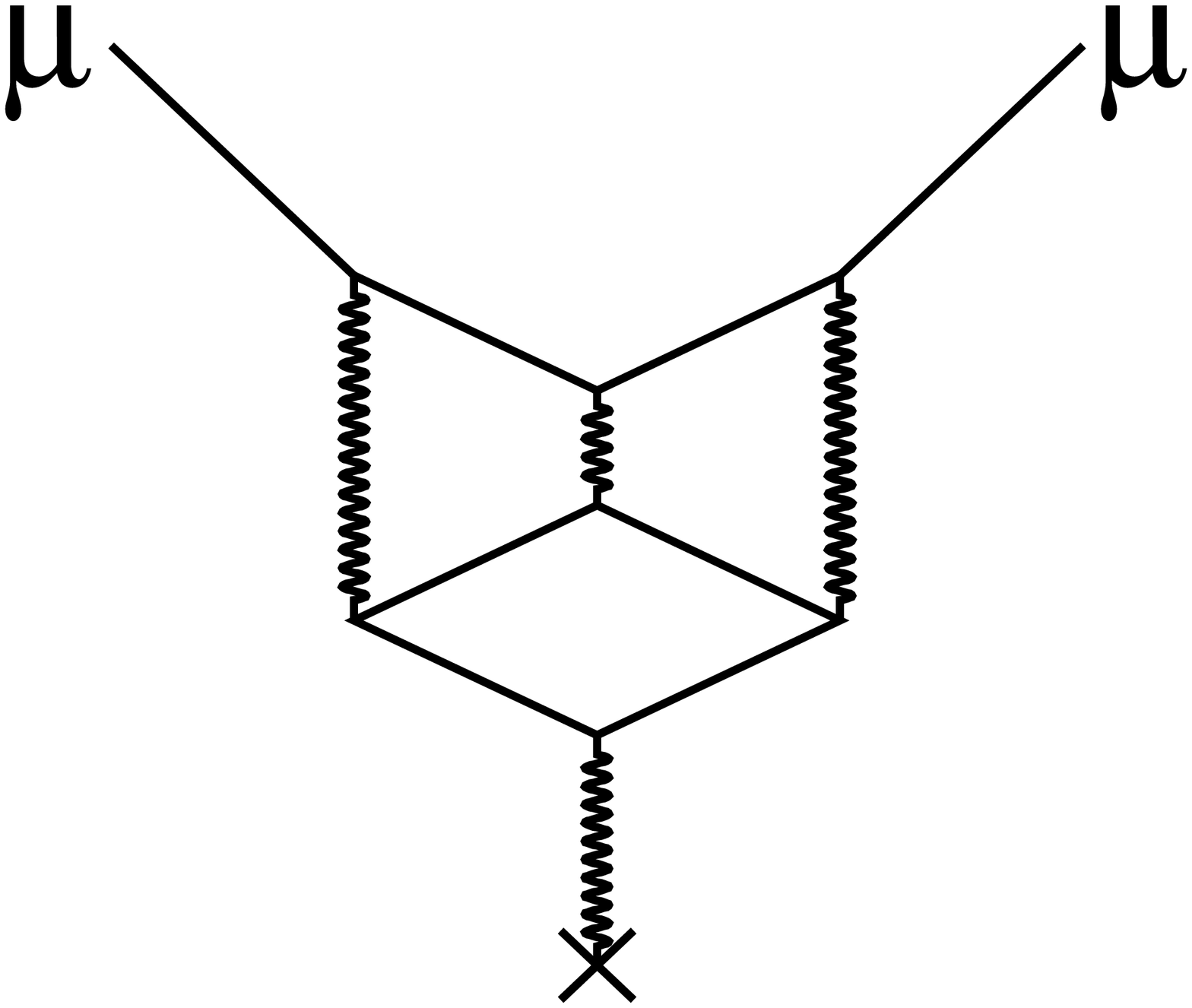}
\caption{Contributions to the $\SM$ calculation of $a_\mu$, from left
to right: QED, weak, hadronic, light-by-light diagrams.}
\label{fig:g-2_contribs}
\end{figure}

The expected value of $a_\mu$ is calculated within the $\SM$ to
similar precision, allowing to test the theory. Since the $\SM$
calculation is sensitive to the full particle content of the theory
through loops of virtual particles, any measured excess could indicate
the existence of new particles not found so far.  Typical diagrams
contributing to the calculation are shown in
Figure~\ref{fig:g-2_contribs}, and are classified as pure QED, weak,
hadronic and light-by-light diagrams.  The pure QED contribution is by
far the dominant term but known extremely
precisely~\cite{Kinoshita:2002ns,Kinoshita:2004wi}.  The
weak~\cite{Davier:2004gb},
hadronic~\cite{Davier:2003pw,Hagiwara:2002ma,Hagiwara:2003da} and
light-by-light~\cite{Davier:2004gb} contributions are all much smaller
but have larger uncertainties.  Adding all contributions, the error on
the $\SM$
calculations~\cite{Davier:2003pw,Hagiwara:2002ma,Hagiwara:2003da} is
dominated by the leading-order hadronic contribution (6-7 units,
depending on calculation), followed by the hadronic light-by-light
contribution (3.5 units).  The measured value turns out to be larger
than the $\SM$ expectation at the level of 2.2 to 2.7 standard
deviations~\cite{Bennett:2006fi}, depending on calculation.

\clearpage

\section{Constraints on the Standard Model}
\label{chap:MSM}

\subsection{Introduction}

The $\SM$ expectations for the various observables discussed so far
depend on the free parameters of the $\SM$ theory, as shown in many
preceeding figures.  This dependence exists already at Born level or
is introduced by radiative corrections involving virtual particles,
and needs to be taken into account in precision calculations.
Conversely, the measurement results can be used to constrain these
free parameters.  For the measurements discussed here, there are five
main $\SM$ parameters on which the predictions depend, conventionally
chosen as: (i) the hadronic vacuum polarisation $\dalhad$ contributing
to the running of the electromagentic coupling constant $\alqed$, (ii)
the coupling constant of the strong interaction, $\alfmz$, determining
QCD corrections, (iii) the mass $\MZ$ of the Z boson , (iv) the mass
$\Mt$ of the top quark, and (v) the mass $\MH$ of the as yet
unobserved Higgs boson.

The hadronic vacuum polarisation is determined in dedicated analyses,
using experimental results only~\cite{bib-JEG2, bib-Burk, bib-BP05},
or also theory constraints introducing some model dependence but
leading to higher precision~\cite{\dalphaQCD,
bib-Troconiz-Yndurain-2004}.  In the following, we use the
experimentally driven result~\cite{bib-BP05}:
\begin{eqnarray}
\dalhad & = & 0.02758\pm0.00035\,,
\label{eq:dalhad:exp:new}
\end{eqnarray}
but in some cases also illustrate the effect of using the more precise
theory-driven result~\cite{bib-Troconiz-Yndurain-2004}:
\begin{eqnarray}
\dalhad & = & 0.02749\pm0.00012\,.
\label{eq:dalhad:qcd}
\end{eqnarray}

\subsection{Z-Pole Results}

Using only the Z-pole measurements of Section~\ref{sec:ew:2f} together
with the hadronic vacuum polarisation, the five $\SM$ parameters are
determined in a $\chi^2$ fit of the theory predictions to the measured
results, taking uncertainties and correlations properly into account.
The fit has a $\chidf$ of 16.0/10, corresponding to a probability of
9.9\%.  The results of the fit are reported in
Table~\ref{tab:msmfit-lep1sld}.

\begin{table}[htbp]
\begin{center}
\renewcommand{\arraystretch}{1.25}
\caption[Results for $\SM$ input parameters from Z-pole measurements]
{Results for the five $\SM$ input parameters derived from a fit to the
Z-pole results and $\dalhad$.  The results on $\MH$, obtained by
exponentiating the fit results on $\LOGMH$, are also shown.}
\label{tab:msmfit-lep1sld}
\begin{tabular}{|c||r@{$\pm$}l||rrrrr|}
\hline
Parameter & \multicolumn{2}{|c||}{Value} 
          & \multicolumn{5}{|c| }{Correlations} \\
          & \multicolumn{2}{|c||}{ }
          & {$\dalhad$} & {$\alfmz$} 
          & {$\MZ$} & {$\Mt$} & {$\LOGMH$}      \\
\hline
\hline
$\dalhad$   &$0.02759$&$0.00035$&$ 1.00$&$     $&$     $&$     $&$     $\\
$\alfmz$    &$0.1190 $&$0.0027 $&$-0.04$&$ 1.00$&$     $&$     $&$     $\\
$\MZ~[\GeV]$&$91.1874$&$0.0021 $&$-0.01$&$-0.03$&$ 1.00$&$     $&$     $\\
$\Mt~[\GeV]$&$173    $&$^{13}_{10}$
                                &$-0.03$&$ 0.19$&$-0.07$&$ 1.00$&$     $\\
$\LOGMH$    &$2.05   $&$^{0.43}_{0.34}$
                                &$-0.29$&$ 0.25$&$-0.02$&$ 0.89$&$ 1.00$\\
\hline
$\MH~[\GeV]$&$111    $&$^{190}_{60}$
                                &$-0.29$&$ 0.25$&$-0.02$&$ 0.89$&$ 1.00$\\
\hline
\end{tabular}
\end{center}
\end{table}

The fitted hadronic vacuum polarisation is nearly unchanged compared
to the input measurement Equation~\ref{eq:dalhad:exp:new}, showing
that the data has only a low sensitivity to it. However, the parameter
is crucial in the determination of the mass of the Higgs boson as
evidenced from the sizeable correlation coefficient of 0.89.  The
fitted strong coupling constant is one of the most precise
determination of this quantity, and in good agreement with other
determinations~\cite{Bethke:2004uy,PDG2004}. The mass of the Z boson
in this $\SM$ analysis is identical to that derived in the
model-independent analysis of Section~\ref{sec:ew:2f}, but now $\MZ$
changed its role to a parameter of the theory, in terms of which all
other observables are calculated.  Even though Z-pole radiative
corrections are sensitive to $\LOGMH$ only, the mass of the Higgs
boson is nevertheless constrained within a factor of two and lies in
the region below $1~\TeV$ as required by theory, showing the
self-consistency of this analysis.

\subsection{The Mass of the Top Quark and of the W Boson}

\label{sec:msm:add:W}

\begin{figure}[p]
\begin{center}
\includegraphics[width=0.495\linewidth]{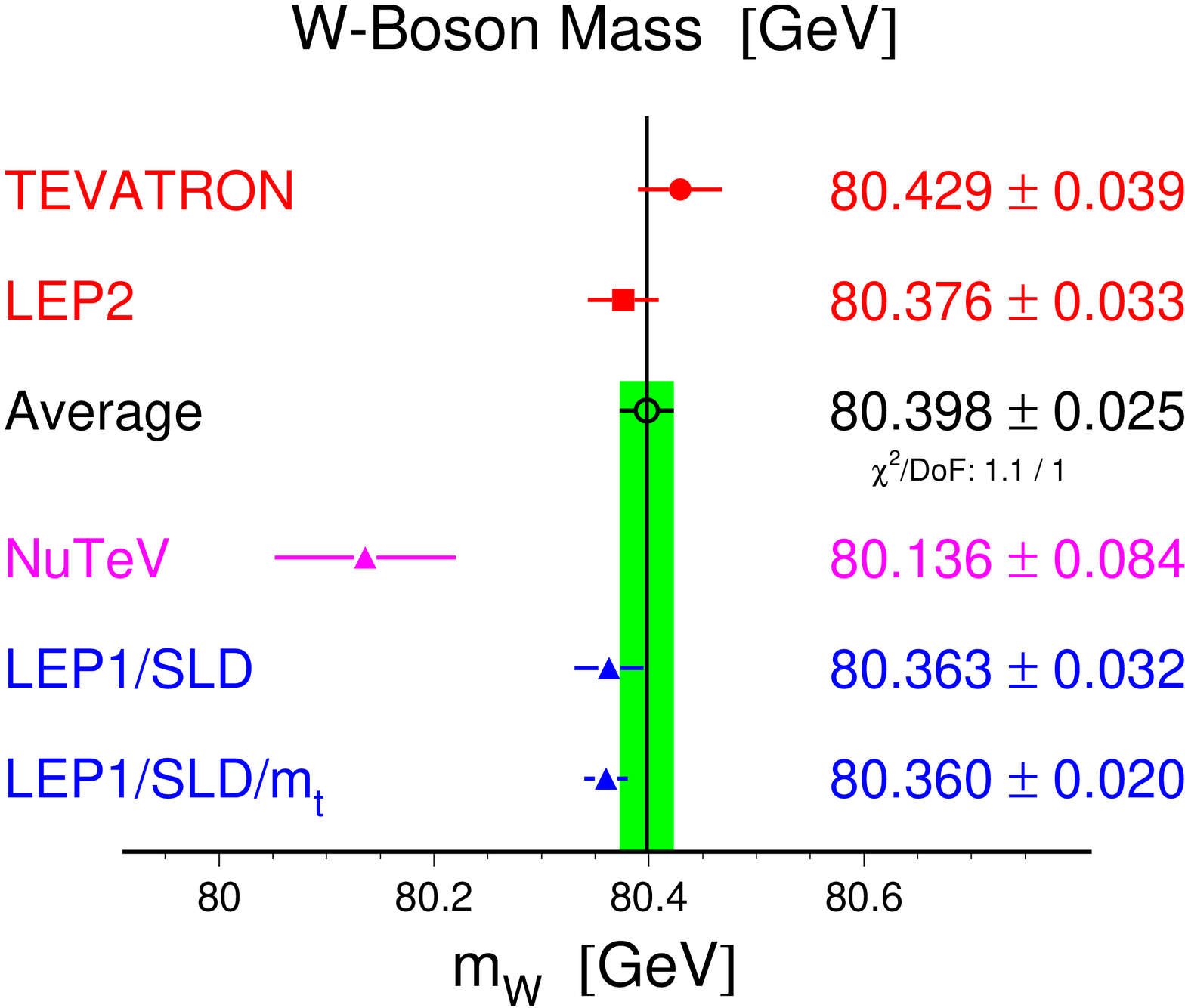}
\hfill
\includegraphics[width=0.495\linewidth]{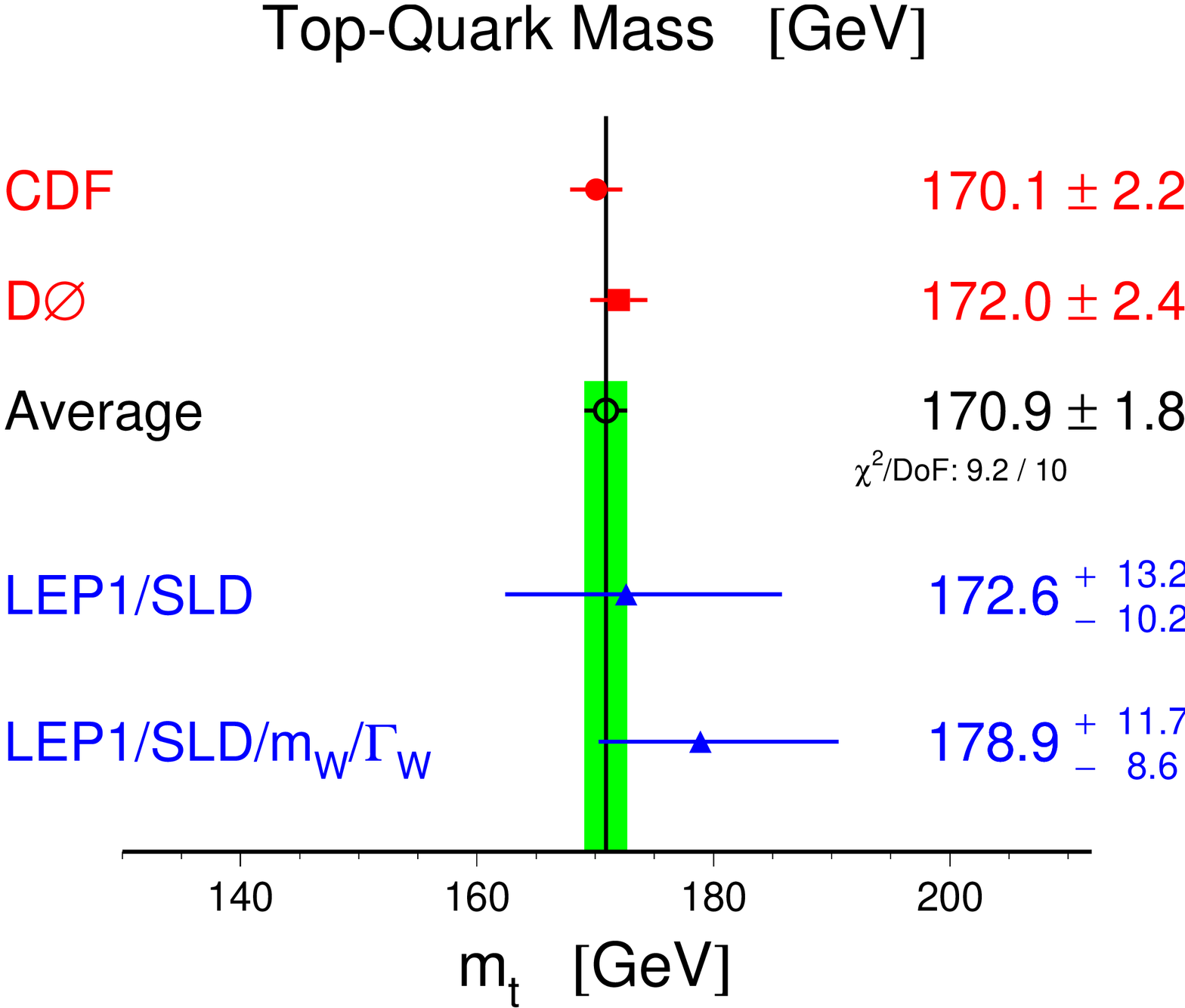}
\caption[Comparison of results on the mass of the W boson and the top
quark] { Left: results on the mass of the W boson, $\MW$. The direct
measurements of $\MW$ at \LEPII\ (preliminary) and at Run-I of the
Tevatron (top) are compared with the indirect determinations (bottom).
The NuTeV result interpreted in terms of $\MW$ is shown separately.
Right: results on the mass of the top quark. The direct measurements
of $\Mt$ at Run-I of the Tevatron (top) are compared with the indirect
determinations (bottom). }
\label{fig:msm:mt} 
\label{fig:msm:mw} 
\end{center}
\end{figure}

The results on the mass and total width of the W boson obtained at
\LEPII\ and the Tevatron and presented in Section~\ref{sec:ew:4f} are
in good agreement, as shown in Figure~\ref{fig:msm:mw}.  The results
are combined assuming no correlations between the two sets of results,
yielding preliminary world averages of:
\begin{eqnarray}
\MW & = & 80.398\pm0.025~\GeV \\
\GW & = &  2.140\pm0.060~\GeV\,.
\end{eqnarray}

Based on the analysis reported in Table~\ref{tab:msmfit-lep1sld}, the
mass of the W boson is predicted to be $\MW = 80.363\pm0.032~\GeV$.
The prediction of $\MW$ is greatly improved when the direct
measurement of the mass of the top quark, Section~\ref{sec:ew:6f}, is
also used in its calculation, with the result: $\MW =
80.360\pm0.020~\GeV$.  This result must be compared with the direct
measurements of $\MW$ performed at the Tevatron and \LEPII\ given
above, $\MW = 80.398\pm0.025~\GeV$.  The different results on $\MW$
are compared in Figure~\ref{fig:msm:mw} (left).  The agreement between
the prediction and the direct measurement is good, constituting an
important and successful test of the electroweak $\SM$.

\begin{figure}[p]
\begin{center}
\vskip -1cm
\includegraphics[width=0.67\textwidth]{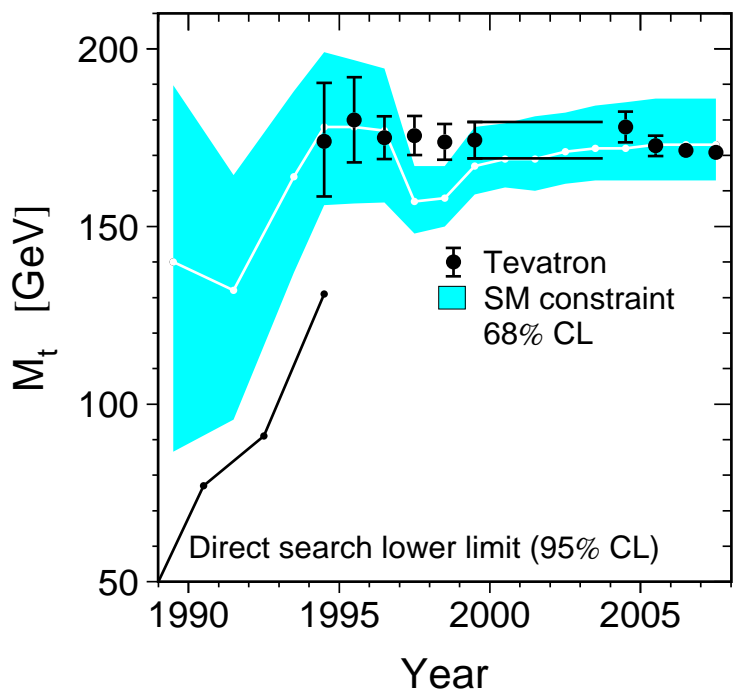}
\vskip -1cm
\caption[Predictions and measurements of $\Mt$.]{Comparison of direct
and indirect determinations of the mass of the top quark, $\Mt$, as a
function of time.  The shaded area denotes the indirect determination
of $\Mt$ at 68\% confidence level derived from the analysis of
radiative corrections within the framework of the $\SM$ using
precision electroweak measurements. The dots with error bars at 68\%
confidence level denote the direct measurements of $\Mt$ performed by
the Tevatron experiments.  Also shown is the 95\%
confidence level lower limit on $\Mt$ from direct searches before
the discovery of the top quark. }
\label{fig:history_mt}
\end{center}
\end{figure}

The mass of the top quark is found with an accuracy of about
$12~\GeV$: $\Mt=173^{+13}_{-10}~\GeV$ as given in
Table~\ref{tab:msmfit-lep1sld}.  This result must be compared with the
direct measurement of $\Mt$ at the Tevatron, with the most recent
combined preliminary result of $\Mt=170.9\pm1.8~\GeV$ as presented in
Section~\ref{sec:ew:6f}.  The good agreement constitutes another
important and successful test of the electroweak $\SM$.  The
prediction of the top-quark mass is improved slightly when the latest
results on the mass and width of the W boson are also used in its
determination: $\Mt=179^{+12}_{-9}~\GeV$.  Historically, this type of
analysis was used to predict the mass of the top quark before it was
discovered at the Tevatron; a historical perspective is illustrated
in~\ref{fig:history_mt}.  The most recent results on $\Mt$ are
compared in Figure~\ref{fig:msm:mt} (right).

\clearpage

\subsection{The Mass of the Higgs Boson}

\begin{figure}[p]
\begin{center}
\includegraphics[width=0.8\linewidth]{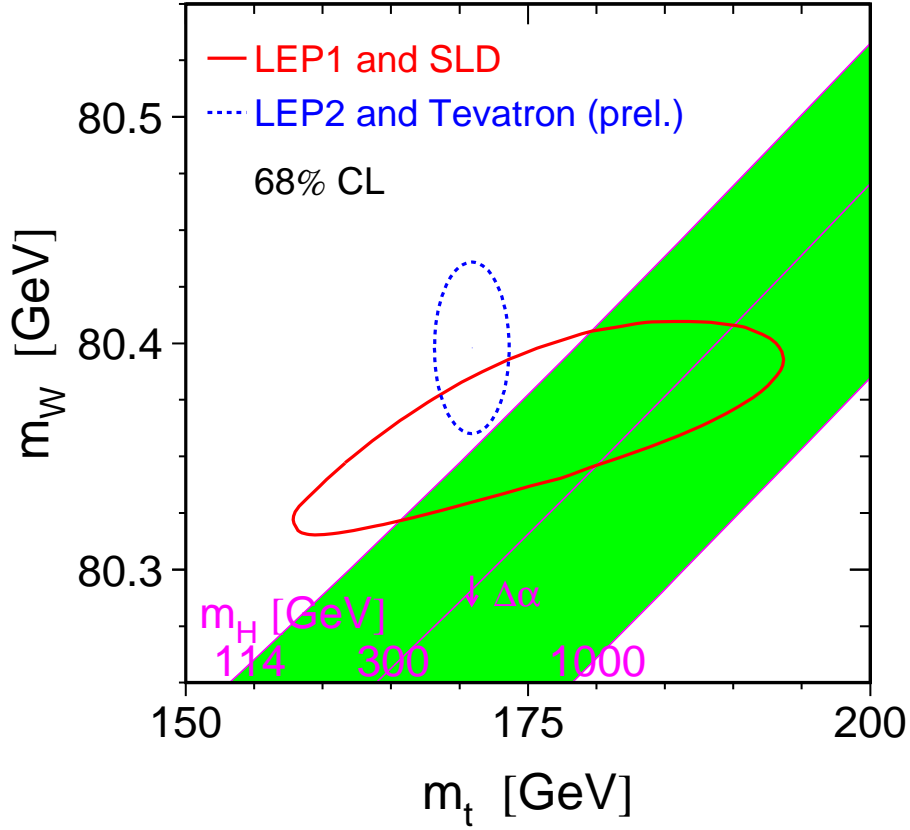}
\vskip -1cm
\caption[Comparison of top-quark and W-boson mass determinations]
{Contour curves of 68\% probability in the $(\Mt,\MW)$ plane.  The
shaded band shows the $\SM$ prediction based on the value for $\GF$
for various values of the Higgs-boson mass and fixed $\dalhad$;
varying the hadronic vacuum polarisation by
$\dalhad=0.02758\pm0.00035$ yields an additional uncertainty on the
$\SM$ prediction shown by the arrow labeled $\Delta\alpha$. }
\label{fig:msm:mt-mw} 
\end{center}
\end{figure}

The predictions of the masses of heavy fundamental particles, such as
top quark and W boson, turned out to be very successful as the
predictions agree well with the direct measurements.  These analyses
clearly demonstrate the predictive power of the electroweak $\SM$ as
well as the precision achieved by the experimental measurements.
Building on this success, the mass of the as yet unobserved Higgs
boson of the minimal $\SM$ will be predicted.  As a first step,
Figure~\ref{fig:msm:mt-mw} shows the comparison between the direct and
indirect determinations of $\Mt$ and $\MW$ in the $(\Mt,\MW)$ plane.
As already discussed, both sets of measurements agree well as shown by
the overlapping contour curves.  Within the $\SM$, $\Mt$ and $\MW$ are
correlated quantities which is driven by the Fermi constant $\GF$.
The exact correlation depends crucially on the mass of the Higgs
boson, which can thus be determined with increased precision from both
sets of results.

\begin{figure}[p]
\begin{center}
\includegraphics[width=0.8\linewidth]{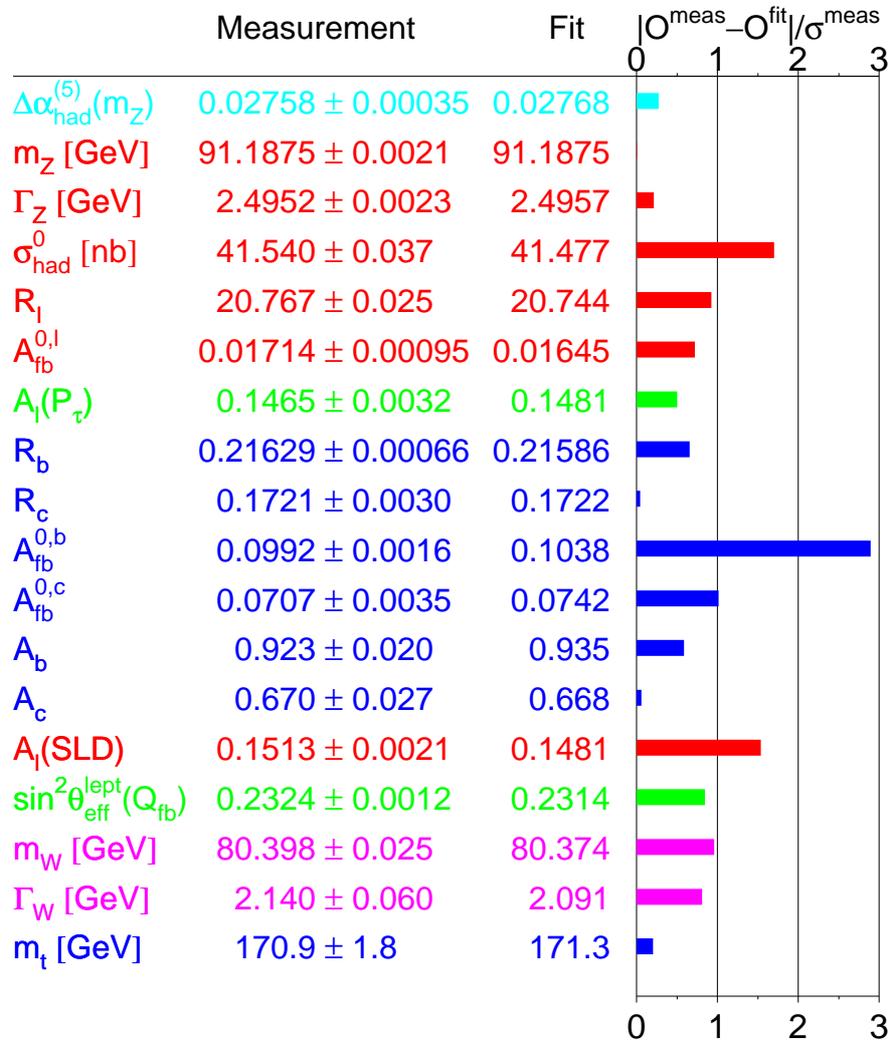}
\caption[Measurements and pulls] {Comparison of the measurements with
the expectation of the $\SM$, calculated for the five $\SM$ input
parameter values in the minimum of the global $\chi^2$ of the fit.
Also shown is the pull of each measurement, where pull is defined as
the absolute value of the difference of measurement and expectation in
units of the measurement uncertainty. }
\label{fig:msm:pulls} 
\end{center}
\end{figure}

Including also the direct measurements of the top quark and of the
mass and width of the W boson, the $\SM$ analysis yields the results
as reported in Table~\ref{tab:msmfit-all}. The $\chi^2$ fit to the
experimental results is 18.2 for 13 degrees of freedom, corresponding
to a probability of 15\%.  The pulls of the measurements relative to
the fit results are listed in Table~\ref{tab:msm:input} and shown in
Figure~\ref{fig:msm:pulls}.  The largest pull occurs for the
forward-backward asymmetry measured in the reaction $\ee\to\bb$ at the
Z pole, another consequence of the effects observed in the b-quark
sector and already discussed in Sections~\ref{sec:Heavy}
and~\ref{chap:Z+coup}.

\begin{figure}[t]
\begin{center}
\includegraphics[width=0.7\linewidth]{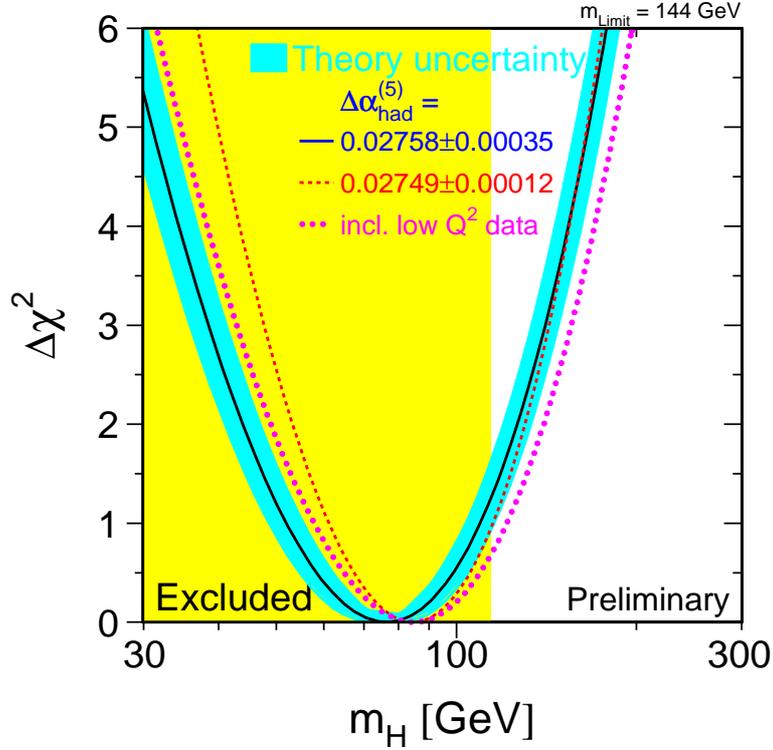}
\caption[The ``blue-band'' plot $\Delta\chi^2(\MH)$]
{$\Delta\chi^2(\MH)=\chi^2_{min}(\MH)-\chi^2_{\min}$ as a function of
$\MH$.  The line is the result of the fit using all 18 results.  The
associated band represents the estimate of the theoretical uncertainty
due to missing higher-order corrections.  The vertical band shows the
95\% confidence level exclusion limit on $\MH$ of $114.4~\GeV$ derived
from the direct search at \LEPII~\cite{LEPSMHIGGS}. The dashed curve
is the result obtained using the theory-driven $\dalhad$ determination
of Equation~\ref{eq:dalhad:qcd}. }
\label{fig:msmfit-chi2} 
\end{center}
\end{figure}

The $\Delta\chi^2(\MH)=\chi^2_{min}(\MH)-\chi^2_{\min}$ curve is shown
in Figure~\ref{fig:msmfit-chi2}.  Compared to the results shown in
Table~\ref{tab:msmfit-lep1sld}, the relative uncertainty on $\MH$ is
decreased by more than a factor of two.  Theoretical uncertainties in
the $\SM$ calculations of the expectations, due to missing
higher-order corrections, are shown as the shaded area around the thin
solid curve.  Including these theoretical uncertainties, the one-sided
95\%~CL upper limit on $\LOGMH$ ($\Delta\chi^2=2.7$) is:
\begin{eqnarray}
\MH    ~ < ~ 144~\GeV \label{eq:sm:mh-exp}\,.
\end{eqnarray}
This limit increases to $182~\GeV$ when the lower limit of
$114~\GeV$~\cite{LEPSMHIGGS} from the direct search for the Higgs
boson at \LEPII, shown as the shaded rectangle in
Figure~\ref{fig:msmfit-chi2}, is taken into account.  The
determination of the limit on the Higgs mass is only marginally
affected by using the theory-driven determination of $\dalhad$,
Equation~\ref{eq:dalhad:qcd}.

The analyses of the high-$Q^2$ results are now used to predict the
values of observables measured in low-$Q^2$ reaction such as those
discussed in Section~\ref{sec:lowQ2}. Measured results and predictions
are compared in Table~\ref{tab:msm:predict}. In general, good
agreement is observed except for the result on the left-handed quark
coupling combination $\gnlq^2$, measured by the NuTeV experiment eight
times more precisely than $\gnrq^2$. The difference to the expected
result is at the level of three standard deviations.  However, the
determination of the Higgs mass and its limit is only marginally
affected when including the low-$Q^2$ results, as visible in
Figure~\ref{fig:msmfit-chi2}.

\begin{table}[p]
\begin{center}
\renewcommand{\arraystretch}{1.25}
\caption[Results for $\SM$ input parameters] {Results for the five
$\SM$ input parameters derived from a fit to the Z-pole results and
$\dalhad$, plus $\Mt$, $\MW$, and $\GW$ from Tevatron Run-I and
\LEPII.  The results on $\MH$, obtained by exponentiating the fit
results on $\LOGMH$, are also shown. }
\label{tab:msmfit-all}
\begin{tabular}{|c||r@{$\pm$}l||rrrrr|}
\hline
Parameter & \multicolumn{2}{|c||}{Value} 
          & \multicolumn{5}{|c| }{Correlations} \\
          & \multicolumn{2}{|c||}{ }
          & {$\dalhad$} & {$\alfmz$} 
          & {$\MZ$} & {$\Mt$} & {$\LOGMH$}      \\
\hline
\hline
$\dalhad$   &$0.02768$&$0.00034$&$ 1.00$&$     $&$     $&$     $&$     $\\
$\alfmz$    &$0.1185 $&$0.0026 $&$ 0.03$&$ 1.00$&$     $&$     $&$     $\\
$\MZ~[\GeV]$&$91.1875$&$0.0021 $&$ 0.00$&$-0.02$&$ 1.00$&$     $&$     $\\
$\Mt~[\GeV]$&$171.3  $&$1.7    $&$-0.01$&$ 0.03$&$-0.02$&$ 1.00$&$     $\\
$\LOGMH$    &$1.88   $&$0.16   $&$-0.54$&$ 0.06$&$ 0.09$&$ 0.39$&$ 1.00$\\
\hline
$\MH~[\GeV]$&$ 76    $&$^{33}_{24}$
                                &$-0.54$&$ 0.06$&$ 0.09$&$ 0.39$&$ 1.00$\\
\hline
\end{tabular}
\end{center}
\end{table}

\begin{table}[p]
\begin{center}
  \renewcommand{\arraystretch}{1.30}
\caption[Overview of results]{ Summary of measurements included in the
  analyses of the five $\SM$ input parameters.  The top 15 results are
  included in the Z-pole and the high-$Q^2$ fit, while the bottom
  three results are only used in the high-$Q^2$ fit.  The total errors
  in column 2 include the systematic errors listed in column 3.  The
  $\SM$ results in column 4 and the pulls (absolute value of the
  difference between measurement and fit in units of the total
  measurement error, see Figure~\ref{fig:msm:pulls}) in column 5 are
  derived from the $\SM$ analysis of all 18 results.  \\
  $^{(a)}$\small{Only common systematic errors are indicated.}\\ }
\label{tab:msm:input}
\begin{tabular}{|ll||r|r||r|r|l|}
\hline
 && {Measurement with}  & {Systematic} & {Standard Model} & {Pull}  \\
 && {Total Error}       & {Error}      & {High-$Q^2$ Fit} & {    }  \\
\hline
\hline
& $\dalhad$~\cite{bib-BP05}
                & $0.02758\pm0.00035$   
                               & 0.00034              & $0.02766\pm0.00035$ & $0.2$ \\
\hline
\hline
&$\MZ$ [\GeV{}] & $91.1875\pm0.0021\pz$
                               &${}^{(a)}$0.0017$\pz$ & $91.1875\pm0.0021\pz$ & $0.0$ \\
&$\GZ$ [\GeV{}] & $2.4952 \pm0.0023\pz$
                               &${}^{(a)}$0.0012$\pz$ &  $2.4957\pm0.0015\pz$ & $0.2$ \\
&$\shad$ [nb]   & $41.540 \pm0.037\pzz$ 
                               &${}^{(a)}$0.028$\pzz$ &  $41.477\pm0.014\pzz$ & $1.7$ \\
&$\Rl$          & $20.767 \pm0.025\pzz$ 
                               &${}^{(a)}$0.007$\pzz$ &  $20.744\pm0.018\pzz$ & $0.9$ \\
&$\Afbzl$       & $0.0171 \pm0.0010\pz$ 
                               &${}^{(a)}$0.0003$\pz$ &  $0.01645\pm0.00023 $ & $0.7$ \\
+& correlation matrix &&&& \\[-2mm]
 & ~~Table~\ref{tab:lsafbresult} &&&& \\
\hline
&$\cAl~(P_\tau)$& $0.1465 \pm0.0033\pz$ & 0.0015$\pz$ & $0.1481\pm 0.0010\pz$ & $0.5$ \\
\hline
&$\cAl$~(SLD)   & $0.1513 \pm0.0021\pz$ & 0.0011$\pz$ & $0.1481\pm 0.0010\pz$ & $1.5$ \\
\hline
&$\Rbz{}$       & $0.21629\pm0.00066$   & 0.00050     & $0.21586\pm0.00006  $ & $0.7$ \\
&$\Rcz{}$       & $0.1721\pm0.0030\pz$  & 0.0019$\pz$ & $0.1722 \pm0.0001\pz$ & $0.0$ \\
&$\Afbzb{}$     & $0.0992\pm0.0016\pz$  & 0.0007$\pz$ & $0.1038\pm 0.0007\pz$ & $2.9$ \\
&$\Afbzc{}$     & $0.0707\pm0.0035\pz$  & 0.0017$\pz$ & $0.0743\pm 0.0006\pz$ & $1.0$ \\
&$\cAb$         & $0.923\pm 0.020\pzz$  & 0.013$\pzz$ & $0.9347\pm 0.0001\pz$ & $0.6$ \\
&$\cAc$         & $0.670\pm 0.027\pzz$  & 0.015$\pzz$ & $0.6684\pm 0.0005\pz$ & $0.1$ \\
+& correlation matrix &&&& \\[-2mm]
 & ~~Table~\ref{tab:14parcor} &&&& \\
\hline
&$\swsqeffl$
  ($\Qfbhad$)   & $0.2324\pm0.0012\pz$  & 0.0010$\pz$ & $0.23138\pm0.00013$   & $0.8$ \\
\hline
\hline
&$\Mt$ [\GeV{}] (Run-I+II~\cite{Mtop-tevewwgWin07})
                & $170.9\pm1.8\pzz\pzz$ &1.5$\pzz\pzz$& $171.3\pm1.7\pzz\pzz$ & $0.2$ \\
\hline
&$\MW$ [\GeV{}]
                & $80.398\pm0.025\pzz$  &             & $80.374\pm0.015 \pzz$ & $1.0$ \\
&$\GW$ [\GeV{}]
                & $ 2.140\pm0.060\pzz$  &             & $2.092\pm0.002  \pzz$ & $0.8$ \\
+& correlation given in &&&& \\[-2mm]
 & ~~Section~\ref{sec:msm:add:W} &&&& \\
\hline
\end{tabular}\end{center}
\end{table}

\begin{table}[htbp]
\begin{center}
  \renewcommand{\arraystretch}{1.30}
\caption[Overview of predictions]{ Summary of predictions for results
 obtained in low-$Q^2$ processes described in
 Section~\ref{sec:lowQ2}, derived from the fit to all
 high-$Q^2$ data.  }
\label{tab:msm:predict}
\begin{tabular}{|ll||r||r|r|l|}
\hline
 && {Measurement with}  & {Standard Model} & {Pull}  \\
 && {Total Error}       & {High-$Q^2$ Fit} & {    }  \\
\hline
\hline
&APV~\cite{QWCs:theo:2003:new}
                &                        &                      &        \\
\hline
&$\QWCs$        & $-72.74\pm0.46\pzz\pz$ & $-72.899\pm0.032\pzz$& $ 0.4$ \\
\hline
\hline
&M\o ller~\cite{E158RunI+II+III}
                &                        &                      &        \\
\hline
&$\swsqMSb$     & $0.2330\pm0.0015\pz$   & $0.23109\pm0.00013$  & $1.3$  \\
\hline
\hline
&$\nu$N~\cite{bib-NuTeV-final}
                &                        &                      &        \\
\hline
&$\gnlq^2$      & $0.30005\pm0.00137$    & $0.30391\pm0.00016$  & $2.8$  \\
&$\gnrq^2$      & $0.03076\pm0.00110$    & $0.03011\pm0.00003$  & $0.6$  \\
\hline
\end{tabular}\end{center}
\end{table}

\clearpage

\section{Summary and Conclusion}

Over the last two decades, a large body of high-precision measurements
became available, testing the $\SM$ of electroweak interactions both
at Born level but also, very importantly, at loop-level.  In general,
the many precise results presented here are in good agreement with the
$\SM$ expectations, thus contraining any hypothetical theories beyond
it.  The notable exceptions, all at the level of about three standard
deviations, are: (i) the (final) forward-backward asymmetry in
$\ee\to\bb$ measured at the Z pole, (ii) the (still preliminary)
leptonic W-decay branching fractions measured in $\eeWW$ at \LEPII,
and (iii) the left-handed quark couplings combination measured in
neutrino-nucleon scattering by the NuTeV experiment.  While the first
two are likely to be of statistical nature, the last one may also be
explained through an improved theoretical understanding of hadronic
physics inside the atomic nucleus.

The precision measurements test the theory at loop level, verifying
the $\SM$ as a renormalisable field theory correctly describing
nature. The data impose very tight constraints on any new physics
beyond the $\SM$.  Any extended theory must be consistent with the
$\SM$ or one or more Higgs doublet models such as super-symmetry.
Masses of heavy fundamental particles are predicted from the analysis
of reactions where these particles are not directly produced but occur
as virtual particles in loops contributing to the process under study.
The predictions of W-boson mass and top-quark mass agree well with the
direct measurements, constituting crucial tests of the electroweak
$\SM$.  In addition, the mass of the as yet unobserved Higgs boson is
constrained, with the result: $\MH<144~\GeV$ at 95\% confidence level.
The direct observation of the Higgs boson and the measurement of its
mass, expected for the next few years, will complete this domain of
the electroweak interaction, but may also lead to new surprises.

\subsection{Prospects for the Future}

The next generation of high-energy particle colliders consists of the
Large Hadron Collider (LHC), and the planned International Linear
Collider (ILC). The LHC, a proton-proton collider with a
centre-of-mass energy of $14~\TeV$, will start operations at the end
of this year (2007). The ILC, a linear electron-positron collider with
a centre-of-mass energy of $500~\GeV$ to $1000~\GeV$, is currently in
the design phase.

The upper limit on the Higgs mass indicated above opens a window of
opportunity for the Tevatron experiments CDF and D\O\ to find the
Higgs boson, but only the LHC is powerful enough to cover the entire
mass range of the $\SM$ Higgs boson allowed by theory - or to find
alternative signs for physics governing electroweak symmetry breaking.
The high centre-of-mass energy also gives the LHC a wide range in
searching for new massive particles beyond the Higgs, such as
supersymmetric particles.  The LHC is a W/Z/top factory, allowing for
high-statistics measurements in any final state with leptons.  In
contrast, the ILC is a precision machine allowing to make highly
accurate measurements in both leptonic and hadronic channels, covering
all possible decay modes of new heavy states in the clean environment
of an $\ee$ collider.  The smaller centre-of-mass energy limits its
mass reach compared to the LHC, but the ILC is essential to be able to
survey and measure precisely all properties of the Higgs boson and
other new particles found at the LHC, setting the stage for detailed
comparisons with theoretical models.

\section{Acknowledgements}

The author thanks the members of the LEP, SLD and Tevatron electroweak
working groups, as well as Guido Altarelli and Tord Riemann, for
valuable and stimulating discussions.

\clearpage

%
%

\section{References}%

\vspace*{-132 pt}
\bibliographystyle{PhysRep}
\bibliography{physrep,ew}

\end{document}